\documentclass{ut-thesis}

\usepackage{graphicx} 
\usepackage{setstack} 
\usepackage{amssymb} 
\usepackage{natbib} 
\usepackage{bbm} 
\usepackage{bbold}
\usepackage[symbol]{footmisc}

\degree{Doctor of Philosophy} \department{Physics} \gradyear{2013} 
\author{Matthew P. Killi} 
\title{Tunable single-particle and many-body physics in graphene}

\begin{document}

\newcommand{\vl}{v_{_L}}
\newcommand{\vc}{\mathbf}
\newcommand{\bk}{{{\bf{k}}}}
\newcommand{\bK}{{{\bf{K}}}}
\newcommand{\cE}{{{\cal E}}}
\newcommand{\bQ}{{{\bf{Q}}}}
\newcommand{\bq}{{{\bf{q}}}}
\newcommand{\br}{{{\bf{r}}}}
\newcommand{\bg}{{{\bf{g}}}}
\newcommand{\bG}{{{\bf{G}}}}
\newcommand{\hbr}{{\hat{\bf{r}}}}
\newcommand{\bR}{{{\bf{R}}}}
\newcommand{\hx}{{\hat{x}}}
\newcommand{\hy}{{\hat{y}}}
\newcommand{\hz}{{\hat{z}}}
\newcommand{\ra}{\rangle}
\newcommand{\la}{\langle}
\renewcommand{\tt}{{\tilde{t}}}
\newcommand{\upa}{\uparrow}
\newcommand{\dna}{\downarrow}
\newcommand{\bS}{{\bf S}}
\newcommand{\vS}{\vec{S}}
\newcommand{\dg}{{\dagger}}
\newcommand{\pdg}{{\phantom\dagger}}
\newcommand{\tphi}{{\tilde\phi}}

\newcommand{\bv}{{{\bf{v}}}}
\newcommand{\cV}{{\cal V}}

\newcommand{\vr}{\mathbf{r}}
\newcommand{\va}{\mathbf{a}}
\newcommand{\vb}{\mathbf{b}}
\newcommand{\pll}{\parallel}
\newcommand{\vR}{\mathbf{R}}
\newcommand{\vK}{\mathbf{K}}

\newcommand{\bp}{{\bf p}}
\newcommand{\py}{{p_y}}
\newcommand{\px}{{p_x}}

\def\re#1{(\ref{#1})} 
\def\Tr{\mathop{\mathrm{Tr}}} 
\def\sgn{\mathop{\textrm{sgn}}} 
\newcommand{\clb}{\color{blue}}
\newcommand{\Ref}[1]{Ref.~\onlinecite{#1}} 
\newcommand{\eq}[1]{Eq.~(\ref{#1})} 
\newcommand{\fig}[1]{Fig.~\ref{#1}} 
\newcommand{\bsigma}{\mbox{\boldmath$\sigma$}} 
\renewcommand{\ni}{\noindent}

\long\def\symbolfootnote[#1]#2{\begingroup%
\def\thefootnote{\fnsymbol{footnote}}\footnote[#1]{#2}\endgroup}

\begin{preliminary}
	
	\maketitle
	
	\begin{abstract}    
	The primary subject of this thesis is graphene and how the rudimentary attributes of its charge carriers, and local moments on its surface, can be directly manipulated and controlled with electrostatic potentials.        
	
	We first consider bilayer graphene subject to a spatially varying electrostatic potential that forms two neighbouring regions with opposite interlayer bias.  Along the boundary, 1D chiral `kink' states emerge.  We find that these 1D modes behave as a strongly interacting Tomonaga-Luttinger liquid whose properties can be tuned via an external gate.                         
	
	Next, we consider superlattices in bilayer graphene.  Superlattices are seen to have a more dramatic effect on bilayer graphene than monolayer graphene because the quasiparticles are changed in a fundamental way; the dispersion goes from a quadratic band touching point to linearly dispersing Dirac cones. We illustrate that a 1D superlattice of either the chemical potential or an interlayer bias generates multiple anisotropic Dirac cones.  General arguments delineate how certain symmetries protect the Dirac points.  We then map the Hamiltonian of an interlayer bias superlattice onto a coupled chain model comprised of `topological' edge modes.  We then discuss the relevance of spatially varying potentials to recent transport measurements.            
	
	This is followed by another study that considers the effect of a magnetic field on graphene superlattices.   We show that magnetotransport measurements in a weak perpendicular (orbital) magnetic field probe the number of emergent Dirac points and reveal further details about the dispersion.  In the case of bilayer graphene, we also discuss the properties of kink states in an applied magnetic field. We then consider the implications of these results with regards to scanning tunnelling spectroscopy, valley filtering, and impurity induced breakdown of the quantum Hall effect.

	Finally, we investigate local moment formation of adatoms on bilayer graphene using an Anderson impurity model.  We construct various phase diagrams and discuss their many unusual features.  We identify regions where the local moments can be turned on or off by applying a external electric fields. Finally, we compute the RKKY interaction between local moments and show how it too can be controlled with electric fields.

\end{abstract}

	 \begin{dedication}   
	\centering
		To Jordan and my family.
	 \end{dedication}        
	\newpage
\begingroup 
\tableofcontents 
\endgroup

\listoffigures 
\chapter*{List of Publications}
\vspace{-1.cm}
\noindent [1] ``{\underline{Tunable Luttinger liquid physics in biased bilayer graphene}}",  M.~Killi, T.-C.~Wei, I.~Affleck, and A.~Paramekanti, {\bf{Physical Review Letters}} {104}, 216406 (2010)
 \\\

 \vspace{-.4cm}
 \noindent [2]  ``{\underline{Controlling local moment
 formation and local moment interactions in bilayer graphene}}",\\
 M.~Killi, D.~Heidarian, and A.~Paramekanti, {\bf New Journal of Physics}  13, 053043 (2011)
 \\
 
 \vspace{-.4cm}
 \noindent [3]  ``{\underline{Band structures of bilayer graphene superlattices}}",
 M.~Killi, S.~Wu, and A.~Paramekanti, {\bf{Physical Review Letters}} 107, 086801 (2011)
 \\
 
 \vspace{-.4cm}
\noindent [4] ``{\underline{Graphene under spatially varying external potentials: Landau 
levels, magnetotransport,}\\
\underline{and topological modes}}",
S.~Wu, M.~Killi, and A.~Paramekanti, ({\bf \textit{Editor's Suggestion}}) {\bf Physical Review B} 85, 195404 (2012)
 \\

 \vspace{-.4cm}
\noindent [5] ``{\underline{Graphene: kinks, superlattices, Landau levels and magnetotransport}}",
 M.~Killi, S.~Wu and A.~Paramekanti, ({\bf \textit{Invited Review Article}}) {\bf International Journal of Modern Physics B}  26, 1242007 (2012)
 \\

 \vspace{-.4cm}
\noindent \symbolfootnote[2]{Not discussed in this thesis.}[6]  ``{\underline{Using quantum quenches to probe the equilibrium current patterns of ultracold atoms} \\ \underline{in an optical lattice}}",
 M.~Killi, and A.~Paramekanti, ({\bf \textit{Rapid Comm.}}) {\bf Physical Review A} 85,  061606(R) (2012)
 \\

 \vspace{-.4cm}
\noindent $^\dag$[7]  ``{\underline{Quantum quenches of ultracold atoms in background synthetic gauge fields}}",
 M.~Killi, S.~Trotzky and A.~Paramekanti ({\textit {in preparation}})

\chapter*{Author Contributions}
This thesis compiles some of the research I, together with various collaborators, have undertaken during my time as a PhD candidate.  It should be noted that some of the discussions and background material presented, particularly in in the Chapter \ref{Chapt:Kink}, recapitulate some of the early foundational work on non-uniformly biased bilayer graphene and the generation of 1D topological modes, and is not necessary original.  What is original material in this chapter is Section \ref{Sect:Kinkchar}, and all the corroborating figures.  Much of the discussion at the end this chapter, in Section \ref{Sect:TTLdisc}, review some of the subsequent work that followed the publication of our paper, \textit{Phys.~Rev.~Lett.} {\bf104}, 216406 (2010), that studied electron interactions in these 1D topological modes. 

This paper is the subject of Chapter \ref{Chapt:TTL} and the original publication was written in collaboration with Tzu-Chieh Wei, Ian Affleck and Arun Paramekanti.  The two external co-authors, Ian Affleck and Tzu-Chieh, were involved in discussions and interpretations of the results.

Chapter \ref{Chapt:SL} contains a study on superlattices in bilayer graphene that is largely composed out of material from our publication \textit{Phys.~Rev.~Lett.} {\bf107}, 086801 (2011). The results in this section were obtained in collaboration with Si Wu and Arun Paramekanti.  I would like to acknowledge that Si Wu was responsible for creating the superlattice dispersion figures and density of states figures in Chapter~\ref{Chapt:SL}.  Si Wu numerically computed the dispersions and density of states using a low energy effective theory, while I substantiated this by calculating the band structure using the full tight-binding Hamiltonian.  All the authors worked out the main analytic results.

The following chapter consists of a related study on the Landau levels, magnetotransport and topological states in monolayer and bilayer graphene when subjected to spatially varying potentials in the presence of a magnetic field.  The material of this chapter derives from the publication \textit{Phys.~Rev.~B.} {\bf 85}, 195404 (2012) that was again completed in collaboration with Si Wu and Arun Paramekanti. The figures in Sections~\ref{section:mono} -- \ref{section:bilayer} on the superlattices were created by Si Wu.  He was also responsible for calculating the transport properties using the Kubo formula. 

Chapter \ref{Chapt:Adatom} discusses local moment formation and RKKY interactions in biased bilayer graphene and is composed out of work published in \textit{New J.~Phys.} {\bf 13}, 053043 (2011). The original paper was written in collaboration with Dariush Heidarian and Arun Paramekanti.  Dariush Heidarian was involved in discussions on the details of calculation and the subsequent results.

\end{preliminary}


\chapter{Introduction}

\begin{figure}
	[tb]
	
	\centering a) 
	\includegraphics[width=0.35
	\textwidth]{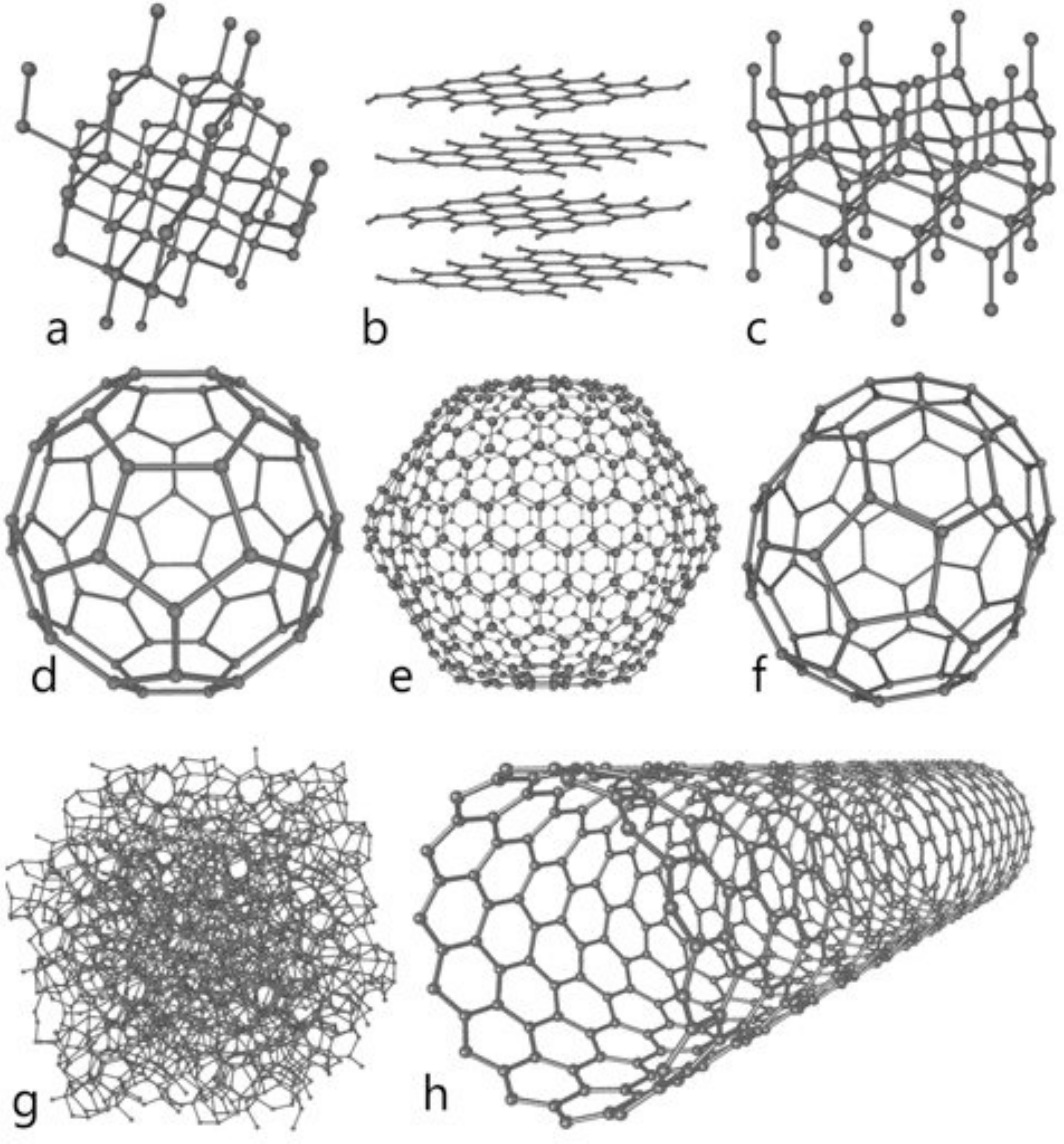}\hspace{1cm}
	b) \includegraphics[width=0.45
	\textwidth]{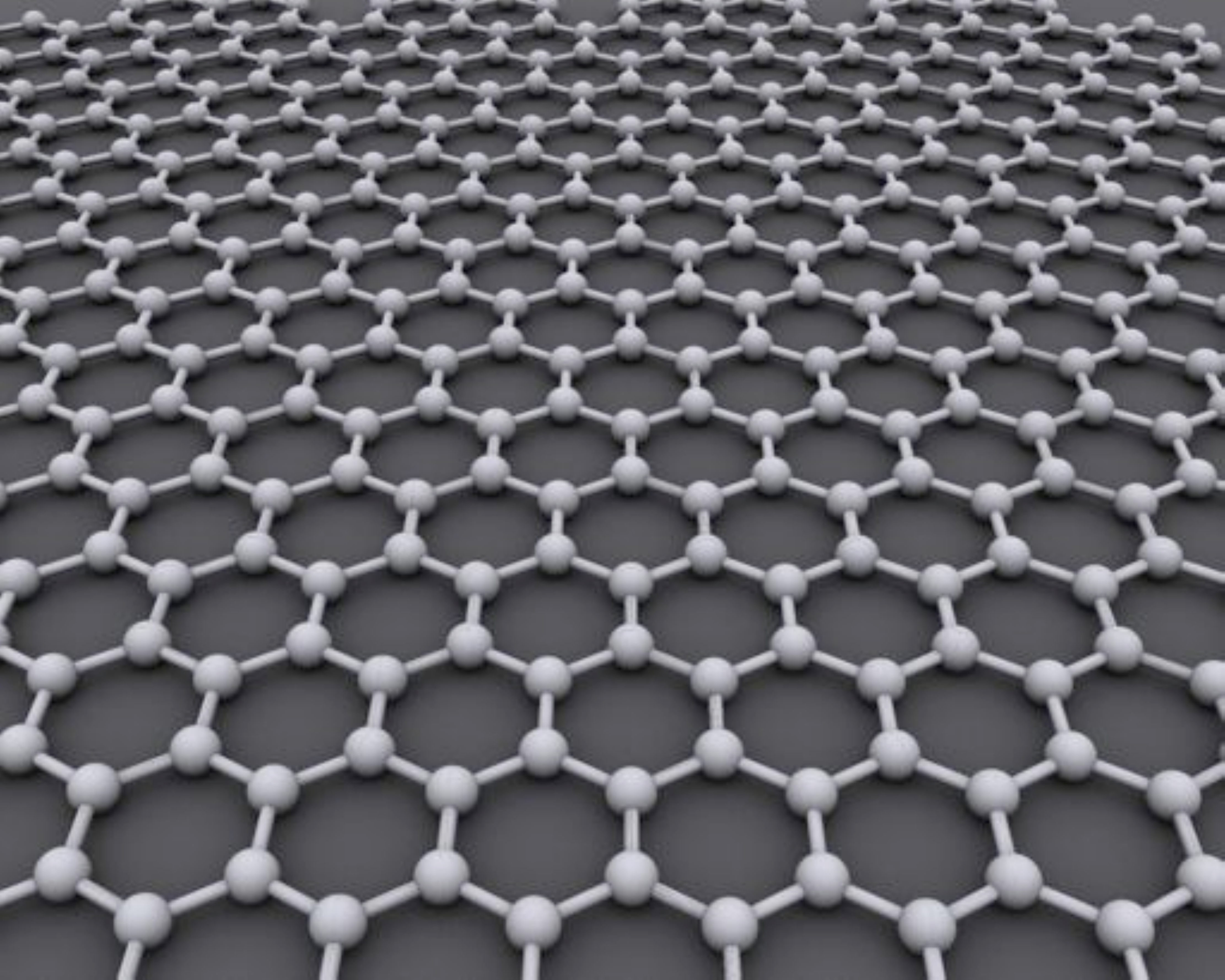}
	\caption{ (a) Various allotropes of carbon. (b) Atomic honeycomb lattice of single layer graphene.  (Images reproduced under the terms GNU Free Documentation Licence)} \label{Fig:Allotropes} 
\end{figure}

Of all the elements, carbon is arguably the most versatile.  Through its unique bonding properties, it forms the backbone of organic matter and is the basis of terrestrial life. In and of itself, carbon can be arranged in a multitude of different ways, forming a vast variety of different allotropes with strikingly disparate physical properties.  Examples of these are shown in Fig.~\ref{Fig:Allotropes}. It is astonishing that the allotropes of carbon span all three spatial dimensions, from 0D Fullerenes, 1D carbon nanotubes, 2D graphene, all the way up to 3D graphite, diamond and Lonsdaleite.  Equally astonishing is where these allotropes have been found, whether it be in ordinary candle soot \citep{Su:2011}, in unearthly fragments from the Canyon Diablo and Goalpara meteorites \citep{Hanneman:1967}, or as far away as 6,500 light years from Earth, within the ancient cosmic dust surrounding a white dwarf in the planetary nebula Tc 1 \citep{Cami:2010}.  Without question, each of these allotropes demonstrate extraordinary physical properties, many of which could be (and are presently being) exploited for innumerable technological applications. However, among all the allotropes of carbon to be discovered, it is graphene that has commanded an unprecedented amount of attention and continues to capture the imagination of the scientific community.

Single layer graphene is considered to be the most elementary allotrope of carbon, as it is the basic building block that constitutes many of the other allotropes, examples of which are shown in Fig.~\ref{Fig:Allotropes} (a).  It can be rolled into nanotubes or buckyballs, layered to form graphite or can simply exist in its freestanding form, as in Fig.~\ref{Fig:Allotropes} (b).  Despite its rudimentary role, of the above listed allotropes it was the last to be discovered. 

This is perhaps not surprising considering many questioned whether graphene's very existence was even theoretically possible. 
It was argued that 2D crystals could not be stable at finite temperature, as this would violate the Mermin-Wagner theorem, which stipulates that continuous symmetries cannot be spontaneously broken at finite temperatures in 2D systems \citep{Mermin:1966}.  As it turns out, the 2D crystal remains intact because of the anharmonic coupling between its bending and stretching modes, which are also responsible for making graphene flexible.  This coupling helps mitigate the detrimental long wavelength fluctuations that would ordinarily cause the sheet to crumple.  In a sense, nature circumvents the Mermin-Wagner theorem by forming relatively static ripples along the surface of the crystal, so that when it is placed on a substrate or suspended by supports it becomes stable \citep{Meyer:2007}. Of course, none of this was known to the early pioneers of graphene research. The stabilizing effect of the anharmonic coupling mechanism was only demonstrated a few years after the initial discovery of single layer graphene in a Monte Carlo study by \citet{Fasolino:2007}.

Through all the contention surrounding the possible existence of graphene, experimentalist continued to attempt to produce fewer and fewer layers of graphene, often with ingenious techniques that required sophisticated instrumentation \citep{Zhang:2005}. The first observation of large freestanding graphene crystals \emph{and} measurements of its electronic properties occurred in 2004 \citep{Novoselov:2004}.  This breakthrough is often heralded as the discovery of a new `wonder material'.  Not long after, Andre Geim and Konstantin Novoselov won the Nobel Prize in Physics in 2010 for their discovery \citep{Geim:2011}.  Ironically, their procedure for extracting the graphene flakes was to repeatedly strip away the layers from bulk graphite using low-budget adhesive tape, a technique colloquially referred to as the `Scotch-tape method'.

The sheer magnitude of this discovery is reflected by the enormous amount of work that has gone into studying graphene.  Since its isolation, there has been a plethora of research spanning both applied and fundamental physics.  Demonstrations of its remarkable electronic and structural properties have been continual, covering the purview of relativistic quantum field theory, general relativity, quantum hall effects, topology, low-dimensional physics and much more.  Aside from this, other prevalent areas of new research, such as topological insulators and Weyl semimetals, were initially seeded by graphene research.  As we approach nearly a decade since its discovery, graphene remains one of the most intensely studied materials and many fascinating, and often unexpected, discoveries occur with almost certain regularity.

\section{Theoretical history: (2+1)D quantum electrodynamics in a pencil}
The first theoretical study of the electronic structure of graphene goes back to the early work of Canadian physicist Phillip Wallace and incredibly predates the discovery of graphene by more than half a century. At the time of this work, he was visiting Nevill Mott at Bristol University and working on the band theory for graphite (essentially stacked layers of graphene), which was historically used as a modulator in nuclear reactors. In his seminal paper \citep{Wallace:1947}, the band structure of bulk graphite was calculated using a tight-binding model.  As a first approximation, he assumed that the interlayer coupling was negligible, thereby indirectly computing the band structure of single layer graphene.

His approximate tight-binding model, which incidentally corresponds to that of a single layer of graphene, describes an atomically thin planar 2D crystal consisting entirely of carbon atoms.  Its crystal structure is that of two interpenetrating triangular lattices that form a regular honeycomb pattern composed of hexagonal rings, which resemble benzene molecules. Each carbon atom is bonded to its three neighbours by hybridized $sp^2$ orbitals that have three lobes.  These strong $\sigma$-bonds form deep valence bands that are electrically inert and are ultimately what make graphene so durable. The main result Wallace reported was that a single layer of graphene has two electronically active half-filled $\pi$-bands that impart it with exotic electronic properties.  These bands form through the hybridization of the remaining $p^z$ orbitals and reflect the unique symmetries of the honeycomb lattice.

\begin{figure}[tb]
	\centering
	 a) \, \includegraphics[width=0.45	\textwidth]{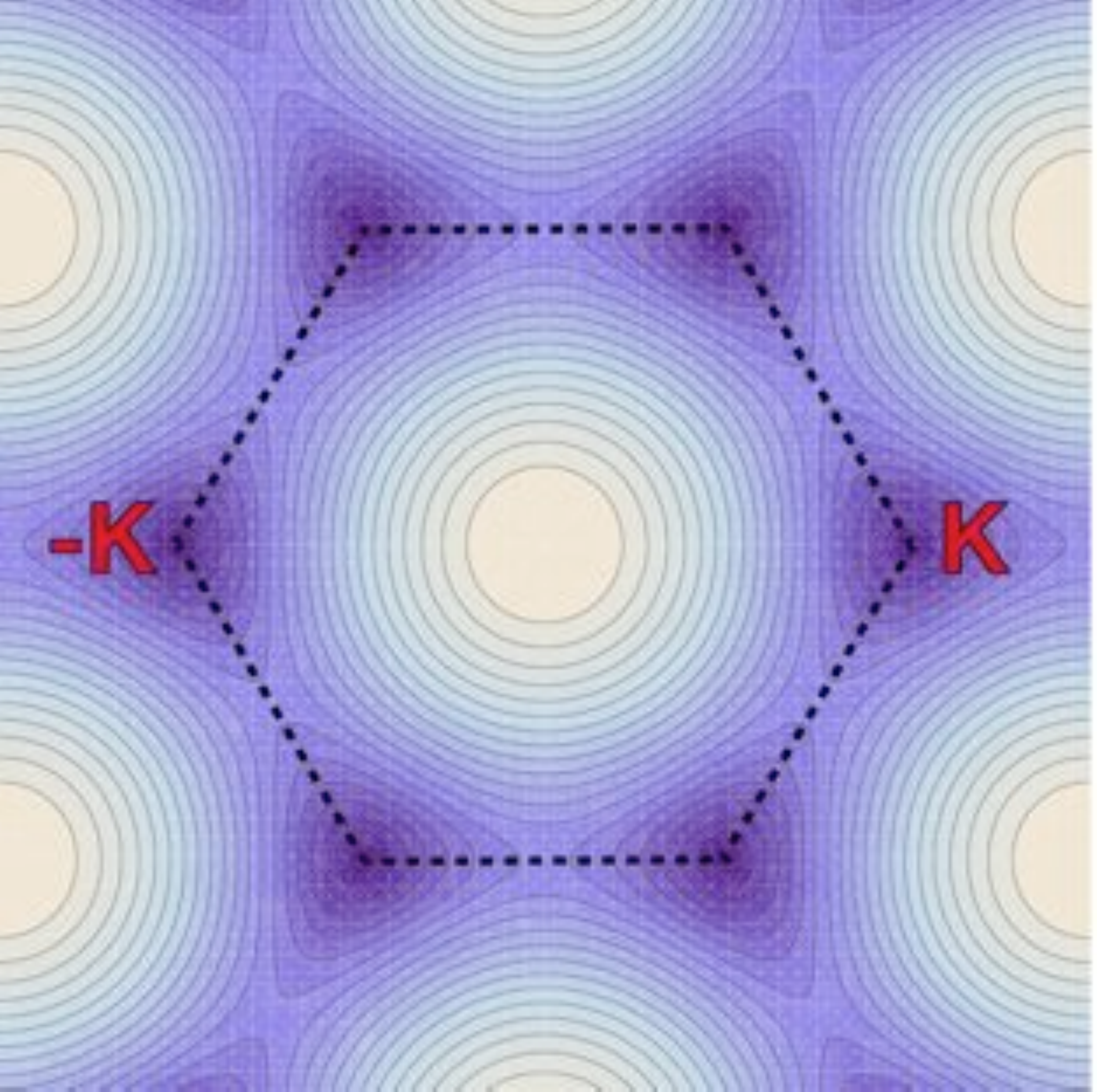} \hspace{1cm}
	 b) \includegraphics[width=0.35\textwidth]{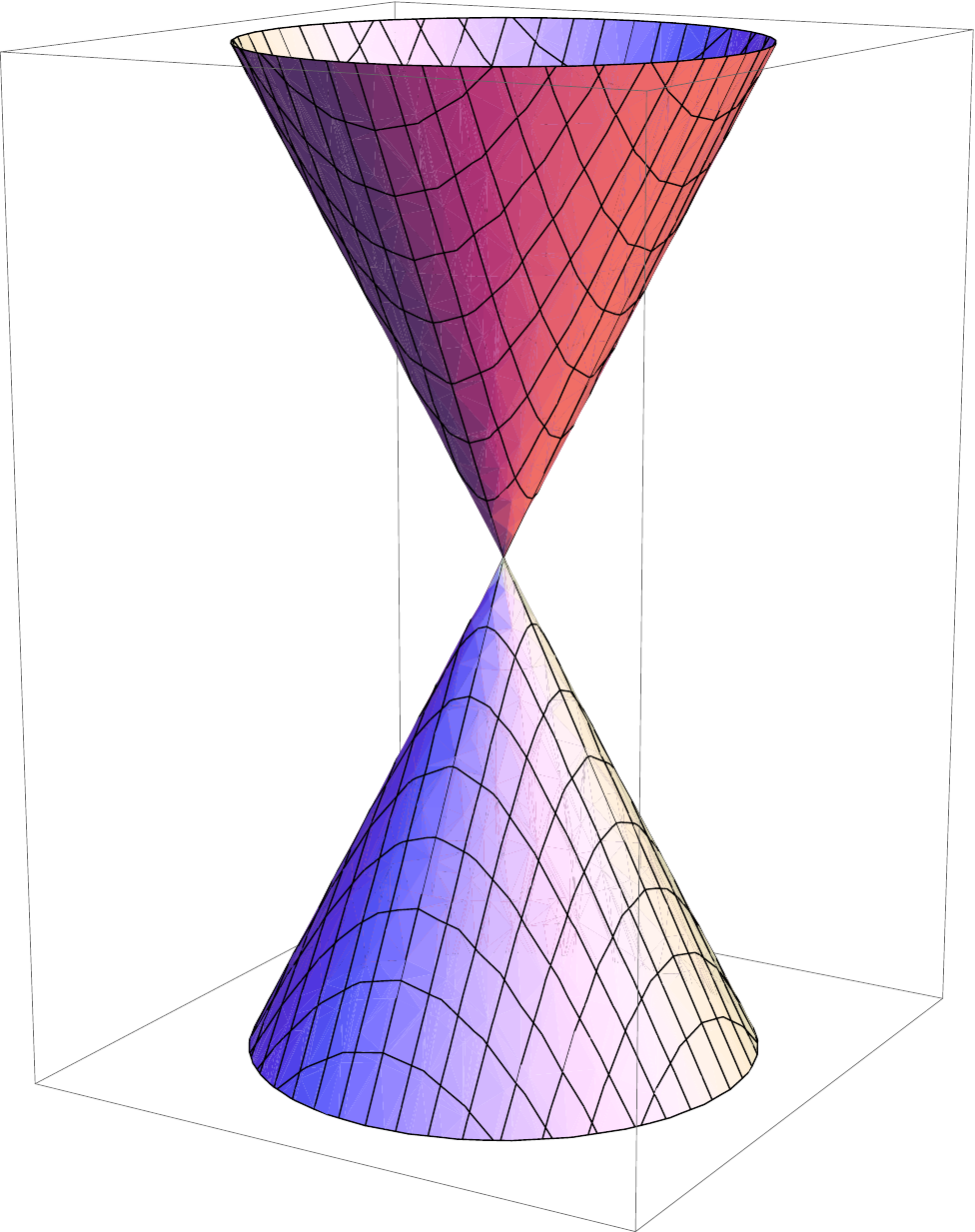} 
	\caption{ (a) Constant energy countours.  Dotted region denotes the Brillouin zone and the two unique $\bK$-points are indicated. (b) Closeup of a Dirac cone about one of the $\bK$-points.} \label{Fig:BZ}
\end{figure}

More specifically, Wallace's calculation demonstrated that if the single layer system is close to (but not at) charge neutrality, there are two separate circular Fermi surfaces.  Each Fermi surface encloses one of the two unique corners of the hexagonal Brillouin zone, which are labeled $\pm \bK$ in Fig.~\ref{Fig:BZ} and referred to as valleys.  As the system approaches the charge neutrality point, the Fermi surfaces shrink continuously and ultimately vanish.  This is all a reflection of the fact the band structure of graphene, shown in Fig.~\ref{Fig:BZ}, supports two linearly dispersing particle-hole symmetric Dirac cones at the $\bK$-points, making graphene semimetallic.  As we shall see, it is the low energy excitations in the vicinity of these two Dirac cones that lie at the heart of much of the perplexing physics observed in graphene.   

Canadian physicist Gordon Semenoff was the first to bring this this point into focus in an article published in \textit{Physical Review Letters} in 1984. In this seminal paper, he made the striking observation that the low energy excitations of graphene, which was still hypothetical at the time, served as a condensed matter analogue to $(2+1)$D dimensional quantum electrodynamics \citep{Semenoff:1984}.  He showed that the Hamiltonian at low energy consists of two decoupled copies of the Dirac equation, which are related by time reversal.  Ignoring the spin of the electrons, the solutions to the Dirac equations have two components, each of which corresponds to the value of the wavefunction on one of the two sublattices.

The spinor-like nature of the wavefunction resembles the internal spin degree of freedom of an electron and is commonly referred to as the pseudospin (note, there is no association with the angular momentum of the state).  Whereas the spin of an electron can in general point in any direction, the pseudospin of a Dirac fermion is restricted to align in one of two directions, depending on whether it is in the conduction or valence band.  If it is in the conduction band, the pseudospin points parallel to quasiparticle's direction of propagation and antiparallel if it is in the valence band. This amounts to the chiral propererty of the Dirac fermions, and is reminiscent of massless neutrinos described by the Weyl equation.  Further details about the tight-binding model, the Dirac equation, and the pseudospin can be found in Section \ref{Section:TBMLG}.

In sum, long before graphene was discovered, Semonoff's paper established the basic theoretical understanding that the low energy quasiparticles of graphene consists of two species of relativistic massless Dirac fermions with a chiral pseudospin degree of freedom.  This, along with later papers (see Ref.~[\citet{Geim:2012}] for an account of the prehistory of graphene), established much of theoretical groundwork for the unconventional electronic and magnetic properties of graphene that result from its Dirac-like quasiparticles.  

It was finally in 2005, one year after the discovery of graphene, over 20 years since Semonoff's article, and nearly 60 years after Wallace's band calculation, two papers appeared in \textit{Nature} that confirmed the existence of Dirac fermions in single layer graphene \citep{Novoselov:2005,Zhang:2005a}.  The history of graphene has therefore followed an often rare sequence of events ---  theoreticians provided predictions \textit{prior} to experimentation, rather than providing explanations for measurements \textit{after} experimentation.

\section{The graphene `gold rush' and beyond}
The question still remains as to why this discovery marked such a watershed in physics research, leading to the so-called `graphene gold rush'. Andre Geim emphasized in his Noble lecture that in actually, it was not so much the discovery of graphene and Dirac fermions that fuelled the flurry of research activity in graphene, but rather the exquisite transport \textit{measurements} \citep{Novoselov:2005,Zhang:2005a} and simplicity of the `Scotch-tape' method used to extract graphene.  Experiments showed that graphene possesses a phenomenally high carrier mobility at room temperature and atmospheric pressure.  This indicated the extraordinary high quality of the exfoliated graphene samples, even at this early stage of development when the fabrication techniques were still quite crude.  Laboratories could therefore produce high calibre graphene samples without the necessity for overly sophisticated and esoteric fabrication techniques, which often require expensive equipment.

The unexpected high quality of graphene, marked by a carrier mobility of over 200 times that of silicon, also caught the attention of industry and applied physicists. It was almost immediately recognized that graphene could be a viable material for high precision electronic devices. At the present time, it appears that the mobility of graphene is not even limited by crystal quality, but rather the imperfections, impurities and phonons of the substrate it is placed on.  This is corroborated by the markedly enhanced mobility in suspended graphene \citep{Du:2008}, and graphene placed on hexagonal boron nitride substrate \citep{Dean:2010,Xue:2011}.

Part of the excitement surrounding graphene also comes from the direct observation of quantum mechanical effects \emph{at room temperature}.  Normally, the observation of quantum mechanical phenomena requires temperatures of a mere few degrees above absolute zero, corresponding to temperatures colder than the boiling point of liquid nitrogen by an order of magnitude.  In contrast, quantum effects, such as integer quantum Hall conductivity, have been observed in graphene samples at much higher temperatures than those needed in typical materials, potentially introducing coherent quantum mechanical effects to room temperature applications \citep{Novoselov:2007}.

Aside from the high quality of graphene samples and their exotic physical and electronic properties, graphene has the significant benefit of being freestanding and exposed --- this makes it well suited for electronic gating and contacting, and surface microscopy.  As a result, the carrier density of graphene, a concomitant of the chemical potential, can be dynamically controlled with an external back-gate.  Moreover, multiple gates can be employed to dope specific regions of the sample in order to form p-n junctions.  It has also been suggested that the electronic structure of atoms absorbed on the surface of graphene can also be controlled with back-gates, opening the door to gate controlled local moment formation \citep{Uchoa:2008,Castro-Neto:2009a,McCreary:2012} and proposals of graphene based chemical sensors \citep{Brar:2011}.  Furthermore, since graphene is essentially an exposed membrane, it is possible to directly deposit and arrange individual atoms on its surface using an atomic force microscope.  The interactions of these atoms with the underlying Dirac fermions could then be studied using scanning tunnelling microscope \citep{Uchoa:2009,Brar:2011}.

Just recently, there have been promising new developments to engineer strain by stretching graphene sheets.  The significance of this is that strain manifests itself in an usual way in graphene ---  it appears to the low energy Dirac quasiparticles in the guise of an effective vector potential \citep{Pereira:2009,Pereira:2009a}.  This can be understood as arising from two effects: i) the geometrical distortion of the Brillouin zone, and ii) the anisotropic change in the nearest neighbour hopping parameters \citep{Guinea:2010}.  Together, these lead to a spatially dependent shift of the Dirac point that, accordingly, can be described as a vector potential, namely $\bk \rightarrow \bk +{\bf \delta k}(\bR)$.  For specific types of strain, the vector potential mimics that of one corresponding to a uniform magnetic field.  Recent experiments on strained graphene show that `bubbled' regions exhibit exceedingly large vector potentials, which would ordinarily require an enormous magnetic field of over $300 T$, far stronger than those created in a laboratory \citep{Levy:2010}. This and another recent experiment \citep{Yan:2012}, also report the presence of pseudo-Landau levels, which puts forth the prospect of observing new physics and entirely new device designs based on strained graphene.

\section{Controlling the defiant}
Although the outlook of graphene research has been opened to a vast expanse of possibility, along with the prospect come the inevitable obstructions that must be overcome.  The Dirac fermions in monolayer graphene are unwieldily defiant quasiparticles and very difficult to tame.  For instance, when these chiral quasiparticles impact a potential barrier at normal incidence, they tunnel through the barrier unobstructed as if it were not even there! Although a fascinating phenomenon in and of itself, it makes it very challenging to corral the particles into confined traps or to guide them along collimated paths.

This perplexing phenomenon is known as the Klein effect \citep{Dombey:1999}.  For massive particles incident on a potential barrier of typical height, the particle reflects back, away from the barrier.  Paradoxically, when the barrier strength is enormously large, of the order of $E\sim mc^2$, it becomes easier for the particle to tunnel through the barrier, as the wavefunction no longer decays exponentially.  Since the Dirac fermions in graphene are massless, the Klein effect is present for arbitrarily small potentials \citep{Katsnelson:2006,Young:2009}.  Prior to graphene, physicists had to contemplate tremendously high energies in order to come across this counter-intuitive phenomenon, such as at event horizon of blackholes \citep{Page:2005} or at the core of heavy atomic systems \citep{Dombey:1999}.

Luckily, potential barriers are not strictly transparent since the transmission of the quasiparticle is sensitive to the angle of incidence.  This angle dependence arises because of the chirality of the Dirac quasiparticle, and is a result of the conservation of pseudospin.  As a corollary to this effect, interfaces between two regions with different carrier density act as if they were an optical medium --- ballistic quasiparticles incident on the interface refract as they pass through it, much like light passing through glass.

Through the judicious use of external gates, physicists have devised many electrical analogues to optical components by harnessing this effect.  These include: lenses to focus the flow of electrons, so-called Veselago lenses similar to those constructed with left-handed metamaterials\citep{Cheianov:2007}, as well as electron guides that operate under the same principles as fibre-optic cables \citep{Williams:2011}.  I suspect that it may even be possible to fabricate an electronic version of an optical invisibility cloak first proposed by my former supervisor, Ulf~\citet{Leonhardt:2006}, and Sir John \citet{Pendry:2006} while I was completing my M.Sc.~at the University of St.~Andrews.  Here, collimated electrons would flow around a cloaked region and emerge back to their original trajectory.

Another often used strategy to contain the Dirac fermions of graphene is to cut the sheets into carefully patterned nanostructures designed to take advantage of geometrical quantum effects.  Using this method, graphene nanoribbons and quantum dots have been fabricated, which demonstrate a whole host of new electronic and magnetic phenomena.  This has lead to the development of `valleytronics' \citep{Rycerz:2007}, interesting edge state physics \citep{Yao:2009,Tao:2011}, a single electron transistor \citep{Ponomarenko:2008} and the realization of room temperature transistor devices \citep{Wang:2008}.

\section{Bilayer graphene: Flatland's new contender}	

Ironically, of all the challenges that researchers face, one the biggest obstacles stems from the very linear dispersing fermions that give rise to the whole slew of interesting phenomena mentioned above.  Many technological applications actually require a material with a functional energy gap between the valence and conducting bands --- in other words, a semiconductor.  Hence, it would be highly desirable to somehow open a bandgap at the Dirac point, thereby transforming graphene from a semimetal to a semiconductor.  In order to do so, graphene must either be altered in such away as to break the sublattice symmetry or translation symmetry in a particular way that hybridizes the two Dirac cones.  Both of these schemes require the daunting task of making microscopic perturbations down at the atomic length scale.

Although experimentalist have claimed to see a bandgap in epitaxial graphene grown on SiC \citep{Zhou:2007}, \citet{Rotenberg:2008} contended that this was not an intrinsic property as the authors claimed, but rather a consequence of the sample being `islanded'.  In another attempt inspired by the prediction of the stability of `graphane' (fully hydrogenated graphene), experimentalists cleverly used the Mo\"ire pattern of graphene/Ir(111) as template to mediate the absorption of hydrogen atoms \citep{Balog:2010}.  This forms a superlattice of confining potentials that appears to open a bandgap.  How useful this method proves to be remains to be seen.

The realization of new graphene-based electronics, integrated circuits and valley/spin-tronics is likely to be contingent on other specific modification, beyond just inducing a bandgap.  For this reason, the goal of many technologically driven researchers is to master control over graphene's properties to a level where it can be tailored to suite specific applications.  While there has been considerable progress in this respect, it is the close cousin of monolayer graphene ---  bilayer graphene --- that appears to be the most amenable.  Together with its multilayer derivatives, bilayer graphene could contain the key that ultimately unlocks the dream of designer-made materials. 

The distinguishing feature of bilayer graphene that makes it more conducive to electronic engineering is the presence of an additional layer with inversion symmetry.  Most importantly, when the layer symmetry is unilaterally broken, a bandgap is generated and bilayer graphene is transformed into a semiconducting state.  In principle, it is quite simple to break the layer symmetry in bilayer graphene because it is \emph{macroscopic}.  By applying an external perpendicular electric field across the sample, obtainable by sandwiching the sample in between a top and bottom gate, a potential bias can be induced between the two layers.  The presence of of such an interlayer bias generates a bandgap at the charge neutrality point that is of the same order as the bias.  The tight-binding model of bilayer graphene, its pseudospin, and the induction of a bandgap is discussed in Chapter \ref{Sect:TBBLG}.

One particularly salient feature of this setup is the direct ability to tune the size of the bandgap by varying the strength of the electric field generated by the two gates.  In addition, by carefully tuning the \emph{average} electrical potential of the two layers, the chemical potential can also be tuned externally and dynamically, just as in backgated single layer graphene samples.  Yet, perhaps the most enticing feature of this method is that it provides a means to spatially vary the interlayer bias and chemical potential by patterning multiple gates in various configurations.            

\section{1D modes in biased bilayer graphene}
In a remarkable study, \citet{Martin:2008} demonstrated that along the interface between two neighbouring regions of bilayer graphene that are biased with opposite parity, there exists 1D conducting modes. The existence of the zero energy modes was argued to be of topological origin.  The author's recognized that the low energy bands possess a non-trivial topological charge about each separate $\bK$-point when the layers are biased.  To the extent that each valley can be taken as independent, the bands about each valley carry a finite Chern number (with opposite sign) if the layer symmetry is broken and the system is in an insulating state.  The key observation is that when the polarity of the bias is reversed, the sign of the Chern numbers changes, indicating that the two insulating states (which have opposite layer polarization) are topologically distinct.  It follows that there exists unidirectional zero energy modes that propagate along any interface between two regions with opposite layer polarization.  Specifically, the number of left moving branches minus the number of right moving branches is given by the difference of the Chern numbers, yielding two right movers in one valley and two left movers in the opposite valley. These 1D `kink' modes are explored in more detail in Chapter \ref{Chapt:Kink}.

One particularly interesting aspect about the kink states is that they are chiral --- the direction of propagation is directly determined by the valley index.  This suggests that currents induced along the 1D channels are valley polarized and potentially useful for valleytronic devices, which harness the extra valley degree of freedom present in graphene.  Although such valley valves and filters have previously been suggested in monolayer graphene, they rely on unadulterated zigzag edges. This presents the daunting task of fabricating graphene ribbons with ultra clean edges, without any dislocation or absorbents.  In this respect, the kink states have a clear advantage; they are present regardless of the details of the potential profile.  Moreover, the ability to generate one dimensional wires with external gates lends the possibility of designing patterned gate configurations that form circuits, which can then be altered \emph{dynamically}.

Aside from technological interest, such kink states also play a prominent role at low energy where electron correlations are presently thought to spontaneously break the `which-layer' symmetry.  Within the layer polarized phase, electron correlations can be treated as a mean-field whose effect is similar to that of an externally induced interlayer bias.  Since the layer symmetry is spontaneously broken, there will be naturally occurring domain walls where kink states very similar to those suggested by \citet{Martin:2008} will form.  As with any spontaneously broken symmetry state, an understanding and characterization of all the possible defects is required for a complete description of both the phase itself and its associated phase transition.

Given the potential technological implications of engineering kink states, and the extensive theoretical and experimental progress made in understanding the strongly correlated phase in bilayer graphene, we further explore the physics of biased induced kink states.  In Chapter \ref{Chapt:TTL}, we go beyond the preliminary single-particle description presented by \citet{Martin:2008} and consider electron interactions.  Electron interactions are shown to play a pivotal and unique role this system, since the single particle description, or more precisely, the Fermi liquid description, breaks down and is no longer valid.  In systems with reduced dimensionality such as this, all excitations are strictly collective.    

\section{Superlattices in bilayer graphene}
Taking into consideration all the novel physics surrounding just a single kink in the interlayer bias, a natural question to ask is \emph{what happens to the low energy modes of the system when multiple kinks and antikinks couple}?  Even if the two valleys remain independent, the topological arguments presented above are no longer valid, so the presence of zero energy modes is not guaranteed.  Consequently, there is nothing known \emph{a priori} about the band structure of a coupled kink/antikink array and it may very well be gapped at low energy.

This same question can also be posed from a very different viewpoint, one where the external potential that forms the kink/antikink array is viewed as a 1D superlattice and bilayer graphene is considered to be a type of semiconducting material.  With this perspective, together with the fact that superlattices in conventional semiconductors have been shown to provide a viable route to band structure engineering \citep{Tsu:2005}, it is natural to ask how general superlattices potentials modify the band structure of bilayer graphene. 

Already, superlattices have been suggested as a promising route to engineer the band structure of \emph{single} layer graphene, paving the way to further new physics and applications. Studies of 1D superlattices in monolayer graphene have demonstrated their remarkable ability to not only regulate the Fermi velocity, but to also generate new Dirac cones    \citep{Park:2008,Park:2008a,Park:2008b,Park:2009,Brey:2009,Barbier:2008,Barbier:2009,Barbier:2010a,Barbier:2010,Killi:2011a,Tan:2011,Arovas:2010,Burset:2011a,Pletikosic:2009,Rusponi:2010}. In the seminal work by \citet*{Park:2008}, they showed that a 1D superlattice in monolayer graphene can lead to strong Fermi velocity renormalization in the direction \emph{transverse} to the superlattice modulation, but surprisingly, leaves the modes with momentum parallel to the interface unchanged.   Thus, the speed at which information can travel depends on the direction of propagation and can be tailored in a continuous fashion.  This counter intuitive phenomenon is a direct consequence of the chirality of the quasiparticles and linear dispersion of the Dirac cones, and has no counterpart in conventional non-chiral 2D systems.  Later, it was shown that superlattices can even lead to nonzero energy Dirac points \citep{Park:2008b} and multiple zero energy Dirac points near each valley. \citep{Park:2009,Brey:2009} These effects again stem from the chirality of the low energy quasiparticles in monolayer graphene.

Significant headway has also been made with experiments.  Many methods to synthesize and generate superlattices in graphene have already been established and new experiments are presently underway.  One such method is to place graphene on specific substrate surfaces whose underlying crystal structure induces a superlattice potential.  For example, a recent set of experiments studied superlattices formed by epitaxial growth of graphene on an Ir(111) surface \citep{Pletikosic:2009,Kralj:2011}.  The lattice mismatch between the metal surface and the crystal structure of graphene produces a Moir\'e pattern that inevitably forms a periodic potential.  Subsequent angle resolved photoemission spectroscopy (ARPES) measurements were able to show clear evidence for the formation of minigaps at the mini Brillioun zone boundaries.  A related experiment by \citet{Rusponi:2010} actually \emph{synthesized} a superlattice on Ir/graphene using the Moir\'e pattern to form metal clusters that self-assemble.  This much stronger potential leads to an even more dramatic modification to the band structure that causes the Dirac cones to become anisotropic.  Yet another viable substrate for superlattice generation is hexagonal boron-nitride (hBN), which is known to induce a superlattice potential because the lattice constants of the two hexagonal crystals are incommensurate.  In perhaps the most remarkable experiment on superlattices to date, \citet{Yankowitz:2012} used scanning tunnelling microscopy to confirm the emergence of entirely new Dirac cones when graphene is placed on a hBN substrate.

As we can see, the progression from theoretical proposals to actual experimentation with superlattices has occurred rapidly for single layer graphene.  In contrast, there has been surprisingly limited theoretical studies on superlattices in bilayer graphene, despite their experimental viability.  Initiated in part by the direct link between an interlayer bias superlattice and the coupling of the kink states, and in response to the lack of theoretical exploration of general superlattice in bilayer graphene, Chapter \ref{Chapt:SL} focuses on the band structure of  bilayer graphene under various superlattice potentials.  Since the quasiparticles of bilayer graphene are also chiral, superlattices are expected to result in highly unconventional modifications to the band structure as well.  In fact, we shall see that the band structures that emerge are actually more profoundly modified than those of single layer graphene and are also much more amenable.

Aside from the above, there is another reason why it is important to understand how bilayer graphene is affected by spatially varying electrostatic potentials.  The presence of charge impurities in proximity to a flake will impose non-uniform potentials that are expected to affect transport, even in biased bilayer graphene.  To see why, it is helpful to review a recent set of experiments that measured the temperature dependence of the resistivity in biased samples. We now temporarily divert our discussion away from `engineered' superlattices to provide this overview and argue that studies on spatially modulated potentials provide an important first step to interpreting the reported measurements.

On theoretical grounds, bilayer graphene is an attractive candidate for transistor applications since it has a tunable gap, which varies in proportion to the electric field perpendicular to the layers \citep{McCann:2006,McCann:2006a}. However, conductance measurements on biased samples do not show the theoretically expected strong suppression of conductance at low temperatures \citep{Zhang:2009}; instead, the measured transport gap is 2 orders of magnitude smaller than the observed optical gap \citep{Oostinga:2008,Xia:2010,Miyazaki:2010}.  It was suggested that the excess conductance arises from `topological' edge states \citep{Li:2011a}, but transport measurements in a Corbino geometry do not support this scenario \citep{Yan:2010}.  This suggests that disorder is playing a predominant role.

Two experimental groups, one at MIT \citep{Taychatanapat:2010} and the other at the University of Maryland \citep{Yan:2010}, reported that the temperature dependence of the resistivity falls within two distinct regimes: (i) A high temperature regime that exhibits activated behaviour, and (ii) a low temperature regime that seems to be dominated by either variable range hopping or nearest neighbour hopping between localized states, as suggested by the MIT group.  In the temperature range between 100 K -- 2 K, the resistivity drops dramatically with temperature and the data is consistent with thermally activated transport with a large energy activation gap, of the order of the perviously observed optical gap, and a second smaller activation gap attributed to an impurity band.  However, at temperatures below 2K, the transport mechanism and the role of disorder becomes more obscure.

At a fixed displacement field, \citet{Taychatanapat:2010} reported that between 2K -- 300mK the temperature dependence of the resistivity data is consistent with either an additional variable range hopping term, $\rho=\rho_{VR}\exp(T_{VR}/T)^3$, or a next nearest neighbour hopping term, $\rho=\rho_{NN }\exp(E_{NN}/T)$, where $\rho_{NN}=\rho_0\exp(2r/a)$, $r$ is the distance between sites and $a$ is the localization length.  However, further measurements showed an unexpected strong exponential dependence of $\rho_{VR}$ on the displacement field, ruling out the variable range hopping scenario (although it is expected to be present at very low temperatures).  Hence, the nearest neighbour hopping model between localized states appears to be the most likely candidate.

This leaves us with the glaring question: \emph{What is the source and character of the localized states responsible for the observed conductance measurements?}  Fits to the nearest neighbour hopping model show that the average distance between localized sites is consistent with the density of charge impurities and indicates that they are somehow responsible for generating the underlying localized states.  Another clue about the localized states is that the localization length extracted from the data is about a nanometer in length and \emph{decreases with the displacement field}.  

The possible role of charge impurities to transport is perhaps not surprising given that they are known to shift the chemical potential locally and form electron/hole puddles, but how is it that they form the underlying localized states responsible for low temperature transport?  One often overlooked effect of charge impurities in bilayer graphene is that they can also strongly modulate the local bandgap.  With this in mind, it is reasonable to expect localized states to be trapped by charge impurities that cause the bandgap to close and reverse sign, in a manner very similar to the kink states explored above.  What is particularly enticing about this scenario is the localization length of such states is expected to decrease with with increasing bias, which is broadly consistent with the results of \cite{Taychatanapat:2010}.

Taken altogether, random charge impurities will form a modulated electrostatic landscape with both chemical potential and bandgap modulations that are expected to not only form subgap states, but also localized states.  Although the primarily focus of Chapter \ref{Chapt:SL} is  on `engineered' superlattices, to the extent that disorder potentials can be decomposed into Fourier components, we also expect to learn something useful about disordered bilayer graphene.

\section{Graphene superlattices in a magnetic field}
Although a  firm understanding of the remarkable restructuring of band structure when bilayer graphene is subjected to 1D superlattices is addressed in Chapter \ref{Chapt:SL}, the question still remains as to how the theoretical band structure can be experimentally verified.  Early experiments aiming to characterize the electronic properties of graphene used quantum Hall measurements to probe its underlying structure and unequivocally demonstrate the relativistic nature of the charge carriers \citep{Zhang:2005a, Novoselov:2007}. In these studies, the Hall conductivity displayed plateaus at atypical values of $\pm 4 (e^2/h) (N+1/2)$, where $N$ is an integer. The origin of these plateaus is well known to be a consequence of the fact that electrons in graphene are governed by a relativistic Dirac Hamiltonian, together with the inherent valley and spin degeneracy. The chiral symmetry of the Dirac Hamiltonian generates a particle-hole symmetric Landau level spectrum -- every positive energy Landau level has a conjugate negative energy Landau level that together are responsible for the positive and negative conductivity plateaus. The Dirac Hamiltonian also supports an additional zero energy Landau level, and leads to the `half-step' offset in the first conductivity plateaus. Finally, the step size between each plateau can be attributed to the presence of four degenerate Dirac cones labelled by different spin/valley indices. Further information about the velocity of the Dirac cone, $v_F$, can also be obtained by measuring the energy gap between the Landau levels, which scales as $v_f \sqrt{B}$ and goes as ${\rm sign}(n)\sqrt{|n+1|}-{\rm sign}(n)\sqrt{|n|}$, where $n$ is the level index. Thus, information about Landau level spectrum and Hall conductivity of a system can provide direct evidence for the existence of Dirac-like quasiparticles and can be used to probe the degeneracy and velocity of the Dirac cones.

Similar measurements of bilayer graphene in the quantum Hall regime are equally valuable.  Again, its Landau level spectrum contains many distinguishing features that unveil the elementary properties of its underlying quasiparticles that derive from the two quadratic band touching points.  Like monolayer graphene, the chiral symmetry of the Hamiltonian is reflected by the  spectrum's particle-hole symmetry.  Similarly, the four-fold degeneracy of the valley and spin degrees of freedom again leads to a step size of $4e^2/h$ in the Hall conductivity, albeit with one crucial exception --- the step size between the first particle-like and hole-like Landau levels is twice the size.  This occurs because there are now $two$ zero energy Landau levels present which can be traced back to the pseudo-spin winding number being $2\pi$ in bilayer graphene.  Aside from this, the magnetic field dependence of the Landau levels is the same as other systems with massive excitations in that it scales linearly.  In contrast to these conventional systems, however, the $n+1$ and $n$ Landau level separation is proportional to ${\rm sign}(n+1)\sqrt{|n+1|(|n+1|-1)}-{\rm sign}(n)\sqrt{|n|(|n|-1)}$ as opposed to being constant. Given the contrasting differences between the monolayer and bilayer, the Landau level spectrum and Hall conductivity can actually be used to distinguish chiral Dirac-like excitations that are massless from chiral Schr\"odinger excitations that derive from linear and quadratic band touching points, respectively.      

Hence, in Chapter \ref{Chapt:BSL} we investigate the electronic properties of both bilayer and monolayer graphene 1D superlattices in the presence of a uniform magnetic field perpendicular to the plane.  Similar to early magnetic field studies of graphene and graphene superlattices \citep{Zhang:2005a, Novoselov:2006, Park:2009}, careful analysis of the Landau levels spectrum and Hall conductivity proffers many potential experimental signatures of the underlying Dirac cones generated by the superlattice. Special attention is also made to electric field modulations where the period length is large and decoupled `topological' modes emerge.  

\section{Local moments on bilayer graphene}    

Another aspect of bilayer graphene that is interesting to explore is how its unusual single particle properties impact the physics of adatoms deposited on its surface, and to what extent the electronic structure of the adatom can be controlled with external gates.  Early work on single layer graphene explored the combination of adatom-graphene hybridization and Hubbard-like interactions on the adatom in the context of local moment formation \citep{Uchoa:2008, Venezuela:2009}, Kondo physics \citep{Uchoa:2010, Zhu:2010, DellAnna:2010, Wehling:2010, Jacob:2010}, RKKY interactions \citep{Vozmediano:2005, Dugaev:2006, Brey:2007, Saremi:2007, Black-Schaffer:2010, Black-Schaffer:2010a, Sherafati:2010}, and adatom positional ordering \citep{Cheianov:2009, Berashevich:2009, Shytov:2009, Abanin:2010}. Many of these early studies show that local moment physics in graphene is expected to be highly non-trivial, largely because of the chirality and linear dispersion of the low energy excitations.  Moreover, much of the physics that is predicted to take place on the adatom may be controlled by tuning the chemical potential with external gates.

It is important to emphasize, however, that at the time of these studies the route to realizing local moments in single layer graphene remained largely unexplored (and almost entirely unexplored for bilayer graphene).  Despite this, the above mentioned studies helped to invigorate experimental efforts to study adatoms, observe their tunable properties and, ideally, witness Kondo physics.  This research also helped initiate many first principles studies that attempt to elucidate the physics of specific candidate adatoms (although this issue still remains controversial at the present time).    

Like the preliminary studies of local moments on single layer graphene, the study presented in Chapter \ref{Chapt:Adatom} to help initiate the exploration of local moment physics in bilayer graphene by studying local moment formation and RKKY interactions. Since the low energy excitations of bilayer graphene are also chiral, and both the chemical potential and the bandgap can be independently tuned in this system, we expect that local moment physics in the bilayer will also proffer many  novel features and will exhibit an even higher degree of tunability.  

Given the nascency of first principle studies at the time of this work, this study aims only to distill some elementary features of tunable local moment physics in an attempt to further entice more detailed studies, first principle calculations and ultimately, experimentation. In time, the physics of adatoms on graphene may be of interest to the nanoscience and quantum computation communities given the possibility to control local moment physics, and adatom-adatom spin and density interactions, by varying the carrier concentration via gating \citep{Wolf:2001}.

\section{Summary of the main results}

\subsubsection{Tomonaga-Luttinger Liquid} 
As a starting point, we considered one of the simplest non-uniform interlayer bias profiles --- that of single interface separating two regions of opposite parity.  Along this interface, localized `topological' edge modes emerge that are analogous to those in quantum Hall systems.  These modes can be thought of as forming chiral 1D quantum wires, since states from opposite valleys are counter-propagating (this follows from the Berry curvature about each valley having opposite sign).  Through the combination of numerical diagonalization and Abelian bosonization, we demonstrated that Coulomb interactions drive these modes into a novel Tomonaga-Luttinger liquid phase.  Two important features of the system were demonstrated.  First, backscattering is largely suppressed by the wide spread of the wavefunctions, making the system relatively robust against disorder.  Second, the velocity of both spin and charge carrying modes, as well as the  Luttinger parameter in the total-charge sector, are directly tunable by changing the strength of the interlayer bias.  Since the Luttinger parameter determines the non-universal power-law decay of correlations in the system, this tunability could be observable in experiments such as scanning tunnelling microscopy.  The discovery of robustness of these modes and tunability has lead to to further studies which explored their possible uses, such as switchable nanowire circuits in gated bilayer graphene.  It has also been suggested that they may be naturally present in disordered bilayer graphene and along domain walls formed in certain spontaneously broken layer symmetry states at low energy.

\subsubsection{Superlattices} In a second related study, we looked at more elaborate external periodic potential profiles that form 1D superlattices in bilayer graphene. Our research showed that the effects of a superlattice are much more dramatic in the bilayer than in the monolayer because the low energy quasiparticles are changed in a fundamental way; the dispersion evolves from a single quadratic band touching point into a pair of linearly anisotropic Dirac cones separated in k-space.  Hence, the low energy chiral `Schr\"odinger-like' quasiparticles metamorphose into massless chiral Dirac fermions.  Remarkably, by varying the parameters of the superlattice both the velocity anisotropy and the distance between the cones can be controlled.  We used a combination of numerical diagonalization and perturbation theory to fully characterize all of the features of the band structure.  In addition, we presented general arguments that symmetry protects the band touching points in the dispersion.  Moreover, we derived a novel tight-binding model that describes the dispersion in terms of a series of coupled topological modes.  This model should be useful in future studies of these topological modes, particularly in understanding their contribution to transport and constructing network models of such edge states.

\subsubsection{Magnetotransport of Graphene Under Spatially Varying Potentials}
Following this, we explored how a uniform magnetic field perpendicular to the plane   affects both mono- and bi- layer graphene exposed to spatially varying potentials. Similar to early magnetic field studies of graphene, careful analysis of the Landau levels spectrum and Hall conductivity proffers many potential experimental signatures of the underlying dispersion.  We identified and studied three different regimes of magnetic field strength:  i) weak, ii) intermediate and ii) strong fields. In the presence of low magnetic fields, the structure of the Landau levels and Hall conductivity contain distinguishing features that reveal the elementary properties of the underlying quasiparticles --- they can not only differentiate between quasiparticles derived from quadratic and linear band touching points, but also the number of different species in the system and their anisotropy.  We found that for intermediate magnetic fields, the Landau levels become dispersive.  An interesting consequence of this is that anisotropy of the diagonal conductivity can be reversed by adjusting the magnetic field strength for 1D superlattices in the monolayer.  When the magnetic field is much stronger than the superlattice strength, the Landau levels of intrinsic graphene emerge.  Through studying the magneto effects, additional insights into the superlattice dispersion are also made.  Finally, we consider gate induced 1D topological kink states in the presence of the magnetic field, and discover an interesting valley symmetry breaking mechanism that results in valley polarized nanowires.

\subsubsection{Local Moments} The last study focuses on local moment formation on plaquette centred adatoms and RKKY interactions between these moments.  Again, the ability to generate a bandgap and control the chemical potential leads to some very interesting tunable properties in the system.  We used an Anderson mean field theory to construct phase diagrams for local moment formation when both the chemical potential and bandgap are varied.  As a consequence of the chiral wavefunctions of bilayer graphene, the coupling between the impurity and the quasiparticles of bilayer graphene has strong momentum and band dependence. This affects many of the details of the phase diagram.  For instance, the self-energy develops a large real part that has nontrivial frequency dependence, which substantially renormalizes the position of the impurity spectral peak.  Of the main results discussed, there were two significant findings. First, local moments could be turned on and off by varying the chemical and interlayer bias of the underlying bilayer graphene substrate. Second, the RKKY interaction strength is not only modified by varying either the chemical potential or the interlayer bias, but the exchange coupling can be made to switch from antiferromagnetic to ferromagnetic by varying both.  
 
\chapter{Graphene Basics}
\section{Tight-binding model of monolayer graphene} \label{Section:TBMLG}

\begin{figure}[tb]
	\centering
	 a) \includegraphics[width=0.45	\textwidth]{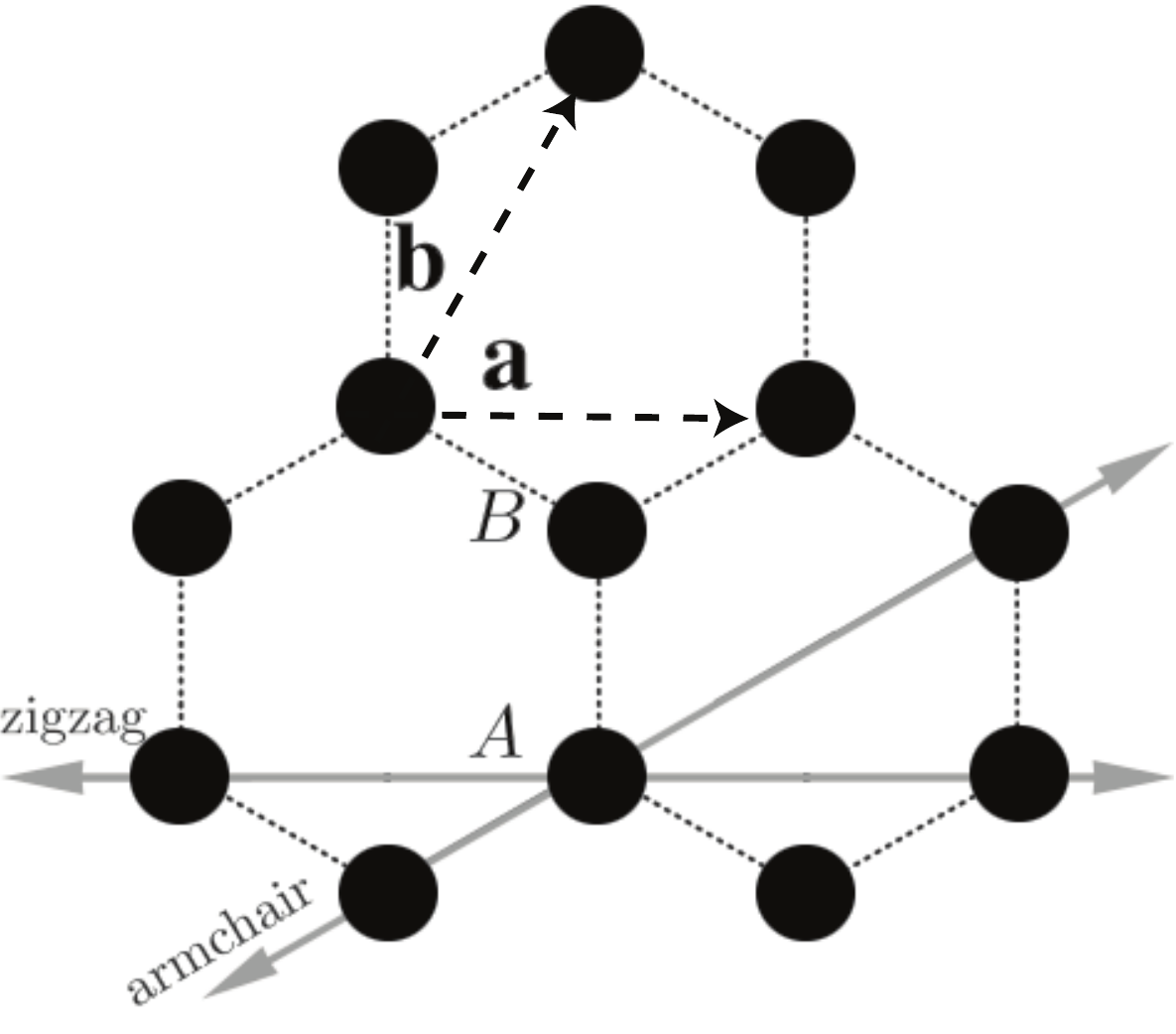}
	 b) \includegraphics[width=0.45\textwidth]{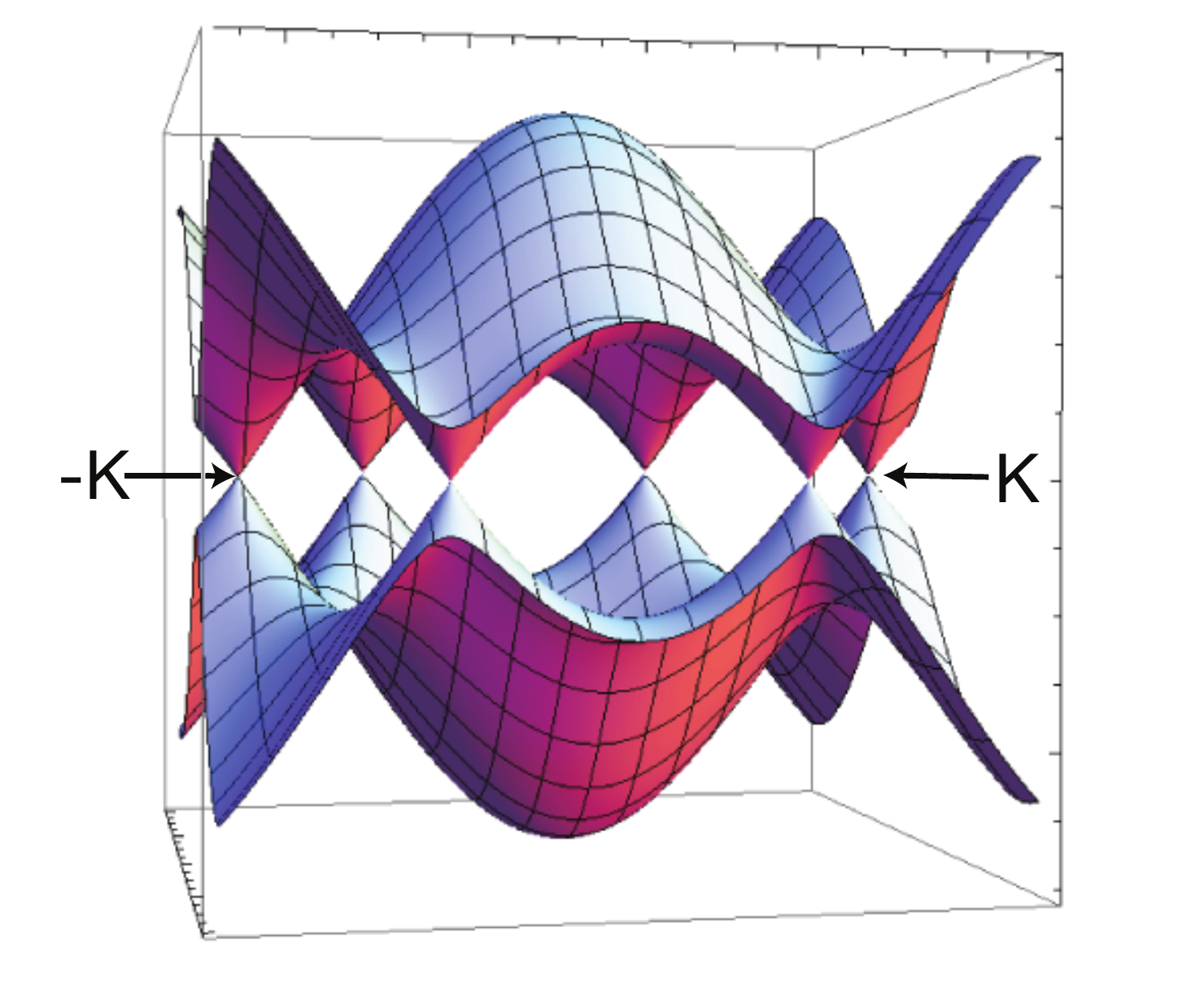} 
	\caption{ (a) Depiction of crystal structure of monolayer graphene and labelling conventions. (b) Full band dispersion of monolayer graphene.}\label{Fig:MLGtb}
\end{figure}

Monolayer graphene is a 2D honeycomb crystal composed entirely of carbon atoms.  Since the honeycomb lattice is strictly not a Bravais lattice, there are two inequivalent sites per unit cell.  The two atoms, labeled $A$ and $B$, form the basis of two interpenetrating triangular lattices spanned by the primitive vectors 
\begin{eqnarray}
	{\bf a}=\left(\phantom{\frac{1}{1}}\!\!\!\!d,0\right), \hspace{0.5cm}  {\bf b}=\left(\frac{1}{2}d,\frac{\sqrt{3}}{2}d\right),
\end{eqnarray}
where $d\sim 2.46$ \AA \, and nearest neighbour distance is $a=d/\sqrt{3}\sim1.42$ \AA.  In Fig.~\ref{Fig:MLGtb}, the two important crystal directions, zigzag and armchair, are also indicated.

Carbon has a valency of 4 electrons, since it contains 6 electrons. The $2s$, $2p_x$ and $2p_y$ orbitals hybridize to form three $sp^2$ orbitals that have three lobes at $120\,^\circ$ angle from each other, which lie in the plane.  The $sp^2$ orbitals of a single carbon atom overlap strongly with each of its three neighbouring carbon atoms' $sp^2$ orbitals.  These orbitals form strong $\sigma$-bonds, which results in the formation of the honeycomb crystal. (Note, the corresponding antibonding band lies at high energy and can be ignored).  This leaves a single electron per site and one remaining $p^z$ orbital.
	
The $p^z$ orbitals have two dumbbell-like lobes orientated perpendicular to the plane and can form $\pi$-bonds.  Since there are two orbitals in each unit cell, they form bonding and antibonding states, which then hybridize to form two bands.  These are referred to as the $\pi$- and $\pi^*$-bands and are together half filled with electrons.

Due to the translational symmetry of the system (and ignoring interactions), the crystal momentum $\bk$ is a good quantum number and the wavefunctions can be written as 
\begin{eqnarray}
	\Psi_s(\bk,\br)=\frac{1}{\sqrt{N}}\sum_{\bR}e^{i\bk\cdot \bR}\, \phi_s(\br-\bR)
\end{eqnarray}
where $s=\{A,B\}$ is the the sublattice index, $\br$ is the position vector, $N$ is the number of unit cells in the lattice, $\bk$ is the crystal momentum, $\bR$ is the position vector of a carbon site and $\phi_s(\br)$ are Wannier functions.  Denoting the transfer integral as
\begin{eqnarray}
	-t&=& \int d \br \, \phi^*_A(\br-\bR_A) \hat{H}  \phi_B(\br-\bR_B)\\ \nonumber
	&\sim&- 3 \, eV
\end{eqnarray}
where $\bR_A$ and $\bR_B$ are the position coordinates of any two neighbouring sites, and by assuming the overlap between next neighbour sites is small [see \citep{McCann:2012} for details], the Hamiltonian can be written as (ignoring spin),
\begin{eqnarray}
	\hat{H}=\sum_\bk \Psi^\dag(\bk) \left(
	\begin{array}{cc}
		0 & f\pdg(\bk) \\
		f^*(\bk)& 0
	\end{array}
		\right) \Psi(\bk),
\end{eqnarray}
where $\Psi(\bk)=\left(\psi^\pdg_{A}(\bk),\psi^\pdg_{B}(\bk)\right)^T$, $\psi_{s}(\bk)$\, $\left(\psi^\dag_{s}(\bk)\right)$ are annihilation (creation) operators, $f(\bk)=-t\sum_{\bf j}e^{i\bk \cdot {\bf \delta}_j}$, and ${\bf \delta}_j$ is one of three vector connecting a site to its neighbour.  Diagonalizing the matrix leads to the dispersion relation $\epsilon(\bk)=\pm |f(\bk)|$, which is plotted in Fig.~\ref{Fig:MLGtb} (b).

Close to half filling and at low energy, the matrix in the Hamiltonian can be expanded about the $\bK$-points at $\pm \bK=(\pm 4 \pi d/3,0)$, to give
\begin{eqnarray}\label{DE}
	 \hat{H}_{\pm \bK}(\bk) = -\hbar v_f (\pm k_x \sigma_x - k_y \sigma_y),
\end{eqnarray}
where $v_f=\sqrt{3}td/2$ is the speed at which the excitations travel (roughly 300 times slower than the speed of light), $k_{x(y)}$ is the $x$($y$) component of the momentum measured from the respected $\bK$-point and the Pauli matrices act on the sublattice degree of freedom.  Equation \ref{DE} is the Dirac equation.  Notice that the two valley Hamiltonians are related by time-reversal symmetry and must have the same dispersion relation in order to preserve the overall symmetries of the system.  The resulting dispersion relation is given by $\epsilon(\bk)= \pm \hbar v_F |\bk|$ and forms two Dirac cones at each $\bK$-point, identical to the one shown in Fig.~\ref{Fig:BZ}.
Solutions to the two valley Hamiltonians are related by time reversal, and so we can just focus on the $+\bK$-point. The wavefunction is given by 
\begin{eqnarray}
	\psi_{+\bK,\, \pm} (\bk)=\frac{1}{\sqrt{2}}	\left( 
	\begin{array}{c}
		e^{-i\theta/2} \\
		\pm e^{i\theta/2}
	\end{array} \right), \label{sol} 
\end{eqnarray}
where $\theta$ is the angle of propagation given by $\cos(\theta)=k_x/|\bk|$ and the $+$($-$) sign corresponds to the positive (negative) energy state. 

 The  explicit form of  Eqn.~\ref{sol} reveals an additional pseudospin degree of freedom associated with two sublattices.  Since the wavefunction is an eigenstate of ${\bf \sigma} \cdot {\bf n}_1 $, where ${\bf n}_1=\Big(\!\cos(\theta),\, \sin (\theta),\, 0\Big)$, the pseudospin is evidently chiral.  Electrons in the conduction band of valley $+\bK$ have their pseudospin pointing parallel to the momentum, and antiparallel to the momentum if they are in the valence band.  Notice that the pseudospin circumvolves once around as an electron is adiabatically evolved around the Dirac point.  This is a direct consequence of $\pi$ Berry phase associated with an \emph{isolated} Dirac point.  It is important to emphasize that the combined Hamiltonian, that is when both valleys are considered, is topologically trivial.  This is in contrast to strong topological insulators where there is an odd number of Dirac points.

\section{Tight-binding model of bilayer graphene} \label{Sect:TBBLG}
Bilayer graphene consists of two stacked copies of monolayer graphene that have a Bernal stacking order. The intrinsic electronic properties deduced from a non-interacting minimal tight-binding model of bilayer graphene share some similarities to those of single layer graphene, albeit with some notable differences.

\begin{figure}[tb]
	\centering a)
	\includegraphics[width=0.45
	\textwidth]{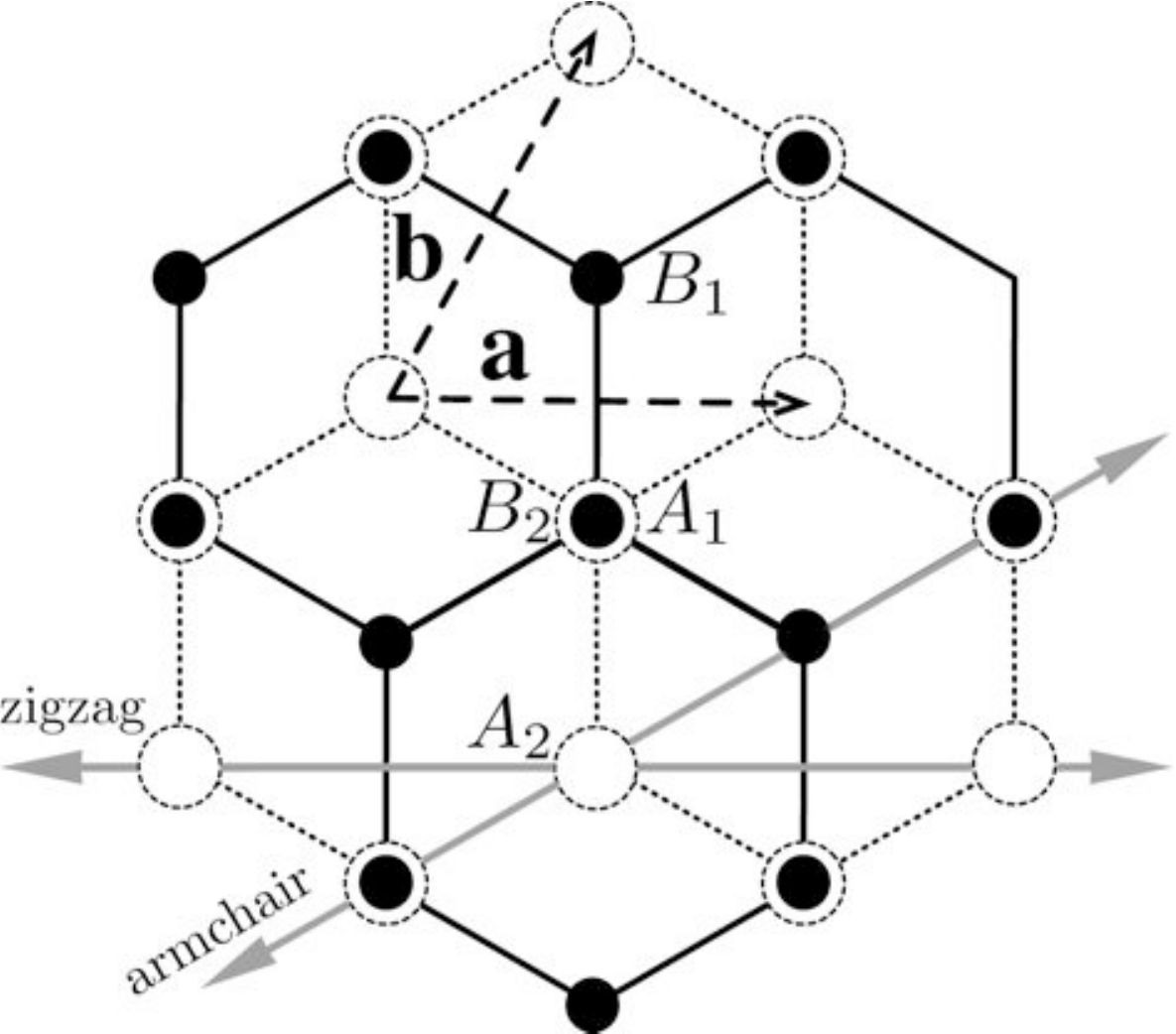} b)
	\includegraphics[width=0.45
	\textwidth]{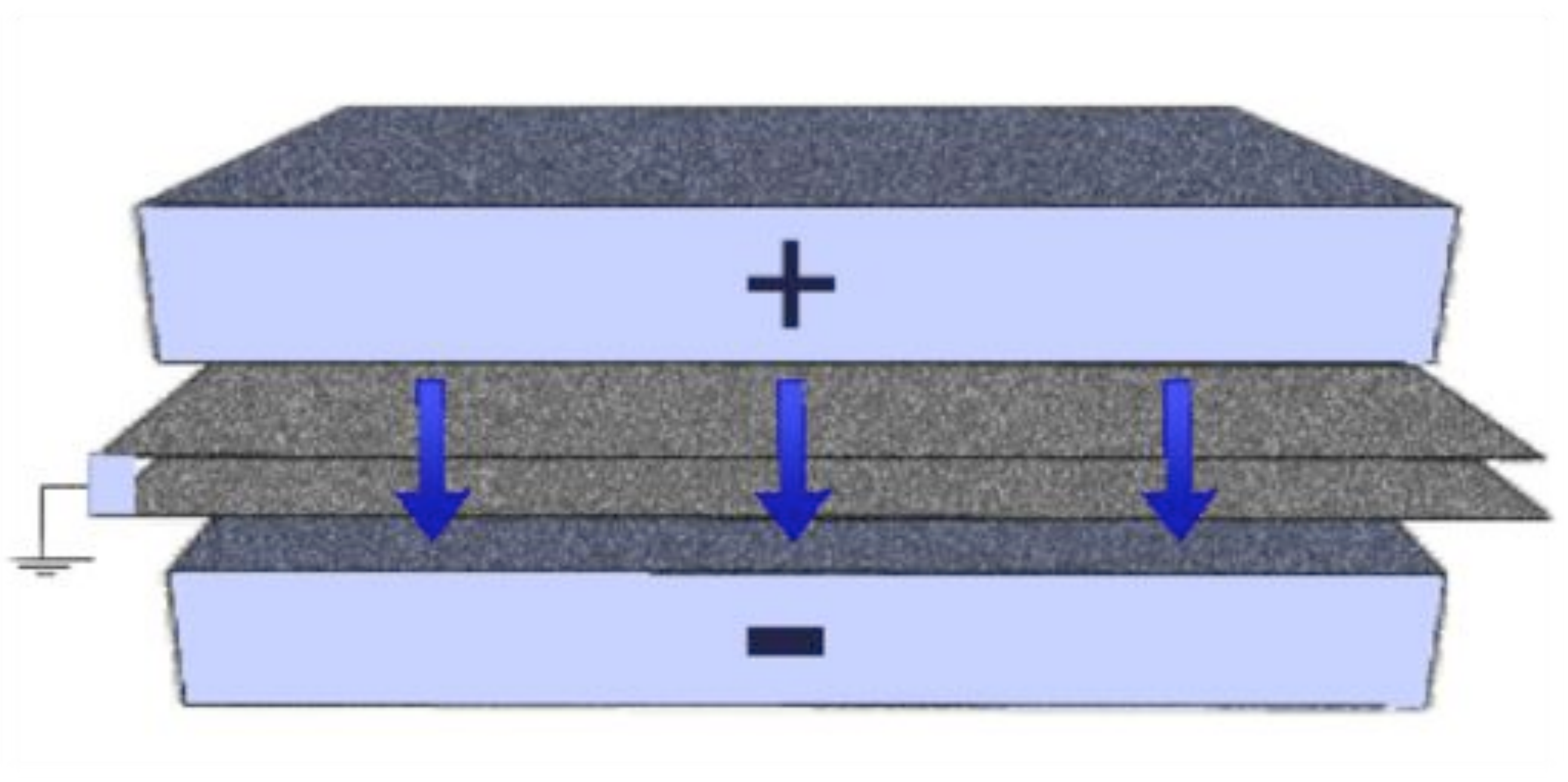} \caption{ (a) Depiction of crystal structure of bilayer graphene and labelling conventions. (b) Schematic diagram of a dual-gate configuration. $V_{\rm T}+V_{\rm B}$ controls the bilayer doping while $V_{\rm T}-V_{\rm B}$ controls the gap via the depicted perpendicular electric field.} \label{Fig:Scheme} 
\end{figure}

To begin, let us establish some of the notation and labelling conventions that will be adhered to throughout the text, unless otherwise noted. A schematic depiction of bilayer graphene is shown in Fig.~\ref{Fig:Scheme} (a). Again, there are two important crystal orientations, the armchair and zigzag direction. The carbon atoms in the bilayer are labeled by a unit cell index $i$, a sublattice index $s=a,b$, and a layer index $\ell=1,2$ labelling top and bottom layers, respectively. The distance between neighbouring carbon atoms in the same layer and on the same sublattice is $d \approx 2.46$ \AA, as it in the monolayer, while the interlayer distance is $d_\perp \approx 3.34$ \AA. The minimal tight-binding model of the electrons in bilayer graphene consists of the same nearest nearest-neighbour hopping amplitude $t \approx 3$ eV between sites on the same layer, and an additional interlayer hopping amplitude $t_\perp \approx 0.15$ eV between sites $(i,s\!=\!a,\ell\!=\!1)$ and $(i,s\!=\!b,\ell\!=\!2)$. Henceforth, we use units where $\hbar\!=\!t\!=\!d\!=\!1$, and set $\bR_i \equiv m \va + n \vb$ where $\va,\vb$ are unit vectors depicted in Fig.~\ref{Fig:Scheme} (a).

Quite generally, spatially dependent on-site potentials, $V_\ell(s,i)$, can also be present in the system. As discussed in the introduction, these potentials can be experimentally imposed through various techniques and displacement fields as high as $2.5$ V/nm can be imposed, and optical bandgaps of the order of $250$ meV have been observed.  For instance, by using a carefully chosen substrate or by patterning external electrical gates on the sample, any number of electrical potential profiles can be attained.  Spatially varying potentials are also expected to manifest inadvertently due to the presence of charged impurities in the surrounding environment. 

\subsection{Uniformly biased bilayer graphene and tunable bandgaps}

Here we consider bilayer graphene in the presence of a single dual-gate shown in Fig.\ \ref{Fig:Scheme} (b) that can be used to establish a homogeneous electrical potential on layer $1$ and layer $2$ independently, given by $V_1$ and $V_2$, respectively. Changing $V_g\equiv V_1-V_2$, and $-\mu \equiv (V_1 +V_2)/2$ allows for both the interlayer bias and the chemical potential to be tuned separately. Just as in the monolayer graphene, the important low energy physics takes place at the charge neutrality point at the $\pm \bK$ points in the Brillioun zone. Thus, for the homogeneous system, an expansion about the two valleys gives the following Hamiltonian,
\begin{eqnarray}
	\label{Eqn:Full}
	&\hat{H}_{\pm {\bf K},\,\sigma} \! \!= \!\!\Psi_{\pm \bK ,\,\sigma}^\dg(\bk)\! \!\left( 
	\begin{array}{cccc}
		\frac{V_g}{2}\!-\!\mu &\!\! v_F (\pm k_x\!-\!ik_y)\!\! & 0 & -t_\perp \\
		\!\!\!\!v_F ( \pm k_x\!+\!ik_y)\!\! & \frac{V_g}{2}\!-\!\mu& 0 & 0\\
		0 & 0 & -\frac{V_g}{2}\!-\!\mu & \!\!v_F (\pm k_x\!-\!ik_y) \!\!\!\!\\
		-t_\perp & 0 & \!\!v_F (\pm k_x\!-\!ik_y) \!\!& -\frac{V_g}{2}\!-\!\mu 
	\end{array}
	\right) \! \!\Psi_{\pm \bK ,\, \sigma}^\pdg(\bk), \nonumber\\
\end{eqnarray}
where $\Psi_{\pm \bK \sigma}^\dg(\bk)=(\psi_{A_1, \sigma}(\bk),\psi_{B_1, \sigma}(\bk),\psi_{A_2, \sigma}(\bk),\psi_{B_2, \sigma}(\bk))$ is the wavefunction composed of the four components corresponding to each atom in the four-site unit cell.

\begin{figure}
	[tb]
	
	\centering a) 
	\includegraphics[width=0.5
	\textwidth]{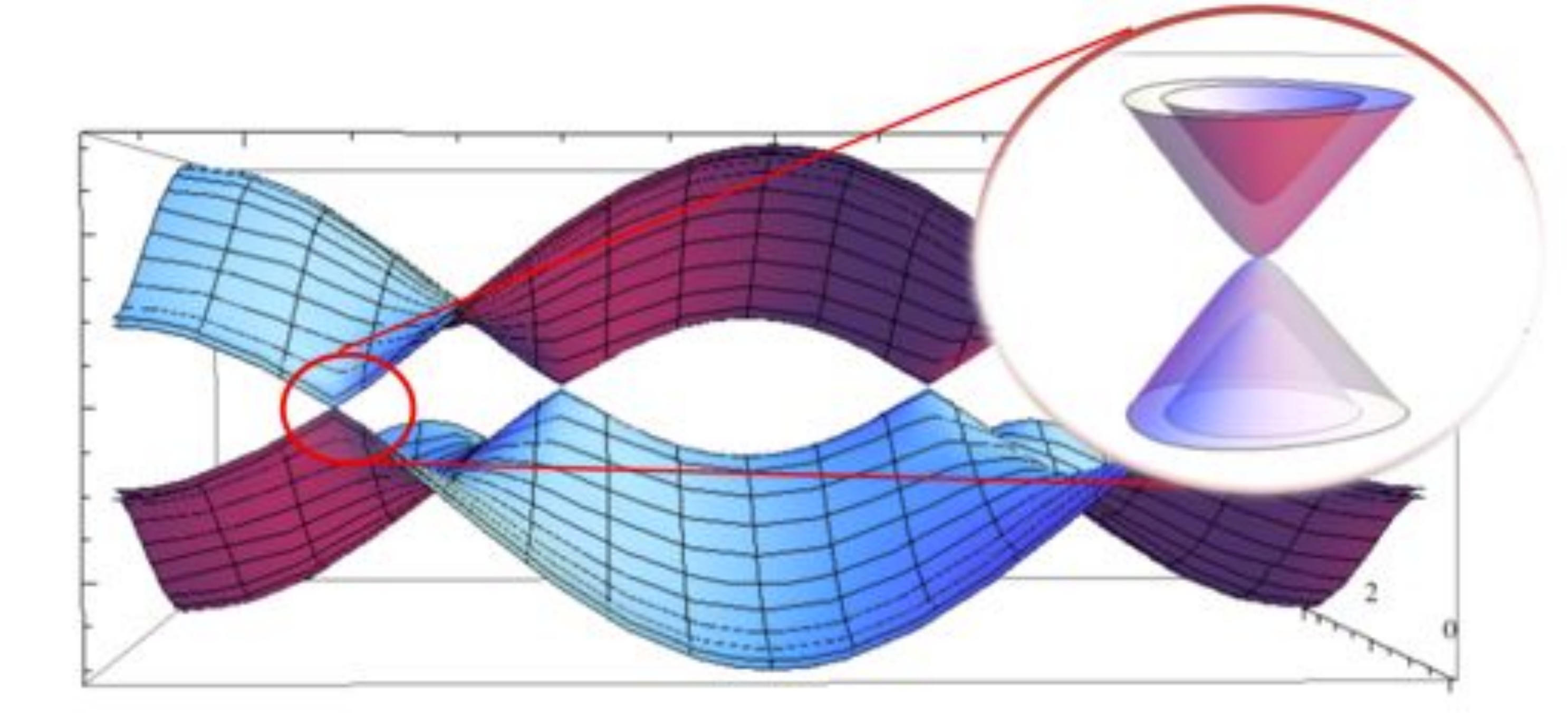} b) 
	\includegraphics[width=0.37
	\textwidth]{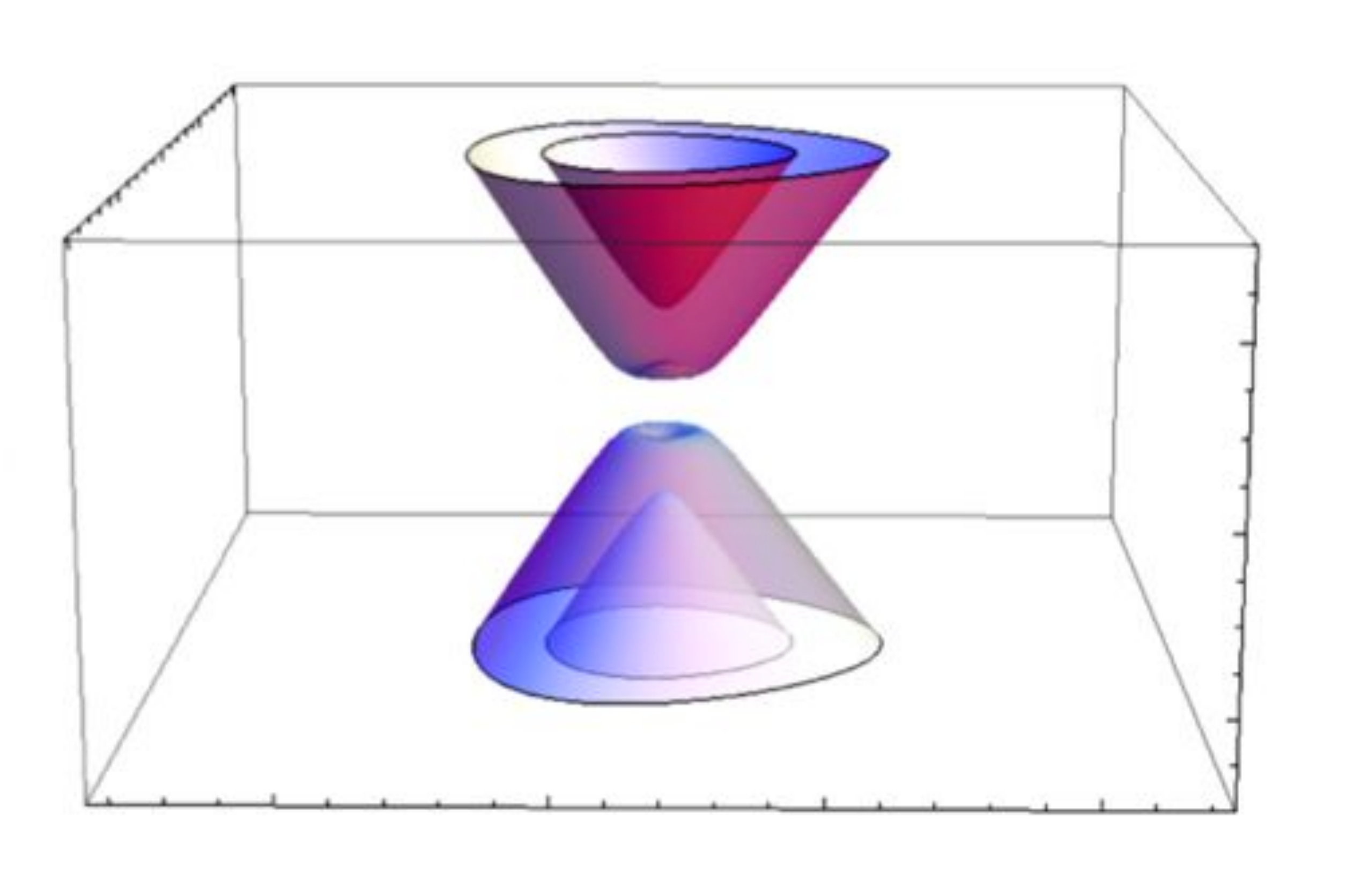}
	
	\caption{ (a) Energy spectrum of full minimal tight binding Hamltionian with $V_g=0$. Inset shows the dispersion about the $+\bK$ point. (b) Energy spectrum with $V_g=0.2t$ close to the $+\bK$ point.} \label{Fig:BBLG} 
\end{figure}

The zero bias spectrum of the full tight-binding model and the spectrum close about the $+\bK$ point is shown in Fig.~\ref{Fig:BBLG} (a).  The most apparent difference between the dispersion of bilayer and monolayer graphene is that instead of two linear Dirac cones at the $\bK$-points, the valence and conduction band touch quadratically.  Thus, while the Fermi surface of both single layer and bilayer graphene vanish at the charge neutrality point, the density of states remains finite for bilayer graphene. In addition to these two bands, there are two high energy bands separated by $ \pm t_{perp}$ from zero energy at the $\bK$-points.

When a bias between the two layers is present, a bandgap that is of the order of the interlayer bias opens up between the low energy conduction and valence band. Notice in the plot on the right panel of Fig.~\ref{Fig:BBLG} there is a `mexican-hat' feature present for larger interlayer biases.

\subsubsection{Low energy two-band model} Further intuition into the formation of a bandgap can be made by considering a low energy model generated by integrating out the two high energy bands.  This is obtained through a Schrieffer-Wolff transformation or using standard  2nd order perturbation theory.  In either case, the virtual hopping processes involving the higher energy bands generates an effective $2 \times 2$ Hamiltonian in terms of the low energy sites on each layer, $A_2$ and $B_1$. The resulting Hamiltonian is valid at energies smaller than $t_{perp}$ and is given by,
\begin{eqnarray}
	\ \label{Eqn:Effective} 
	\hat{H}_{\pm \bK \sigma}^{eff}\!\!=\!\!\tilde{\Psi}^\dg_{\pm \bK \sigma}(\bk) \left( 
	\begin{array}{cc}
		 V_g/2 -\mu & - (\pm k_x - i k_y)^2/2m^* \\
		-(\pm k_x + i k_y)^2/2m^* & - V_g/2 - \mu 
	\end{array}
	\right) \tilde{\Psi}^\pdg_{\pm \bK \sigma}(\bk) 
\end{eqnarray}
where $m^*\equiv t_{\perp}/(2 v_F^2)\sim 0.04 m_e$ is the effective mass of the quasiparticles, and the wavefunction now has two components given by $\tilde{\Psi}^\pdg_{\pm \bK \sigma}(\bk)=\Big(\psi_{A_2 \pm \bK \sigma}(\bk),\, \psi_{B_1 \pm \bK \sigma}(\bk) \Big)$.

It is extremely useful to rewrite this Hamiltonian as $H_{\pm \bK \sigma}^{eff}=\tilde{\Psi}^\dg_{\pm \bK \sigma}(\bk)\, {\bf \sigma} \cdot {\bf f}_{\pm \bK} (\bk) \, \tilde{\Psi}^\pdg_{\pm \bK \sigma}(\bk)$, where 
\begin{eqnarray}
{\bf f}_{\pm \bK} (\bk) = -\Big(\!\left(k_x^2-k_y^2\right)/2m^*,\, k_x k_y/2m^*,\,  V_g/2 \Big).
\end{eqnarray}
Here, ${\bf f}(\bk)$ can be thought of as a pseudo-Zeeman field acting on the pseudospins in momentum space.  In this form, many of the interesting properties of the Hamiltonian and solutions to the wavefunctions are much more transparent.
The two valley dispersion relations are again equal due to symmetry and are easily determined to be $\varepsilon(\bk)=\pm \sqrt{(V_g/2)^2 + (\bk^2/2m^*)^2} - \mu$.  Note that at this order, the effective Hamiltonian does not capture the `mexican-hat' feature in the electronic spectrum that is seen in the full tight-binding model. 

In terms of the eigenfunctions, the wavefunctions also have a two component structure similar to monolayer graphene, but with an important difference.  Consider first the system without a potential bias and the corresponding solutions to the wavefunction about the $+\bK$-point
\begin{eqnarray}
	\psi_{\pm} (\bk)=\frac{1}{\sqrt{2}}	\left( 
	\begin{array}{c}
		e^{-i\theta} \\
		\pm e^{i\theta}
	\end{array} \right), \label{sol2} 
\end{eqnarray}
where $\theta$ is still the angle of propagation given by $\cos(\theta)=k_x/|\bk|$ and the $+$($-$) sign corresponds to the positive (negative) energy state.  Notice the absence of the factor of $1/2$ in the argument.  This implies that the Berry phase associated with a single quadratic band touching point is $2\pi$.  The wavefunctions are also eigenstates of the operator $\sigma \cdot {\bf n}_2$, where ${\bf n}_2=\Big(\! \cos(2 \theta), \,\sin(2 \theta),\, 0\Big)$, meaning the pseudospin lies parallel or antiparallel to the unit vector ${\bf n}_2$, where ${\bf n}_2$ winds twice as an electron circumvolves about the band touching point.

Now consider the wavefunctions when there is an interlayer potential bias.  The potential bias breaks the $Z_2$ layer symmetry and can be though of as a mass generating term.  This can also be understood in terms of the pseudospin.  If you view the the vector field ${\bf f}(\bk)$ as momentum dependent pseudo-Zeeman field acting on the pseudospin,  the field picks up a z-component when an interlayer bias is present.  The result is that that the pseudospin lifts off the XY plane and becomes canted, implying the crystal is layer polarized.  At the $\bK$-point, the pseudo-Zeeman field points entirely in the z-direction and the pseudospin is fully polarized along the z-axis.  This gives two states that are `Zeeman split' by an amount $\Delta$, corresponding to the bandgap.

\part{Graphene Under Spatially Varying External Potentials}\label{Part:Kink} 

\chapter{Preliminaries: 1D Topologically Confined Kink States} \label{Chapt:Kink} 
\subsubsection{The material in the section is largely based on the article M.\ Killi, T.-C.\ Wei, I.\ Affleck and A.\ Paramekanti  {\textit{Phys.~Rev.~Lett.} {\bf104}, 216406 (2010)}.}

There are a few fundamental concepts regarding bilayer graphene that will be heavily relied upon in Part \ref{Part:Kink} of the thesis. The aim of this preliminary chapter is to briefly outline some of the pertinent physics of bilayer graphene in order to establish a firm foundation in which later chapters build upon. Here, we review the \textit{non-interacting} properties of the 1D kink modes that arise in non-uniformly biased bilayer graphene.  An understanding of the basic properties of these modes is essential for Chapter \ref{Chapt:TTL}, as it involves an extensive study of interactions between between electrons in these kink states. Later, in Chapters \ref{Chapt:SL} and \ref{Chapt:BSL} we will again find that a sound understanding of the properties of these 1D modes is imperative to developing a deep and intuitive understanding of the band structure of 1D electric field superlattices.

\section{The properties of non-interacting kink states} \label{Sect:Kink}
\begin{figure}
	[tb]
	
	\centering a) \, 
	\includegraphics[width=0.4
	\textwidth]{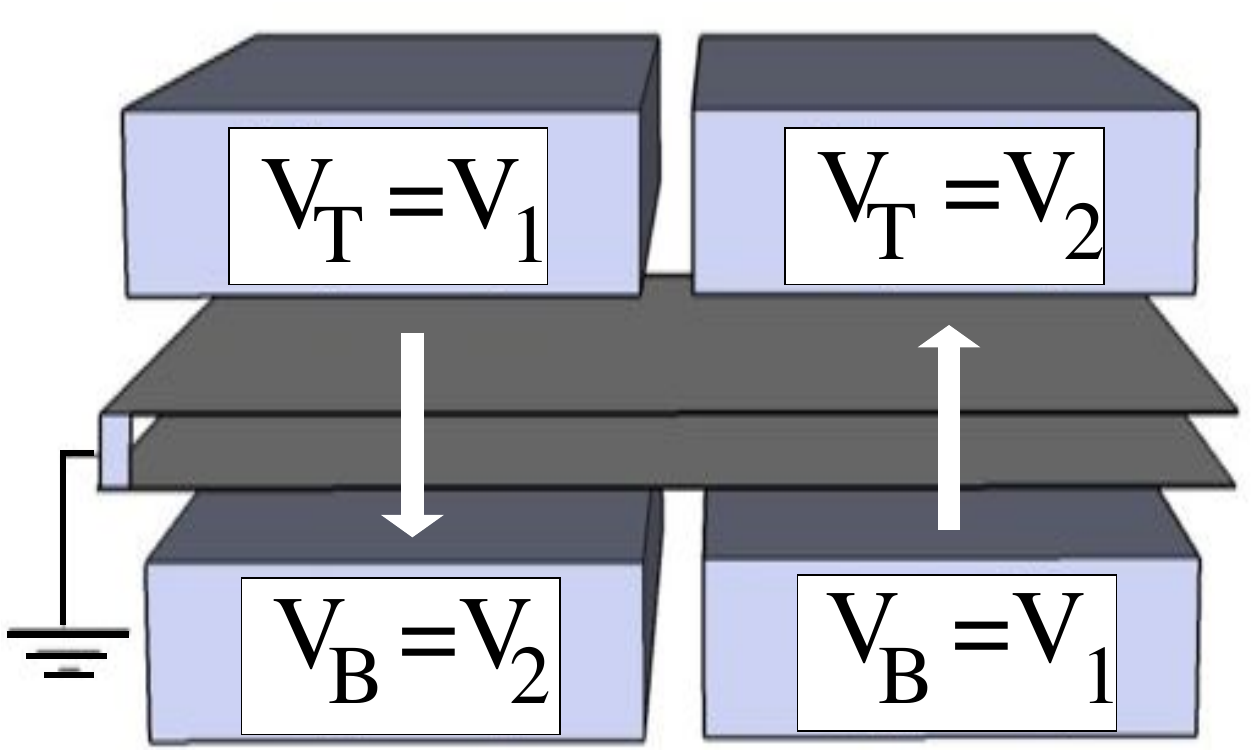} b) \, 
	\includegraphics[width=0.4
	\textwidth]{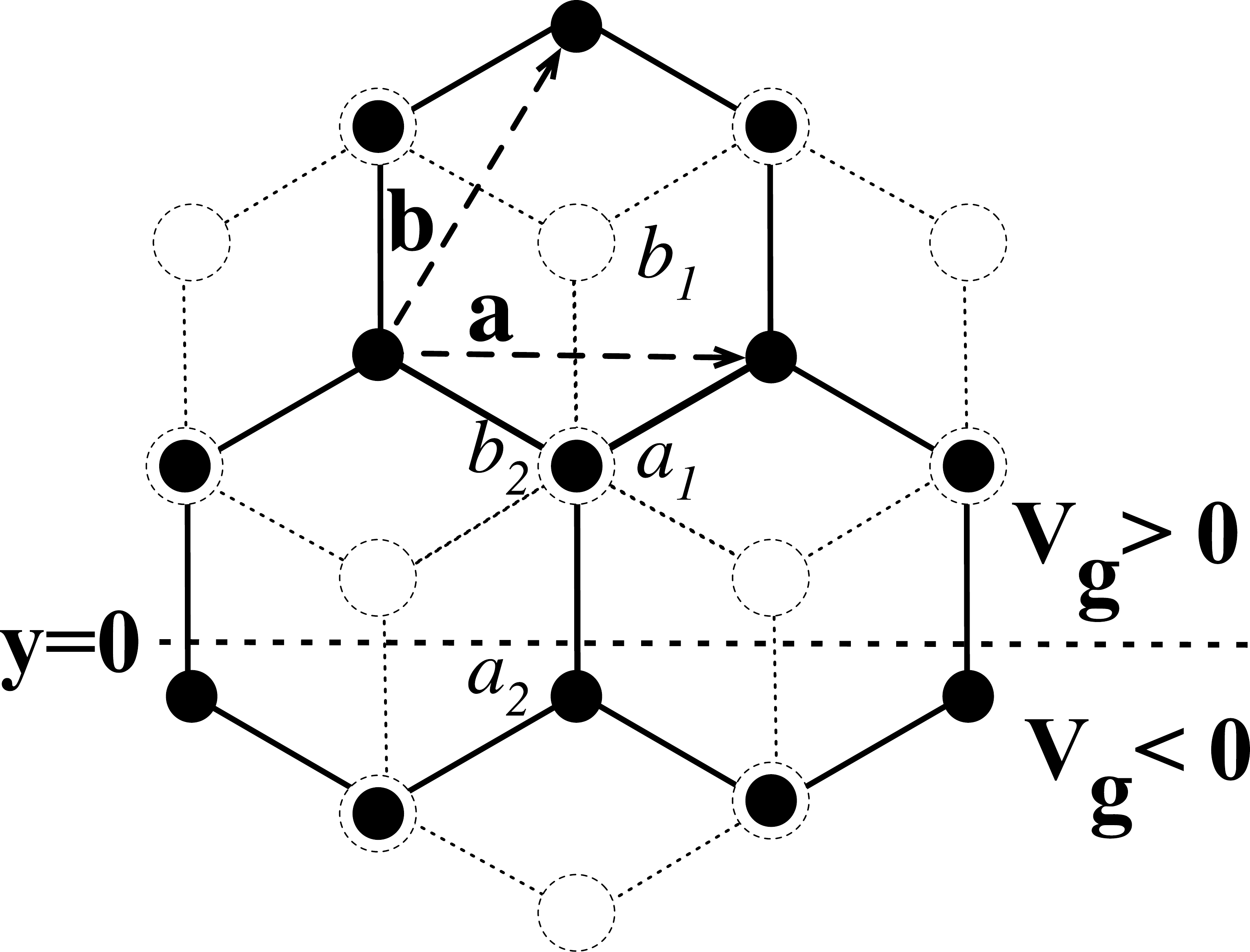}
	
	\caption{ (a) Schematic diagram of a pair of external dual-gates that generate the interlayer bias to reversal along the interface between them. (b) Structure of bilayer graphene with a `bias kink' in $V_g=V_{\rm T}-V_{\rm B}$ at $y=0$.} \label{Fig:Schematic} 
\end{figure}

Having covered the homogeneously biased system, we now consider the non-homogenous biased system originally proposed by \citet*{Martin:2008}, where two neighbouring regions of bilayer graphene are biased with opposite parity, as shown schematically in Fig.\ \ref{Fig:Schematic} (a). In this geometry, gapless 1D dimensional modes are expexted to emerge at the interface where the biase reverses sign (see also \citet{Yao:2009}). 

In order to describe the effect of the external gates as depicted in Fig.~\ref{Fig:Schematic}, the potential term is written as, 
\begin{eqnarray}
	-\frac{1}{2} \sum_{\vR,\sigma} \left(-1\right)^{\ell}V_g (y_{n,s,\ell}) \, \hat{n}_{\vR,\sigma}, 
\end{eqnarray}
where $\hat{n}_{\vR,\sigma}$ is the electron number operator for spin-$\sigma$ at the site labeled by the set of indices $\vR=(m,n,s,\ell)$. Here, we have assumed that the bias $V_g(y)$ depends only on the $y$-coordinate, which is determined by $(n,s,\ell)$, and is $m$-independent so that translational symmetry is preserved along $\va$. We assume $V_T+V_B$ is chosen to fix the bilayer chemical potential at the mid-gap. For later convenience, we set $\vR=(m,\vr)$, where $\vr\equiv(n,s,\ell)$. For simplicity, we assume the gates induce a potential with a step profile $V_g(y)= {\rm sign}(y) \,V_g $ since the dominant effect of a potential can be shown to come from the strength of the bias and not the details of its spatial profile.\footnote{For a bias kink much broader than $ v_f/ \sqrt{V_g t_\perp}$ additional (non-topological) modes may appear in the gap \citep{Martin:2008}} 

Before we explicitly solve the Hamiltonian, a heuristic understanding of the solution can be made. For a general potential profile with $V_g(y>0)=-V_g(y<0)$ and $V_g( y\to\pm\infty) = \pm V_g$, the bulk region far from $y=0$ has a gap $\Delta \approx V_g$. On both sides of the interface, there is a mass generating term proportional to $\sigma_z$, however, they have \textit{opposite signs}. Hence, it is expected that zero energy 1D `kink' modes exist that are confined to the interface but are free to propagate along it. One may question this overly simplistic picture, and rightfully so. Strictly speaking, the total Berry curvature of the Hamiltonian --- that is one that extends around the entire Brillioun zone, encompassing both the $\pm \bK$ points --- has a trivial topological charge of zero. It is only when each valley is taken as independent can the system be thought of as having topological character analogous to that which is present in the quantum Hall effect. Remarkably, as discussed at the end of this chapter, even in the absence of rigorous topological protection, these 1D conducting channels are remarkably robust to disorder, orientation and other potentially parasitic effects.

\subsection{Soliton wavefunctions and mid-gap dispersion} \label{Sect:Kinkchar} 
\begin{figure}
	[h] \centering a) 
	\includegraphics[height=5.cm]{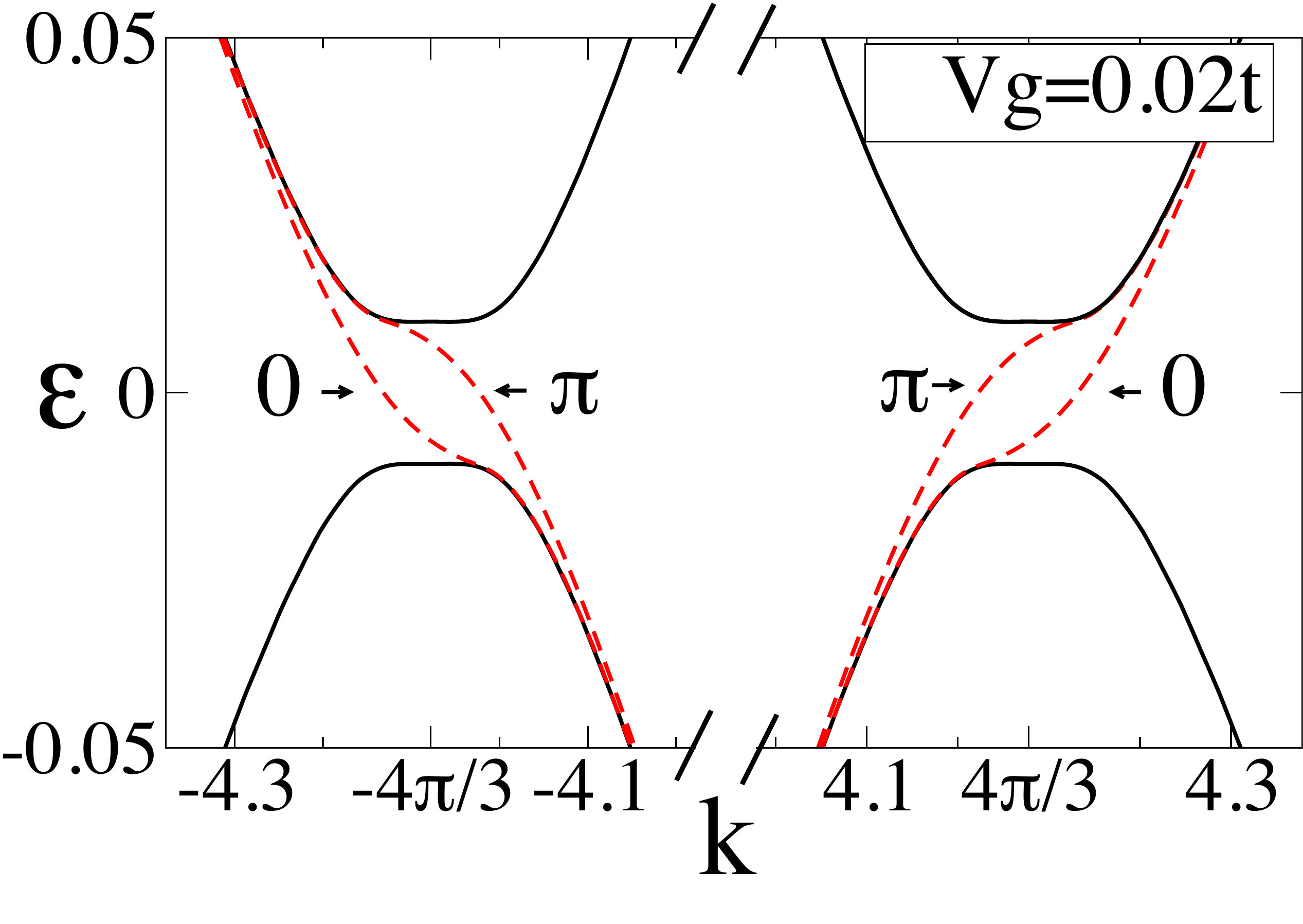} \hspace{.2cm} b)
	\includegraphics[height=5.cm]{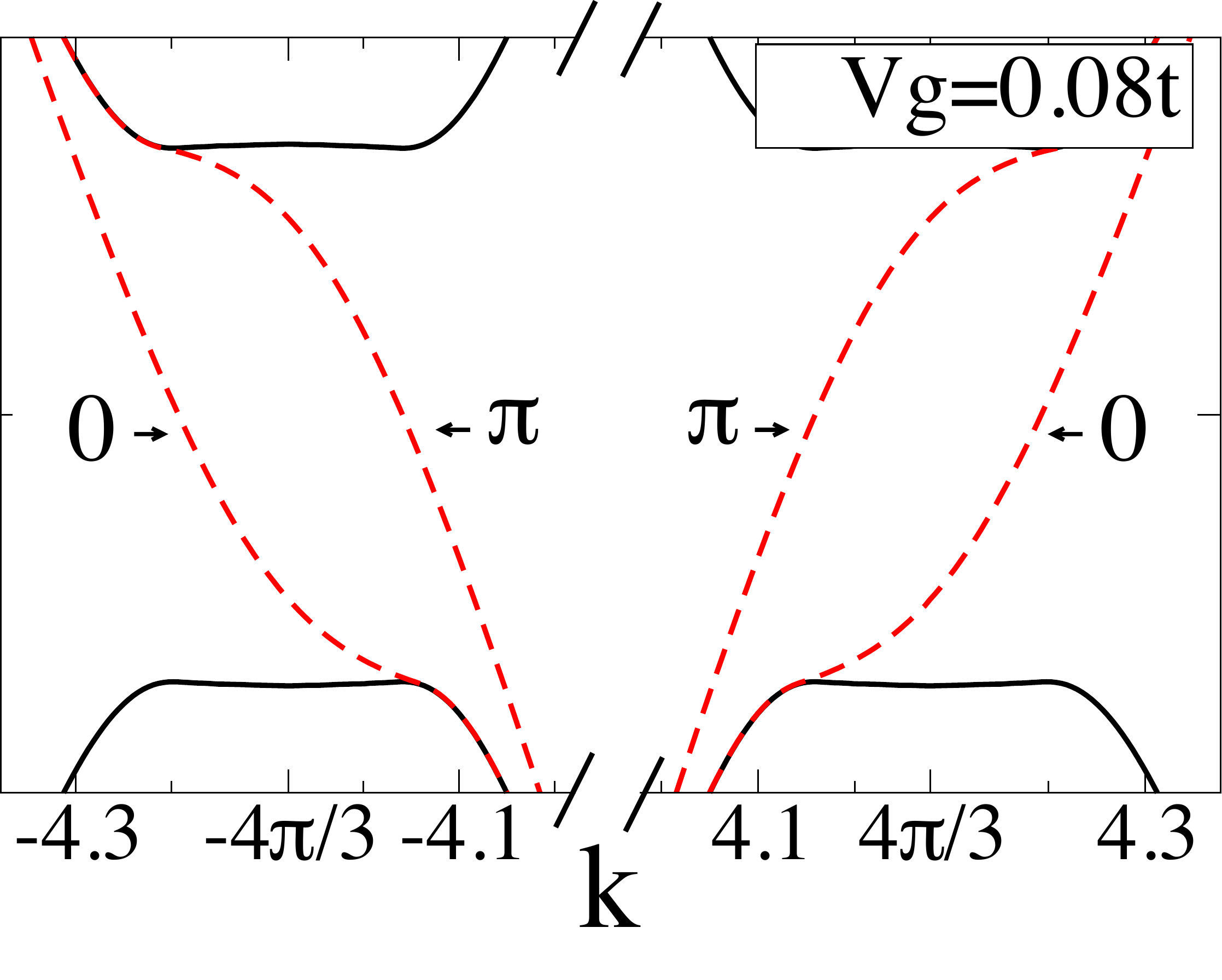}
	\caption{Dispersion about the K-points with (a) $V_{g}=0.02t$ and (b) $V_{g}=0.08t$. Edge-mode bands are indicated by the labelled arrows and bulk-states by the hatched region.} \label{Fig:Dispersions}
	
\end{figure}

Turning back to the full Hamiltonian, it can be solved by Fourier transforming along the direction parallel to the interface, with momentum labelled $k$, and then numerically diagonalizing the matrix for each $k$. Numerical results for various experimentally viable gate biases \citep{Zhang:2009} are shown in Fig.~\ref{Fig:Dispersions}.  The spectrum shows two right (left) movers for each spin at the $\bK$ (-$\bK$) point, consistent with the findings of \citet{Martin:2008}. Drawing an analogy with two leg Hubbard ladders \citep{Fabrizio:1992,Balents:1996,Chudzinski:2008}, we refer to the high/low energy bands as the $\pi$/$0$-bands. 
As seen qualitatively from each figure, and by comparing Fig.~\ref{Fig:Dispersions} (a) with (b), the Fermi velocity of the two bands are equal and they both change significantly with the bias voltage. 

\begin{figure}
	[tb] \centering a) 
	\includegraphics[height=4.9cm]{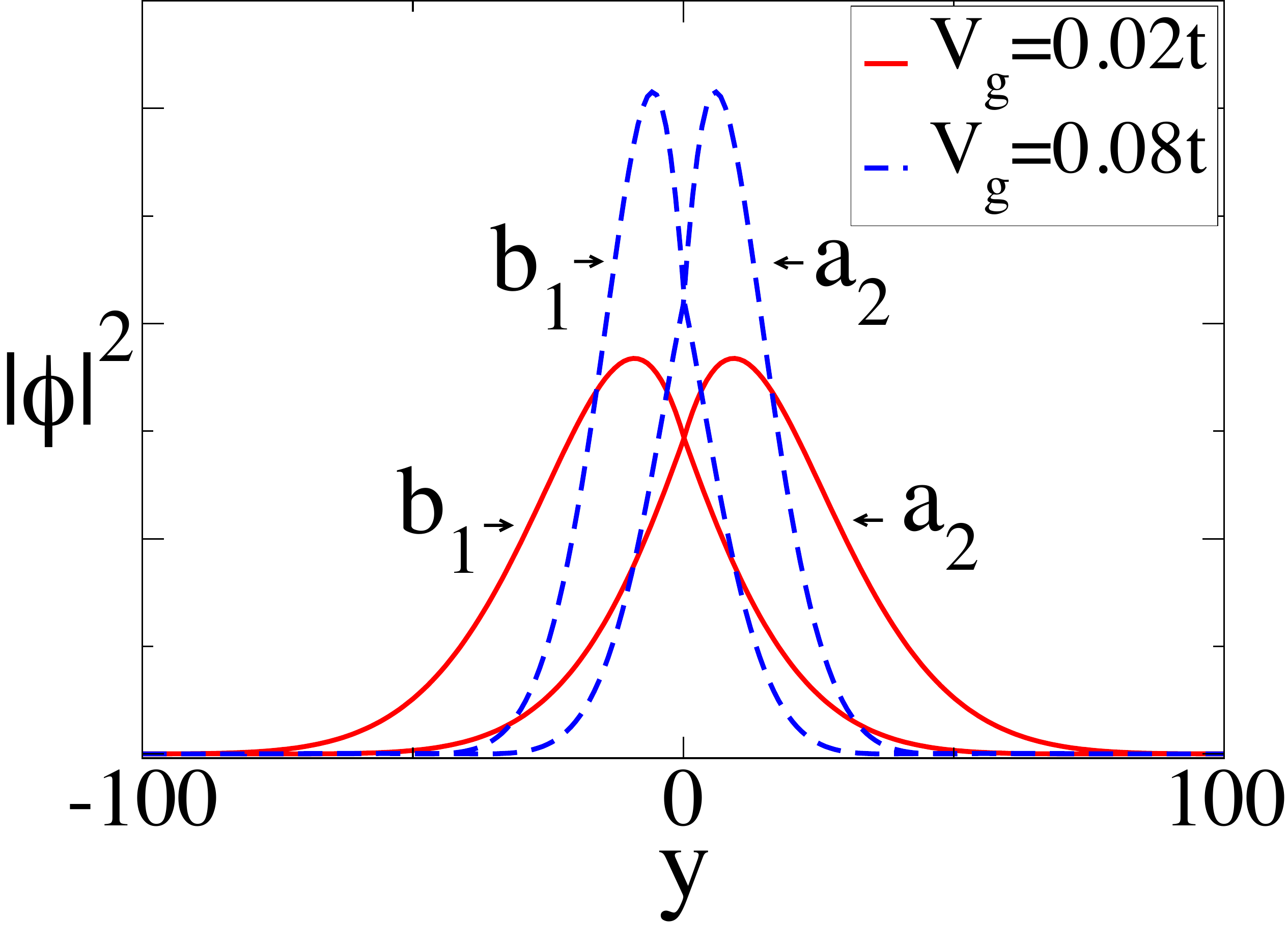} \hspace{0.2cm} b) 
	\includegraphics[height=4.9cm]{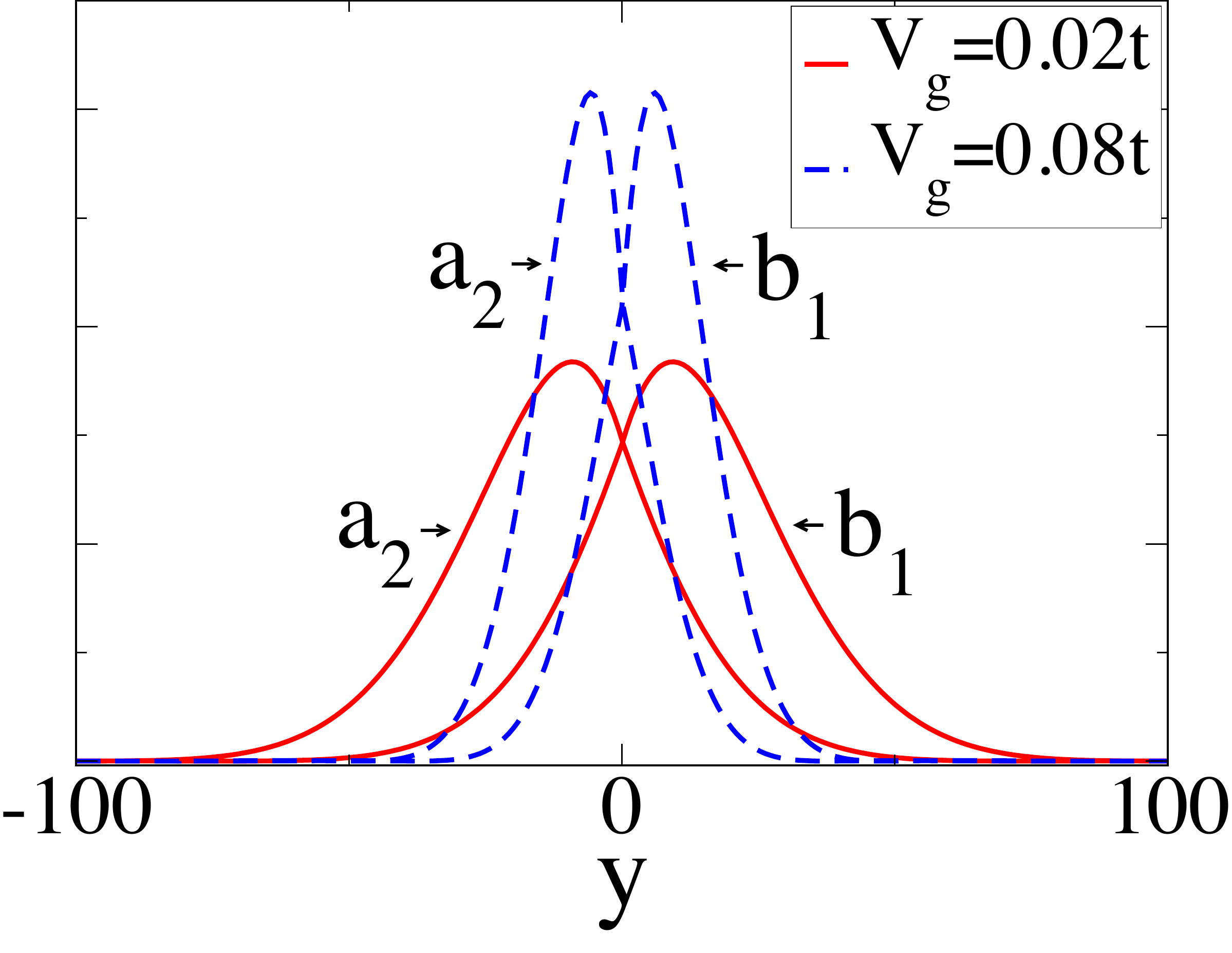} \caption{The modulus square of the zero-energy wavefunction of (a) the $0$-band and (b) the $\pi$-band on the dominate sites $a_2$ and $b_1$ at $V_g=0.02t$ and $V_g=0.08t$ (independent of being a right or left mover).} \label{Fig:Spread} 
\end{figure}

Figure \ref{Fig:Spread} shows the plots of the modulus square of the zero-energy wavefunctions in the $0$-band and $\pi$-band at high and low gate voltage. We see explicitly from these figures that the wavefunctions are localized to the interface and that the spread of the wavefunction transverse to the interface direction also varies significantly with gate bias. With respect to the overall width of the wavefunction, a simple scaling analysis suggests that the wavefunction width should go as
\begin{eqnarray}
 l\sim (m^* V_g)^{-1} \sim \left(\frac{t}{\sqrt{V_g t_{\perp}}}\right)a.
\end{eqnarray}

Note that the zero-energy wavefunction of the $\pi$-band is related by interchanging the layers and reflecting about the interface so that wavefunction `leans' into the region of positive potential instead. Both this fact and the equality of the Fermi velocities of the two bands ($V_{F}^{0/\pi}=V_{F}$), are a consequence of the symmetry that relates the two bands by an inversion about the $K$ (-$K$) point at low energies \citep{Martin:2008}.

\chapter{Tunable Tomonaga-Luttinger Liquid in Bilayer Graphene} \label{Chapt:TTL}
\subsubsection{The material in this section is largely based on the article M.\ Killi, T.-C.\ Wei, I.\ Affleck and A.\ Paramekanti \textit{Phys.~Rev.~Lett.} {\bf104}, 216406 (2010).}

There are a few specific questions concerning electron interaction effects in kink states that we address in this chapter.  First, \emph{do electron interactions induce strong backscattering processes that suppress the ability to transport excitations in the system}?  The answer to this question is essential to properly evaluate the potential functionality of kink states in electronic devices and valleytronic circuits.  Second, although interactions are known to drive 1D systems into a Tomonaga-Luttinger liquid phase, \emph{what is the precise nature of this Tomonaga-Luttinger liquid state?}  The properties of the Tomonaga-Luttinger liquid phase or more specifically, the extent to which the velocity and Luttinger parameter are renormalization, are expected to strongly depend on the various details and nuances of this unique system. Moreover, given the fact that these 1D conducting states are not topologically protected, interactions could also gap-out various modes in the system.  The final question we are concerned with is \emph{to what extent does variation of the interlayer bias strength affect the properties of the kink state when interactions are included}?  

To explore these questions, we consider the kink states within the context of the original setup proposed by \citet{Martin:2008}, as discussed in the previous chapter. We find that the low energy wavefunctions of the 1D modes have a broad spread in the direction transverse to the interface --- this leads to the dominance of the forward scattering part of the Coulomb interaction between electrons, in a manner akin to large radius carbon nanotubes \citep{Kane:1997}. Within an abelian bosonization framework, the forward scattering processes are shown to drive the system into a strongly interacting Tomonaga-Luttinger liquid phase \citep{Giamarchi:2004}.
\begin{figure}
	[tb]
	
	\centering a) \, 
	\includegraphics[width=0.4
	\textwidth]{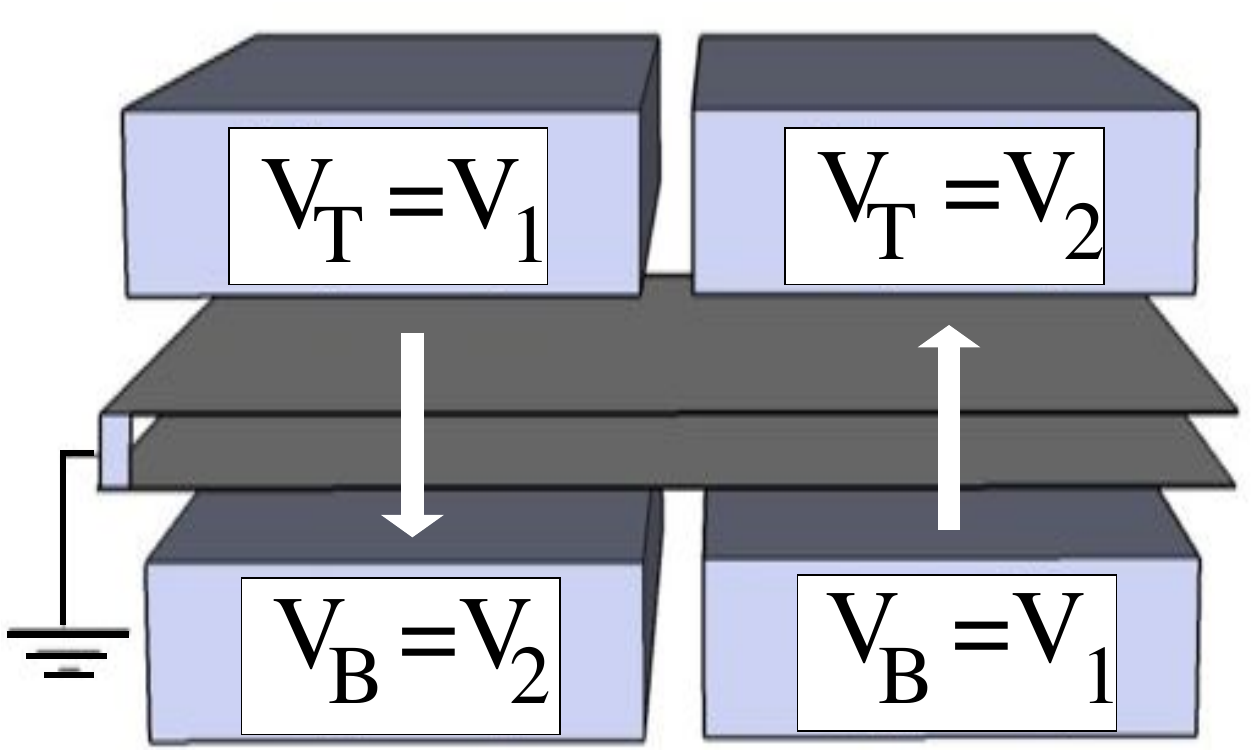} b) \, 
	\includegraphics[width=0.4
	\textwidth]{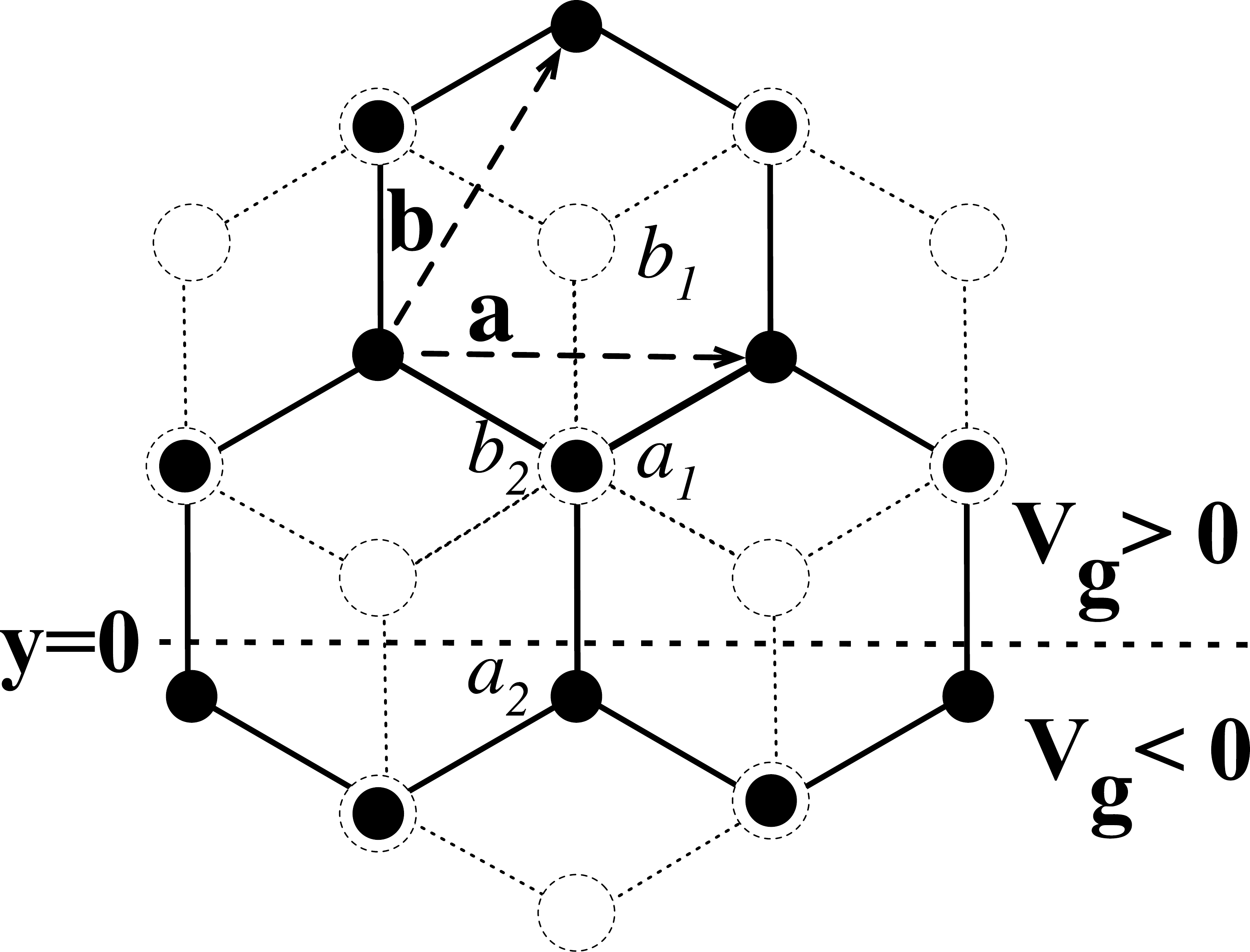}
	
	\caption{ (a) Schematic diagram of external gates. $V_{\rm T}+V_{\rm B}$ controls the bilayer doping while $V_{\rm T}-V_{\rm B}$ controls the gap via the depicted perpendicular electric field. (b) Structure of bilayer graphene with a `bias kink' in $V_g=V_{\rm T}-V_{\rm B}$ at $y=0$.} \label{schematic} 
\end{figure}
Remarkably, we find the Luttinger parameter of this liquid is tunable by adjusting the gate potential. This results from two competing effects: (i) An increased bias causes further confinement of the wavefunctions to the interface, enhancing the effect of interactions; (ii) An increase in the bias increases the Fermi velocity of the low-energy modes, suppressing the effect of interactions relative to the kinetic energy. The net result is that the Luttinger parameter in the total charge channel, $K_{c +}$, can be varied between $0.15$-$0.2$ by increasing the bias over an experimentally accessible range. At the same time the Luttinger parameter in the transverse charge channel, $K_{c-} \approx 0.63$, is relatively independent of the bias. We thus show that gated bilayer can realize a tunable Tomonaga-Luttinger liquid. Such band structure and wavefunction tuning of Luttinger liquids has been suggested in a few other systems recently --- in cold atomic gases \citep{Zhai:2005, Moritz:2005}, in magnetic waveguides in graphene \citep{Hausler:2008}, in carbon nanotubes in crossed electric and magnetic fields \citep{DeGottardi:2009,DeGottardi:2010}, and in gated topological insulators \citep{Yokoyama:2010}. Such a Luttinger liquid with dominant forward scattering is also expected to arise at charge density wave domain walls in single layer graphene, where the charge density wave involves a weak sublattice density modulation induced by an appropriate substrate \citep{Semenoff:2008}, in non-uniformly biased multilayer graphene with inversion symmetry, and along domain walls present in the spontaneously polarized state thought to persist at low temperatures in pristine bilayer graphene. 

\section{Effective 1D Hamiltonian} To derive the effective low-energy 1D hamiltonian, we assume a suitable energy cutoff that is smaller than the bulk gap and focus on those single particle states that lie within this energy window and are confined to the 1D interface region. To do this, we first expand the local field operators, $\hat{\Psi}_{\vR,\sigma}$,  in the complete basis, \begin{eqnarray}
\hat{\Psi}_{\vR,\sigma} = \frac{1}{\sqrt{L}} \sum_{k, \alpha} e^{i k m d}\varphi^{\alpha}_{k}(\vr) \, \hat{\psi}^{\alpha}_{k \sigma},
\end{eqnarray}
where $\varphi^{\alpha}_{k}(\vr)$ is the wavefunction of the state in band-$\alpha$ with momentum $k$. We then restrict the bands to the set $\alpha=\{0,\pi\}$ and consider only momenta in the vicinity of the four Fermi points, $\pm k^{\pi}_{F}$ and $\pm k^{0}_{F}$. An additional simplification is made by neglecting the small momentum dependence of the wavefunctions since $\varphi^{\alpha}_{\pm k^{\alpha}_{F}+q}(\vr) \approx \varphi^{\alpha}_{R/L}(\vr)$ for small momenta $q$, where $\varphi^{\alpha}_{R/L}\!(\vr)$ is the zero-energy wavefunction at $\pm k_{F}^{\alpha}$. In doing so, the $\vr$-dependence of the wavefunction can be separated to yield the low energy field operators projected to the 1D subspace via 
\begin{equation}
	\label{field} \hat{\Psi}_{\vR,\sigma} \approx \sum_{r=\pm, \alpha=\{0,\pi\}} \varphi^{\alpha}_{r}(\vr) \, e^{i r k^{\alpha}_{F} x}\, \hat{\psi}^{\alpha}_{r \sigma} (x), 
\end{equation}
where $r$ is the label R/L for left/right movers and takes the values $+/-$ in the expression, and $\psi^{\alpha}_{r \sigma}(x)$ are slowly varying field operators exclusively dependent on the position along the interface (which we now denote by the continuous variable $x=m d$).

\begin{figure}
	[t] \centering
	
	\includegraphics[width=4.5cm]{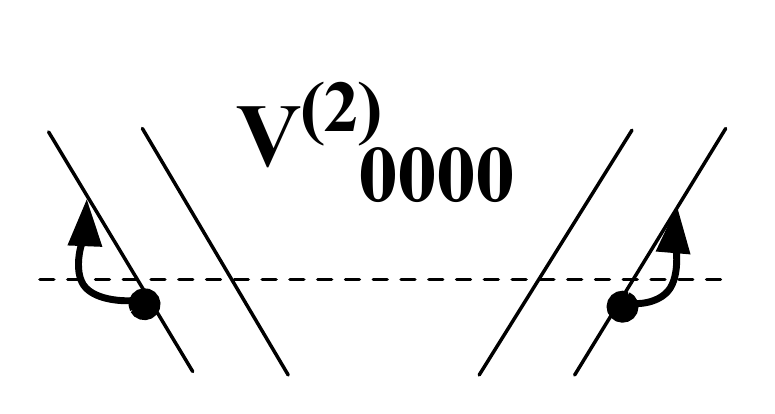}\hspace{0cm} 
	\includegraphics[width=4.5cm]{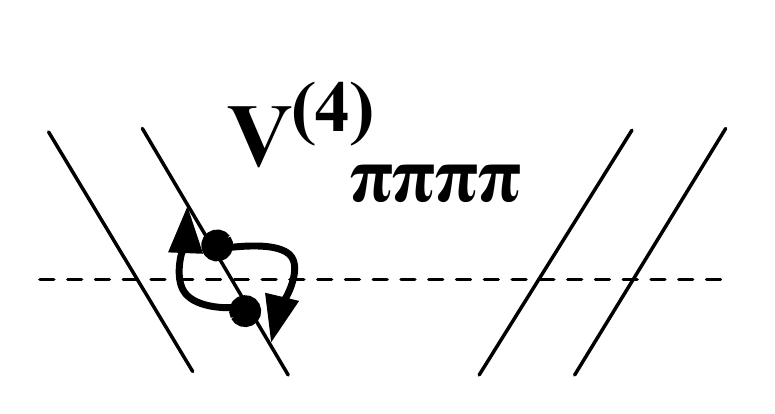}\hspace{0cm} 
	\includegraphics[width=4.5cm]{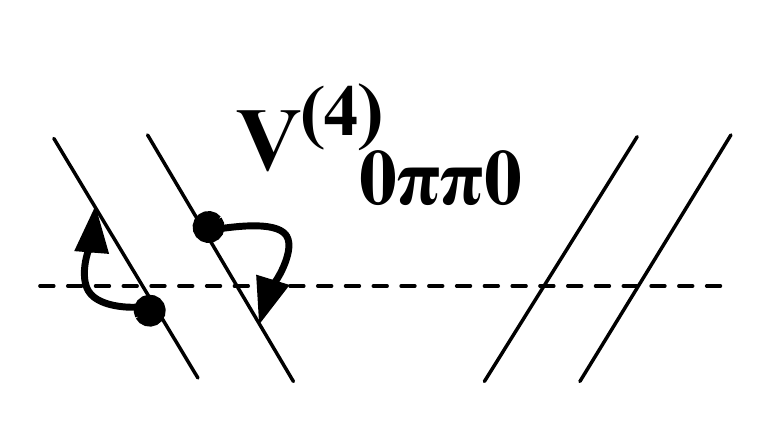}\\
	\vspace{-.2cm} 
	\includegraphics[width=4.5cm]{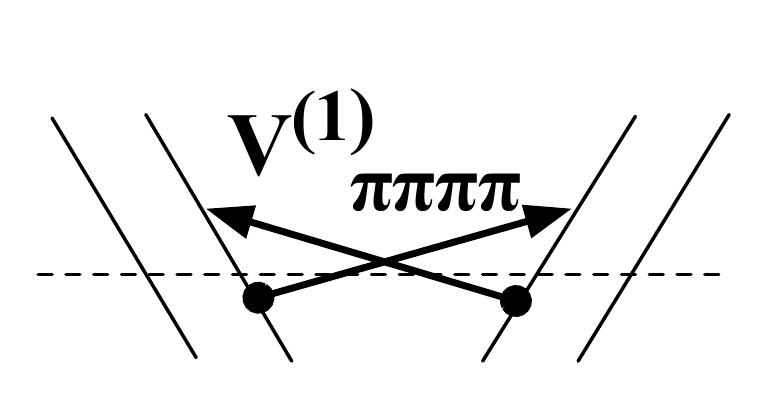}\hspace{0cm} 
	\includegraphics[width=4.5cm]{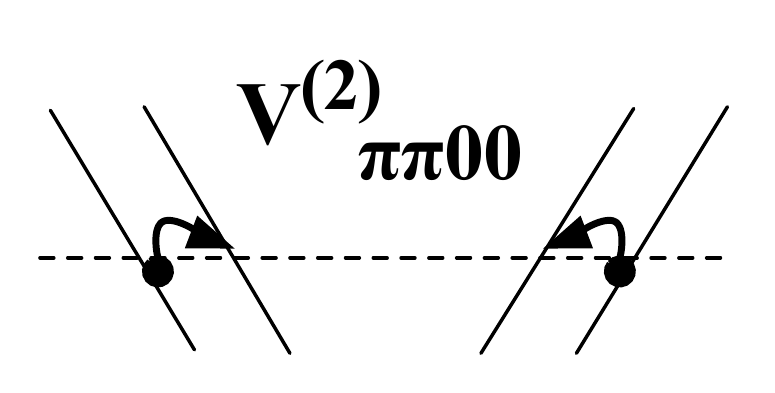}\hspace{0cm} 
	\includegraphics[width=4.5cm]{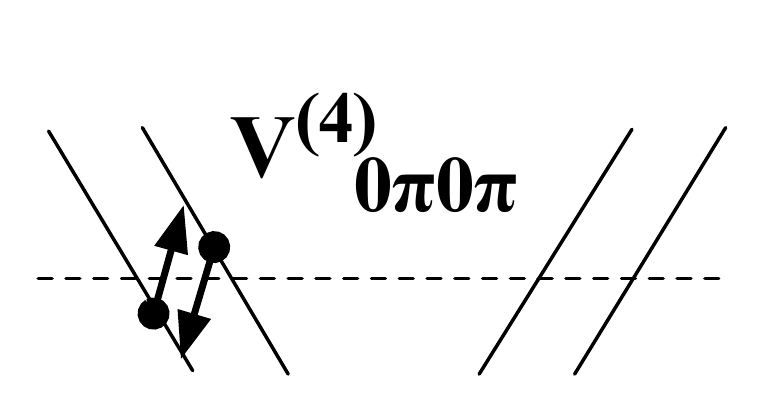}
	
	\caption{\label{scattering}Examples of various scattering processes.}
	
\end{figure}

We now rewrite the entire Hamiltonian in terms of operators in the reduced 1D subspace. The free part is simply linearized to give
 \begin{eqnarray}
	\sum_{|q|<\Lambda} \sum_{r \alpha \sigma} rq V_{F} \hat{\psi}^{\alpha\dag}_{r \sigma} (q) \hat{\psi}^\alpha_{r \sigma} (q).
\end{eqnarray}
The effective interaction between fermions in the 1D channel is obtained by a straightforward substitution of Eqn.\ (\ref{field}) into the Coulomb term, \begin{eqnarray}
\frac{1}{2}\sum_{\sigma \sigma'} \sum_{\mathbf{R R'}} \Psi^{\dag}_{\vR,\sigma} \Psi^{\dag}_{\vR',\sigma'} U(\vR,\vR') \Psi_{\vR',\sigma '}\Psi_{\vR,\sigma},
\end{eqnarray}
followed by a summation over $\vr$. This gives rise to various scattering terms, many of which are rapidly oscillating and can be dropped. The effective Hamiltonian obtained contains many terms of the general form 
\begin{eqnarray}
	\label{effective} \frac{V^{(i)}_{\alpha \beta \gamma \delta}}{2}\sum_{x} \hat{\psi}^{\alpha \dag}_{r_{1} \sigma} \! (x) \hat{\psi}^{\beta \dag}_{r_{2} \sigma'} \! (x) \hat{\psi}^{\gamma}_{r_{3} \sigma'} \!(x) \hat{\psi}^{\delta}_{r_{4} \sigma} \!(x). 
\end{eqnarray}
Here, $V^{(i)}_{\alpha \beta \gamma \delta} \equiv V^{(i)}_{\alpha \beta \gamma \delta} (r_{1} k_{F}^{\alpha}- r_{4}k_{F}^{\delta}\!)$ is the Fourier component of the effective 1D potential 
\begin{eqnarray}
	\tilde{V}^{(i)}_{\alpha \beta \gamma \delta}(x\!- \! x') \!=\! \! \sum_{\vr, \vr'} \! U\!(\vR,\vR') {\varphi^{\alpha}_{r_{1}}}^{\!*} \!(\vr) {\varphi^{\beta}_{r_{2}}}^{\!*} \! (\vr') {\varphi^{\gamma}_{r_{3}}} \! (\vr') {\varphi^{\delta}_{r_{4}}} \! (\vr). 
\end{eqnarray}
The effective interaction Hamiltonian contains all terms of the form in Eqn.\ (\ref{effective}) that have a combination of $R/L$ and band indices that conserve (crystal) momentum. The index $i$ classifies the scattering processes using standard g-ology notation \citep{Haldane:1981,Penc:1990,Sorella:1992,Fabrizio:1992,Balents:1996,Chudzinski:2008} (see Fig.\ \ref{scattering}): $i=1$ refers to backscattering, $i=2$ to forward-scattering involving both right and left movers, and $i=4$ to forward-scattering involving only right or only left movers. The number of distinct processes is greatly reduced by the fact that all interband scattering terms with parallel spin merely renormalize the coefficients of a corresponding intraband term with parallel spin. 

This Hamiltonian is qualitatively similar to that obtained for Hubbard ladders \citep{Fabrizio:1992,Balents:1996,Chudzinski:2008} with two significant differences. First, both the wavefunctions and $V_{F}$ are sensitive to changes in the applied gate bias. The modified wavefunctions alter the distance dependence of the effective Coulomb interaction, while the change in $V_{F}$ adjusts the relative interaction strength parameterized by the fine-structure constant $\alpha \rightarrow \alpha \frac{c}{V_{F}}$. Second, we note that the long-range nature of the Coulomb interactions, together with the large spread of the low energy wavefunctions, causes the small momentum forward scattering processes to dominate. This is reminiscent of large radius single wall carbon nanotubes, where the extension of the wavefunctions around the tube radius suppresses the bare backscattering \citep{Kane:1997}. We have checked that the bare values of these backscattering and interband scattering terms are very small, consistent with this argument. For instance, $V^{(1)}_{0000}/V^{(2)}_{0000} \sim 10^{-3}$ and $V^{(4)}_{0\pi0\pi}/V^{(2)}_{0000} \sim 10^{-2} $ at $V_g=0.02t$, so that such processes are expected to be important only at extremely low energy and temperature. For the remainder of the chapter, we focus on the low energy regime where backscattering can be ignored.

\section{Bosonization} Using the standard abelian bosonization procedure \citep{Giamarchi:2004}, we introduce the bosonic field $\hat{\phi}_{\alpha \sigma}(x)$ and the phase $\hat{\theta}_{\alpha \sigma}(x)$ whose spatial derivative $
\partial_{x} \hat{\theta}_{\alpha \sigma} = \hat{\Pi}_{\alpha \sigma}$ is conjugate to $\hat{\phi}_{\alpha \sigma}(x)$. Fermion operators can be represented in terms of these boson fields via 
\begin{eqnarray}
	\hat{\psi}^{\alpha}_{r \sigma}(x) \sim e^{i \left( r \hat{\phi}_{\alpha \sigma}(x) - \hat{\theta}_{\alpha \sigma}(x) \right)}.
\end{eqnarray} 
It is a simple matter to rewrite the density-density interactions in the boson representation using the two relations,
\begin{eqnarray}
	\partial_{x} \hat{\phi}_{\alpha \sigma} &=& - \pi \left( \hat{\rho}_{R \alpha \sigma} +
	\hat{\rho}_{L \alpha \sigma} \right) \\
	\partial_{x} \hat{\theta}_{\alpha \sigma} &=& \pi \left( \hat{\rho}_{R \alpha \sigma} - \hat{\rho}_{L \alpha \sigma} \right).
\end{eqnarray}
 In addition, the symmetry between the bands allows us to lighten our notation by defining $V_{A}\!\equiv \!V^{(2)}_{\alpha \alpha \alpha \alpha}\!=\!V^{(4)}_{\alpha \alpha \alpha \alpha}$ and $V_B \! \equiv \! V^{(2)}_{\alpha \bar{\alpha} \bar{\alpha} \alpha} \!=\!V^{(4)}_{\alpha \bar{\alpha} \bar{\alpha} \alpha}$.

This leads to the Hamiltonian, 
\begin{eqnarray}
	H_{1}= \frac{1}{2 \pi} \int \! dx (
	\partial_{x} \Phi)^{T} \hat{u} \cdot \hat{K}^{-1} (
	\partial_{x} \Phi) + (
	\partial_{x} \Theta)^{T} \hat{u} \cdot \hat{K} (
	\partial_{x} \Theta), 
\end{eqnarray}
with 
\begin{eqnarray}
	\hat{u} \cdot \hat{K}^{-1} &=& V_{F} \, \mathbbm{1} +\frac{V_{F}}{2 \pi} \left(
	\begin{array}{cccc}
		g_{A} & g_{B} & g_{A} & g_{B}\\
		g_{B} & g_{A} & g_{B} & g_{A}\\
		g_{A} & g_{B} & g_{A} & g_{B}\\
		g_{B} & g_{A} & g_{B} & g_{A} 
	\end{array}
	\right) \\
	\hat{u} \cdot \hat{K} &=& V_{F} \, \mathbbm{1}. 
\end{eqnarray}
Here $g_{A/B} \equiv (2V_{A/B})/ V_{F}$, and $\Phi=(\phi_{0 \uparrow}, \phi_{\pi \uparrow}, \phi_{0 \downarrow}, \phi_{\pi \downarrow})^{T}$ with a similar definition for $\Theta$.

\begin{figure}
	[t] \centering 
	\includegraphics[height=5.9cm]{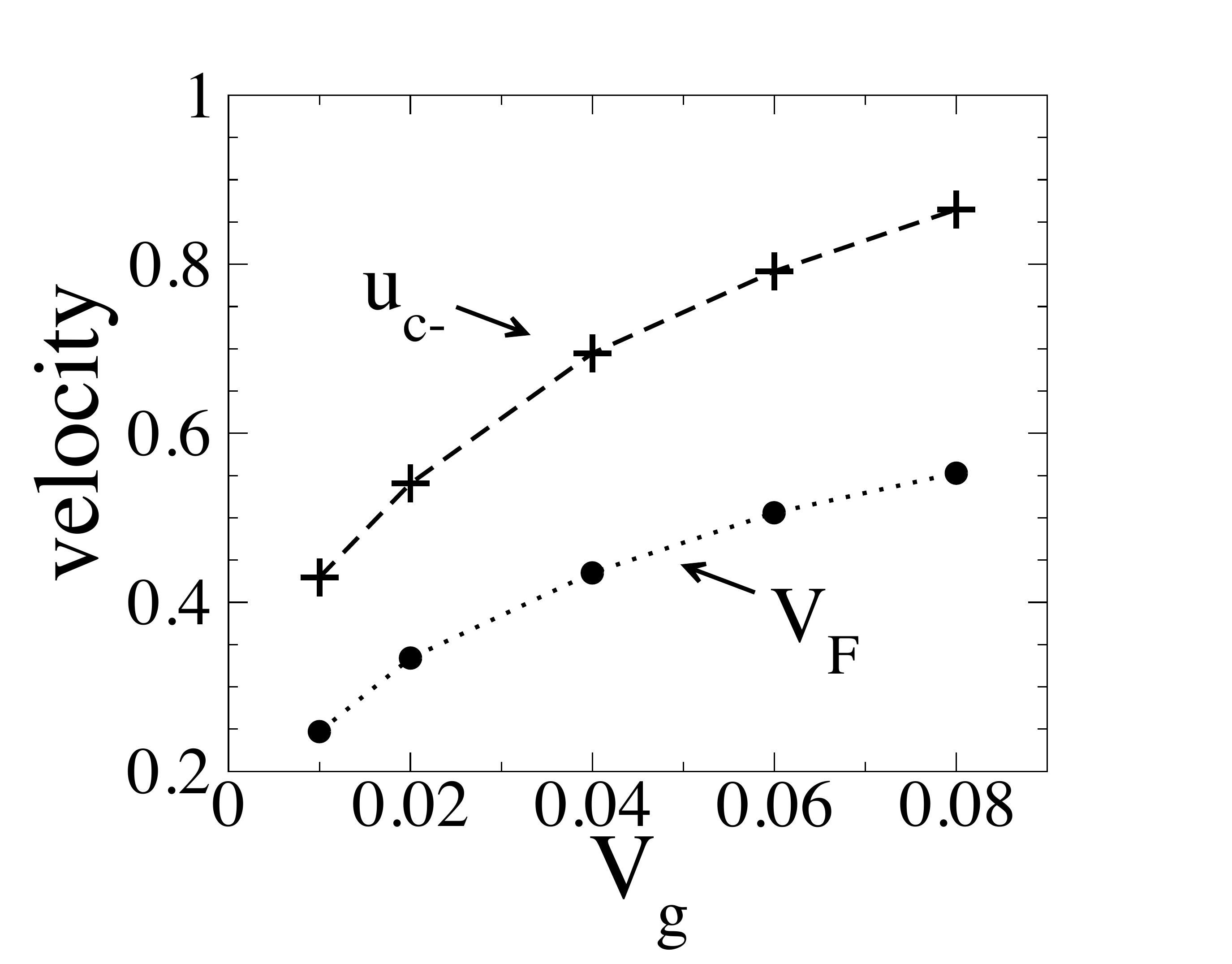}\hspace{0.2cm} 
	\includegraphics[height=5.9cm]{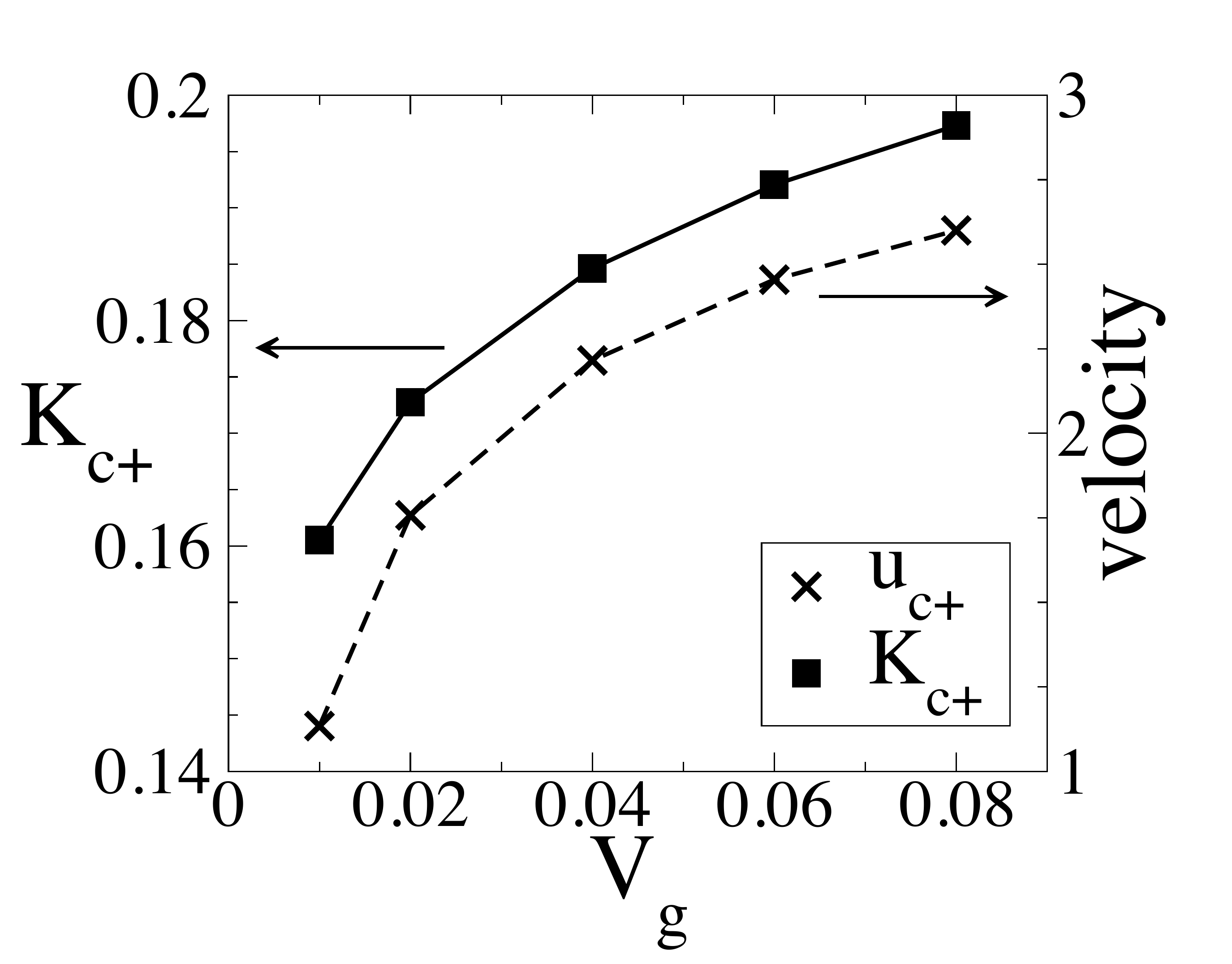} \vspace{-0.4cm} \caption{Fermi velocity $V_F$ and mode velocities $u_{c\pm}$ in the charge sector (in units of $td$), and the Luttinger parameter of total charge sector as a function of $V_{g}$. We assume a screening length of $1000 d \approx 0.3\mu m$ for the Coulomb interaction and a short distance cutoff of 0.5d.} \label{parameters} \vspace{0 cm} 
\end{figure}

This Hamiltonian is diagonal in the total/transverse basis defined via $\phi_{\nu \pm}=\phi_{\nu 0} \pm \phi_{\nu \pi}$, where $\nu$ labels the spin ($s$) or charge ($c$) sector and $\phi_{c(s) \alpha}=\phi_{\alpha \uparrow} \pm \phi_{\alpha \downarrow}$. In this basis, the spin and charge sectors decouple. The spin modes are unaffected by interactions, the Luttinger parameters $K_{s \pm} \!=\! 1$ and the velocities $u_{s \pm} \!= \!V_F$. The charge modes have renormalized velocities and nontrivial Luttinger parameters, given by 
\begin{eqnarray}
	u_{c \pm} & = V_{F} \left(1+ y_{c \pm}\right)^{\frac{1}{2}} \\
	K_{c \pm} &= \left (1+y_{c \pm} \right)^{-\frac{1}{2}}, 
\end{eqnarray}
where $y_{c \pm} = 2 (V_A \pm V_B)/ \pi V_{F} $. At the Gaussian level, the only effect of the interactions is thus to strongly modify $K_{c \pm}, u_{c \pm}$. Fig.(\ref{parameters}) shows these parameters plotted for various gate voltages. As seen from the figure, $K_{c+}$ can be tuned significantly by the external bias; by contrast, $K_{c-} \approx 0.63$ (not shown) is relatively bias independent.

\section{Observable consequences} The strong interactions in the charge channel lead to three different velocities for the spin ($V_F$) and charge ($u_{c\pm}$) modes in the Luttinger liquid. Mapping out these dispersing modes, as has been done in semiconductor heterostructures \citep{Auslaender:2005}, appears to be challenging in the biased bilayer graphene system. A more accessible signature of the Luttinger liquid physics is the energy dependence of the single particle density of states. We expect $n(\epsilon) \sim \epsilon^{\alpha}$, with $\alpha > 0$. Such a suppression of the density of states is expected to lead to a tunneling conductance $G \sim T^\alpha$ (for voltages $e V \! \ll \! k_B T$) or a nonlinear differential conductance $dI/dV \sim V^\alpha$ (for $e V \! \gg \! k_B T$). We find $\alpha_{\rm bulk} = \frac{1}{8} (K_{c+}+K_{c+}^{-1}+K_{c-}+K_{c-}^{-1} - 4)$ and $\alpha_{\rm edge} = \frac{1}{4}(K_{c+}^{-1}+K_{c-}^{-1} - 2)$, so that bias dependent tunneling exponents are expected to be observed. Various charge density, spin density and superconducting pair correlators are expected to show power-law decays in this intermediate energy Luttinger liquid regime. We find that charge and spin density wave operators at $2 k_F^0$ and $2 k_F^\pi$ are most strongly enhanced by interactions in this regime, decaying along the interface as $|x-x'|^{-(2+\!K_{c+}\!+\!K_{c-}\!)/2}$, with the precise lattice scale modulation pattern in $(n,s,\ell)$ being determined by the prefactors set by the wavefunctions $\varphi^{\alpha}_{R/L}(\vr)$ from Eq.(\ref{field}). Modulations at $k_F^0\!\pm\!k_F^\pi$ are subdominant.

\section{Discussion} \label{Sect:TTLdisc} We have shown that bilayer graphene in a suitable gate geometry can realize a tunable Luttinger liquid with four gapless modes. We attribute the persistence of the gapless modes to the strong suppression of backscattering, which makes this system an attractive candidate for electronic device applications.  At very low energy (or temperature), however, we expect that such processes will gap out all sectors except the $c+$ channel, which should still exhibit Luttinger liquid physics. This is expected to break down once backscattering and interband scattering terms become important; since the bare values of these interactions are small and that they are all marginal, and thus flow slowly, there is an intermediate energy window where the physics discussed above should be observable.

Turning to the effect of additional interlayer hopping terms ($\gamma_3$) between the $a_2$ and $b_1$ sites, we have checked that this renders $V_F^0 \neq V_F^\pi$. This asymmetry is small for moderate bias voltages; further, interband scattering tends to equalize the velocities \citep{Penc:1990,Chudzinski:2008} so that small velocity asymmetries are expected to be unimportant. We have assumed that the Fermi level is tuned, via $V_T+V_B$, to be precisely in the middle of the gap - small deviations that tend to slightly dope the interface states while leaving the bulk gap intact will not qualitatively alter the physics discussed here. While the `domain wall' modes discussed here are topologically protected \citep{Martin:2008} independent of the precise bias profile transverse to the wire, disorder along the wire direction will lead to backscattering. Backscattering is mitigated by the wavefunction spread, as is discussed in the conclusion of the thesis, but we expect it will lead to insulating behavior at very low energy \citep{Kane:1992}. Fourier transform scanning tunneling spectroscopy would then be useful to uncover the underlying Luttinger liquid physics \citep{Kivelson:2003}.
 
\chapter{Band Structure of Bilayer Graphene in Inhomogeneous Potentials} \label{Chapt:SL}
\chaptermark{Band Structures of Bilayer Graphene Superlattices}
\subsubsection{The material in this section is largely based on the article M.\ Killi, S.\ Wu A.\ Paramekanti \textit{Phys.~Rev.~Lett.} {\bf107}, 086801 (2011).}


In this chapter, we go beyond the simple non-uniform potentials discussed in the previous chapter and study the band structures of bilayer graphene superlattices, arising from periodic modulations of the chemical potential and the bias, using an effective low energy Hamiltonian. Our main results are the following. (i) Although the minimal model of bilayer graphene has {\it quadratic} band touching points, we find, remarkably, that a weak 1D chemical potential modulation leads to the generation of {\it linearly} dispersing massless Dirac fermions with a tunable and anisotropic velocity. These Dirac fermion excitations are robust and rely on the chiral nature of the bilayer graphene quasiparticles. Beyond a critical modulation amplitude, these Dirac modes get gapped out. (ii) An electric field superlattice is shown to support linearly dispersing massless Dirac fermions and finite energy Dirac points, which survive even for strong modulations. We provide a picture for these modes within a novel coupled chain model of `topological' edge states. (iii) For 2D superlattices, we show that for chemical potential and electric field superlattices the quadratic band touching points are protected for symmetric superlattices with $C_4$ or $C_6$ symmetry. (iv) We compute the density of states for {\it biased} bilayer graphene with superimposed 1D potential modulations, and find a plethora of subgap modes, which we argue are important for understanding transport data. While our results on 1D superlattices overlap with work on Kr\"onig-Penney models \citep{Barbier:2010a,Barbier:2010}, our analysis provides simpler insights, highlights the role of the quasiparticle chirality, and is applied here to more general potential profiles as well as to 2D superlattices.  For the below numerical calculations, we take the superlattice length to be $60d\sim 20$nm, which lies between the currently experimentally accessible in gated systems ($\sim 100$ nm  (private commumincation)) and  corrugated systems ($ \sim 3$ nm  \citep{Yan:2012a}).

\section{ Effective Hamiltonian approach} The low energy Hamiltonian for Bernal-stacked bilayer graphene can be obtained by expanding its minimal tight binding spectrum near one of the Brillouin zone corners ($\bK$ points) \citep{McCann:2006}. When the bias (i.e., interlayer potential difference) is not too large, $|\Delta|\ll t_{\perp}$, we find ${\mathcal H}=\psi^{\dagger}\hat{H}\psi$ \citep{McCann:2006}, where 
\begin{eqnarray} \label{Hred}
	H&=&H_{\rm kin}+H_{\rm SL} \\      \nonumber
	 {H}_{\rm kin}&=&-\frac{v_F^2}{t_{\perp}}\left( 
	\begin{array}{cc}
		0 & (\pi^\dagger)^2 \\
		\pi^2 & 0 
	\end{array}
	\right)
	\\                      \nonumber  
	H_{\rm SL}&=&
	\left( 
	\begin{array}{cc}
		V_1({\bf x}) & 0 \\
		0 & V_2({\bf x}) 
	\end{array}
	\right), 
\end{eqnarray}
and $\psi^{T}=(a_{\bf x},b_{\bf x})$, with $a$ ($b$) being the electron operator on the top (bottom) layer. Here, $\pi \! =\! -i 
\partial_x \! + \! 
\partial_y$, $v_F \! =\! \sqrt{3}td/2 \! \approx \! 10^6$~m/s is the Fermi velocity, $t \! \approx \! 3$~eV is the nearest neighbor hopping integral, $d \! \approx \! 2.46$ \AA~is the distance between neighboring atoms on the same sublattice, $V_{1,2}$ are the potentials on each layer, and $t_{\perp} \!\approx \! 0.15 t$ is the interlayer coupling. Unless stated, we set $t\!\!=\!\!d\!\!=\!\!1$. We will ignore inter-valley scattering assuming the potentials are varying slowly on the scale of $d$, so that identical physics is expected around the other valley (at $-\bK$). Such an approach has been successfully used to study superlattices in monolayer graphene \citep{Park:2008a,Park:2008b,Park:2009}.

To diagonalize $H_{\rm kin}$, we Fourier transform and then make a unitary transformation $a_{\bf p} \!=\! (\alpha_{\bf p} \!+\! \beta_{\bf p})/\sqrt{2}$, $b_{\bf p} \!=\! {\rm e}^{2 i \theta_{\bf p}} (\alpha_{\bf p} \!-\! \beta_{\bf p})/\sqrt{2}$, where $\cos\theta_{{\bf p}}\!=\!p_x/p$ and $p\!=\!\sqrt{p_x^2+p_y^2}$. This leads to 
\begin{eqnarray}
	H_{\rm kin} = 
	\sum_{\bf p}\left(\varepsilon_{e}({\bf p})\beta^{\dagger}_{\bf p} \beta^\pdg_{\bf p} +
	 \varepsilon_{h}({\bf p})\alpha^{\dagger}_{\bf p}\alpha^\pdg_{\bf p}\right).
\end{eqnarray}
 Here $\varepsilon_{e,h}({\bf p})\!=\!\pm p^2/2 m^*$ are energies of electron (hole) states, with an effective mass $m^* \!\equiv\! t_{\perp}/(2 v_F^2)$. This minimal model supports quadratic band touching points at $\pm \bK$.

When $V_{1,2}({\bf x})$ are periodic, we can also Fourier transform the superlattice potential to obtain 
\begin{eqnarray}
	H_{\rm SL}=\sum_{{\bf p},{\bf G}}\Psi^{\dagger}({\bf p}) W_{{\bf p},{\bf G}} \Psi({\bf p}-{\bf G}),
\end{eqnarray}
where 
\begin{eqnarray}
	\label{sl} W_{{\bf p},{\bf G}}\!\!=\!\!\frac{1}{2} \left( 
	\begin{array}{cc}
		\! V_1({\bf G})\!+\!V_2({\bf G}){\rm e}^{2i\theta} & V_1({\bf G})\!-\!V_2({\bf G}){\rm e}^{2i\theta} \! \\
		\! V_1({\bf G})\!-\!V_2({\bf G}){\rm e}^{2i\theta} & V_1({\bf G})\!+\!V_2({\bf G}){\rm e}^{2i\theta} \! 
	\end{array}
	\right)\!\!, \label{W} 
\end{eqnarray}
with $\Psi^\dg({\bf p})\!=\!(\alpha^\dg_{\bf p},\beta^\dg_{\bf p})$ and $\theta \!\equiv\! \theta_{{\bf p}-{\bf G}}\!-\!\theta_{{\bf p}}$ is the angle between momenta ${\bf p}\!-\!{\bf G}$ and ${\bf p}$. Our aim is to understand the band structures of superlattices described by $H_{\rm kin}+H_{\rm SL}$. We will study 1D superlattices with period $\lambda$ along $\hat{y}$, so that the reciprocal lattice vectors, $\{\bG\}$, are integer multiples of $\bQ=(0,2\pi/\lambda)$, and the mini Brillouin zone (MBZ) boundaries are at $p_y=\pm \pi/\lambda$. We will also study 2D superlattices.

\section{1D electric field superlattice} 

An electric field superlattice corresponds to $V_1(x,y)=-V_2(x,y)=U(x,y)$. Solving for the resulting band structure, we find that it depends sensitively on the modulation type. To illustrate this, we consider a periodic potential, with $U(y)=2 U (1-w/\lambda)$ for $0 \leq y < w$, and $U(y)=-2 U w/\lambda$ for $w \leq y < \lambda$. We have set the average potential on each layer to be zero. If $w=\lambda/2$, the resulting {\it symmetric} superlattice is found to support a pair of anisotropically dispersing massless Dirac fermions at zero energy at $(\pm p^*_x,0)$. In addition, it supports a Dirac point at nonzero positive (as well as negative) energies at $(0,\pi/\lambda)$ (or equivalently $(0,-\pi/\lambda)$). However, an {\it asymmetric} superlattice, with $w \neq \lambda/2$, leads to a gap for all these Dirac fermions. More generally, we find that if the superlattice potential commutes with a generalized parity operator, ${\cal P}$, which corresponds to $y \to -y$ followed by exchanging the two layers of bilayer graphene, then these gapless Dirac points survive. Breaking ${\cal P}$ leads to gaps.      

\begin{figure}
	[t] \centering 
	\includegraphics[width=.45 
	\textwidth]{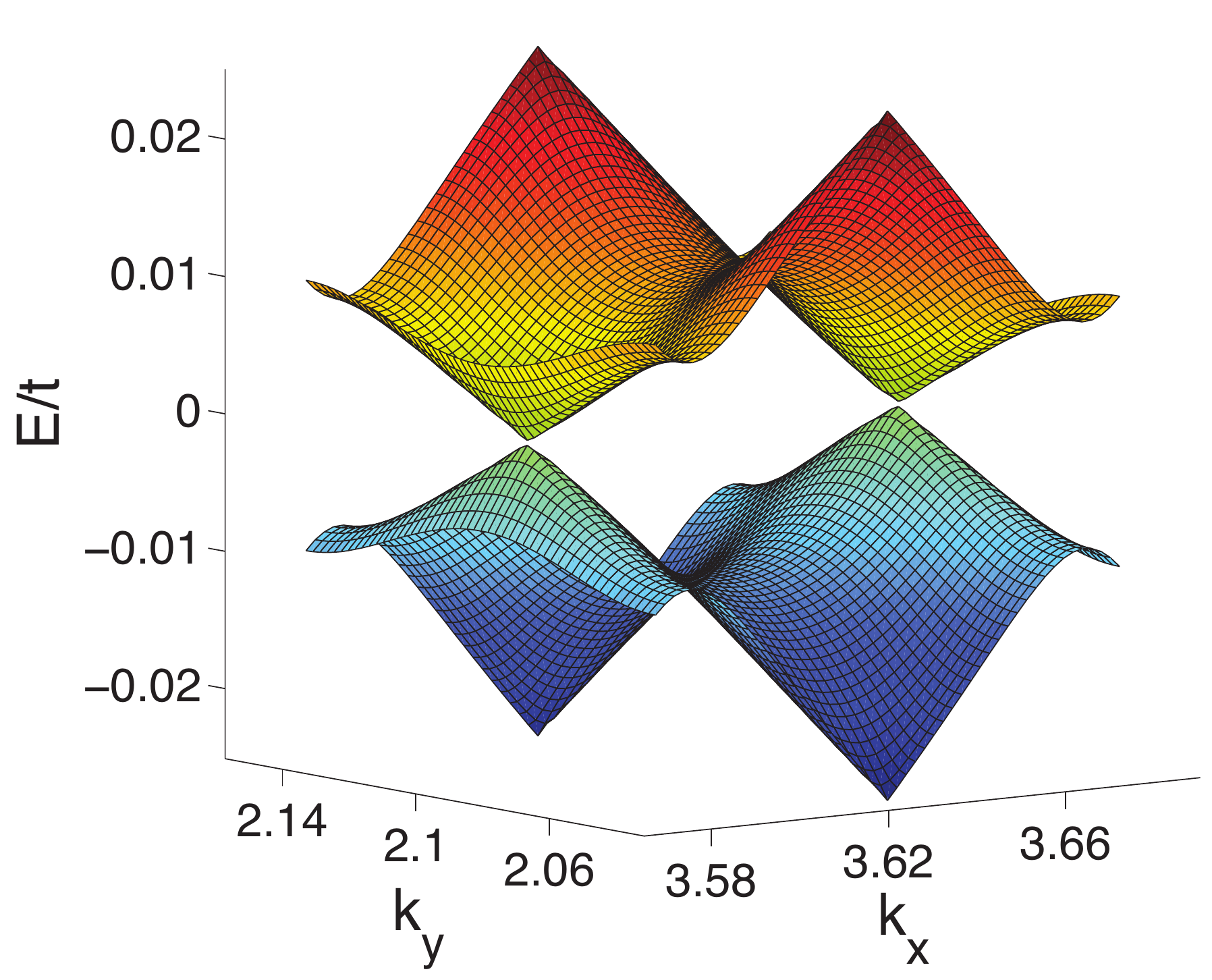} 
	\includegraphics[width=.45 
	\textwidth]{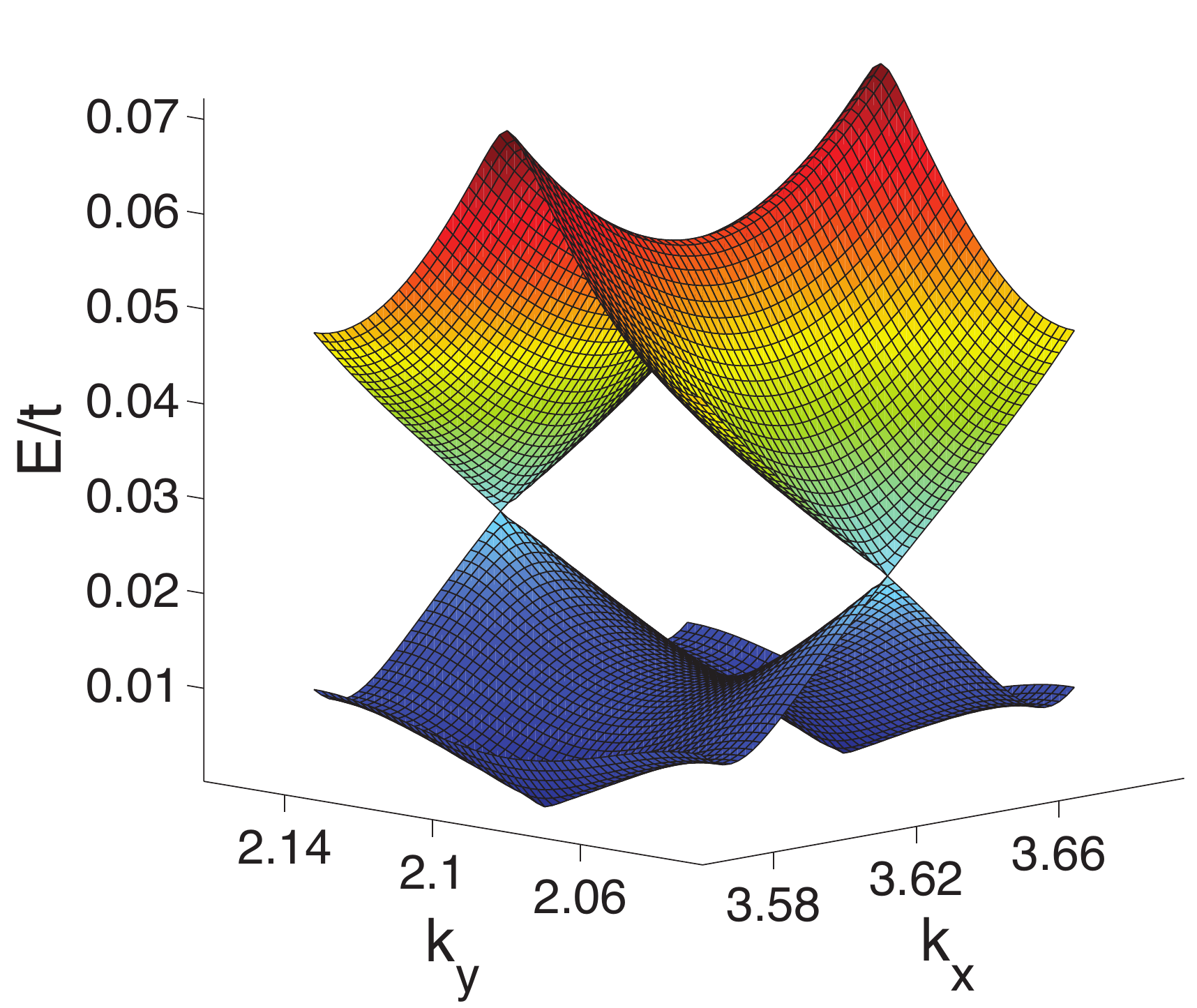} \caption{Energy spectrum for a 1D {\it symmetric} (see text) electric field superlattice with $\lambda=60d$ and $U=0.03 t$, showing a pair of zero energy massless Dirac fermions at $(\pm p^*_x,0)$ and a nonzero energy Dirac point at $(0,\pm \pi/\lambda)$.} \label{fig3} 
\end{figure}

\subsection{An array of coupled kink and antikink wires} 

\begin{figure}
	[tb] \centering 
	\includegraphics[width= 0.45 
	\textwidth]{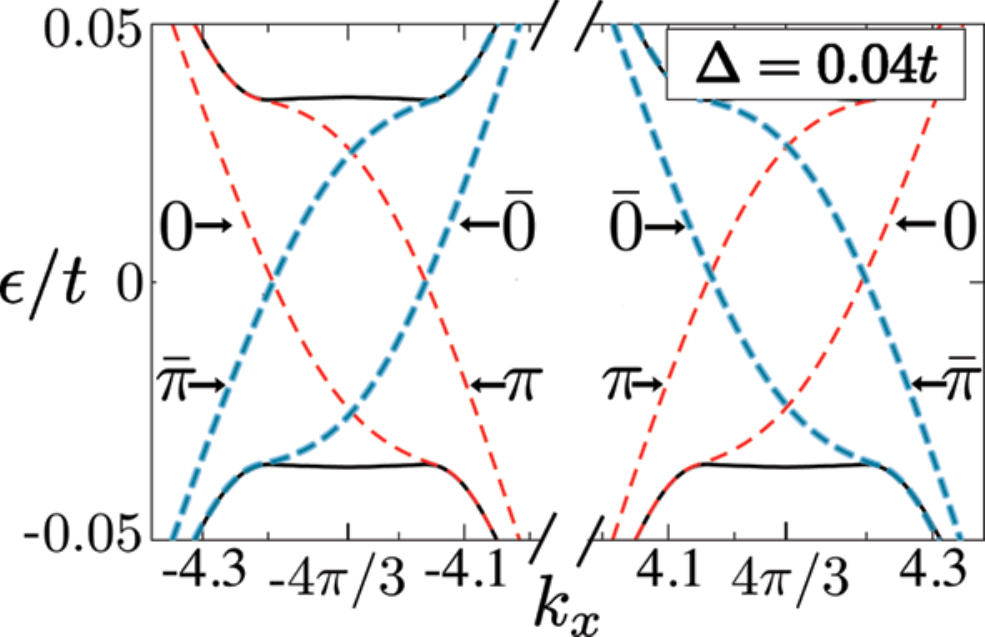} \quad \quad 
	\includegraphics[width=0.35 
	\textwidth]{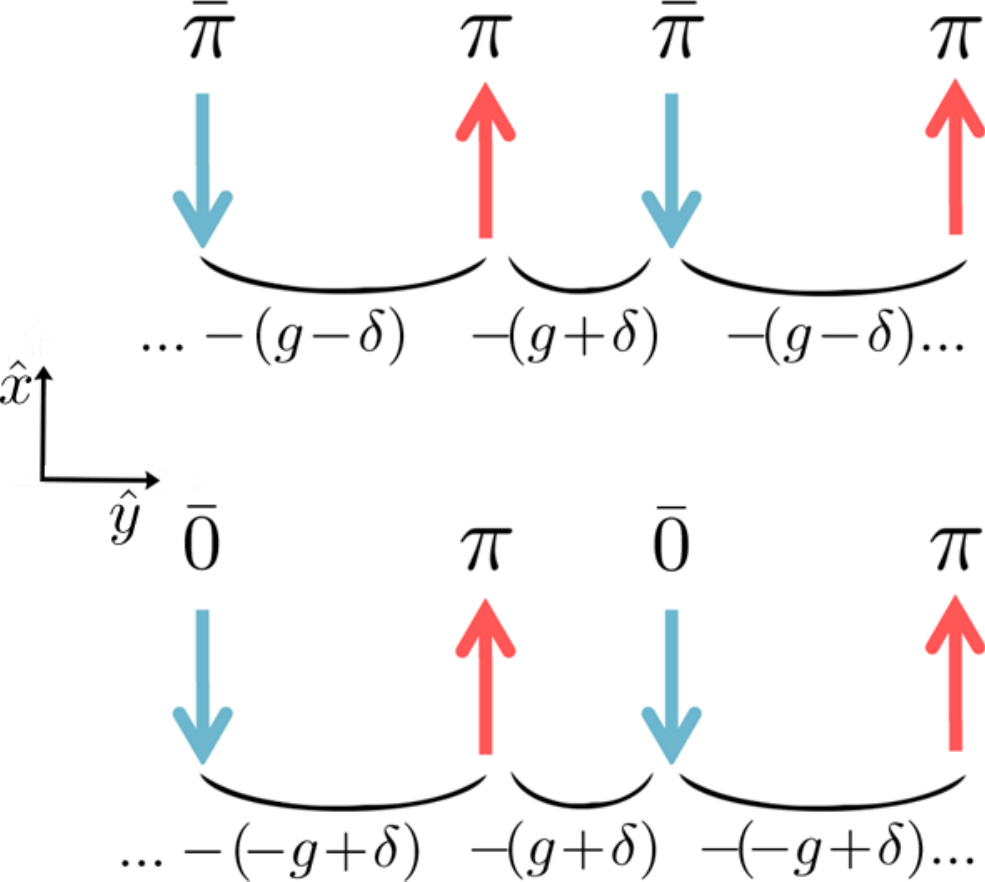} \caption{ Left: Spectrum of isolated kink (thin, red) and antikink (thick, blue). Higher (lower) energy modes are labelled $\pi$ ($0$) at a kink and as $\bar{\pi}$ ($\bar{0}$) at an antikink. Right: Schematic of hopping between the $\pi-\bar{\pi}$ and $\bar{0}-\pi$ states, where $g$ is the average and $\delta$ the variation of the magnitude of the hopping parameter.} \label{wiredispersion} 
\end{figure}

A simple route to understanding these results that leads to other interesting predictions is to view the superlattice as a periodic array of `kinks' and `antikinks' where a kink (antikink) corresponds to where the electric field flips from pointing up (down) to pointing down (up). A single such kink/anti-kink in the bias is well understood \citep{Martin:2008,Killi:2010,Xavier:2010,Li:2011a} and was discussed thoroughly in Chapter \ref{Chapt:Kink}. In the absence of interactions a kink (anti-kink) supports a pair of right-moving (left-moving) `topological' edge states near the ${\bf K}$ point for each spin. By time-reversal, these right and left movers get interchanged at the $-{\bf K}$ point. These modes are depicted in Fig.~\ref{wiredispersion}. (Although these modes were suggested to be topologically protected, we reiterate that they are not truly stable against disorder; nevertheless disorder induced backscattering is weak (see Chapter \ref{Chapt:Conclusion} and \citet{Li:2011a}.) At a kink, we denote the higher (lower) energy edge state as $\pi$ ($0$), while we denote these states as $\bar{\pi}$ ($\bar{0}$) at an antikink. Hence, there are four points at each valley where a kink and anti-kink mode cross: two of these occur at zero-energy ($\pi$-$\bar{0}$ and $\bar{\pi}$-$0$ crossings), and two of them occur at nonzero energy ($\pi$-$\bar{\pi}$ and $0$-$\bar{0}$ crossings). We will show below that these crossing points evolve into massless Dirac fermion modes in the MBZ of the superlattice. In order to see this, we construct a tight-binding model of such coupled `topological' kink states.

\subsubsection{Symmetry properties} In order to do so, we exploit the observation that the Hamiltonian with the single kink (or antikink) potential is invariant under ${\cal P}$, since $\mathcal{P^{\dag}}H(y)\mathcal{P}\!=\!\sigma_x H(-y) \sigma_x \!=\! H(y)$. Physically, ${\cal P}$ is the combined operation of layer inversion and reflection about a kink (or antikink). The $0/\bar{0}$ states are {\it even} under ${\cal P}$, while the $\pi/\bar{\pi}$ states are {\it odd} under ${\cal P}$ \citep{Martin:2008}. Explicitly, the wavefunctions take the form 
\begin{eqnarray}
	\left( 
	\begin{array}{cc}
		f(y) \\
		g(y) 
	\end{array}
	\right)=\left( 
	\begin{array}{cc}
		f(y) \\
		f(-y) 
	\end{array}
	\right),\left( 
	\begin{array}{cc}
		f(y) \\
		-f(-y) 
	\end{array}
	\right), 
\end{eqnarray}
with corresponding eigenvalues of $+1$ and $-1$ of the operator $\mathcal{P}$, respectively. Knowing the symmetry properties of the kink and antikink states allows us to construct a reduced Hamiltonian that describes the hybridization between neighboring edge modes about the four band crossing points.

For the case of the zero energy band crossing points (between the $\pi$- and $\bar{0}$-modes or the $\bar{\pi}$- and $0$-modes at either K-point), the soliton wavefunctions of the kink wire have $opposite$ $\mathcal{P}$ symmetry to the soliton wavefunctions of the two neighbouring anti-kink wires. The hopping between neighboring `wires' along $\hat{y}$ is then between states that have opposite velocities (since it is between a kink and an antikink edge state) and it is between a `p-wave' like state (${\cal P}$-odd) and an `s-wave' like state (${\cal P}$-even). In contrast, for the case of the finite band crossing points (between the $\pi$- and $\bar{\pi}$-modes or the $\bar{0}$- and $0$-modes at either K-point), the soliton wavefunctions of the kink wire have the $same$ $\mathcal{P}$ symmetry as the soliton wavefunctions localized to its two neighbouring anti-kink wires. Hence, the hopping along these `wires' along $\hat{y}$ is again between states with opposite velocities but are now either both `s-wave' or `p-wave', depending on if band crossing is in the valence or conducting band. As we will now show, the transfer integrals describing the hopping between neighbouring wires along the array is dependent on whether the parity of the coupled modes is the same or opposite. 
\begin{figure}
	[tb] \centering 
	\includegraphics[width=0.9 
	\textwidth]{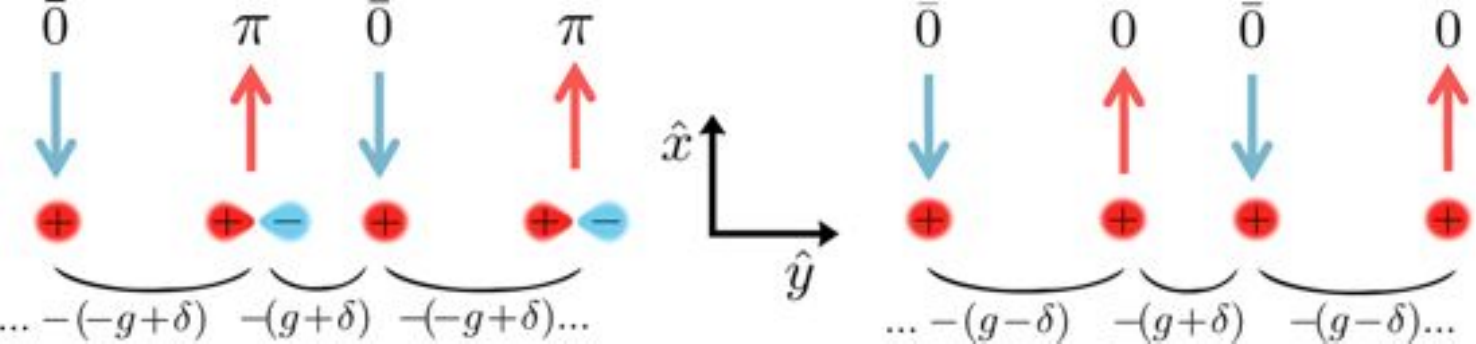} \caption{ Variation of the hopping parameter along the wire array. Left: Hopping between zero energy modes of opposite parity. Right: Hopping between finite energy modes with the same parity. Shape, orientation and sign of the wavefunctions are completely schematic and serve only to be illustrative of effect of parity on the hopping integrals.} 
\end{figure}
\subsubsection{Transfer integrals} For concreteness, let us consider the region where the $\pi$- and $\bar{0}$-bands cross at the K point for a symmetric modulation with $w=\lambda/2$. Let us set the $y=0$ point to be an anti-kink wire. Again, we know that the anti-kink $\bar{0}$-modes have wavefunctions of the form 
\begin{eqnarray} 
	\Psi^{\bar{0}}(y)=(w(y),w(-y))^T 
\end{eqnarray}
while its two neighbours' $\pi$-modes have wavefunctions of the form 
\begin{eqnarray}
\Psi^\pi(y+\lambda/2)=\left(
\begin{array}{c}v(y+\lambda/2) \\
	-v(-y-\lambda/2)
\end{array}\right)
\end{eqnarray}
and
 \begin{eqnarray}
\Psi^\pi(y-\lambda/2)=\left(
\begin{array}{c}v(y-\lambda/2)\\-v(-y+\lambda/2)
	\end{array}\right),
\end{eqnarray}
respectively. For simplicity, we will assume that the wavefunction overlap is finite only in the region between the wires, as this assumption does not effect our main result. 

The transfer matrix that determines the hopping parameter between the central wire and its left neighbour is then 
\begin{eqnarray}
	g_1=\int^0_{-\lambda/2} \Psi^{\bar{0} \dag}(y) \, H(y) \, \Psi^\pi(y+\lambda/2) \, \mathrm{d}y, 
\end{eqnarray}
and between its right neighbour 
\begin{eqnarray}
	g_2=\int^{\lambda/2}_{0} \Psi^{\pi \dag}(y-\lambda/2) \, H(y) \, \Psi^{\bar{0}}(y) \, \mathrm{d}y. 
\end{eqnarray}

Inserting $H(y)= \sigma_x H(-y) \sigma_x$ and changing $y \rightarrow -y$ in the expression for $g_2$, gives us the relation 
\begin{eqnarray}
	g_1=-g_2^*\equiv |g|e^{i \theta} 
\end{eqnarray}
between alternate bonds. The same relationship holds for the other zero energy band crossing point. However, repeating this calculation for the finite energy band crossing points (where the wavefunctions have the same parity) yields the corresponding relation 
\begin{eqnarray}
	g_1'=g_2^{*}{'}\equiv t=|g'|e^{i \theta} . 
\end{eqnarray}

Without loss of generality, in both cases the hopping parameter be assumed to be real, as it is always possible to remove the phase factor by a simple gauge transformation. Hence, we have deduced from very general symmetry arguments that when $w=\lambda/2$ the hopping parameter between the $0$- ($\bar{0}$-) and $\pi$- ($\bar{\pi}$-) modes of neighbouring wires alternates sign along the chain, and is uniform between $0$- ($\pi$-) and $\bar{0}$- ($\bar{\pi}$-) modes of neighbouring wires.

If we generalize this case where $w\neq\lambda/2$, the separation between neighbouring wires is unequal and the magnitude of the hopping parameter will begin to alternate along the bonds. Again considering the $\pi$- $\bar{0}$-band crossing at the K point and taking the the wire at larger $y$ to be further than to the anti-kink wire than the other neighbour, $g_{1}=-\left(-g+\delta\right)$ and $g_{2}=-\left(g+\delta\right)$, where $g>0$ is the average magnitude of the hopping between neighbouring wires and $\delta>0$ is the deviation. Alternatively, for the finite energy band crossing points, the hopping parameter can be shown to be $g_{1}=-\left(g-\delta\right)$ and $g_{2}=-\left(g+\delta\right)$.

\subsubsection{Reduced tight-binding Hamiltonian} Using the index $n$ to label the wires, the interchain hopping parameter will then alternate as $(-1)^n g$ for equally spaced wires and as $g+\delta, -g+\delta$ (with $\delta < g$) if pairs of wires are closer to each other. Linearizing the dispersion at the crossing point, and letting $v_0$ denote the velocity of the linearized modes, 
\begin{eqnarray}
	H(\px)=&v_0&\sum_n \left((-1)^n (p_x-p^*_x) c^{\dg}_{\px n}c_{\px n}\right) \nonumber \\
	&-&\sum_n (g (-1)^n + \delta) \left(c^{\dg}_{\px n}c_{\px n+1} + h.c.\right) 
\end{eqnarray}
where $p^*_x$ is the location of the $\pi-{\bar 0}$ crossing point in the single kink/antikink problem, and $c_{\px n}$ annihilates an electron on wire $n$ with momentum $\px$. Let $\xi(p_x)\equiv v_0 (p - p^*_x)$. Fourier transforming, we find $H(\px)=\sum'_\py \Psi^\dg(\py) {\bf \sigma} \cdot {\bf h}(\px) \Psi(\py)$, where ${\bf h}(\px)=\left(\xi(p_x) , -2 g \sin(\py), -2 \delta \cos(\py)\right)$, with $\Psi(\py) =(c_\py \, c_{\py+\pi})^T$, and $\sum'_\py$ runs over the MBZ. The dispersion is thus \begin{eqnarray}
E=\pm \sqrt{\xi^2(p_x) + 4 \delta^2 \cos^2(\py) + 4 g^2 \sin^2(\py)}.
\end{eqnarray}
 Consequently, when $w=\lambda/2$, and the Hamiltonian commutes with ${\cal P}$, we have $\delta=0$ and a Dirac cone is generated at $(p^*_x, 0)$, consistent with numerical results. When $w \neq \lambda/2$, the Hamiltonian breaks ${\cal P}$ --- we then have $\delta \neq 0$, which leads to a gap $4\delta$. Similar arguments hold for the other zero energy band crossing points. The velocity of the Dirac fermions is highly anisotropic and depends on $g$ --- this can be controlled by tuning the superlattice period and amplitude.

The above analysis can also be repeated for the nonzero energy ($0$-$\bar{0}$ and $\pi$-$\bar{\pi}$) crossings; in the {\it symmetric} case, $w=\lambda/2$, we find Dirac cones at $(0,\pm \pi/\lambda)$ on the MBZ. Once again, a modulation with $w\neq\lambda/2$ results in a finite $\delta$ and opening of band gap.

Interestingly, just as in polyacetylene, a domain wall between a gapped region with $w >\lambda/2$ and a gapped region with $w <\lambda/2$ leads to new subgap soliton modes. Since each kink/antikink is itself like a domain wall, these should be viewed `solitons in a soliton lattice'!

\section{1D chemical potential superlattice} 

\begin{figure}
	[t] \centering 
	\includegraphics[width=.45 
	\textwidth]{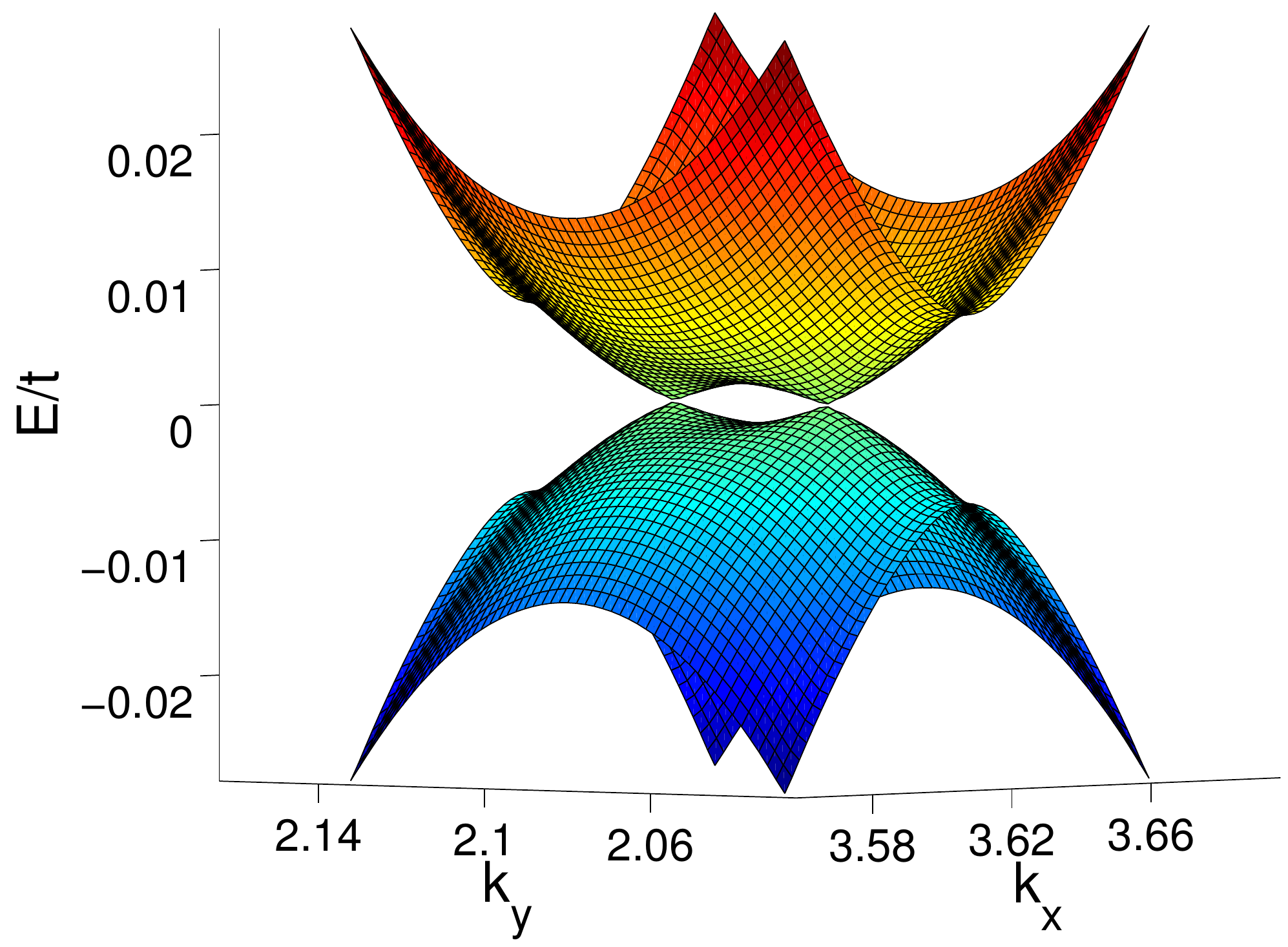} 
	\includegraphics[width=.45 
	\textwidth]{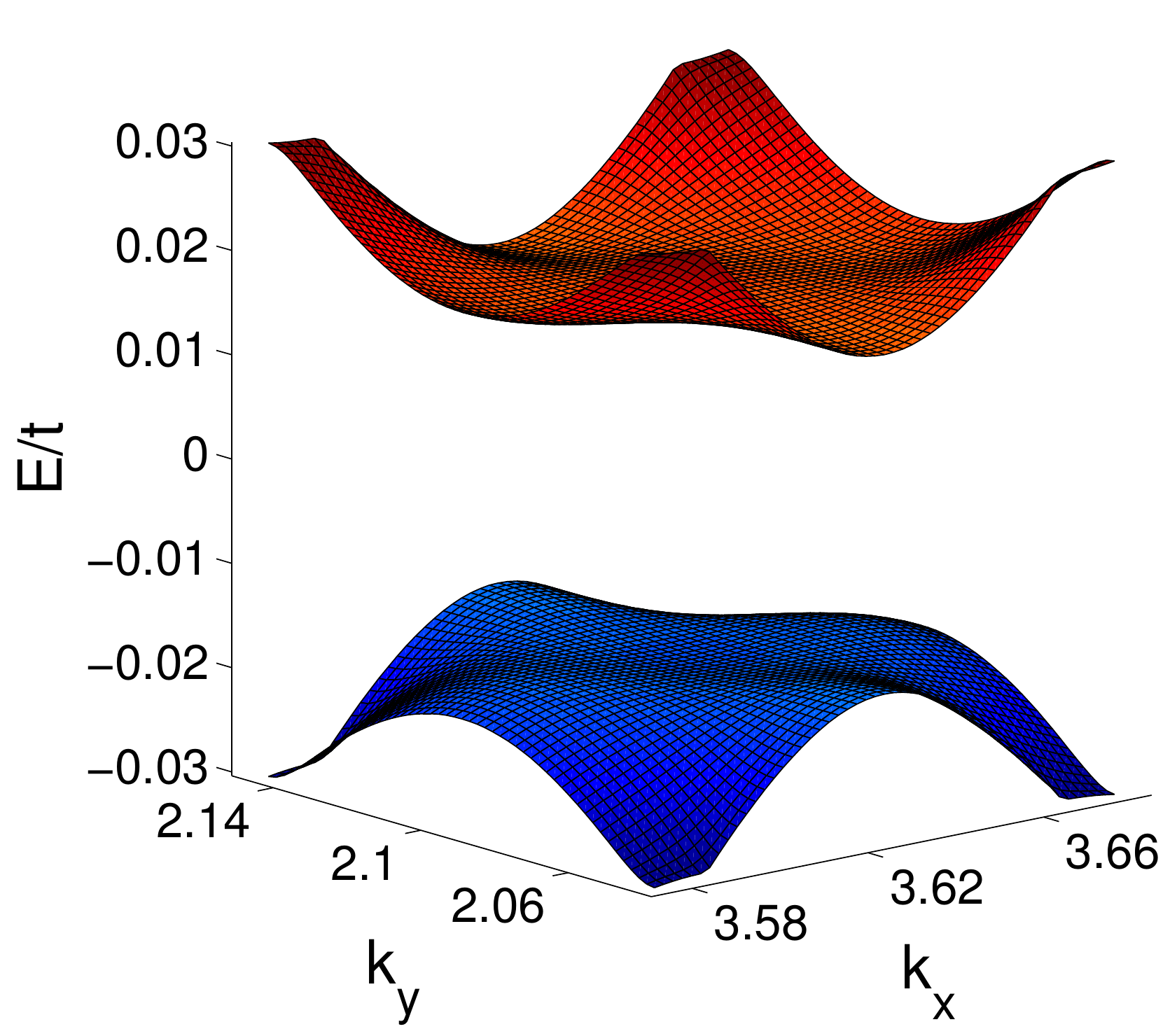} \caption{Energy spectrum for a 1D superlattice with step-like chemical potential modulation of amplitude $U$. We set $\lambda=60d$, with (a) $U=0.01t$ showing two Dirac nodes split along $\hat{y}$ near $\bK$, and with (b) $U=0.04t$ showing a full gap.} \label{fig1} 
\end{figure}

Imposing a periodic potential $V_{1}(x,y)=V_2(x,y)=U(x,y)$ corresponds to a chemical potential modulation. Numerically solving for the band structure of a periodic 1D modulation using the above effective Hamiltonian, we find a pair of zero energy Dirac points in the MBZ in the vicinity of each valley. This is shown in Fig.~\ref{fig1} for a periodic step-like potential with (i) $U(x,y)=U$ for $0 \leq y < \lambda/2$ and (ii) $U(x,y)=-U$ for $\lambda/2 \leq y < \lambda$. With increasing $U$, these Dirac points move away from each other along $\hat{y}$. Beyond a critical modulation amplitude a full gap opens up.

The existence of two Dirac cones at each valley is deeply rooted in the chiral nature of the low energy bilayer graphene quasiparticles, which causes the matrix elements of Eqn.~\ref{sl} to depend on the scattering angle $\theta$. For states with momenta parallel to the modulation direction, $\theta=0$ or $\pi$, the off-diagonal matrix elements vanish; the electron and hole states then decouple, but electron-electron and hole-hole mixing is allowed. However, in an extended zone scheme, all such electron (hole) states within the first MBZ only mix with electron (hole) states of higher (lower) energy, and so the energy of these states will be globally shifted down (up). This results in two level crossings along the modulation direction, which are {\it protected by the chirality of the low energy bilayer graphene quasiparticles}. If this electron-hole decoupling was true for all momenta, we would see the two parabolic bands crossing on a full circle in the MBZ, but going to momenta $(\delta p_x, p_y)$ leads to electron-hole mixing linear in $\delta p_x$; this results in an avoided level crossing and the robust emergence of two Dirac cones in the MBZ. 

The location and velocity anisotropy of Dirac cones, as well as the critical modulation amplitude to gap them out, can be predicted using perturbation theory in $U(\bG)$. The second order energy correction of states with ${\bf p}=(0,p_y)$ is 
\begin{eqnarray}
	\Delta E^{(2)}({\bf p})=\sum_{n\neq 0}{|U(n{\bf Q})|^2}/\left( {\varepsilon_{e,h}({\bf p})-\varepsilon_{e,h}({\bf p}+n{\bf Q})}\right). 
\end{eqnarray}
Since ${\varepsilon_{e}({\bf p})<\varepsilon_{e}({\bf p}+n{\bf Q})}$ while ${\varepsilon_{h}({\bf p})>\varepsilon_{h}({\bf p}+n{\bf Q})}$ in the MBZ, this correction is always negative (positive) for electron (hole) states, as expected. Thus, the two bands will intersect and cross linearly at momenta $(0,\pm p^*_y)$.

\subsection{Location of the Dirac points}

For small bias strengths, the location of Dirac point, $(0,p^*_y)$, can be estimated by setting the absolute value of the second order energy correction equal to the free electron energy, 
\begin{eqnarray}
	\label{E} \frac{p^{*2}_y}{2m^*}=2m^* \sum_n|U(n{\bf Q})|^2 \! &\!&\! \! \! \! \left(\frac{1}{n^2Q^2+2p^*_y nQ}\right. +\left. \frac{1}{n^2Q^2-2p^{*}_ynQ}\right). 
\end{eqnarray}
Equation \ref{E} can then be expanded in the $p_y/Q \ll 1$ limit to give 
\begin{eqnarray}
	\label{E2} p^{*2}_y \approx \frac{2m^{*2}\lambda^{2}}{\pi^{2}} \sum_n \! &\!&\! \! \! \!\frac{|U(n{\bf Q})|^2}{n^2}, 
\end{eqnarray}
or retaining only the fundamental harmonic 
\begin{eqnarray}
	\label{E3} p^*_y \approx \frac{\sqrt{2}m^{*} |U({\bf Q})| \lambda}{\pi}. 
\end{eqnarray}

For the case of a square potential where $|U(n \bQ)|=2U/n \pi$ for odd $n$ and keeping only the fundamental harmonic, 
\begin{eqnarray}
	\label{p} p^*_y \approx 2\sqrt{2}m^{*} U \lambda / \pi^{2}. 
\end{eqnarray}
Alternatively, Eq.\ \ref{E2} can be computed explicitly through the relation $\sum_{n_{odd}}1/n^4=\pi^4/96$ to give $p^*_y \approx \pm \frac{\sqrt{3}m^*U \lambda}{6}$. A comparison of the two expressions confirms that higher order harmonic corrections are indeed small $(\sim 1\%)$ and can be ignored.

\subsection{Anisotropy of the Dirac cones}

Also of interest is the degree of anisotropy of the group velocity about the Dirac cone. Along the $\bp=(0,p_y)$ direction, the above perturbation theory indicates that $v_y=p^*_y/m^* \approx\frac{\sqrt{2}\lambda |U({\bf Q})|}{\pi}$. To calculate the velocity along the $p_x$ direction, we perform a degenerate perturbation theory for an electron and hole states with momentum $\bp=( p^*_y \, \theta, p^*_y)$ for a small angle $\theta$ while retaining only the leading order harmonic. For such states, the off-diagonal terms in the Hamiltonian reduce to, 
\begin{eqnarray}
	W_{\bp, \bp-\bQ}= \left( 
	\begin{array}{cc}
		U(\bQ) & \frac{iU(\bQ)}{2}\sin (2 \theta_{\bp, \bp-\bQ}) \\
		\frac{iU(\bQ)}{2}\sin(2\theta_{\bp, \bp-\bQ}) & U(\bQ) 
	\end{array}
	\right). 
\end{eqnarray}

Given the small finite momentum along the $p_x$ direction, the originally pure electron states get mixed with hole states and vice versa. To leading order in $\theta$, 
\begin{eqnarray}
	|\Psi^e_\bp \rangle =|\alpha_\bp \rangle &+&\frac{U(\bQ)|\alpha_{\bp+\bQ} \rangle}{\epsilon_e(\bp)-\epsilon_e(\bp+\bQ)} +\frac{U^*(\bQ) |\alpha_{\bp-\bQ} \rangle}{\epsilon_e(\bp)-\epsilon_e(\bp-\bQ)} \nonumber \\
	&-&\frac{i\theta U(\bQ) |\beta_{\bp+\bQ}\rangle}{\epsilon_e(\bp)-\epsilon_h(\bp+\bQ)} - \frac{i\theta U^*(\bQ) |\beta_{\bp-\bQ} \rangle}{\epsilon_e(\bp)-\epsilon_h(\bp-\bQ)} \nonumber 
\end{eqnarray}
and, 
\begin{eqnarray}
	|\Psi^h_\bp \rangle =|\beta_\bp \rangle &+&\frac{U(\bQ)|\beta_{\bp+\bQ} \rangle}{\epsilon_h(\bp)-\epsilon_h(\bp+\bQ)} +\frac{U^{*}(\bQ) |\beta_{\bp-\bQ} \rangle}{\epsilon_h(\bp)-\epsilon_h(\bp-\bQ)} \nonumber \\
	&-&\frac{i\theta U(\bQ) |\alpha_{\bp+\bQ}\rangle}{\epsilon_h(\bp)-\epsilon_e(\bp+\bQ)} - \frac{i\theta U^*(\bQ) |\alpha_{\bp-\bQ} \rangle}{\epsilon_h(\bp)-\epsilon_e(\bp-\bQ)}. \nonumber 
\end{eqnarray}
Since, $\langle\Psi^e_\bp | H |\Psi^e_\bp \rangle = \langle\Psi^h_\bp | H |\Psi^h_\bp \rangle=0$ to first order in $\theta$ and are degenerate, we must compute the off-diagonal matrix elements $\langle\Psi^e_\bp | H |\Psi^h_\bp \rangle$ of the Hamiltonian. To leading order in $\theta$, 
\begin{eqnarray}
	\langle\Psi^e_\bp | H |\Psi^h_\bp \rangle \!\! \! \!&=&\!\!\!\! -2i\theta |U(\bQ)|^2 \left( \frac{1}{\epsilon(\bp)-\epsilon(\bp+\bQ)}\right. \nonumber +\frac{1}{\epsilon(\bp)-\epsilon(\bp-\bQ)} +\frac{1}{\epsilon(\bp)+\epsilon(\bp+\bQ)} \nonumber \\
	&+&\!\left. \frac{1}{\epsilon(\bp)+\epsilon(\bp-\bQ)}\right), 
\end{eqnarray}
where $\epsilon(\bp)\equiv\epsilon_e(\bp)=-\epsilon_h(\bp)$. In the small $\bp$ limit, the matrix element reduces to 
\begin{eqnarray}
	\langle\Psi^e_\bp | H |\Psi^h_\bp \rangle=-\frac{16i m^* \theta |U(\bQ)|^2 }{|\bQ|^2}. 
\end{eqnarray}
This give the solution $\epsilon_\bp= \pm 16 m^* \theta |U(\bQ)|^2 / |\bQ|^2$ to the perturbed Hamiltonian is then, from which the velocity can be calculated. Using the relation $p^*_y \, \theta=p_x$ for small the theta, 
\begin{eqnarray}
	\epsilon(p_x)&=&\frac{16m^*|U(\bQ)|^2 p_x}{p^*_y \, |\bQ|^2} \nonumber \\
	\rightarrow &v_x&=2\sqrt{2}\lambda |U({\bf Q}|/\pi 
\end{eqnarray}
Hence, the anisotropy of the velocity in the Dirac cone is predicted to be $v_x/v_y = 2$ for small U, which is again remarkably consistent with the numerical results.

\subsection{Critical interaction strength} 

From numerical results consistent with the perturbative results above, the Dirac points are seen to move out towards the mini-zone boundary with increasing $U$. Once these Dirac points reach the MBZ boundary, Bragg scattering between them opens up a full gap. Using the approximation provided for $p^{*}_{y}$ in Eq.\ \ref{E3}, it is also possible to make a crude estimate of the critical potential, $U_c$, before the onset of a bandgap. The critical potential strength, $U_c$, can be determined from the condition that the Dirac point $p^*_y=p_c=\pm Q/2$. Although the above perturbative result is not strictly valid in this regime, it provides a crude estimate of 
\begin{eqnarray}
	|U_c(\bQ)| \approx \frac{\pi^2}{\sqrt{2}m^*\lambda^2}. 
\end{eqnarray}
In the case of a step potential with $\lambda=60d$, $U_{c}\approx \frac{\pi^3}{2\sqrt{2}m^*\lambda^2} \approx 0.03t $, which is reasonably close to the observed numerical value of $0.02t$.

\section{2D superlattices} We have also considered 2D chessboard like superlattices with fourfold rotation symmetry. For both types of 2D superlattices, chemical potential or electric field, the quadratic band touching point remains intact when the superlattice potential is `symmetric', $ V_{1,2}(x+\lambda/2,y)=V_{1,2}(x,y+\lambda/2)=-V_{1,2}(x,y). $ This is consistent with the fact that no Dirac points can be generated in a way that conserves both topological charge and $C_4$ (or $C_6$) symmetry \citep{Sun:2009}. For asymmetric superlattices, higher order corrections lead to modifications to the energy spectrum at the {\bf K}-point. For chemical potential superlattice, the charge neutrality point shifts slightly in energy, due to higher order effects which reflect particle-hole symmetry breaking. For electric field superlattices, breaking generalized parity opens a small gap at the ${\bf K}$-point. 

\section{Experimental implications} Our work demonstrates that superlattice modulations in bilayer graphene can generate new Dirac fermion modes. Such modes are perturbatively stable to interaction effects, and could be experimentally explored by suitable choice of substrates. Disorder will also lead to such bias and chemical potential modulations, albeit in random fashion. One source of such fluctuations is the presence of charged impurities, embedded in the underlying substrate (SiO$_2$) or, in the case of suspended bilayer graphene, in the residue of the etching/washing process. Such impurities are expected to locally shift the charge neutrality point, and to suppress or enhance the bandgap depending on the relative sign of the bias and the impurity electric field. If the impurity lies close to the surface it can locally reverse the parity of the interlayer bias leading to `topological' subgap modes. Another source of superlattice fluctuations is rippling \citep{Ishigami:2007, Bao:2009}, which would modulate the electric field perpendicular to the bilayer at the ripple wavelength. 
\begin{figure}
	[tb] \centering 
	\includegraphics[width=0.45 
	\textwidth]{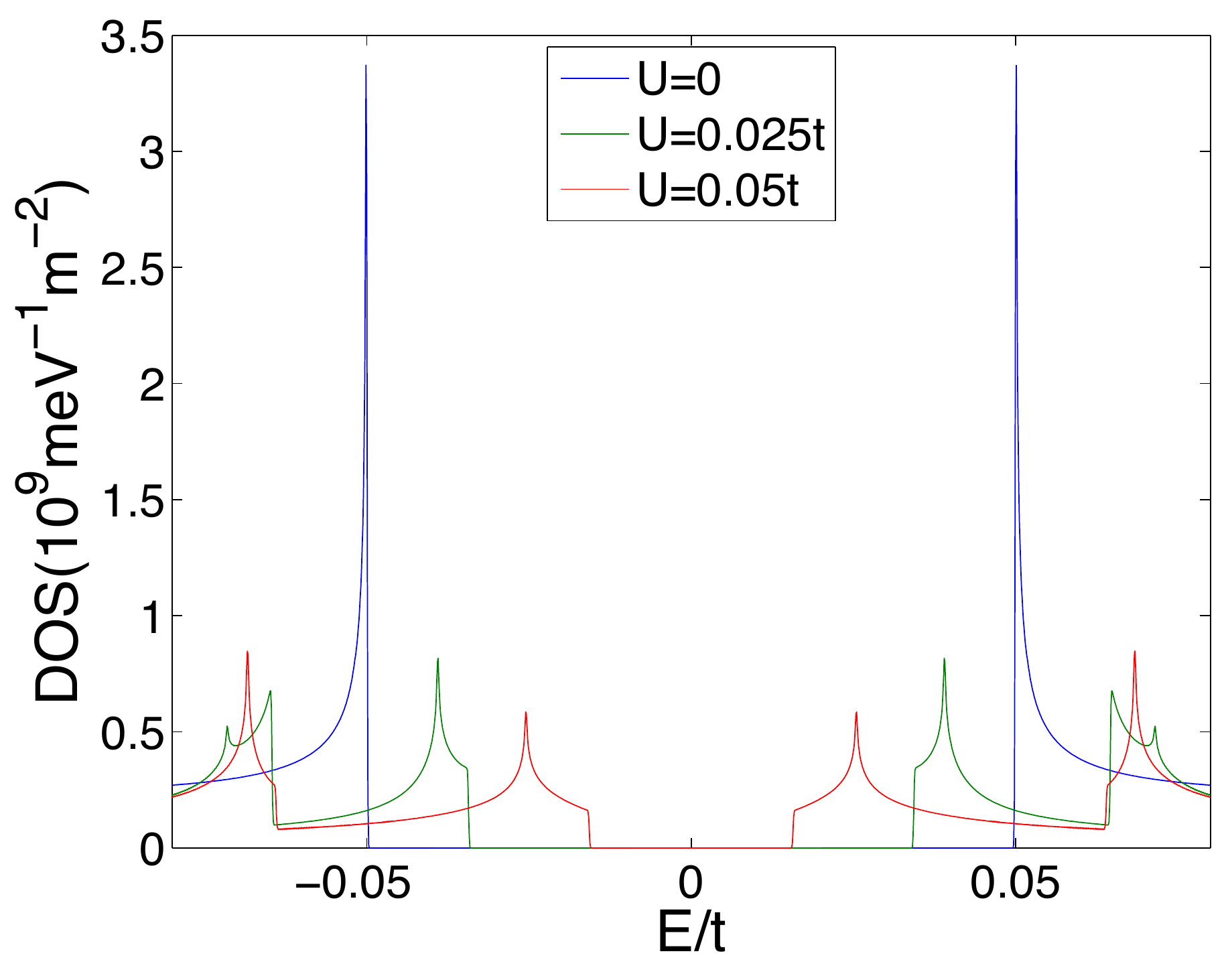} \quad 
	\includegraphics[width=0.45 
	\textwidth]{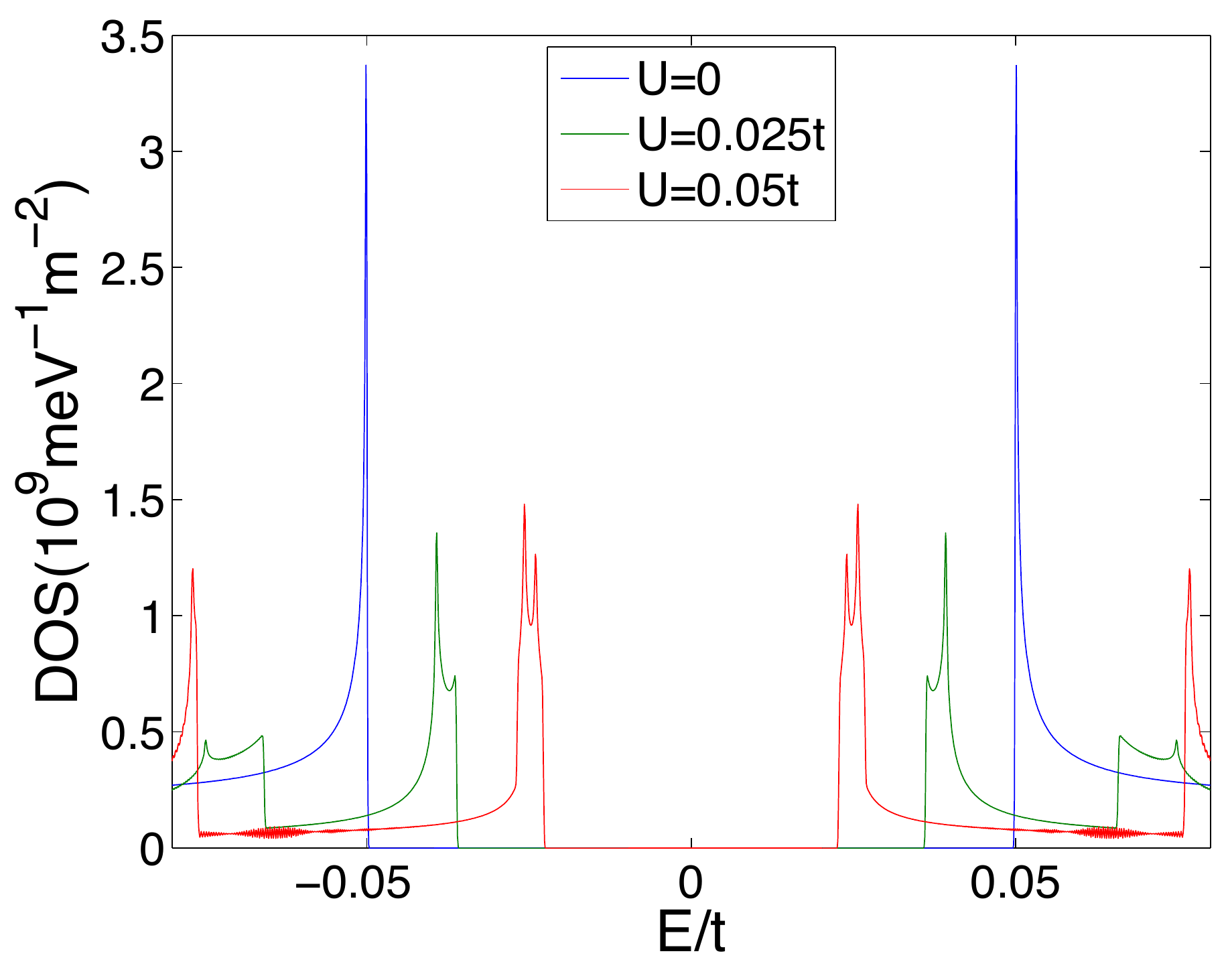} \caption{Density of states for bilayer graphene subject to a uniform bias of $\Delta=0.1t$ and various chemical potential (left) and electric field (right) superlattices with period $\lambda=60d$.} \label{fig4} 
\end{figure}

As a starting point to understanding the expected role of chemical potential and electric field fluctuations, Fig.\ \ref{fig4} shows plots of the density of states of a biased superlattice with periodic 1D modulations. In the absence of a superlattice, the DOS diverges as $1/\sqrt{E}$ at the gap edge arising from the $\sim p^4$ dispersion of modes near the gap edge. We find that both chemical potential or bias modulations, cause low energy subgap modes states in this system. For chemical potential modulations, the subgap states are due to the local shift in the charge neutrality point. At finite temperature, regions with a slightly shifted charge neutrality point will have thermally activated `electron-hole' puddles that contribute to transport. For bias modulations, weak modulations locally enhance/suppress the bandgap, while strong modulations form `topological' states in the bulk along interfaces where the field reverses sign \citep{Martin:2008,Killi:2010, Xavier:2010,Li:2011a}. The energy of these `topological' midgap states $decreases$ for large and dilute fluctuations, as the overlap between edge mode wavefunctions is reduced.

{\it Random} potential fluctuations will have two important effects not captured in our study of periodic modulations. First, it will cause the low energy density of states to broaden, causing further suppression of the bandgap predicted by the periodic modulation. Second, dilute localized `topological' states induced in the bulk by strong random electric field modulations due to charged impurities will contribute to transport --- this is broadly consistent with the temperature dependence of the resistance in biased bilayer graphene \citep{Yan:2010,Taychatanapat:2010,Miyazaki:2010}. 
 
\chapter{Graphene Under Spatially Varying Potentials: Landau Levels, Magnetotransport, and Topological Modes} \label{Chapt:BSL}
\chaptermark{Landau Levels, Magnetotransport, and 1D Modes}
\subsubsection{The material in this section is largely based on the article S.\ Wu, M.\ Killi and A.\ Paramekanti  \textit{Phys.~Rev.~B.} {\bf 85}, 195404 (2012).}

In this chapter, we examine the electronic properties of both bilayer and monolayer graphene 1D superlattices in the presence of a uniform magnetic field perpendicular to the plane.  Inspired by early magnetic field studies of graphene and graphene superlattices \citep{Zhang:2005a, Novoselov:2006, Park:2009}, we show that careful analysis of the Landau levels spectrum and Hall conductivity provides many experimentally accessible signatures of the underlying Dirac cones generated by the superlattices explored in the previous chapter. In addition,  we also carefully consider electric field modulations where the period length is large and decoupled `topological' modes emerge. Although our superlattice results are most relevant to graphene with chiral low energy excitations, the effects explored in this chapter may also find interesting counterparts in low density two dimensional electron gasses in semiconductor heterostructures, in which a honeycomb pattern has been successfully imprinted, with a possibility of other kinds of superlattice patterns imprinted in the future \citep{Singha:2011}.          

We obtain the following key results for 1D superlattices in monolayer graphene and bilayer graphene in a perpendicular (orbital) magnetic field. (1) We show that transport in a weak magnetic field can not only be used to probe for the presence of extra Dirac points, but may also be used, in clean systems, to obtain further information about their velocities via the energy and degeneracy of higher Landau levels. (2) We show that a moderate magnetic field in monolayer graphene leads to a dramatic reversal of the transport anisotropy generated by the superlattice potential alone, an effect arising from the superlattice induced dispersion of the zero energy Landau level. This field tunable transport anisotropy may find useful applications in monolayer graphene superlattices. For example, one can use the change of the anisotropy as an on/off switch and even perform bit or gate operations with these superlattices. However, this field tunable transport anisotropy is found to be absent in bilayer graphene. (3) We consider the Landau levels in bilayer graphene in the presence of a uniform bias, and a kink in the bias that leads to 1D topologically bound states, and the coupling between such modes in an array of kinks. We also discuss the real space structure of the Landau level wave functions and the local density of states which is expected to be relevant to scanning tunneling spectroscopy (STS) studies such as those described in \citet{Connolly:2012,Rutter:2011}. (4) Finally, we consider possible implications of these results for valley filtering and breakdown of the quantum Hall effect.

For realistic superlattice period, $\lambda=170\, d \sim 48$ nm  with $d=2.8$ \AA\, being the nearest neighbor lattice constant, the field strengths corresponding to weak and intermediate magnetic field are about 0.1 T and 8 T, respectively.  The weak field regime can be brought to more accessible levels by shorting the superlattice to about a $15$ nm, which has been experimentally observed in corrugated samples \citep{Yan:2012a}.  Also, for the superlattice period of $\lambda=170 d$, the minimum superlattice potential required to generate three Dirac points in one valley is about $200$ meV. All these values are experimentally viable and the interesting physics described in points (1) and (2) above should be accessible.

The organization of this chapter is as follows: Section \ref{section:mono} discusses 1D superlattices in monolayer graphene in a magnetic field. We begin this section with a brief description of the band dispersion of 1D superlattices in the absence of a magnetic field.  In the next part, Section \ref{section:LL}, we introduce the low energy effective Hamiltonian and then identify the following three magnetic field regimes that lead to very different electronic dispersions: (i) weak, (ii) intermediate and (iii) high fields. We present the Landau level dispersions of these three regimes in Section~\ref{section:weak}. We then present the diagonal and Hall conductivity in Section~\ref{section:conductivity}. We find that the diagonal conductivity shows strong anisotropy reversal and the Hall conductivity no longer exhibits plateaus when the magnetic field is tuned from weak to intermediate strength. 

Section \ref{section:bilayer} of this paper examines 1D superlattices in bilayer graphene and begins with a description of the low energy theory that describes both chemical potential and electric field superlattices. We discuss the Landau level dispersion and transport for 1D chemical potential and electric field superlattices in Sections~\ref{section:chemical} and \ref{section:electric}, respectively.

Then in the Section~\ref{section:realspace}, we consider the real space picture of interlayer modulations with dilute kinks and antikinks that form decoupled soliton modes. We first provide an overview of the Landau level spectrum of uniformly biased bilayer graphene in Section~\ref{section:uniform} and then compare it to the spectrum generated when there is a single isolated antikink in the interlayer bias in the beginning of Section~\ref{section:kink}. Also in this section, we present the tunneling current and provide key signatures that could identify the presence of the kink in scanning tunnelling measurements and suggest how the occurrence of disorder induced kinks may contribute to the breakdown of the quantum Hall effect. Following the discussion of a single isolated interlayer bias kink, we discuss a periodic array of kink and antikinks in Section~\ref{section:array}. We argue that for $\ell_B<\lambda$, it is only necessary to consider a single kink-antikink pair and observe that regardless of the magnetic field strength, the low energy physics is governed by the soliton modes, which remain robust. Employing a simple low energy effective theory to describe the soliton modes in a magnetic field, we gain further insights into the low energy physics of the superlattice and discuss why the zero energy Landau levels are robust from this perspective. Finally, in Section~\ref{section:valley} we consider the coupling between the soliton states generated by a kink-antikink pair and how this leads to a single 1D one-way conducting band with definite valley index in each kink separately. 

\section{Monolayer graphene superlattices}\label{section:mono} 
\begin{figure}
	[tb] 
	\includegraphics[width=.47
	\textwidth]{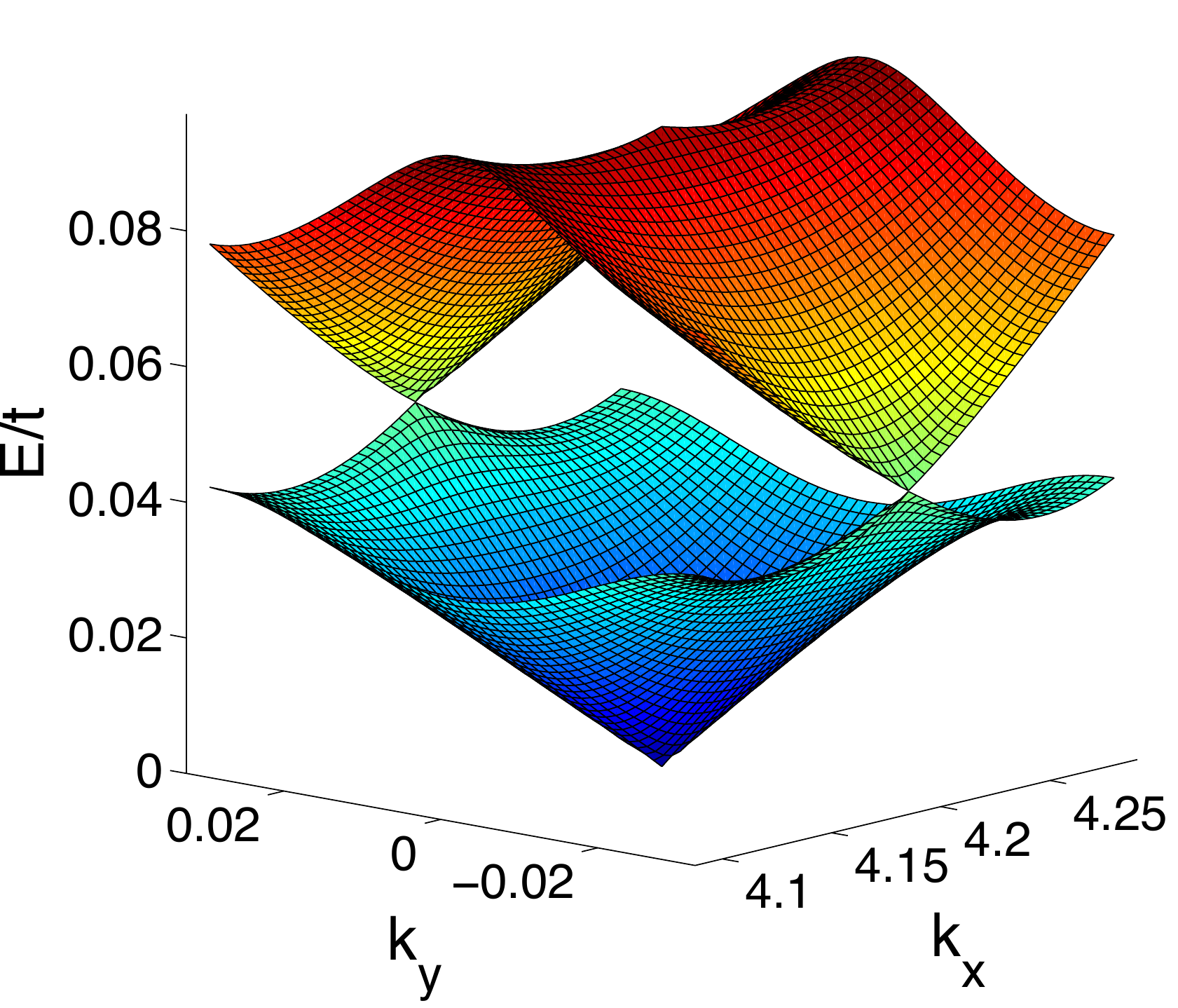} 
	\includegraphics[width=.47
	\textwidth]{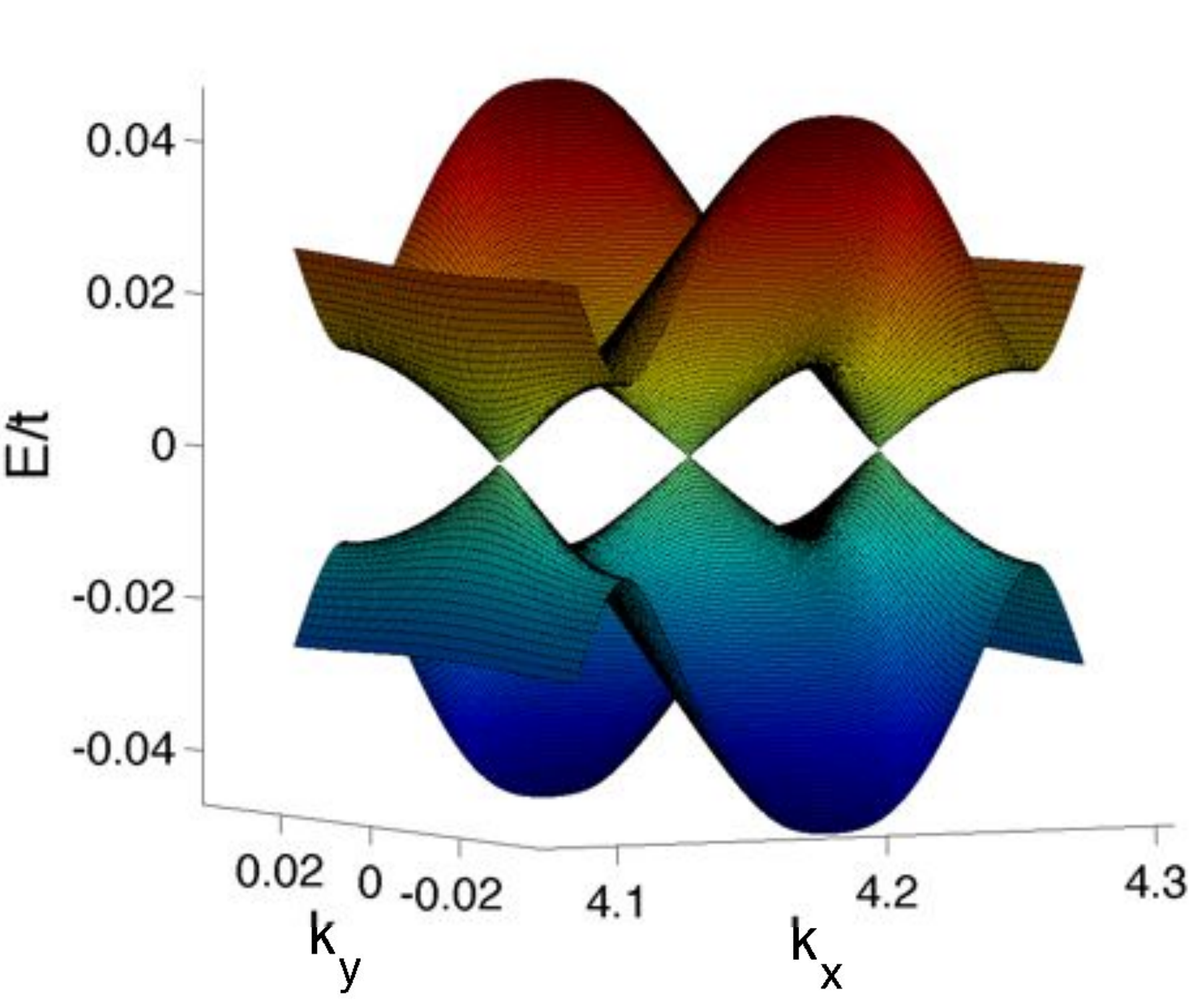} \caption{Energy spectrum of monolayer graphene superlattices with different (dimensionless) superlattice strengths, $\tilde{U}=1$ (left) and $\tilde{U}=3$ (right). For $\tilde{U}=1$, there is only one anisotropic zero energy Dirac cone present in the spectrum, while for $\tilde{U}=3$, two additional Dirac cones are generated in the direction perpendicular to superlattice. For even stronger superlattice potential, more Dirac cones will emerge in pairs.  The superlattice period length is $\lambda=60d$ in both figures.} \label{GSL} 
\end{figure}

\subsection{Band structure of monolayer graphene superlattices}

It is useful to briefly review the effect of the superlattice potential on the low energy spectrum near $\bK$ at zero magnetic field \citep{Park:2009,Brey:2009,Barbier:2010a,Barbier:2010} before we present the superlattice spectrum in a nonzero magnetic field (see Fig. \ref{GSL}). For $U=0$, where $U$ is the superlattice strength, the spectrum consists of a single Dirac point at $\bK$, with an isotropic velocity at low energy. The first effect of turning on a nonzero $U$ is to make this Dirac cone anisotropic, with $v_y = v_f$, but $v_x < v_f$. For $U = 4\pi v_f/\lambda$, with $\lambda$ as the superlattice period, one finds $v_x = 0$. Further increasing $U$, with $4\pi v_f/\lambda < U < 8\pi v_f/\lambda$, leads to {\it three} zero energy Dirac points: one of these continues to be located at $\bK$, while two new Dirac points emerge, which are symmetrically split off from the $\bK$ point along the $\pm x$-direction. All three Dirac cones have anisotropic velocities \citep{Barbier:2010a}. It is convenient to measure the strength of the superlattice potential in terms of the dimensionless parameter $\tilde{U}$, defined by $U \equiv (2 \pi v_f/\lambda) \tilde{U} $, so tha  $0 < \tilde{U} < 4$ leads to the different spectra with additional Dirac cones. Further increasing $\tilde{U}$ leads to even more complex spectra with even more Dirac points. 
                          
\subsection{Landau level spectrum}\label{section:LL}

In this section and for the remainder of this chapter, we will ignore electron spin as the Zeeman splitting of the Landau levels is negligible. As detailed in Section~\ref{Section:TBMLG}, the low energy Hamiltonian of monolayer graphene is given by a $2\times 2$ matrix at each valley, $ \hat{H}_0 \! = \! v_f (s \pi_x \sigma_x - \pi_y \sigma_y)$, where pseudospin $\sigma_z=\pm 1$ labels the two sublattices, while the two (decoupled) valleys at $\pm \bK=\pm 4\pi \hat{x}/3$ are labelled by $s \! = \! \pm 1$. Here, $v_f \! = \! \sqrt{3}td/2$ is the isotropic Fermi velocity, where $d=\sqrt{3}a \! = \! 2.84$ \AA\, is the primitive lattice spacing, $a$ is the nearest carbon-carbon distance, and $t \! = \! 3$ eV is the transfer integral. (We set $\hbar \! =\! 1$ and $d=1$ for convenience.)
In a uniform perpendicular magnetic field, $\pi_j \! = \! -i \nabla_j \! - \! e A_j$; for ${\bf B} \! =\! - B \hat{z}$, and in the Landau gauge, the vector potential ${\bf A} \! = \! By\hat{x}$.

Diagonalizing $\hat{H}_0$, we find the Landau level spectrum $\varepsilon_n={\rm sgn}(n)\sqrt{|n|}\omega_c$, where $\omega_c \! =\! \sqrt{2}v_F/\ell_B$, with $\ell_B \! =\! 1/\sqrt{eB}$. For $s=+1$ (i.e., at valley ${\bK}$), the $n \! \neq \! 0$ eigenfunctions are given by 
\begin{equation}
	\phi_{n,k,+} (x,y)=\frac{e^{ i k x}}{\sqrt{2 L}}\left( 
	\begin{array}{c}
		\psi_{|n|,k}(y) \\
		- {\rm sgn}(n) \psi_{|n|-1,k}(y) 
	\end{array}
	\right), \label{eigen1} 
\end{equation}
where $L$ and $k$ are the system length and electron momentum deviation from $\bK$, both along the $x$-direction, while for $n=0$, 
\begin{equation}
	\phi_{0,k,+} (x,y)=\frac{e^{ikx}}{\sqrt{L}} \left( 
	\begin{array}{c}
		\psi_{0,k}(y) \\
		0 
	\end{array}
	\right). \label{eigen2} 
\end{equation}
Here, $\psi_{n,k}(y)$ is the n-th eigenstate of a (shifted) 1D harmonic oscillator, 
\begin{eqnarray}
	\psi_{n,k}(y)&=&\frac{1}{\sqrt{2^{n}n!\sqrt{\pi}\ell_B}}{\rm exp} \left[-\frac{1}{2}\left(\frac{y-y_0}{\ell_B}\right)^2\right] \times H_{n}\left(\frac{y-y_0}{\ell_B}\right), 
\end{eqnarray}
centered at $y_0 \! = \! k\ell_B^2$, and $H_n$ are Hermite polynomials. For $s\!=\!-1$ (i.e., at $- \bK$), the eigenfunctions are given by $\phi_{n,k,-}(x,y) \!=\! -i \sigma_y \phi_{n,k,+}(x,y)$. The entire low energy Landau levels of monolayer graphene are thus $\phi_{n,k,\pm} (x,y) {\rm e}^{\pm i K_x x}$.

We now introduce a periodic 1D chemical potential modulation $V(y)$, with period $\lambda \! \gg \! d$, and study its effect on the Landau levels.  We consider the limit where the disorder is weak enough so that the mean-free path is larger than $\lambda$.  The set of eigenfunctions $\phi_{n,k,s} (x,y) {\rm e}^{ i s K_x x}$, with $s=\pm 1$, form a convenient basis to study the superlattice Hamiltonian in a magnetic field. (This basis choice is different from the one used by Park, {\it et al.}, and allows us to numerically access a wide range of magnetic fields.) Due to momentum conservation along the $x$-direction, the superlattice Hamiltonian is diagonal in $k$. Further, for $\lambda \gg a$, intervalley scattering is strongly suppressed. We will therefore assume that the two valleys stay completely decoupled. (We focus below on valley ${\bK}$ with $s\!=\! +1$; we expect identical physics around valley $-{\bK}$.) With this approximation, the only effect of the superlattice potential is therefore to induce Landau level mixing.

To be concrete, we choose an explicit form for the superlattice potential. For simplicity, we set $V(y) \! =\! \frac{U}{2}\cos\left(\frac{2\pi y}{\lambda}\right)$.  The results obtained from this specific superlattice potential can easily be generalized to other superlattice profiles by including multiple Fourier components, under the condition that the potential does not vary rapidly over the atomic scale. For numerical computations, the following integral identity\footnotetext[1]{URL 18.17.25 and 18.17.26 from the Digital Library of Mathematical Mathematical Functions at http://dlmf.nist.gov/.} for harmonic oscillator states proves useful: 
\begin{eqnarray}
	&& \int_{-\infty}^{\infty} dy \cos\left(\frac{2\pi y}{\lambda}\right)\psi_{n,k}(y) \psi_{n+m,k}(y) \nonumber \\
	= && \left(\frac{n!}{(n+m)!}\right)^{1/2} \!\! \left(\frac{\sqrt{2}\pi\ell_B}{\lambda}\right)^m \!\! \exp\left(\!-\frac{\pi^2\ell_B^2}{\lambda^2}\! \right) \nonumber\\
	&&\times L_n^{(m)}\left(\frac{2\pi ^2\ell_B^2}{\lambda ^2}\right) \cos\left(\frac{2\pi y_0}{\lambda}+\frac{m\pi}{2}\right), \label{identity} 
\end{eqnarray}
where $m \geq 0$, and $L_n^{(m)}$ is the generalized Laguerre polynomial. Using this identity, we simplify and numerically compute the matrix elements of the full Hamiltonian, retaining up to 3000 Landau levels. With this Hamiltonian, the spectrum of the 1D superlattice  in a magnetic field can be numerically calculated. 

Since the superlattice introduces an additional length scale other than the magnetic length, it is necessary to consider various regimes of magnetic field strength.   Within the continuum limit, the results below do not change when $\lambda$ is scaled as long as the corresponding magnetic field and superlattice strength is similarity rescaled.  It should be noted that lattice effects will take place for short periods and when the orientation of the superlattice is along the armchair direction where the Dirac cones are no longer separated in momentum space \citep{Pal:2012a}.  We identified three  regimes for $\tilde{U} \sim {\cal O}(1)$ that lead to qualitatively different electronic structure:

(i) {\bf Weak field}: This regime corresponds to having $\omega_c \ll U$, where the Landau level spacing is much smaller than the superlattice amplitude, so that $2 \pi \ell_B/\lambda \gg 1$. In this regime, the magnetic field may be viewed as effectively `probing' the zero field superlattice excitations.

(ii) {\bf Intermediate field}: In this regime, $\omega_c \sim U$, which means $2\pi \ell_B/\lambda \sim 1$, so that the superlattice potential and the magnetic field have to be treated on equal footing.

(iii) {\bf Strong field}: Here, $ \omega_c \gg U$ or, equivalently, $2\pi \ell_B/\lambda \ll 1$. In this regime, the superlattice potential only weakly perturbs the Landau levels of pristine graphene.

\begin{figure}
	[h!] \centering 
	\includegraphics[width=.9 
	\textwidth]{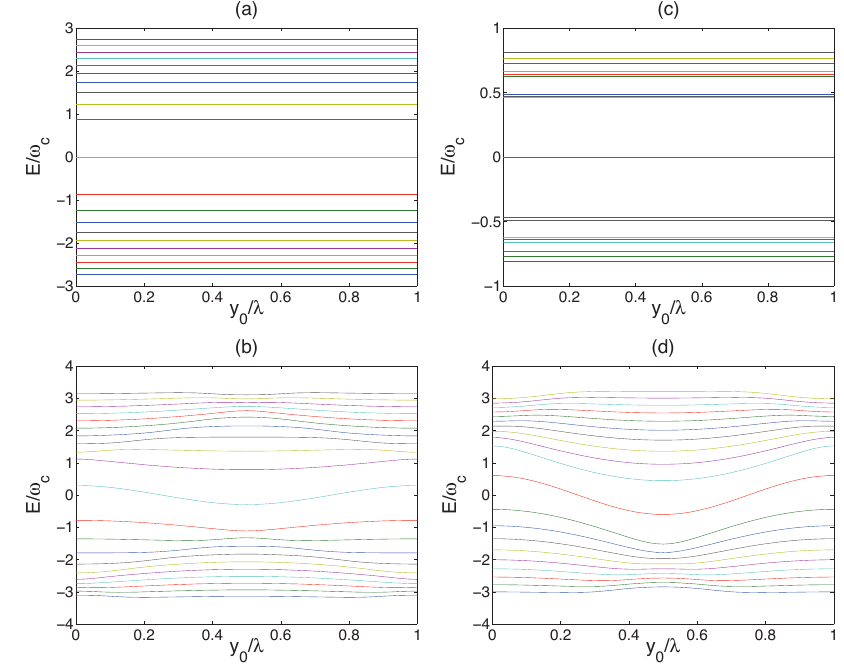} \caption{Landau levels of monolayer graphene superlattice for different (dimensionless) superlattice strengths $\tilde{U}$, and magnetic fields $B$. The spectrum is shown for weak field ($\ell_B=2\lambda$, top panels) and intermediate field ($\ell_B=0.2\lambda$, bottom panels). Left panels (a,b) correspond to $\tilde{U}=1$ which supports a single anisotropic zero energy massless Dirac fermion. Right panels (c,d) correspond to $\tilde{U}=3$ which supports three zero energy massless Dirac fermions with anisotropic velocities --- the weak field zero energy Landau level thus has three times as many states for $\tilde{U}=3$ as it does for $\tilde{U}=1$, while the $n=\pm 1, \pm 2$ levels have degeneracy splitting in weak field due to the Dirac fermions having two different mean velocities. For $\ell_B \ll \lambda$ (not shown), the Landau levels closely resemble that of pristine graphene. See text for a detailed discussion of the Landau level structure.} \label{GSL_LL} 
\end{figure}

Fig.~\ref{GSL_LL} shows the spectrum of the graphene superlattice in different field regimes for superlattice strengths $U\!\!=\!\!2\pi v_f/\lambda$ (or $\tilde{U}=1$) and $6\pi v_f/\lambda$ (or $\tilde{U}=3$). This allows us to contrast the behaviour of the spectrum of the superlattice in a magnetic field without or with extra Dirac points being present at zero field, and to explore consequences for quantum Hall physics and transport.

\subsubsection{Weak field regime} \label{section:weak}

When the magnetic field is weak, $\ell_B=2\lambda$ (top panels in Fig. \ref{GSL_LL}), we find that the energy spectrum barely depends on the value of $k$, or equivalently, $y_0$. This is due to the fact that when magnetic length $\ell_B$ is larger than the superlattice period $\lambda$, the matrix elements of the Hamiltonian do not depend on the center of the Landau wavefunctions, and therefore produce flat bands that are $k$ independent. In other words, in this regime the magnetic field may be viewed as effectively `probing' the structure of the zero field superlattice dispersion and forms Landau levels that depend on the details of the superlattice spectrum at low energy (i.e.~the Dirac cones).

For $\tilde{U}=3$, the low energy spectrum of the superlattice contains three anisotropic Dirac points at zero energy, generating zero energy Landau levels with three times the degeneracy of that when $\tilde{U}=1$. Further, the Dirac cone centred at $\bK$ has a slightly different average velocity $\sqrt{v_x v_y}$ compared to the other two cones that are symmetrically split off from $\bK$ along $\pm \hat{x}$. This unique cone results in the Landau levels at nonzero energy becoming weakly split, as is most clearly seen for the first two excited levels (at positive or negative energy, i.e., with $n=\pm 1, \pm 2$). We numerically determined $v_x$ and $v_y$ for each of the three Dirac points and found good agreement between the energy levels obtained on this basis of having Dirac fermions with two different average velocities, and those obtained numerically.

At higher energies, $E/\omega_c \gtrsim 2$ for $\tilde{U}=1$ or $E/\omega_c \gtrsim 1$ for $\tilde{U}=3$, the spectrum begins to deviate from this simple behaviour expected for a linear Dirac spectrum. This deviation is expected from the curvature of the dispersion, which appears upon going beyond the linearized approximation.

\subsubsection{Intermediate field regime} \label{section:inter}

At intermediate fields when $\ell_B=0.2\lambda$, the spectrum at low energy can most simply be understood as resulting from the superlattice potential invoking a strong dispersion to the Landau levels. In simple terms, if we assume that the state labelled by momentum $k$, or equivalently position $y_0$, has an energy that is modulated by the superlattice potential, we expect a periodic modulation of of the level with period $\lambda$ and an amplitude proportional to the superlattice amplitude $U$. The features of the low energy Landau levels, $n=0,\pm 1, \pm 2$, as seen from the lower panels in Fig.~\ref{GSL_LL}, are consistent with this scenario, as can be seen by the periodic modulation tracing the $\cos(2\pi y/\lambda)$ form of the superlattice potential and the modulation for $\tilde{U}=3$ being roughly thrice as strong as the modulation for $\tilde{U}=1$. We also observe that for $\tilde{U}=3$ the low energy Landau levels overlap with each other. This has nontrivial effect on the DC conductivity, as shown in the following subsection. For higher energy Landau levels, the energy spectrum maintains a periodic modulation but no longer retains the simple cosine form. As the energy gets higher, the distribution of Landau levels becomes more dense and the energy difference between two adjacent levels is now comparable to the matrix element of superlattice potential. Here, a simple first order perturbation correction is not enough to account for the dispersion and second order perturbation from adjacent levels must be taken in account.

\subsubsection{High field regime} \label{section:high}

For very strong magnetic fields, the intrinsic Landau level structure of graphene is recovered. Here, only one zero energy level exists at the Dirac point (ignoring valley and spin degeneracy), and the energy gap follows the familiar square root relation. With such a strong magnetic field, the superlattice is merely a small perturbation and can only give rise to a slight modulation of the Landau levels following our previous argument for intermediate field. From a perturbative point of view, the energy corrections up to first order are given by 
\begin{equation}
	\Delta E^{(1)}=\int dy \phi_n^*(k,y)V(y)\phi_n(k,y), 
\end{equation}
which, upon using the identity (\ref{identity}), depends sinusoidally on the center position of Landau wavefunctions. Thus, even in a strong magnetic field the energy spectrum is not strictly dispersionless, but again traces the superlattice modulation. However, it should be kept in mind that the ratio of the amplitude of this modulation to the Landau level spacing, $U/\omega_c$, is extremely small in the high field regime. This dispersion, though small, can give rise to interesting magnetoresistance oscillations known as Weiss oscillations, in addition to the usual Shubnikov-de Hass oscillations \citep{Matulis:2007}. \citet{Matulis:2007} showed that in comparison to two-dimensional electron gases with a parabolic dispersion relation, the Weiss oscillations that are expected in graphene superlattices are much more pronounced and are more robust against temperature damping than in typical materials. This can be attributed to the different Fermi velocities of Dirac and normal electrons at the same chemical potential \citep{Matulis:2007}. Similarly, Weiss oscillations in bilayer graphene chemical potential superlattices have also been discussed by \citet{Tahir:2008}.

\subsection{Diagonal and Hall conductivity in monolayer graphene superlattices} \label{section:conductivity}

Once we have the eigenvalues and eigenfunctions for the superlattice in a perpendicular magnetic field, both the AC and DC conductivities can be calculated directly by the Kubo formula \citep{Akkermans:2007}, 
\begin{eqnarray}
	\sigma_{ij}(\omega)&=&\frac{iv_F^2}{\lambda\ell_B^2}\int_0^{\lambda}dy_0\sum_{\alpha,\beta} \frac{f(E_{\alpha})-f(E_{\beta})}{E_{\alpha}-E_{\beta}}\nonumber\\
	&&\times \frac{\langle\alpha k |v_i|\beta k\rangle\langle\beta k |v_j|\alpha k\rangle} {E_{\alpha}-E_{\beta}-\omega-i\gamma}. \label{Kubo} 
\end{eqnarray}
Here, we have set $\gamma=10^{-3}\times \omega_{c}(B=1 \rm T)$ to be the Landau level broadening and measure the conductivities in units of $e^2/h$, $E_{\alpha}(y_0)$ and $|\alpha k\rangle$ are the $\alpha$-th eigenvalue and the corresponding eigenstate of the system, which can be expanded in the basis of $|nk\rangle$, with $y_0=k\ell_B^2$ and $\phi_n(k,y)=\langle y|nk\rangle$ being the Landau level wavefunctions for pristine graphene. Continuing, $f(E)$ is the Fermi-Dirac distribution, and $v_i$ is the velocity operator in the $\hat{i}$-direction given by $v_i=v_F\sigma_i$, where $\sigma_i$ is the corresponding Pauli matrix. Using Eq. (\ref{eigen1}) and (\ref{eigen2}), we can obtain the expression for the matrix elements of velocity operators from 
\begin{eqnarray}
	\langle m k |v_x| n k\rangle &=& v_F\big[{\rm sgn}(m)\delta_{|m|-1,|n|}
	 +{\rm sgn}(n)\delta_{|m|,|n|-1}\big], 
\end{eqnarray}
and 
\begin{eqnarray}
	\langle m k |v_y| n k\rangle &=& -iv_F\big[{\rm sgn}(m)\delta_{|m|-1,|n|}
	 -{\rm sgn}(n)\delta_{|m|,|n|-1}\big], 
\end{eqnarray}
by expanding $|\alpha k\rangle$ into $|nk\rangle$. Note that $\langle \alpha k|v_y|\alpha k\rangle=0$ is always true for any state. This is because the wavefunction is always localized in the $\hat{y}$ direction.
\begin{figure}
	[!h] \centering 
	\includegraphics[width=.9
	\textwidth]{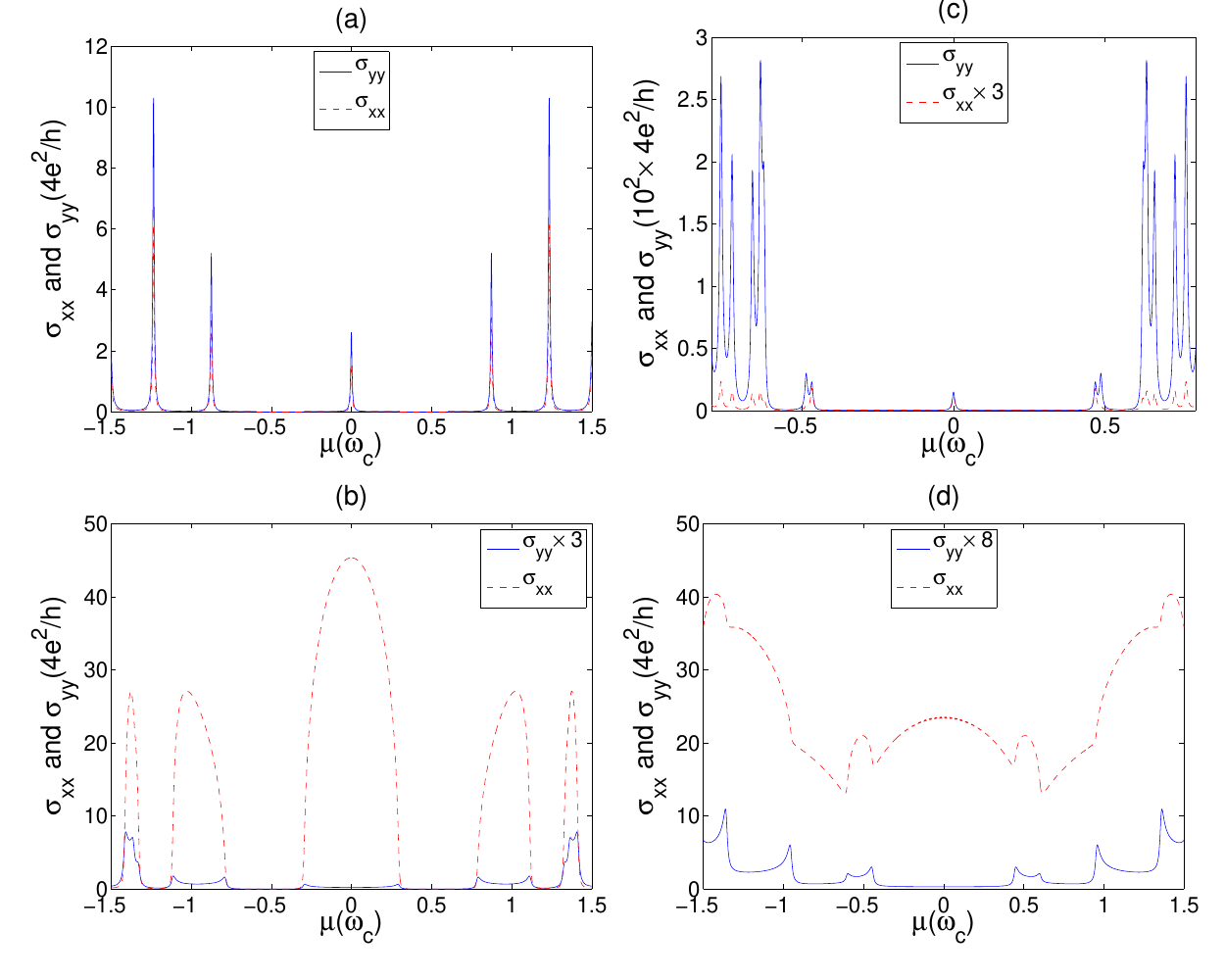} \caption{ Diagonal DC conductivities of monolayer graphene superlattice for different (dimensionless) superlattice strengths $\tilde{U}$, and magnetic fields $B$. The conductivity is shown for weak field ($\ell_B=2\lambda$, top panels) and intermediate field ($\ell_B=0.2\lambda$, bottom panels). Left panels (a,b) correspond to $\tilde{U}=1$, and right panels (c,d) correspond to $\tilde{U}=3$. The conductivities show strong anisotropy when magnetic field strength is varied --- for weak field (a,c), $\sigma_{yy}$ is larger than $\sigma_{xx}$, which is a consequence of the Fermi velocity renormalization in the absence of magnetic field; for moderate field (b,d), the anisotropy is reversed, since $\hat{v}_x$ acquires intra-Landau level contributions, as explained in the text. For $\ell_B \ll \lambda$ (not shown), results for pristine graphene are recovered and the transport becomes isotropic.} \label{diagonal_SLG} 
\end{figure}

Results for the DC diagonal conductivities as a function of chemical potential are shown in Fig. \ref{diagonal_SLG}. This figure was constructed by setting the frequency, $\omega$, to zero in Eq. (\ref{Kubo}), so that only the real part of the conductivity tensor is non-zero. In a weak magnetic field, the conductivities show strong anisotropy with $\sigma_{yy}$  larger than $\sigma_{xx}$. This is a direct consequence of the Fermi velocity renormalization in the absence of a magnetic field (see Fig. \ref{diagonal_SLG} (a) and (c)). The major contribution to the diagonal conductivities derives from the off-diagonal matrix elements, $\langle \alpha k|v_i|\beta k\rangle$ with $\alpha\neq\beta$, since $\langle \alpha k|v_y|\alpha k\rangle=0$ and $\langle \alpha k|v_x|\alpha k\rangle\simeq 0$ due to the flat band structure. Since the matrix elements of $v_y$ are always larger than those of $v_x$, this gives rise to the observed anisotropy in a weak field. 

When the magnetic field strength is at intermediate strengths, the conductivity still displays anisotropy, but now with $\sigma_{xx}$ significantly larger than $\sigma_{yy}$ (see Fig. \ref{diagonal_SLG} (b) and (d)). Here, $v_x$ has acquired a diagonal matrix element, $\langle \alpha k|v_x|\alpha k\rangle= 
\partial E_{\alpha}(y_0=k\ell_B^2)/ 
\partial k\neq 0$ because the energy spectrum has become dispersive and $v_y$ still lacks a sizable contribution. Notice the positions of the conductivity peaks of $\sigma_{yy}$  correspond exactly to the minimum and maximum of the energy band, where the density of states is the largest. For $\sigma_{xx}$, however, the conductivity is at a minimum at the band edge because the average of the velocity operator, $\langle v_x \rangle$, is zero at these energies. Therefore, the intra-Landau level contribution to $\sigma_{xx}$ will be the smallest at the band edge and thus leads to the dips in the conductivity. For a weak superlattice potential, $\sigma_{xx}$ drops to zero when there are no overlapping Landau levels.  However, in a strong superlattice, $\sigma_{xx}$ always shows a dispersive transport property. In a strong magnetic field (not shown here), the Landau levels structure of pristine graphene is recovered and the conductivities become isotropic (see, for example \citet{Ferreira:2011a}). In this case, the superlattice is merely a perturbation to the magnetic field and thus should have only a minor effect on the magnetotransport properties.

\begin{figure}
	[!h] \centering 
	\includegraphics[width=.9 
	\textwidth]{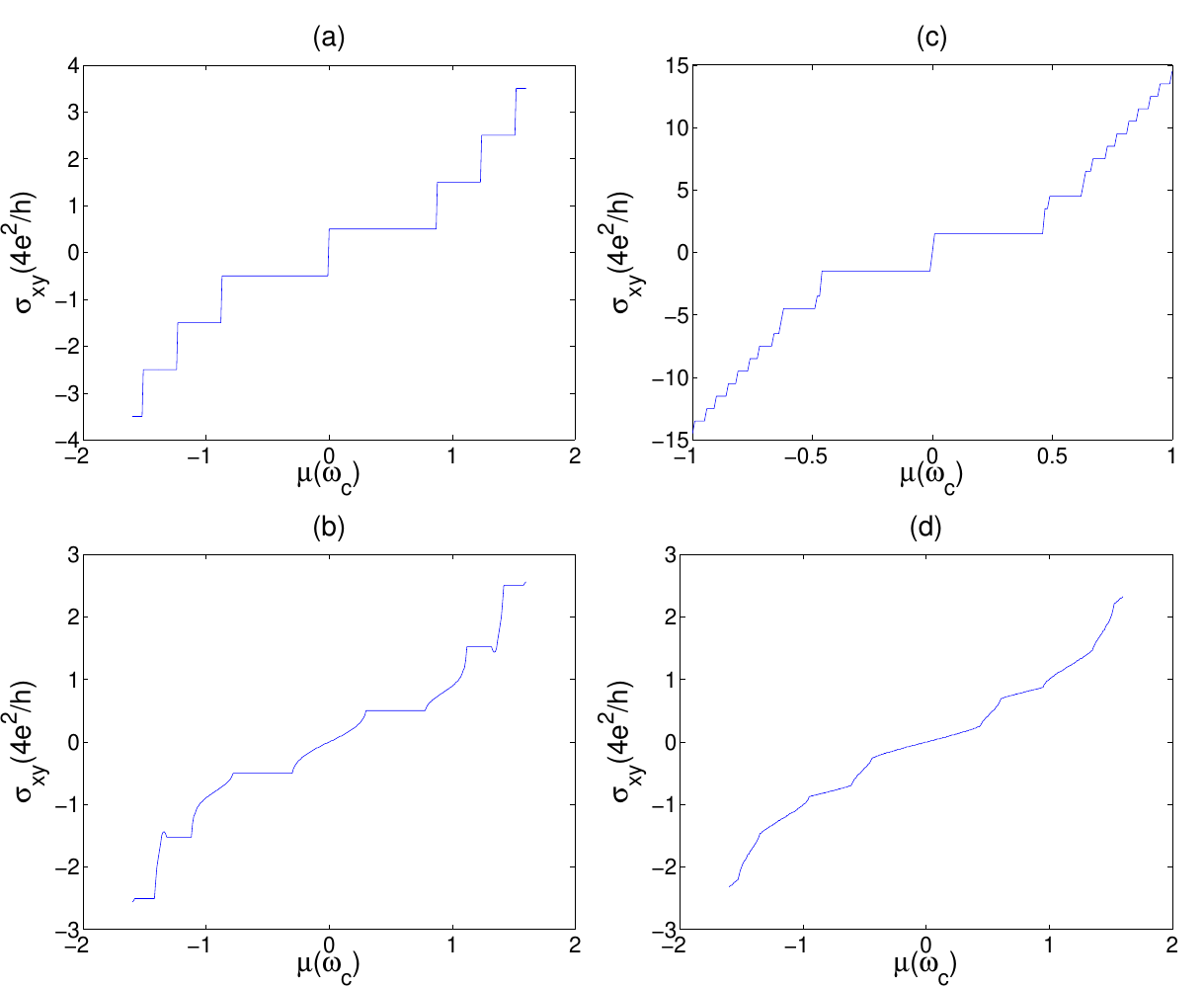} \caption{ The DC Hall conductivity of a monolayer graphene superlattice for different (dimensionless) superlattice strengths $\tilde{U}$, and magnetic fields $B$. The conductivity is shown for weak field ($\ell_B=2\lambda$, top panels) and intermediate field ($\ell_B=0.2\lambda$, bottom panels). Left panels (a,b) correspond to $\tilde{U}=1$, and right panels (c,d) correspond to $\tilde{U}=3$. For weak field (a,c), the Hall conductivity shows well-defined plateaus, as a consequence of the nearly flat energy bands. For an intermediate field (b,d), the bands become dispersive and the Hall conductivity no longer shows step-like structure. However, for a weak superlattice (b), the energy bands are not fully overlapped so that the Hall conductivity still shows small plateaus when the chemical potential falls between two bands.  In this case, the value of $\sigma_{xy}$ changes by one between adjacent steps, as expected from Dirac physics. For $\ell_B \ll \lambda$ (not shown), result for pristine graphene is recovered.} \label{Hall_SLG} 
\end{figure}

The computation of the DC Hall conductivity is shown in Fig. \ref{Hall_SLG}. For a weak magnetic field (Fig. \ref{Hall_SLG} (a) and (c)), the Hall conductivity shows well defined plateaus that reflect the nearly flat energy bands in the spectrum. The values of the Hall conductivity in the vicinity the Dirac points, taking into account of spin and valley degeneracies, are $\pm 1/2(4e^2/h)$ in a weak superlattice ($\tilde{U}=1$) and $\pm 3/2(4e^2/h)$ in a strong superlattice ($\tilde{U}=3$). The latter result resembles the anomalous half integer quantum Hall effect in pristine graphene, but the Hall conductivity triples due to the presence of three Dirac points. Moving away from the Dirac point, we observe quantum Hall plateaus with higher conductivities, and the value increases by one each time the chemical potential crosses another Landau level. For an intermediate magnetic field (Fig. \ref{Hall_SLG} (b) and (d)), there are no longer well defined plateaus, again because of the dispersive energy spectrum. However, for a weak superlattice potential, the Landau levels do not overlap with each other. In this case, if the chemical potential falls between two Landau levels, a small plateau will manifest with the same value expected from Dirac physics. Once again, when the magnetic field becomes strong enough, the Landau level structure for pristine graphene is restored, just as before, and the Hall conductivity shows the the anomalous half integer quantum Hall plateaus.

\begin{figure}
	[h] \centering 
	\includegraphics[width=.9
	\textwidth]{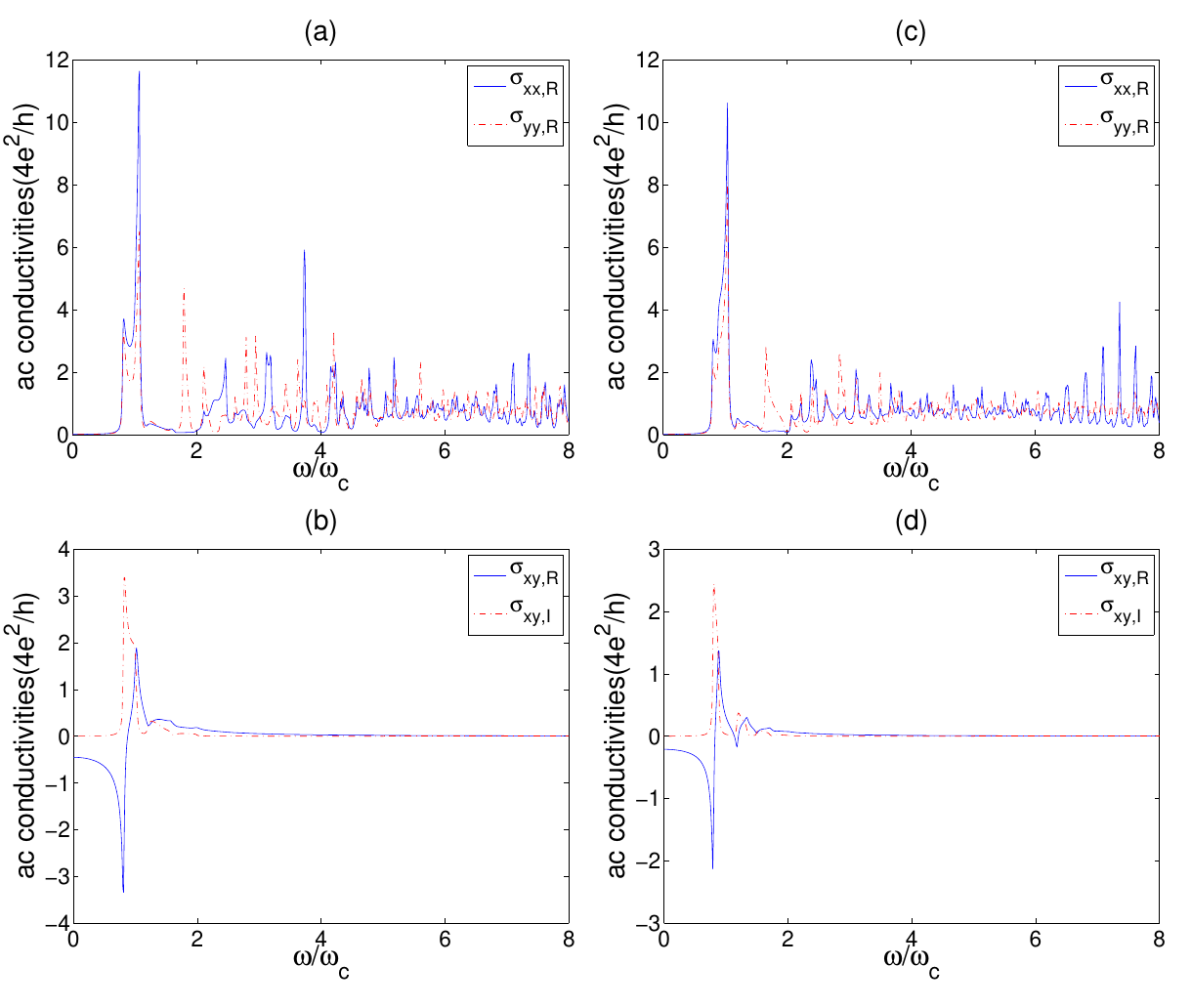} \caption{ The AC conductivity of a monolayer graphene superlattice for different (dimensionless) superlattice strengths $\tilde{U}$, in  an   intermediate magnetic field, $\ell_B=0.2\lambda$, with $\mu=0.2\omega_c$. Left panels (a,b,c) correspond to $\tilde{U}=1$, and the right panels (d,e,f) correspond to $\tilde{U}=3$.} \label{ac_SLG} 
\end{figure}

Fig. \ref{ac_SLG} shows the AC conductivities of graphene superlattices in an intermediate magnetic field. For weak and strong magnetic fields, the results are similar those of pristine graphene \citep{Ferreira:2011a} since, in both cases, the Landau levels are nearly flat and the real part of the conductivities show strong peaks when the photon frequencies exactly correspond to the energy differences between two Landau levels. In an intermediate magnetic field, the results are complicated by the dispersion of Landau levels. At low frequencies, there can be optical transitions in a range of photon energies, and the real part of the diagonal conductivities is at a maximum along the band edge where the density of states is also at a maximum. At higher frequencies, the Landau levels become less dispersive so that sharp peaks begin to emerge. These results can be linked with graphene's unusual magnetooptical properties, for instance, the observed giant Faraday rotation \citep{Ferreira:2011a, Crassee:2011}. While the anisotropy in the diagonal conductivities can lead to anisotropic rotation angles for incident waves with different polarization plane, this effect is actually quite small and hard to observe experimentally.

\section{Bilayer graphene superlattices}\label{section:bilayer}
\begin{figure}
	[!htb] \centering 
	\includegraphics[width=.38
	\textwidth]{L60w30U001m.pdf} 
	\includegraphics[width=.38
	\textwidth]{L60w30U004m.pdf}\\
	\includegraphics[width=.38
	\textwidth]{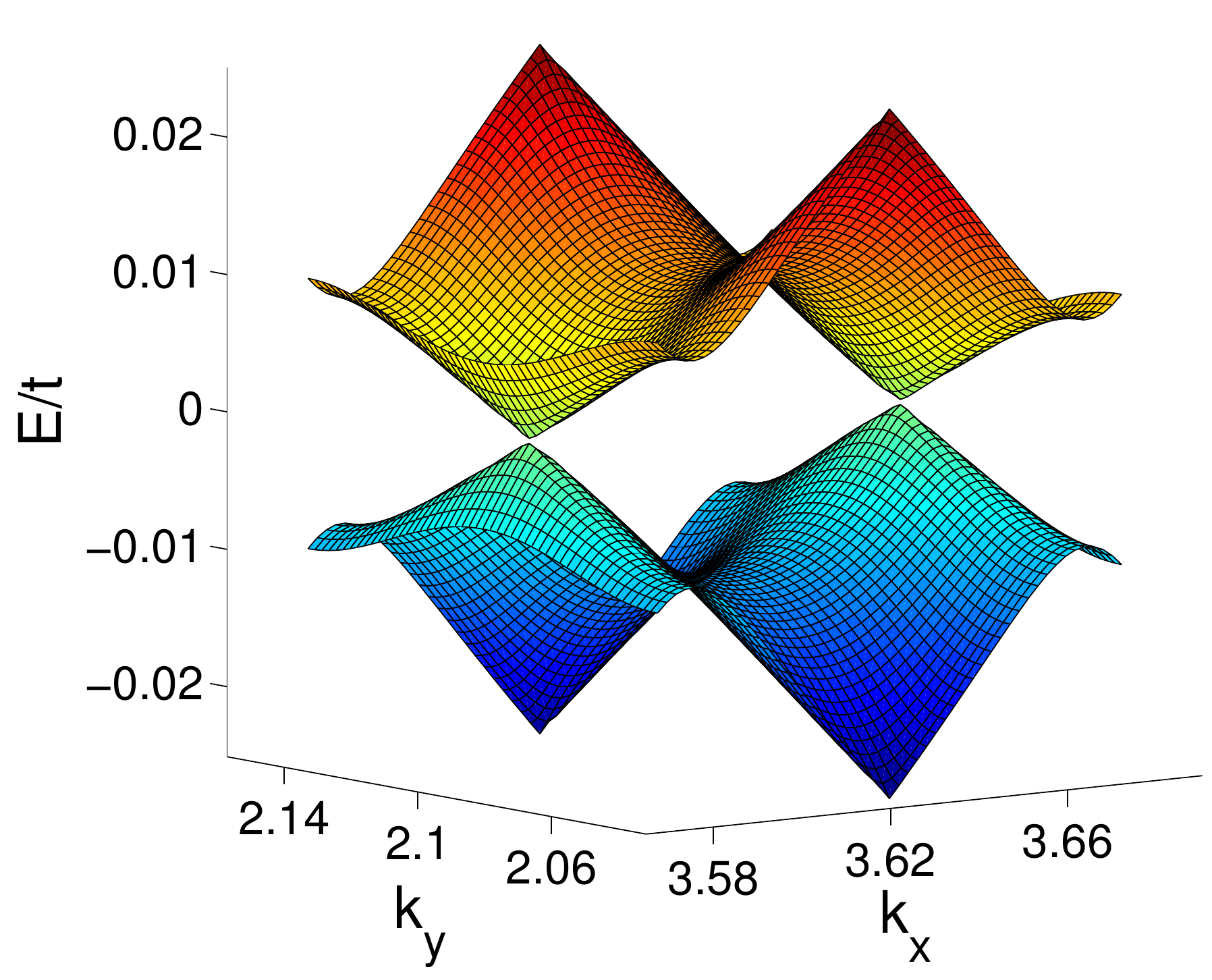} 
	\includegraphics[width=.38
	\textwidth]{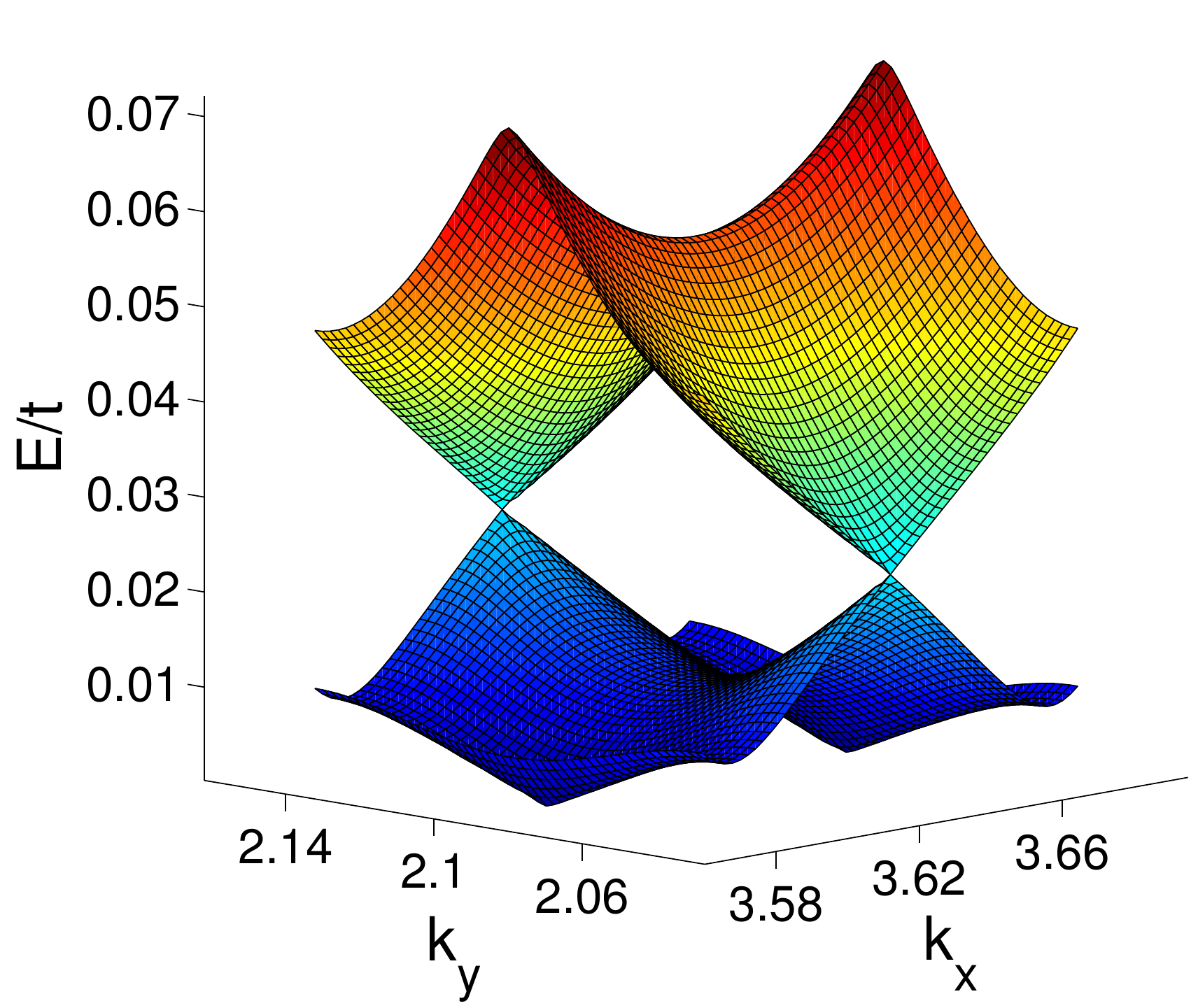} \caption{Band structures for different types of bilayer graphene superlattices with $\lambda=60d$. In a chemical potential superlattice, when the superlattice strength is weak ($V_0=0.01t$, top left), two anisotropic Dirac cones are generated in the spectrum. When the superlattice strength exceeds a critical value, the spectrum becomes gapped ($V_0=0.04t$, top right). In an electric field superlattice, however, there are always two Dirac cones at zero energy ($V_0=0.04t$, bottom left). In addition to the zero energy Dirac cones, there are finite energy Dirac cones (bottom right).} \label{BLGSL} 
\end{figure}

\subsection{Band structure of bilayer graphene superlattices}

Before presenting the results for bilayer graphene superlattices in a perpendicular magnetic field, we briefly review the band structures of bilayer graphene superlattices in the absence of a magnetic field. In bilayer graphene, there are two distinct types of superlattices, namely, chemical potential and electric field superlattices. Other superlattices can be thought as a superposition of these two fundamental types, and can be similarly studied.

In a chemical potential superlattice, the superlattice potentials are exactly the same on each layer. As has been shown by us in previous work \citep{Killi:2011a}, the electron and hole states are completely decoupled along the superlattice direction. Therefore, the superlattice correction to the energy spectrum pushes electron (hole) states down (up) and causes the two quadratic bands cross each other inside the mini Brillouin zone (MBZ). However, in a generic direction, this decoupling is absent and level anti-crossing gives rise to a gapped spectrum, which increases linearly as we alter the angle with respect to the superlattice modulation direction. This then induces the appearance of two Dirac cones. Further increasing  the superlattice strength, the two band crossing points move outside MBZ and the spectrum becomes gapped (see top figures of Fig. \ref{BLGSL}).

In an electric field superlattice, the situation is drastically different. Now, the superlattice potentials on the two layers are exactly opposite in sign, which can effectively be viewed as a periodic arrangement of potential kinks and anti-kinks. Earlier results have indicated that two topological 1D modes will be confined to each kink or anti-kink \citep{Killi:2010,Martin:2008,Xavier:2010}. These topological modes propagate perpendicular to the superlattice direction and have opposite velocities at a kink and an antikink. When one has such modes arranged periodically, they will couple with each other and zero energy and finite energy Dirac cones will appear, as explained by an effective ``wire'' model in \citet{Killi:2011a} (see bottom figures of Fig. \ref{BLGSL}).

\subsection{Magnetic properties of bilayer graphene superlattices}

From the previous section, we can see that the phenomenon resulting from the existence of Dirac points is most manifest when the magnetic field is weak compared to the superlattice strength. A very special feature of bilayer graphene superlattices is that anisotropic Dirac points can be generated in the energy spectrum, which replace the original quadratic band touching points. Our numerical results also indicate that in an intermediate magnetic field, the Landau levels are only slightly dispersive. Therefore, in this section, we will focus on the properties of bilayer graphene superlattices in weak magnetic fields using a two-band effective Hamiltonian. We hope that our results can provide means to the determination and characterization of these anisotropic Dirac points. 
\begin{figure}
	[tb] \centering 
	\includegraphics[width=.85 
	\textwidth]{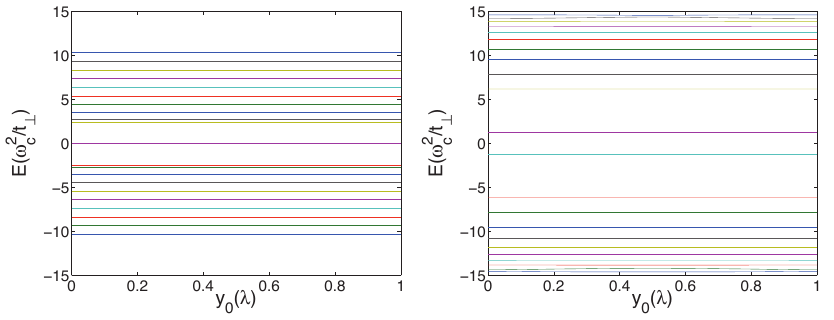} \caption{Energy spectrum of bilayer graphene subject to a chemical potential superlattice and a weak perpendicular magnetic field ($\ell_B=2\lambda$). For both panels, $\lambda=60d$. In the left panel, $V_0=0.01t$, two anisotropic Dirac points are generated in the spectrum in the absence of magnetic field, and the zero energy Landau levels are doubly degenerate. In the right panel, $V_0=0.04t$, no Dirac points are present, and no zero energy Landau level exists. In both panels, the energy bands are nearly flat. For both intermediate fields (not shown here), the energy bands are all slightly dispersive, in contrast to    superlattices where the energy bands overlap. This can be seen from the diagonal conductivities in Fig. \ref{BLGSL_diagonal} } \label{BLGSL_LL} 
\end{figure}

The low energy physics of bilayer graphene near the Brillouin zone corners at $\pm \bK$ is, in the presence of a perpendicular magnetic field, described by the following two-band effective Hamiltonian, 
\begin{equation}
	\hat{H}_0=-\frac{v_F^2}{t_{\perp}}\left( 
	\begin{array}{cc}
		0 & (s \pi_x + i\pi_y)^2 \\
		(s \pi_x - i\pi_y)^2 & 0 
	\end{array}
	\right), 
\end{equation}
with $s=\pm 1$ corresponding to $\pm \bK$ as in the case of monolayer graphene. Similar to the single layer case, we have replaced the momentum operator with its canonical counterpart to take into account of the magnetic field effect. In the following, we will work with the same gauge choice, $A=By\hat{x}$. For $s=+1$, the eigenvalues and the corresponding eigenvectors of the above Hamiltonian are 
\begin{eqnarray}
	&&\varepsilon_n={\rm sgn}(n)\sqrt{|n|(|n|-1)}\omega_c^2/t_{\perp},\nonumber\\
	&&\phi_{n,k,+}(x,y)=\frac{e^{i k x}}{\sqrt{2 L}} \left( 
	\begin{array}{c}
		\psi_{|n|,k}(y) \\
		- {\rm sgn}(n)\psi_{|n|-2,k}(y) 
	\end{array}
	\right), \label{Eq10} 
\end{eqnarray}
with $|n| \geq 2$. In addition, there are two zero energy solutions, 
\begin{eqnarray}
	&&\varepsilon_1=0,\ \ \ \ \ \phi_{1,k,+} (x,y)=\frac{e^{ i k x}}{\sqrt{L}}\left( 
	\begin{array}{c}
		\psi_{1,k} (y) \\
		0 
	\end{array}
	\right), \nonumber\\
	&&\varepsilon_0=0,\ \ \ \ \ \phi_{0,k,+} (x,y)=\frac{e^{ ikx}}{\sqrt{L}}\left( 
	\begin{array}{c}
		\psi_{0,k} (y) \\
		0 
	\end{array}
	\right). 
\end{eqnarray}
For $s=-1$, the corresponding eigenvectors are given by $\phi_{n,k,-} (x,y) = \sigma_x \phi_{n,k,+} (x,y)$. The full low energy Landau level wavefunctions thus take the form $\phi_{n,k,\pm} {\rm e}^{\pm i K_x x}$. and these serve as a good basis to study the magnetic field effect of bilayer graphene superlattices. 

The superlattice can be modelled by assuming spatially varying potential profiles on each layer, 
\begin{equation}
	\hat{H}_{sl}=\left( 
	\begin{array}{cc}
		V_1({\bf x}) & 0 \\
		0 & V_2({\bf x}) 
	\end{array}
	\right), 
\end{equation}
where different combinations of $V_1$ and $V_2$ give rise to different types of superlattice, e.g., $V_1({\bf x})=V_2({\bf x})$ for chemical potential superlattices and $V_1({\bf x})=-V_2({\bf x})$ for electric field superlattices. In the following, we will only consider 1D superlattices, where for certain choices of the superlattice strength, various anisotropic Dirac points can be generated in the spectrum.

\subsubsection{Chemical potential superlattices} \label{section:chemical}

For 1D chemical potential superlattices, we take $V_1({\bf x})=V_2({\bf x})=V(y)=\frac{V_0}{2}\cos(2\pi y/\lambda)$. In previous work \citep{Killi:2011a}, we have shown that for 1D chemical potential superlattice, two anisotropic Dirac points derive from the original quadratic band touching points as long as the superlattice strength is smaller than a critical value. This results from the chiral symmetry, which protects the level crossing in the superlattice direction and results in level anti-crossings in the other directions. Once the superlattice strength increases beyond this critical value, Bragg reflection will open up a gap at the energy spectrum at the mini Brillouin zone boundary. This behaviour has also been found by \citet{Tan:2011}. 
\begin{figure}
	[!h] \centering 
	\includegraphics[width=.9 
	\textwidth]{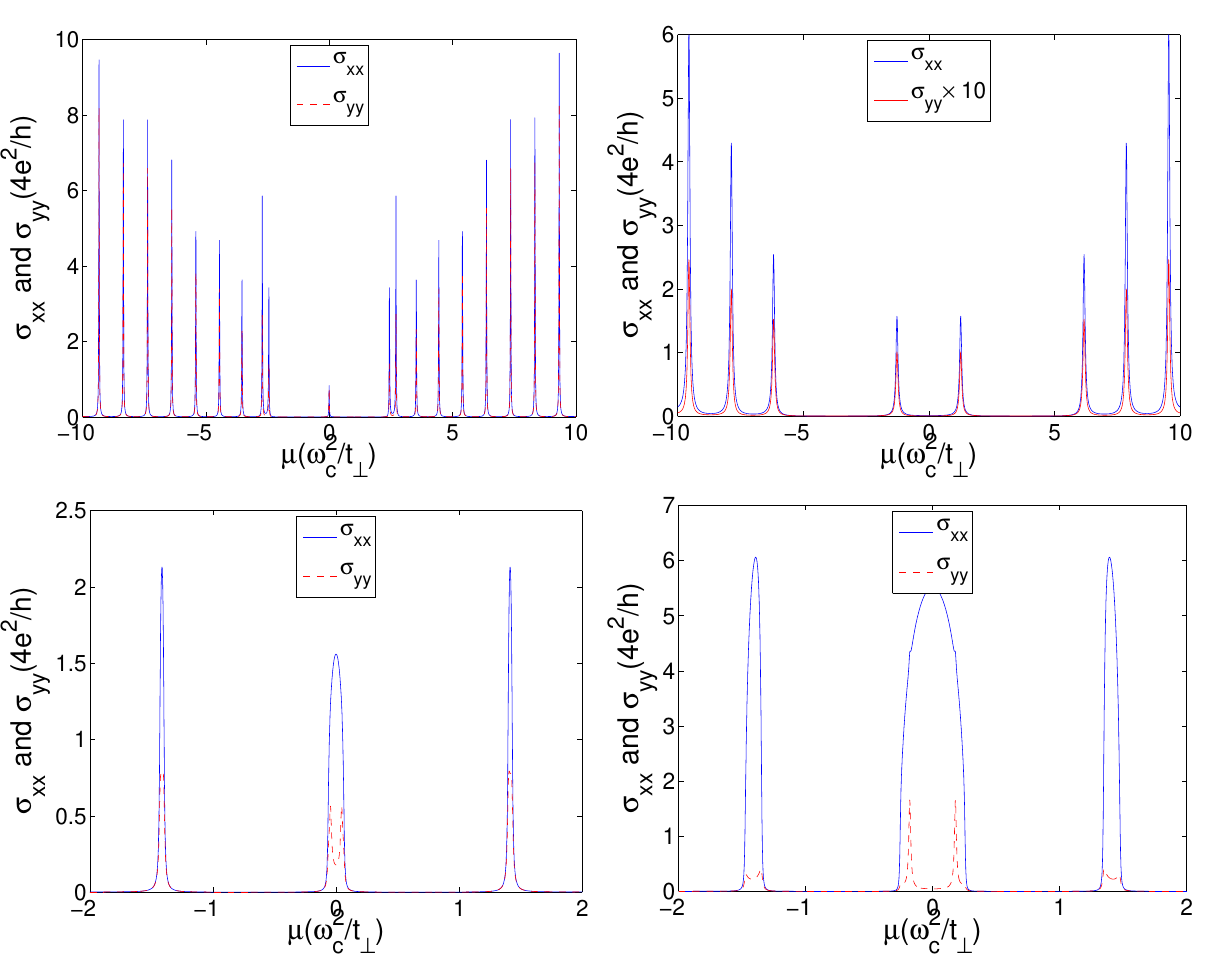} \caption{Diagonal DC conductivities of bilayer graphene superlattice for different superlattice strengths $V_0$, and magnetic fields $B$, with fixed $\lambda=60d$. The conductivity is shown for weak field ($\ell_B=2\lambda$, top panels) and intermediate field ($\ell_B=0.2\lambda$, bottom panels). Left panels (a,b) correspond to $V_0=0.01t$, and right panels (c,d) correspond to $V_0=0.04t$. The conductivities show anisotropy, where $\sigma_{xx}$ is always larger than $\sigma_{yy}$, in contrast to anisotropy reversal in monolayer graphene superlattices. } \label{BLGSL_diagonal} 
\end{figure}

Results for the energy spectrum of chemical potential superlattices in a weak magnetic field are shown in Fig. \ref{BLGSL_LL}. In the left panel, we have chosen $V_0=0.01t$, where $t=3$ eV is again the intralayer nearest neighbour transfer integral. For this value of superlattice potential, our previous calculation has shown that there will be two Dirac points generated in the mini Brillouin zone. Here, similar to the single layer case, the energy spectrum shows a flat band structure. At the Dirac point, there are two nearly degenerate zero energy levels.  Above and below the Dirac point, we can also observe two sets of doubly degenerate energy levels. These are the $n=\pm 1$ Landau levels of each anisotropic Dirac cone. Moving further away from the Dirac point, no such degenerate levels are present. This is because the emergent Dirac cones only have a linear dispersion at rather low energy. In the right panel, $V_0=0.04t$, an energy gap opens up and there exists non-zero energy levels near the Dirac point. Energy bands in an intermediate field regime (not shown here) are only slightly dispersive, where the amplitudes of the dispersion are extremely small compared to the Landau level spacings. 

Fig. \ref{BLGSL_diagonal} shows the DC diagonal conductivity of a chemical potential superlattices. Similar to the single layer case, we see anisotropy in the conductivity. However, there is no anisotropy reversal as the magnetic field strength is tuned. In a weak field, the transport anisotropy is determined by the anisotropic Dirac cones. From the results found in \citet{Killi:2011a}, $v_x \simeq 2v_y$ for the emergent Dirac cones. Therefore, the conductivity in the $\hat{x}$ direction, $\sigma_{xx}$, should be larger than $\sigma_{yy}$ in a weak magnetic field. For an intermediate magnetic field the average velocity in the $\hat{x}$ direction is none zero, $\langle \hat{v}_x\rangle \neq 0$,  due to the dispersion of the energy bands. On the other hand, $\langle \hat{v}_y\rangle$ is always equal to zero. This implies that $\sigma_{xx}$ will acquire intra-LL contributions, while $\sigma_{yy}$ is mainly determined by inter-LL contributions and is therefore small compared to $\sigma_{xx}$. 

\begin{figure}
	[tb] \centering 
	\includegraphics[width=.55 
	\textwidth]{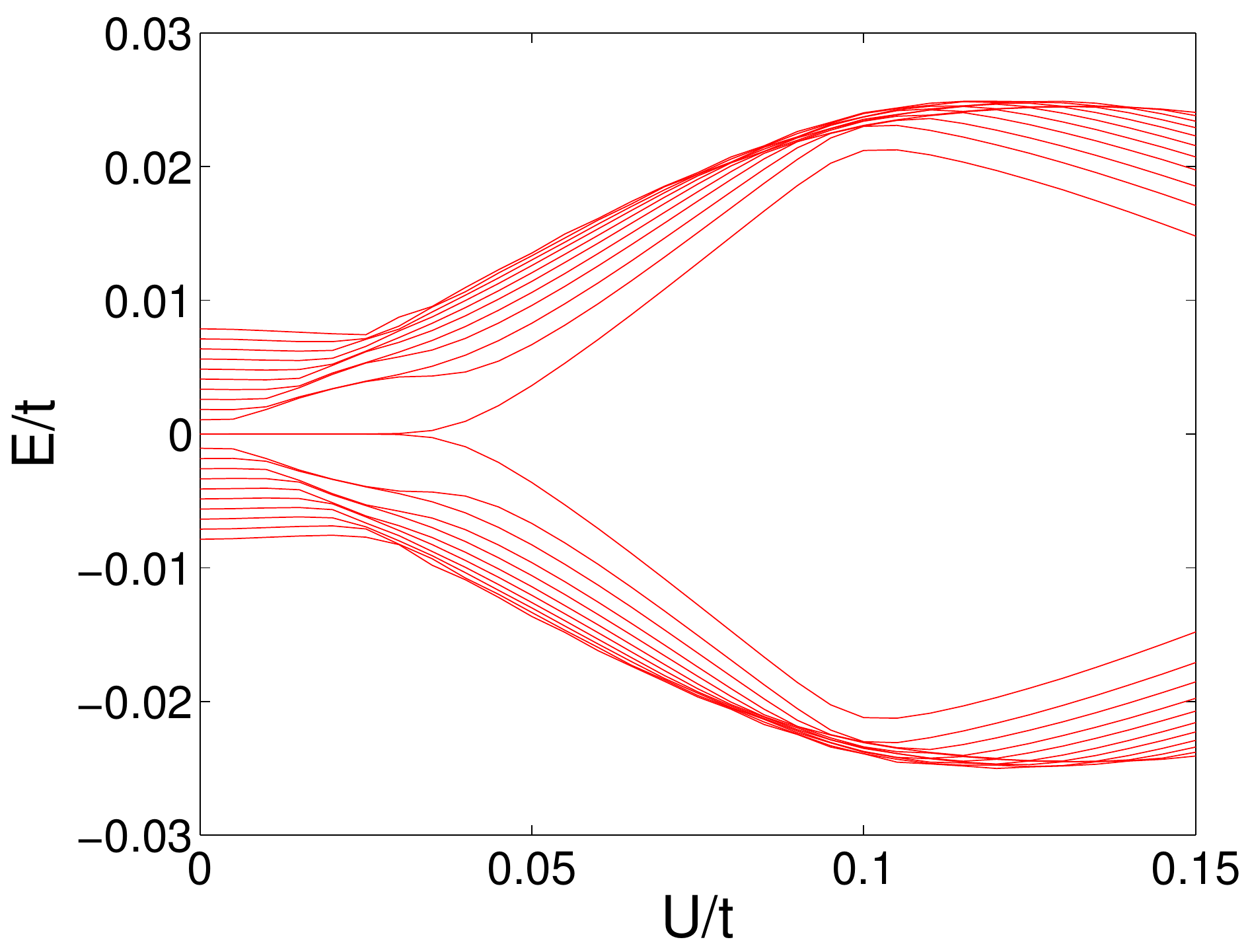} \caption{Evolution of low lying energy levels in a chemical potential superlattice as a function of superlattice potential strength $U$, with $\ell_B=2\lambda$, $y_0=0$, $\lambda=60d$. } \label{evolution} 
\end{figure}

We can further demonstrate how the flat energy levels evolve as the strength of the superlattice potential varies, as shown in Fig. \ref{evolution}. Here, we have fixed the magnetic field strength by setting $\ell_B=2\lambda$ and also fixed the center position of the wavefunctions to be at $y_0=0$. As the superlattice potential tends to zero, where the problem reduces to pristine bilayer graphene in a perpendicular magnetic field, we can see the energies of the Landau levels follow the pattern of Eq. (\ref{Eq10}). As the superlattice potential increases, the physics gradually becomes dominated by anisotropic Dirac cones, which can be seen from the appearance of doubly degenerate levels at nonzero energies.

This crossover from non-relativistic to relativistic behaviour can be qualitatively understood by considering the competition between the characteristic energy scales in these two regimes. In pristine bilayer graphene, the low energy excitations are massive, with an effective mass $m^*=t_{\perp}/2v_F^2$. In a magnetic field $B$, the characteristic energy scale is the cyclotron frequency, $\omega_c^{\prime}=eB/m^*c=1/m^*\ell_B^2$. On the other hand, the anisotropic Dirac points generated by superlattice have anisotropic Fermi velocities $v_y=\sqrt{2}\lambda|U({\bf Q})|/\pi$ and $v_x=2v_y$, where ${\bf Q}=\hat{y}2\pi/\lambda$ \citep{Killi:2011a}. Therefore, the characteristic energy scale associated with these Dirac points is $\omega_c=\sqrt{2v_xv_y}/\ell_B$. We expect to see a smooth crossover as these two energy scales are comparable to each other. A rough estimate shows that the crossover should occur around $U\sim 0.002t$, which is quite close to the value that is observed.

Further increasing the superlattice potential, the doubly degenerate zero energy levels become gapped and all the levels are pushed away from Dirac point. Surprisingly, at rather strong superlattice potentials, $U\sim 0.22t$, zero energy Landau levels appear again, and all of the higher energy levels become doubly degenerate. This phenomenon can be understood from the result of Tan {\it et al} \citep{Tan:2011}. As  shown there, when the superlattice potential is strong enough for a chemical potential superlattice, anisotropic Dirac cones will emerge and this naturally leads to the zero energy Landau level in the presence of a magnetic field. According to \citet{Tan:2011}, for certain values of the superlattice potential, there can be four Dirac points in the spectrum (although this is not present in our calculation). Of course, for stronger superlattice potentials, the two-band description of bilayer graphene is no longer valid. Despite this, we have reproduced this result in a four-band effective Hamiltonian approach. In a strong chemical potential superlattice, there exists a four-fold degenerate zero-energy Landau levels, and the degeneracy reduces to two upon further increasing the superlattice strength.

\subsubsection{Electric field superlattice}\label{section:electric}

\begin{figure}
	[tb] \centering 
	\includegraphics[width=.9 
	\textwidth]{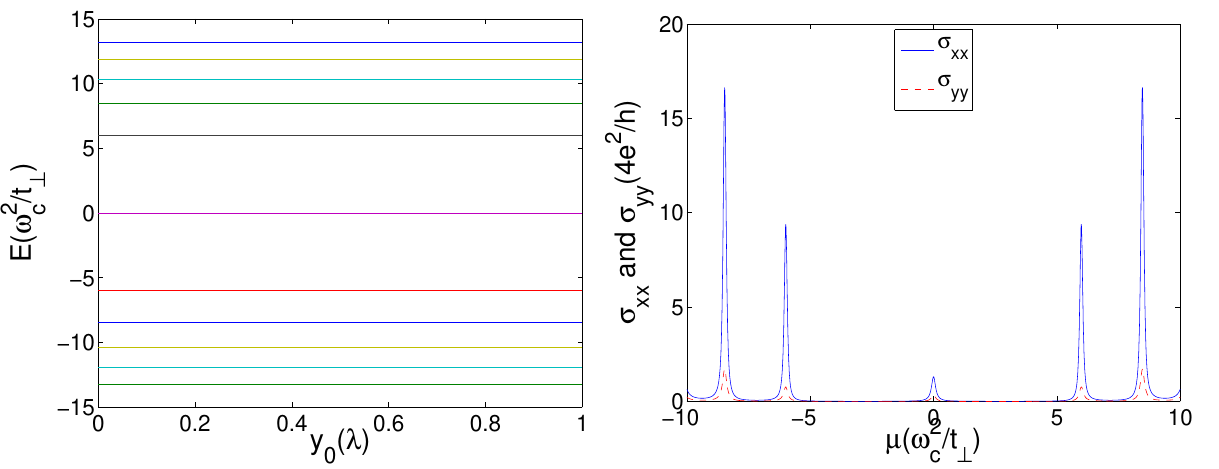} \caption{Energy spectrum and DC diagonal conductivities of electric field superlattice, in a weak magnetic field, with $\ell_B=2\lambda$, $U=0.04t$, $\lambda=60d$. Results for intermediate magnetic fields are similar to those of chemical potential superlattices and are not shown here.} \label{delta} 
\end{figure}

For 1D electric field superlattices, we will take $V_1({\bf x})=-V_2({\bf x})=V(y)=\frac{V_0}{2}\cos(2\pi y/\lambda)$. Following the same procedure as in the previous section, we have calculated the energy spectrum and the DC conductivities for an electric field superlattice in a magnetic field, shown in Fig.~\ref{delta}. 

\begin{figure}
	[tb] \centering 
	\includegraphics[width=.55 
	\textwidth]{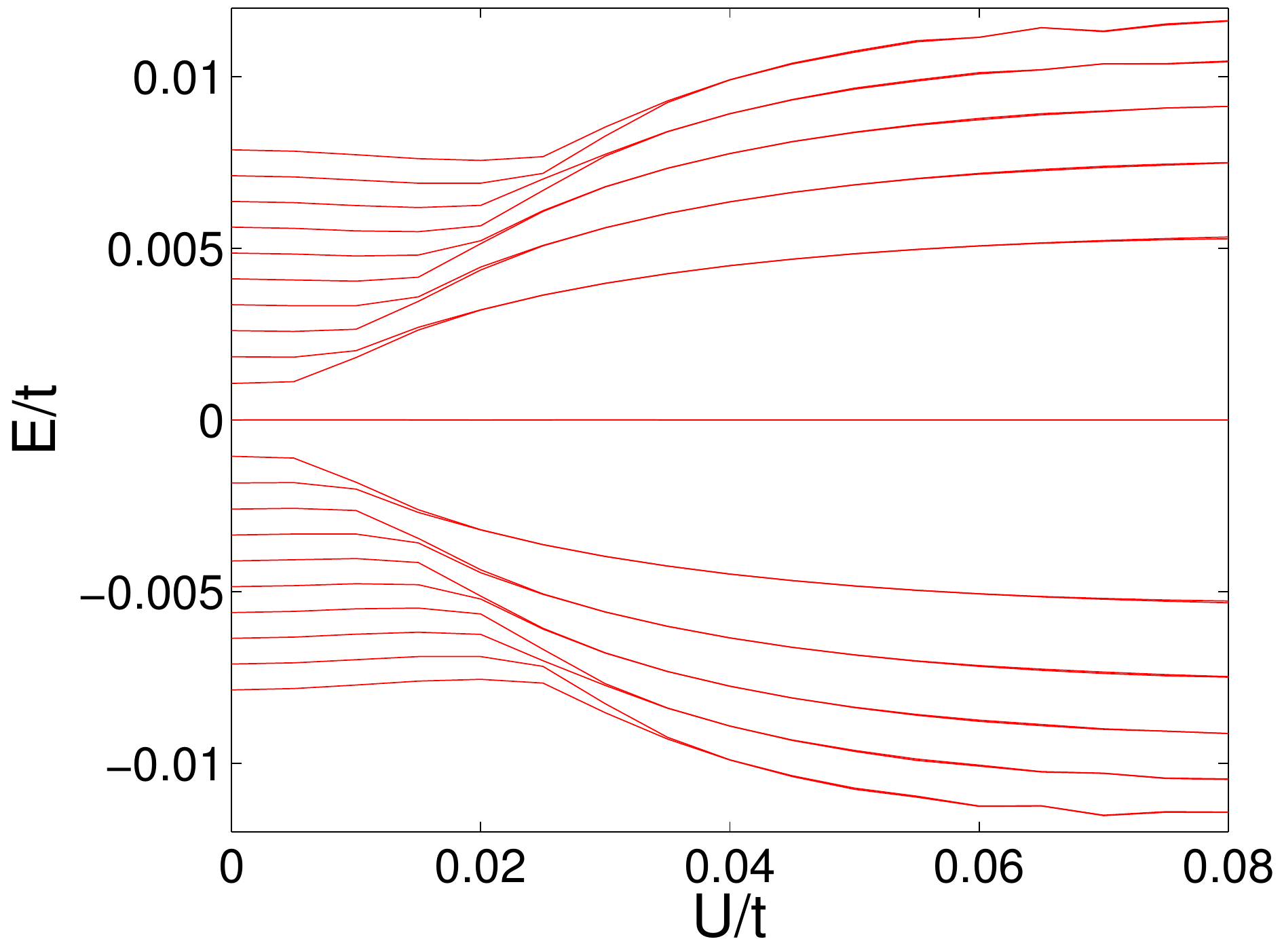} \caption{Evolution of low lying energy levels in an electric field superlattice as a function of superlattice potential strength $U$, with $\ell_B=2\lambda$, $y_0=0$, $\lambda=60d$.} \label{evolution_eSL} 
\end{figure}

In a 1D symmetric electric field superlattice, there are always two zero energy Dirac points present in the spectrum, which results from the coupling of 1D zero modes at kink/antikink of the superlattice potential profile. Earlier, we have developed an effective ``wire'' model to describe these Dirac points and demonstrated that they are topological stable if a generalized ``inversion'' symmetry, $V(y+\lambda/2)=-V(y)$ ({\it i.e.}, flipping the electric field and shifting $y$ by $\lambda/2$), is preserved \citep{Killi:2011a}. This implies that when a weak magnetic field is applied, doubly degenerate zero energy levels should appear at the Dirac point. Indeed, as it can be seen from Fig. \ref{delta} and \ref{evolution_eSL}, these two levels always stay at zero energy, independent of the superlattice potential. Moreover, for the superlattice potential we choose, $V_0=0.03t$, nearly degenerate levels appear even at nonzero energies (Fig. \ref{delta}). These correspond to the Landau levels derived from anisotropic Dirac points, up to $n=\pm 4$.

From Fig. \ref{evolution_eSL}, it is more clear that at strong superlattice potential, the physics is strongly dominated by the Dirac points, where higher energy levels become doubly degenerate and resemble the higher Landau levels of the Dirac cones. When the superlattice potential is weak, equally spaced Landau levels are recovered, as in the chemical potential superlattice, which also indicates a non-relativistic to relativistic crossover at a certain superlattice strength. Notably different from the chemical potential superlattice case, the relativistic behaviour survives to higher energies as the superlattice potential increases, which means the linear approximation description of Dirac cones works in a larger energy range. This is consistent with earlier results \citep{Killi:2010}. Consider a potential profile with a kink of bias reversal. When the bias is large, the conduction and valence bands are widely separated and a pair of linearly dispersed zero modes will traverse the gap. Now, when these kinks and antikinks are arranged periodically, Dirac cones will result from the coupling of these modes, and the Fermi velocity along the superlattice direction inherits its value from the freestanding zero mode \citep{Killi:2011a}. Therefore, as the superlattice potential increases, the energy range where the Dirac cone approximation is valid also increases. This leads to the robust relativistic physics at large superlattice potential. 

\section{Real space picture for interlayer bias modulations in bilayer graphene} \label{section:realspace}

As suggested in an early study by the present authors \citep{Killi:2011a}, it is extremely useful to view an interlayer bias modulation as establishing a series of `kink' and `antikink' modes along the zero-lines where the interlayer bias reverses polarity. Within the effective $2\times 2$ low-energy description of bilayer graphene, the interlayer bias can be regarded as a mass generating term and the zero-lines represent boundaries where the sign of the mass changes. As consequence, along a single isolated kink (antikink), a pair of two right (left) moving `topological' modes emerge at $\bK$ and corresponding two left (right) moving modes emerge at the other $-\bK$ \citep{Martin:2008, Jung:2011a, Qiao:2011, Killi:2011a}. The typical length scale over which these kink-antikink modes are confined, $l$, varies from $\sim100d$ for $V_g=0.01t$ to $<50$ lattice sites for $V_g=0.1t$.

Accordingly, an interlayer bias modulation can be thought of as a periodic array of coupled kink/antikink solitons. With this viewpoint, and using the symmetry properties of the kink and antikink modes, \citet{Killi:2011a} derived an effective low-energy hamiltonian describing how anisotropic Dirac cones precipitate precisely at the band crossing point between the kink and antikink modes when the soliton modes overlap and couple. With this same perspective, further intuition into the quantum Hall effect in the superlattice system can be obtained by studying the properties of 1D kink states in a magnetic field.

Aside from its relevance to superlattices, a study into the properties of the kink/antikink states also provides valuable insights into disordered bilayer graphene samples in the quantum Hall regime. Various sources of disorder are known to generate a random electrostatic potential landscape. Throughout these samples, 1D kink-states are expected to percolate along precipitous fluctuations that reverse the parity of interlayer bias. If the percolation networks are well extended, these kink states would contribute to the breakdown of the quantum Hall effect \citep{Connolly:2012}. Recent STS measurements of bilayer graphene samples in the quantum Hall regime also indicate that even when the sample is uniformly biased by external gates, the disorder is still strong enough to locally reverse the polarity of the interlayer bias \citep{Rutter:2011, Abergel:2011a}. These recent measurements provide an additional motivation to study the properties of kink states in a magnetic field and, specifically, to examine how they modify the tunneling current.

In this section, we consider interlayer bias modulations in quantum Hall regime from the perspective of the kink-antikink modes. We study the dilute limit where the kink and anti-kink states are well separated and confined to the zero-lines (i.e.~$\lambda \gg l$). Within this regime, the low energy states of the system are completely dominated by midgap soliton modes. We start by first reviewing the effects of a magnetic field on uniformly biased bilayer graphene in Section \ref{section:uniform} and then non-uniformly biased bilayer graphene with a single isolated anti-kink in the potential profile in Section \ref{section:kink} . In addition to studying the dispersion relation, we examine how the tunneling current is modified in the proximately of a kink. Next, we consider an array of decoupled kink and anti-kink modes in Section \ref{section:array} . For magnetic fields such that $\ell_B < \lambda$, it is only necessary to understand the dispersion relation of an isolated kink/anti-kink pair as this is sufficient to describe the entire band structure. This allows us to consider a zigzag ribbon of width $N_{\rm cell}$ with one period of modulation (i.e.~a pair of kink anti-kink states) and implement a Peierl's substitution in the full tight-binding Hamiltonian. For weaker magnetic fields such that $\ell_B \gtrsim \lambda$, it becomes necessary to include a larger number of kink/anti-kink states to describe the high energy features of this system. Despite this, further insights into the low energy modes can be made by combining the knowledge obtained from the study a single kink/anti-kink pair and by using a simple low energy effective theory (see \citet{Zarenia:2011a}). Aside from this, in Section \ref{section:valley} we demonstrate that a coupled pair of kink/anti-kink states opens asymmetric bandgaps in each valley that can be precisely controlled using a magnetic field. This effect may provide a viable route to developing a switchable valley filter.

\subsection{Uniform bias} \label{section:uniform} 

Although it has been discussed extensively in the literature, we review and emphasize a few important points about the Landau level structure in uniformly biased bilayer graphene. Details about the Landau level structure in biased bilayer graphene can be found in \citet{McCann:2006, McCann:2006a, Castro:2007, Pereira:2007a, Nakamura:2008, Mazo:2011a, Nakamura:2009, Nakamura:2009a}; here we only briefly summarize some of the features as they pertain to our current discussion. 
\begin{figure}
	[!h] \centering 
	\includegraphics[width=0.9 
	\textwidth]{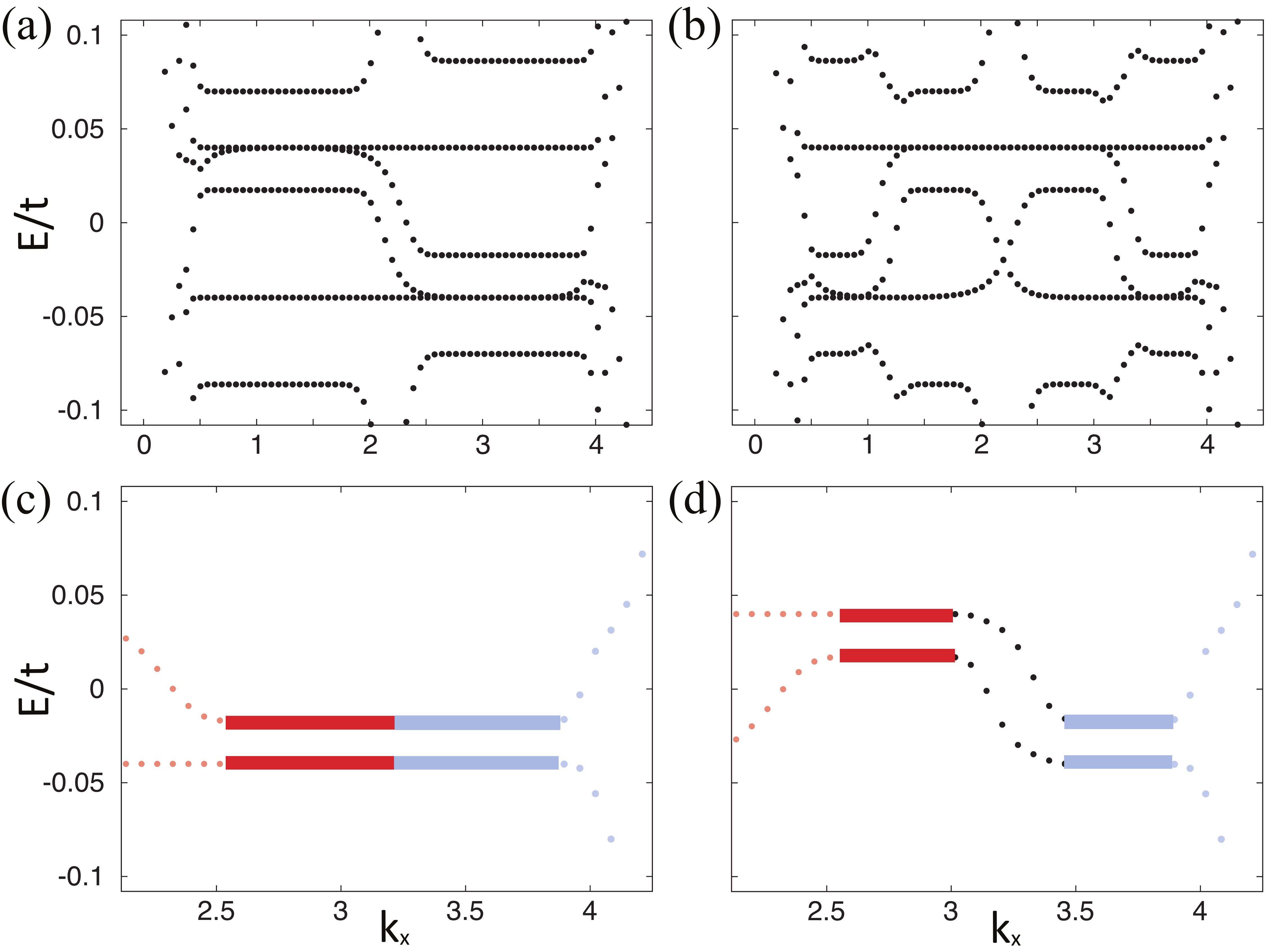} \caption{Landau level spectrum of bilayer graphene strip in the presence of (left column) a uniformly interlayer bias and (right column) with an anti-kink bias profile with $V_g=0.08t$, $N_{cell}=300$. (c,d) n=0 and n=1 Landau levels. Solid red (blue) lines are bulk states localized in the upper (lower) half of the sample and red (blue) dotted lines are the upper (lower) edge states. Black dotted lines connecting the bulk Landau levels are anti-kink states. We have chosen a strong magnetic field of $B=120$ T in order to make the relevant features easily discernible and note that the spectrum is qualitatively identical at lower magnetic fields, although the splitting of the $n=0$ and $n=1$ Landau levels is small.} \label{KinkEdge} 
\end{figure}

In the Landau gauge that preserves translation symmetry along the zigzag direction, inspection of Fig.~\ref{KinkEdge} (a) shows that there are two distinct energy level spectrums in the dispersion relation, one about each valley. Each energy level consists of two well-defined regions: a flat, non-dispersive region consisting of bulk Landau levels and dispersive edge states \citep{Mazo:2011a, Nakamura:2009} (note, the edge state dispersion for the $n=0$ and $n=1$ are unique in that they contained non-dispersive regions, as seen in Fig.~\ref{KinkEdge} (a,c)). The wavefunctions of the non-dispersive Landau level states are well confined within the bulk and, starting from states on the far left side of the level and ending on the right side, move from one edge of the sample to the other.

A comparison of the spectrum about the \bK-points clearly shows that the interlayer bias breaks the valley degeneracy of the Landau level spectrums \citep{McCann:2006a, Nakamura:2008}. Although the valley degeneracy is broken, the states in each valley are connected by a symmetry that relates the low energy Hamiltonian of opposite $\bK$-points. If we define the operator $\cal{P}$ to correspond to the interchanging of $A-B$ sublattices combined with layer exchange (i.e. $A_1\leftrightarrow B_2$ and $B_1\leftrightarrow A_2$), it is a simple matter to show that ${\cal P}^\dag H_\bK(- V_g){\cal P}=H_{\bK'}(- V_g)$. Thus, upon interchanging the sites labelling, the quasiparticles in each valley are governed by the same low energy Hamiltonian but with the effective relative parity of the interlayer bias reversed, leading to the observed valley degeneracy breaking.

Inspection of the band structure shown in Fig.~\ref{KinkEdge} (c) reveals that in the presence of an interlayer bias, the zero energy $n=0$ and $n=1$ Landau levels are shifted to finite energy and their degeneracy is lifted. Note, the energy shift of these two states is positive in one valley an negative in the other. A second crucial observation about the $n=0$ and $n=1$ Landau states (and to a lesser extent the higher energy states) is that the wavefunctions are strongly localized to one of the two layers determined by its valley index. Such layer polarization is directly observed in STS measurements and information about the Landau level spectrum, valley index and local interlayer bias can be obtained \citep{Rutter:2011}. Below, we discuss further how the presence of a kink and anti-kink affects the tunneling current and gives rise to clear signatures observable in STS measurements.

\subsection{Single kink or antikink in the magnetic field} \label{section:kink}

In the absence of a magnetic field, an isolated kink in the interlayer bias generates two unidirectional dispersing subgap bands in one valley and two oppositely dispersing bands in the other valley related by time reversal. An anti-kink generates similar low energy bands, although the velocity of these modes is reversed in each valley (see Fig.\ \ref{Fig:Dispersions} and \citet{Martin:2008} for details). 

When a large magnetic field is introduced, the energy spectrum resembles that of the Landau level structure of uniformly bias bilayer graphene, but with further distinctive features. Consider the energy spectrum about a single valley. Comparison of Fig.~\ref{KinkEdge} (c) and (d) shows that instead of each energy level having a single non-dispersive bulk Landau level, each Landau level breaks into two flat regions shifted in energy, which are connected through a dispersive mode. States in these two flat regions correspond trivially to the bulk Landau states that would have been generated by having a uniform interlayer bias of the same parity as the local bias. Hence, these two regions are composed entirely of Landau states that are well localized in the bulk and respond only the local interlayer bias. As in the previous case, far way from the ${\bK}$-point the bulk states evolve into edge states, and are again sensitive only to the local interlayer bias at that edge.

However, it is for the $n=0$ and $n=1$ Landau levels that the presence of the kink states are directly observed. Starting from a bulk state in the $n=0$ ($n=1$) Landau level on one side of the interface and tracking the states to the other side of interface (moving left to right in momentum space), we observe the following sequence. The bulk $n=0$ ($n=1$) Landau level states begin to continuously evolve into dispersive kink states as the guiding center approaches the zero-line and then emerge from the interface on the other side as bulk states in the $n=1$ ($n=0$) Landau level.

Although other dispersive states connecting the high energy bulk Landau levels on either side of the interface are also present, they are not of topological origin. These states are simply Landau level states that become dispersive as their wavefunction begins to overlap with the region of spatially varying potential. Hence, the properties of these states are inherently sensitive to the the magnetic field and the details of potential profile. In contrast, the kink-states are of topological origin and so persist even in the absence of a magnetic field. Due the strong confined of these states, the properties of these states are remarkably robust and insensitive to the magnetic field (see \citet{Zarenia:2011} for more details).

\begin{figure}
	[!h] \centering 
	\includegraphics[width=0.8 
	\textwidth]{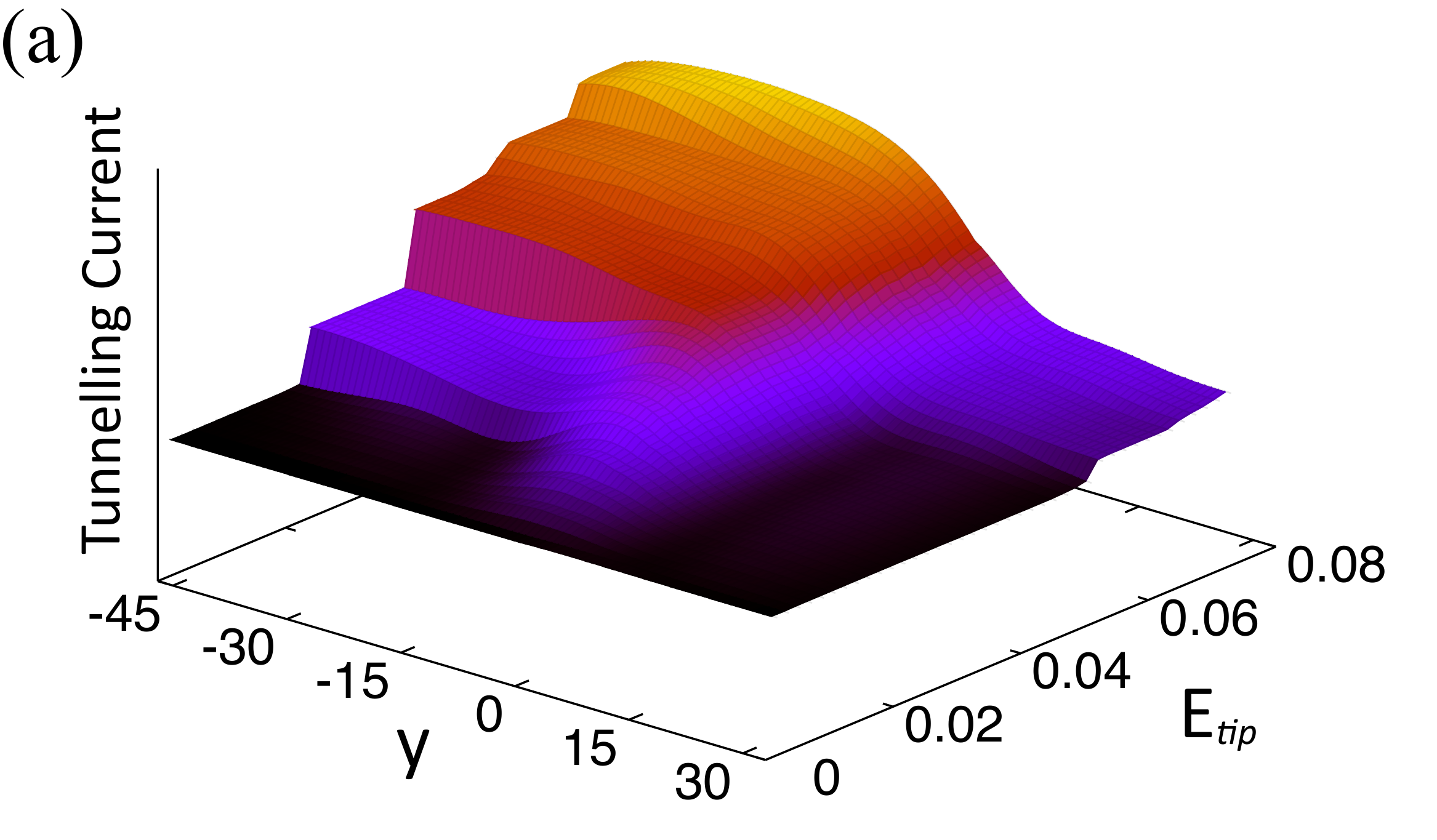} \\
	\vspace{0.1cm} 
	\includegraphics[width=0.9 
	\textwidth]{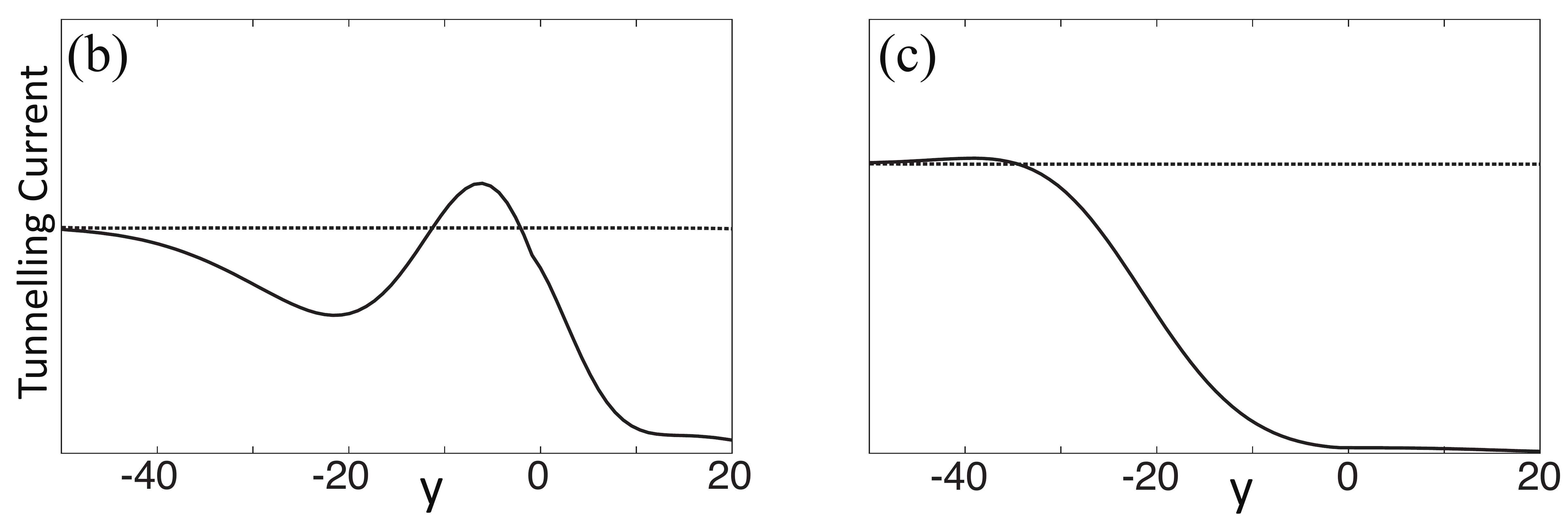}
	\caption{Tunneling current along the $y$ direction (in units of $d$) in the proximity to a kink in the interlayer potential bias with $|V_g=0.08t|$ and large $B=80$ T to avoid extraneous finite size of affects. (a) Tunneling current across a kink for states as function of the sample-to-tip bias, $E_{tip}$, and $\mu=0.0$. (b) Cut where $E_{tip}$ is between the $n=0$ and $n=1$ Landau level ($E_{tip}=0.025t$). Shift of weight towards the interface is a signature of the kink modes. (c) Tunneling current with $E_{tip}$ above $n=0$ with $\mu$ in between the $n=1$ and $n=0$ Landau level ($E_{tip}=0.042t$ and $\mu=0.037t$). Evidence of a kink state is given by the sharp suppression of the tunneling current.} \label{LDOS} 
\end{figure}
The tunneling current of the $A_2$ sites in the vicinity of a kink is shown in Fig.\ \ref{LDOS}. It was computed by allowing electrons to tunnel within an energy window between the chemical potential $\mu=0$ and the sample-to-tip bias $E_{tip}$, and is proportional to the integrated density of states between $\mu$ and $E_{tip}$. Away from the interface, sharp Landau level plateaus easily distinguish the Landau level spectrum. The marked suppression of the tunnelling current on the other side of the interface where the parity of the bias is reversed is indicative of the strong layer polarizability discussed previously. Only when the $E_{tip}$ lies above the $n=2$ Landau level is there any weight on the $A_2$ sublattice in this region, consistent with the low energy theory.

When $E_{tip}$ lies between the chemical potential and the $n=1$ Landau level, there is notable enhancement of the tunneling current close to the interface. A cut along $E_{tip}=0.025t$ shown in Fig.\ \ref{LDOS} (b) shows signatures that this enhancement is due to the presence of strongly confined kink states. Namely, the confinement modes to the zero-line manifests as a transfer of weight from the bulk to the interface and results in the combined downturn in the bulk and sudden enhancement in the tunnelling current close to the interface. Evidence of the strong localized states can also be seen when $E_{tip}$ lies just above the $n=0$ Landau level and $\mu=$ lies just below. Here, the tunnelling into the kink states is suppressed by Pauli blocking, leading to a sharp reduction in the tunnelling current close to the interface. In contrast, at higher energies the reduction in the tunnelling current is a very gradual rolloff across the interface, showing no indication of localized interface modes.

Evidence demonstrating the robustness of these modes in the presence of magnetic field suggest these modes may play an important role in establishing percolation networks responsible for the break down of the quantum Hall effect \citep{Connolly:2012}.

\subsection{Array of kinks and antikinks} \label{section:array}

\begin{figure}
	[tb] \centering 
	\includegraphics[width=0.8
	\textwidth]{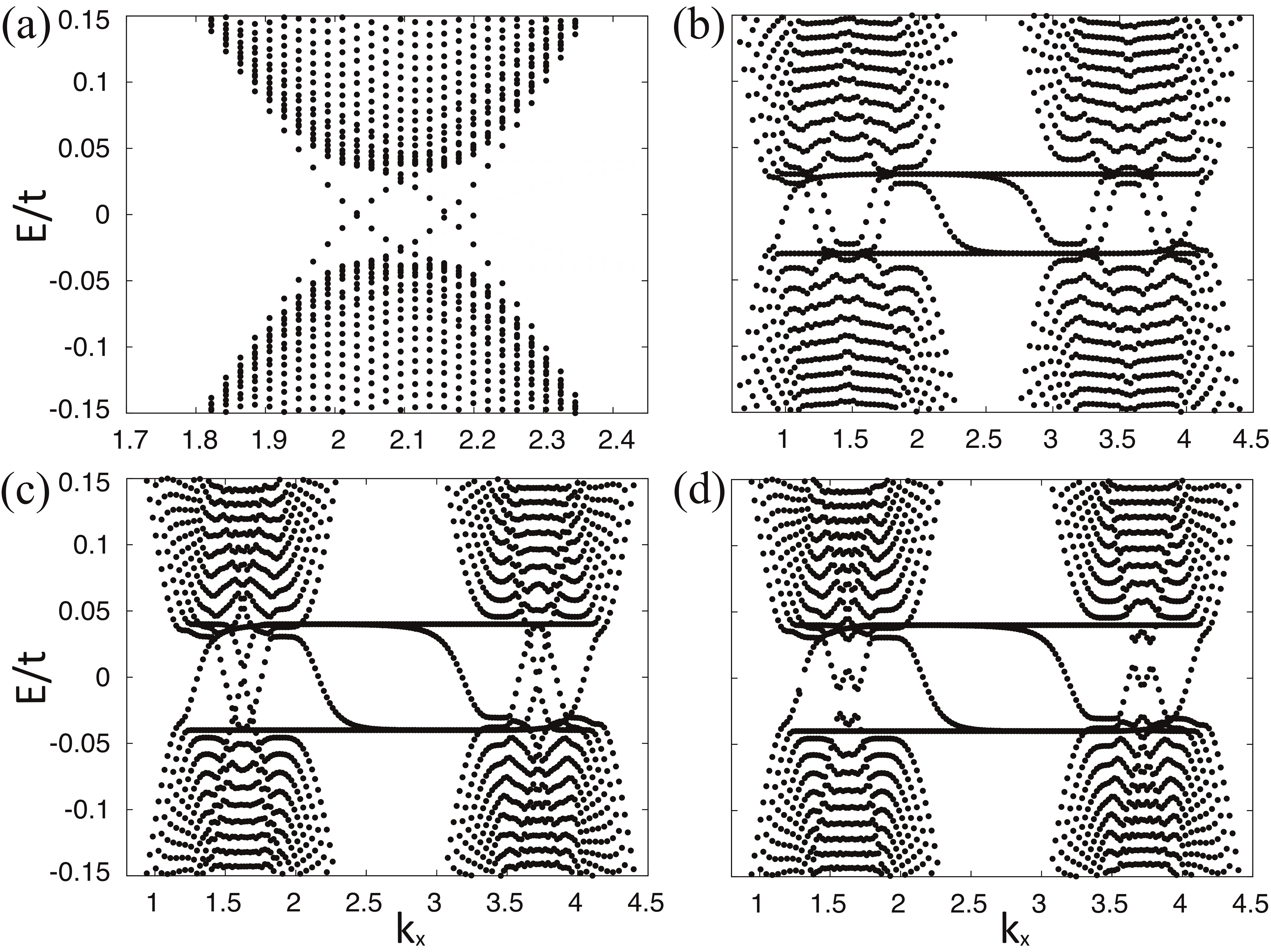} \caption{Dispersion of bilayer graphene in the presence of kink/anti-kink with $|V_g|=0.08t$. (a) decoupled modes with $B=0$ T, (b) decoupled modes with $\lambda_{sep}=100d$ and $\ell_B=17d$, (c) $\lambda_{sep}=40d$ and $\ell_B=17d$, and (d) $\lambda_{sep}=20d$ and $\ell_B=17d$. Note, the large spread of the wavefunctions leads to coupling of the zero energy modes for $\lambda_{sep} < 40d$ and the qualitative features remain robust for similar $\ell_B/\lambda_{sep}$.} \label{Pair} 
\end{figure}

We begin by first considering the dilute limit where the kink and anti-kink soliton modes are well separated and decouple. In the the regime where $\ell_B < \lambda$, the Landau level states are well localized and respond only to their local environment. Hence, the dispersion relation is expected to consist of a series of bulk regions -- identical to those generated by a uniform bias with alternating bias parity -- connected by intermediate kink/anti-kink states. This viewpoint is confirmed by inspecting the dispersion relation of a single kink/anti-kink pair (which, neglecting the spurious edge states, compose the unit cell) shown in Fig.\ \ref{Pair} (b).

In the opposite regime where $\ell_B > \lambda$, the dispersion relation in Fig.\ \ref{Pair} (c) indicates the bulk Landau levels become dispersive in general. Here, the wavefunctions are spread over a wide region encompassing the kink/anti-kink of the potential profile, and thereby altering their properties. In contrast, with respect to the low energy kink/antikink states the magnetic field again does little more than shift their position in momentum space; the mode velocity is robust and the wavefunctions are rigid. Only at higher energies, close to the Landau level energy levels are the dispersive kink states modified.

In sum, the effect of a increasing magnetic field at low energy is to shift the kink and anti-kink modes relative to each other (the anti-kink modes move right the kink modes move left in momentum space). Since the mode velocity of the both the kink and anti-kink are reversed in each valley, this shift pushing the band crossing points up in energy in one valley and down in the other. In this way, a magnetic field can be used to control the band crossing points between modes of adjacent wires.

Further insights into the robustness of the low energy modes present in the continuum model of the 1D interlayer bias superlattice studied above can also be made by employing a simple low energy theory inspired by the kink/anti-kink viewpoint. In the dilute limit and in the absence of a magnetic field each kink (anti-kink) generates an identical copy of the low energy modes about each valley. At each \bK-point in Fig.~\ref{Pair}~(a), the two modes for each kink (antikink) can be linearized about zero energy (see \citet{Killi:2010} for more details). An effective low energy hamiltonian of the superlattice about one of the \bK-points can be written as 
\begin{eqnarray}
	H(\px)=v_0\sum_n\big{[}(-1)^n (p_x-p^*_1) c^{\dg}_{\px n}c_{\px n} \nonumber \\
	+(-1)^n (p_x-p^*_2) f^{\dg}_{\px n}f_{\px n}\big{]} 
\end{eqnarray}
where $v_0$ is the velocity of the modes about zero momentum, and $c^{\dg}_{\px n}$ ($f^{\dg}_{\px n}$) are operators that create electrons on wire $n$, with momentum $p_x$ about momentum close to one of the two zero energy crossing points labelled $p^*_{i}$. (Note, the low energy theory of the opposite valley is identical except that the velocity of the kink/anti-kink modes are reversed.) Representing a magnetic field by the Landau gauge field ${\bf A}=-By\hat{x}$, electrons hopping along a given wire $n$ see an average constant gauge field given by ${\bf A_n}=-B\lambda n$. In a magnetic field, the low energy effective Hamiltonian becomes 
\begin{eqnarray}
	H(p_x)=v_0\sum_n\big{[}(-1)^n (p_x+n\lambda/\ell_B^2-p^*_1) c^{\dg}_{p_x n}c_{\px n} \nonumber \\
	+(-1)^n (p_x+n\lambda/\ell_B^2-p^*_2) f^{\dg}_{p_x n}f_{p_x n}\big{]}. 
\end{eqnarray}
These results demonstrate that the full dispersion of the superlattice consists of multiple copies of a single kink-antikink pair dispersion repeated every $\lambda/\ell_B^2$ in momentum space, consistent with the numerically computation of the dispersions relations provided above. Notice, that the modes in a given wire crosses modes in the two neighbouring wires symmetrically about zero energy. If the magnetic field is weak, the kink modes of wire $n$ cross with the anti-kink modes of wire $n\pm1$ at an energy of $\pm v_0 \frac{\lambda}{2\ell_B^2}$. If the modes couple, level repulsion will cause the kink/antikink modes of the neighbouring wire to mix and split, but because the bandcrossing is symmetric about zero energy and each the each mode couples to both neighbouring modes equally (for a symmetric superlattice), a zero energy band remains intact.

\subsection{Valley filter} \label{section:valley} 
In studying a kink/anti-kink single pair in the coupled regime, we observe a remarkable effect not observed in a previous study \citep{Zarenia:2011} of the kink states that used a low energy model that did not capture the valley degeneracy lifting. As the wavefunctions of the kink and antikink modes begin to overlap, a bandgap opens at the band crossing points. Since the size of the bandgap is determined by the degree of overlap between the wavefunctions, it can be narrowed by either increasing the kinks-antikink separation or by reducing the interlayer bias, which causes the wavefunctions to spread further from the interface. Furthermore, as the magnetic field strength increases, the bandgap shifts to positive energy in one valley and negative energy in the other due to the valley asymmetry in the band crossing point. Remarkably, for a single coupled kink/anti-kink pair an energy window opens where only unidirectional modes are present along each kink and anti-kink (the direction of flow can be flipped by reversing the magnetic field). Moreover, all the conducting modes have identical valley indices and are exclusively conducting.  Hence, only valley polarized currents can flow through the wires, thereby acting as a valley filter similar to those suggested along edge states 	\citep{Rycerz:2007} and line defects \citep{Gunlycke:2011, Wright:2011} when the chemical potential is tuned to lie in the bandgap.

\section{Summary and discussion} 

In this chapter, we studied magnetic properties of monolayer graphene and bilayer graphene under various types of 1D external potentials. For monolayer graphene, we identified three regimes of magnetic field strengths that generate distinctive features in the Landau level dispersion and transport properties. Under a weak magnetic field, monolayer graphene superlattices exhibit zero energy Landau levels whose degeneracies are identical to the number of Dirac points present in the spectrum. At higher energies but still within the linear range of the spectrum, differences between the Dirac cones cause the Landau level degeneracy to be lifted. Therefore, measurements that carefully determine the degeneracy of the Landau levels can be used to probe and characterize the underlying anisotropic Dirac cones. We further showed that the diagonal conductivities show strong anisotropy, with conductivity in the superlattice direction larger than that in the transverse direction. Interestingly, the anisotropy can be reversed for an intermediate magnetic field, where the Landau levels become dispersive. This field tunable anisotropy may find interesting device applications such as in switching or in resistive bits.

We then considered two types of superlattices for bilayer graphene and again showed that analysis of the Landau level spectrum provides clear signatures of the underlying Dirac cones. However, the effects of a magnetic field were shown to be rather different in the bilayer superlattice systems than on the monolayer superlattice system. Although the diagonal conductivity of both superlattices also have strong anisotropy like in the monolayer, there is no field tunable anisotropy reversal. Furthermore, regardless of the type of superlattice there are distinct crossovers from nonrelativistic physics to relativistic Dirac physics as the superlattice strength is tuned. In the case of a chemical potential modulation, the zero energy Landau level vanishes at a critical superlattice strength where the Dirac cones becomes gapped. As for the electric field superlattice, two zero energy Landau levels are always present irrespective of the strength of the magnetic field superlattice strength or magnetic field. This demonstrates a remarkable robustness of the Dirac points inherited from their topological origin.

We have also studied how a magnetic field affects bilayer graphene subject to a uniform bias and with bias reversing kinks in the potential profile. The motivation here was to gain a real space perspective of the Landau level wavefunctions in the quantum Hall regime. Together with a low energy model, this study further established an intuition into the properties of the electric field superlattice. Moreover, it provided valuable insights into the topological kink states present in disordered bilayer graphene and suggests how that they may contribute to the breakdown of the quantum Hall effect. Finally, we proposed that a pair of coupled kink-antikink modes subject to a magnetic field could serve as a possible route to fabricate a switchable one-way valley filter.
 
\chapter{Part I: Closing Remarks} \label{Chapt:Conclusion}     
Throughout Part \ref{Part:Kink} of this thesis,  we explored the multifaceted effects of non-uniform potentials on graphene --- these afford the possibility to directly engineer the rudimentary attributes of the charge carriers, such as the speed at which they travel, their mass, and indeed in some cases, their dimensionality and very nature; massive `Schrodinger-like' particles can be transmuted into massless `Dirac-like' particles.  We will also explored how a magnetic field affects the modified dispersions and can be used to create further interesting and potentially useful effects. 

Moving foreward, there has been considerable theoretical work on the 1D kink states introduced in Chapter \ref{Chapt:Kink}, and on the superlattices discussed in Chapters \ref{Chapt:SL} and \ref{Chapt:BSL}. Since our study on electron interactions, which was the topic of Chapter \ref{Chapt:TTL}, new studies on kink states have considered the effects of disorder \citep{Qiao:2011}, the orientation of the domain wall \citep{Jung:2011a,Xavier:2010,Nunez:2011}, a magnetic field \citep{Zarenia:2011a,Zarenia:2011,Wu:2012,Killi:2012}, spin-orbit coupling \citep{Klinovaja:2012}, and the presence of kink states in a quantum hall ferromagnetic phase \citep{Huang:2012,Mazo:2012}.  Although the exploration of superlattices in bilayer graphene is still in its infancy, there has already been follow-up research on the physics discussed in Chapters \ref{Chapt:SL} and \ref{Chapt:BSL}.  New studies have examined crystal lattice effects \citep{Pal:2012a}, the screening of the electric field that results from the electron transfer between the layers \citep{Tan:2011}, as well as 1D Dirac gap superlattices in monolayer graphene \citep{Maksimova:2012}.

On the experimental side, there has been equally progressive research regarding both kink states and superlattice physics. One particularly noteworthy development comes from a fresh experiment carried out by Yacoby's group at Harvard that, for the first time ever, demonstrated the ability to locally control bandgaps in bilayer graphene \citep{Allen:2012}.  This is, of course, highly pertinent to the progression of both kink state and superlattice fabrication.  Other new methods for assembling specific patterns of non-uniform potentials have also been established \citep{Sun:2011}.

Aside from modified bilayer graphene, there is now experimental evidence that suggests the existence of kink-like states in a gapped insulating phase in bilayer graphene \citep{Bao:2012}.  Results from this experiment indicate that a strongly correlated phase precipitates in suspended samples of exceptional quality.  In this phase, conductivity measurements are most consistent with a scenario in which there are naturally occurring domain walls that support percolating conduction channels akin to kink states. Consequently, this has rejuvenated much interest in kink states and how they couple. 

The current aim of this final chapter is not so much to draw to conclusion the topics of Part I, but rather to attempt to open and reinvigorate future research. Hence, in this closing chapter we survey some recent developments and exciting new proposals, and also present a few open-ended questions to the reader.  But first, we take this opportunity to address some important outstanding issues at length.

First and foremost, we attend to the paramount issue of disorder, as it has the potential to make any attempt at observing 1D physics in the kink states moot.  We address this issue by pointing to some pivotal studies that report compelling evidence that the 1D modes are far more robust than na\"ive expectations, and we provide new insights into their results.  We also touch on orientation effects. Upon addressing these issues, we then highlight some fascinating new proposals surrounding the physics of biased induced kinked states and superlattices, and close our discussion by presenting a few forward-looking questions and suggestions for future research.  With these closing remarks, we hope to support the conviction that 1D kink states and superlattices are soon to be experimentally realized, highly relevant to current experiments, and remain an enticing avenue worthy of future exploration. 


\section{Effects of disorder and orientation on kink states}

The significance of disorder in 1D systems cannot be overstated.  Even without considering electron interactions, the transport properties of 1D conductors with few conducting channels are incredibly sensitive to disorder.  Any residual disorder in the system, either intrinsic or extrinsic, will ultimately cause the conducting electrons to localize within a characteristic length scale, which is short for conventional nanowires.  Beyond this length, diffusive transport sets in and the resistance increases exponentially.

Compared to conventional nanowires that form from geometrical confinement, the 1D kink modes in bilayer graphene are unique in that they derive from the topology of the Hamiltonian. Therefore, disorder is expected to affect transport through the kink states differently.  Hence, a separate investigation specific to the kink states is required.  A few paramount questions regarding the issue of disorder (disregarding electron interactions) have recently been explore:  \emph{Over what length can the conduction considered to be ballistic} and \emph{does disorder induce backscattering between the two valleys}, or correspondingly, \emph{is the valley index preserved}?  

Naively, it would appear that the 1D kink states may not be robust conducting channels, since they are not topologically protected as they are in topological insulators, where the zero energy states occur at the Kramer's points and are protected by time-reversal symmetry. However, \citet*{Qiao:2011} recently carried out an extensive study into various potentially detrimental mechanisms. Through numerical conductance calculations and by examining the local density of states, they showed the low energy states are remarkably resilient to both short and long range disorder.

In their study, short range disorder was modelled by adding random potentials on each site.  The strength of the potential was sampled from a distribution of energies within a given range.  Surprisingly, even when the disorder strength was comparable to the bandgap, the transport remained ballistic.  Even for finite ranged disordered potentials, the conductance hardly degraded.  Their results suggest that ballistic transport may be possible from tens to over hundreds of microns.  

Guided by our early study on electron interactions discussed in Chapter \ref{Chapt:TTL} \citep{Killi:2010}, they found the overall robustness of the kink states can also be attributed to the sharp suppression of backscattering due to large spread of wavefunction.  This is reminiscent of the ballistic transport found in carbon nanotubes \citep{White:1998}, where the disorder averages out over the diameter of the tube.  Both of these systems have a clear advantage over quantum wires formed by geometrical confinement ---  the width of the wavefunction is naturally large and can be tuned \emph{without} introducing further conducting channel, which simultaneously enhance backscattering.  In other words, the number of low energy backscattering channels is fixed, and only the matrix elements and the proximity to the high energy bulk bands are affected. 

Using Fermi's golden rule, \citet{Qiao:2011} determined the localization length to be proportional to $\sim E_g^2\, M/W^2$, where $E_g$ is the bandgap, $M$ is proportional to the width of the wire and $W$ is proportional to average strength of the disorder. Given the scaling argument presented in Section \ref{Sect:Kinkchar}, which showed that $M$ is proportional to $\sim V_g^{-\frac{1}{2}}$, the localization length approximately scales as $E_g^{\frac{3}{2}}$.  For comparison, the localization length of carbon nanotubes scales linearly with the circumference of the tube and, although still relatively long when compared to GaAS/AlGaAs wafers, is a few microns.  

To much surprise, the current along the kink states remains ballistic and valley polarized even when a wire is `bent' around a corner, which is in stark contrast to geometrically confined nanowires in GaAs/AlGaAs heterostructures.  However, it is important to keep in mind that a study by \citet*{Nunez:2011} did show that trigonal warping terms in the Hamiltonian can lead to substantial modifications to the midgap bands when the interlayer bias is small and becomes more sensitive to the wire orientation. (It should also be noted that exclusively along the armchair orientation a small bandgap opens between the 1D bands.)  Remarkably, similar kinks states were also shown to emerge when the potential forms a ring \citep{Xavier:2010}.  Taken together, these studies substantiate the viability of using these kink states as robust quantum wires that could potentially transmit valley polarized currents for valleytronic applications.

\section{Recent developments}
\subsubsection{Kink states} 
    
Another important point that we now emphasize is that the formation of kink states is not restricted to bilayer graphene and much of the physics discussed so far also applies to other systems.  Theoretical proposals have shown the presence of similar kink states in gated multilayer graphene \citep{Jung:2011a}, substrate induced domain wall in single layer graphene \citep{Semenoff:2008}, and also boron nitride sheets \citep{Jung:2012} and twisted bilayer graphene \citep{Kindermann:2012b}.  For few layer systems, it was shown that as long as the stacking order is inversion symmetric, 1D kinks states precipitate along interfaces where the sign of the inversion symmetry-breaking term changes.  The number of zero energy modes at each valley is again determined by its Chern number, which is directly related to the number of layers. 

Other interesting and entirely new phenomena have also been suggested to take place within the kink modes.  \citet*{Klinovaja:2012} recently consider similar non-uniform interlayer biases in bended bilayer graphene flakes that form `half' carbon nanotube.  The local curvature serves to enhance the spin orbit interaction, which results in helical spin modes  that have the potential to support Majorana fermions when contacted with an s-wave superconductor.  In regards to electron interactions, \citet*{Huang:2012} suggested that the kink states form a distinctive charge density pattern in the $\nu=0$ Landau level.  This provides a hallmark signature of the quantum hall ferromagnetic phase that was recently proposed in the $\nu=0$ Landau level.

As we can see, the topologically confined kink modes proposed by \citet{Martin:2008} have the potential to host a wide range of exceptional phenomena.  While much of this physics is very enticing, it all remains theoretical at this time.  However, substantial improvements in sample quality and the overall resilience of these states suggest that it is simply a matter of time before these predictions become reality.  Above all else, the recent report in \emph{Nature}, which provided the first demonstration of microscopic bandgap control in bilayer graphene, is certainly cause for excitement \citep*{Allen:2012}.  The sophistication of this technology is already at a level where clean gate-defined quantum dots have been fabricated, hinting that other gate-defined devices, such as the kink states and superlattices, lies in the visible horizon.
  
\subsubsection{Superlattices}
There also continues to be much theoretical interest in the 1D superlattices described in Chapters \ref{Chapt:SL} and \ref{Chapt:BSL}, and experiments on them are currently underway (private communication).  \citet*{Tan:2011} performed a corroborative study on 1D chemical potential superlattices in bilayer graphene and obtained results consistent with those presented in Chapter \ref{Chapt:SL}.  In addition, they investigated the effects of the charge transfer between the layers and subsequent electron screening by implementing a self-consistent Hartree-Fock theory.  Their results indicate that the emergent Dirac cones do not become gapped until a much higher critical superlattice strength than expected.

Another interesting development comes form a study by \citet{Maksimova:2012} that investigated a 1D Dirac gap superlattice in \emph{single layer graphene} similar to that discussed by \citet{Ratnikov:2009}.  Here, the local bandgap is induced by breaking the $Z_2$ sublattice symmetry, and the superlattice can be thought of as a periodic array of domain walls in a charge density wave existing on the A-B sublattices.  In some sense, this is the single layer version of the interlayer bias superlattice discussed in Chapter \ref{Chapt:SL}.  Even though this system can also be thought of as an array of kink-like states, there are no zero energy modes when the superlattice is weak, in sharp contrast to bilayer graphene.  This strongly reenforces the fundamental role that the generalized parity operator plays in the formation of the emergent Dirac cones in bilayer graphene.  It would be interesting to study this system from the `kink state' point-of-view and explicitly contrast it to the related model in bilayer graphene, or perhaps do similar studies for multilayer graphene.   

\section{Future directions}

While we anticipate the experimental realization of kink states and the implementation of 1D superlattice potentials, there a number of fascinating questions well worth exploring and some important issues that require further investigation.  Seeing that much of our focus has been on electron interactions in kink states, we first mention a few unresolved problems surrounding the Tomonaga-Luttinger liquid phase in Chapter \ref{Chapt:TTL}, and also some persistent issues concerning the combined effect of disorder with electron interactions.  We then turn our discussion over to superlattices.

Although the physics of Chapter \ref{Chapt:TTL} is expected to be valid over a sizeable energy window, it does not describe the ground state properties of the system.  In particular, the backscattering terms become increasingly relevant at low temperatures and cannot be neglected.  Thus, a full description of the phase diagram must include the various sine terms and a full renormalization group treatment is required.  Depending on which of the sine terms lead the renormalization flow, some of the phases will become pinned causing certain modes to become massive.

Another issue that has yet to be carefully understood pertains to the velocity asymmetry of the subgap modes, which occurs in the extended tight-binding model due to trigonal warping.  While preliminary (unpublished) studies show that although the velocity of the bands flow to a common value, it flows slowly.  Since the renormalization of the sine terms are also marginal, it is unclear if the velocity asymmetry can be ignored.  Once the effect of the velocity asymmetry is properly accounted for and which modes are gapped, the exact properties of the ground state of the kink state can then be determined.    

Besides the ground state properties of the kink states, there remains to be a careful investigation into the effect of disorder that also incorporates electron interactions.  This is particularly a concern for 1D systems, as oscillations in the charge density, known as Friedel oscillations, occur close to an impurity and electron interactions cause the impurity strength to strongly renormalized at low energy.  A similar renormalization is also expected to be highly relevant for kink states, except the renormalization may occur more slowly since the initial suppression of backscattering may help mitigate the flow.  Only when interactions are included can the low temperature localization length be determined.  Even at higher energies, electron interactions are expected to enhance the localization of charge carriers beyond the predictions of \citet{{Qiao:2011}}

Let us now close by posing a possible extension that would serve as a nice segue between the kink states and superlattice physics of Part I.  It would be interesting to explore how the transmission through a finite series of kink and antikinks crosses over to the band structure of the full superlattice.  Even at the single particle level, such as study would be important to experimentalists, since the natural progression of experiments is to start with transmission studies through multiple junctions before ultimately forming a bulk superlattice.  Likewise, the crossover between multiple p-n junctions to a chemical potential superlattice would also be useful.  In this case, the Fabrey-Perot type resonances would crossover to bulk band structure effects.  Finally, the multiple kink/antikink array may also serve as a potential platform to study the crossover from 1D Tomonaga-Luttinger liquid physics to 2D physics.  It is particularly conducive because of the high degree of tunability in the system, as the overlap between the wavefunctions can be controlled via the gate potential. This just a small sample of the many interesting questions that remain and we expect that once 1D superlattices and kink states have been experimentally realized, entirely new and unforeseen questions will surface.

\part{Local Moment Adatoms in Bilayer Graphene}\label{Part:Adatom} 
\chapter{Controlling Local Moment Formation and Local Moment Interactions in Bilayer Graphene} \label{Chapt:Adatom} 
\chaptermark{Controlling Local Moments in Bilayer Graphene}
\subsubsection{The material in this section is largely based on the article M.\ Killi, D.\ Heidarian and A.\ Paramekanti \textit{New J.~Phys.} {\bf 13}, 053043 (2011).}

In this chapter, we study adatoms on the surface of bilayer graphene. We examine local moment formation on the adatoms and how these effects can be controlled by tuning the chemical potential and applying perpendicular electric field \citep{Killi:2011}.
\begin{figure}
	[tb] \centering a)
	\includegraphics[width=.4 
	\textwidth]{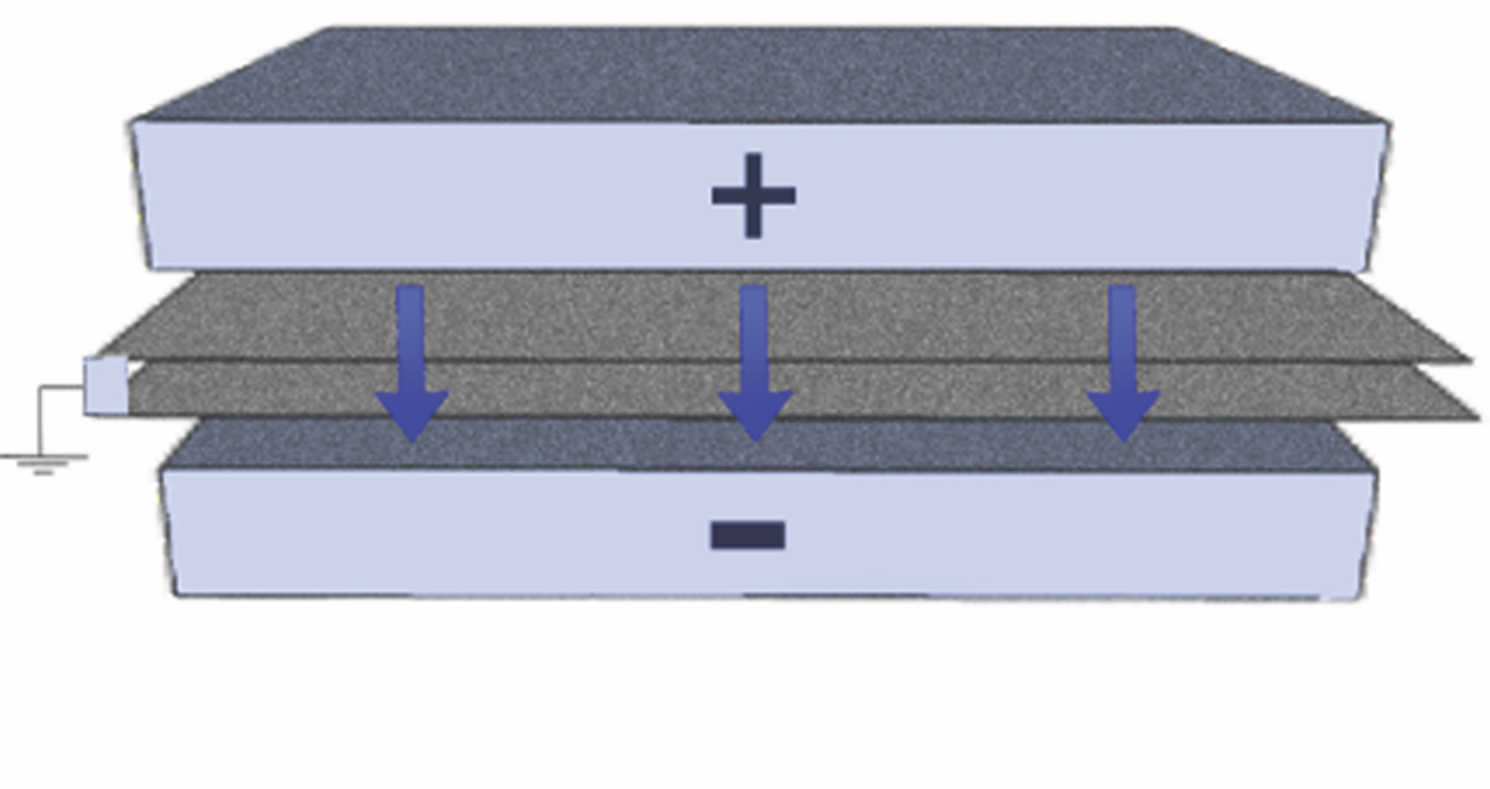} b)
	\includegraphics[width=.4 
	\textwidth]{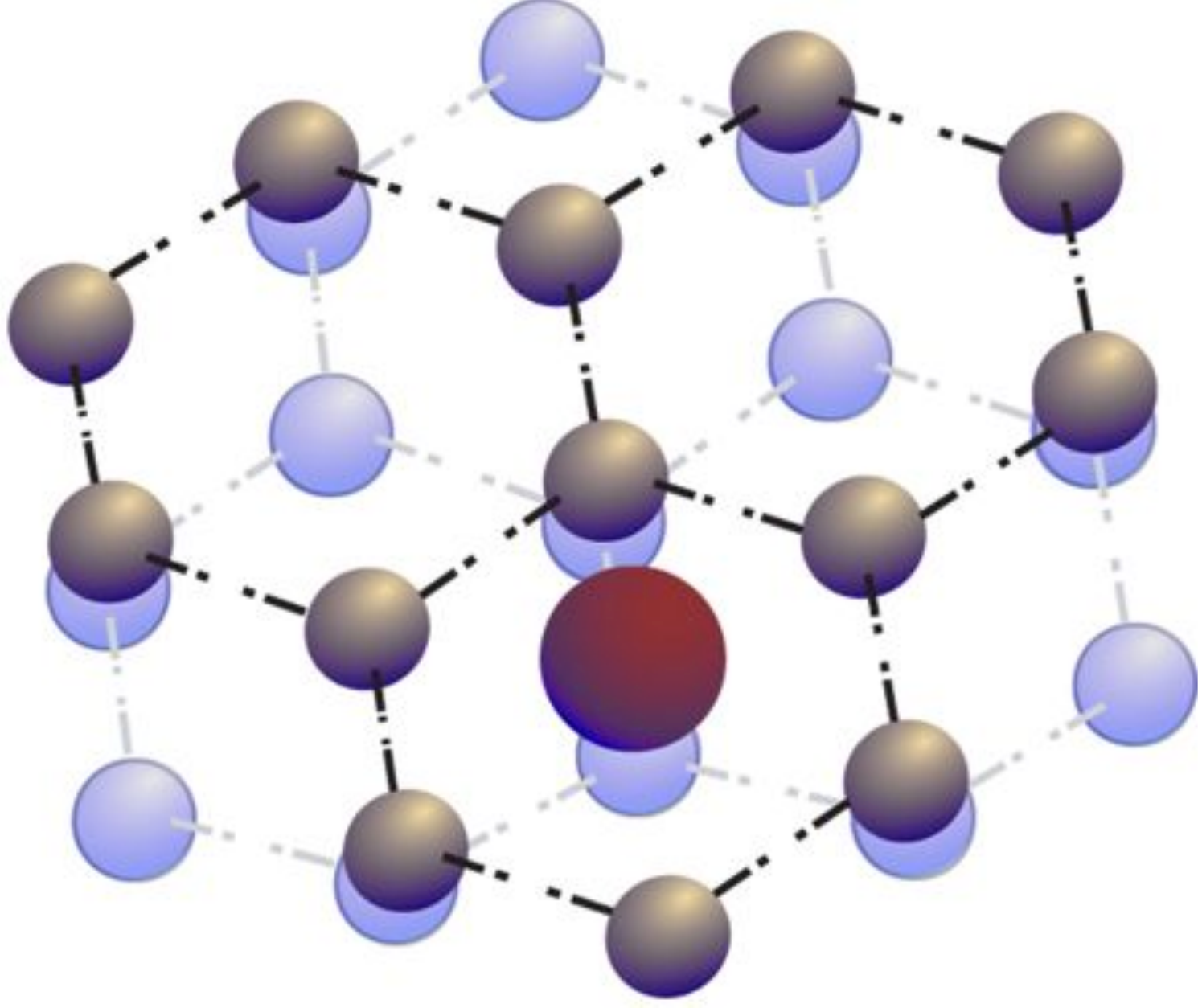} \\
	c)
	\includegraphics[width=0.8 
	\textwidth]{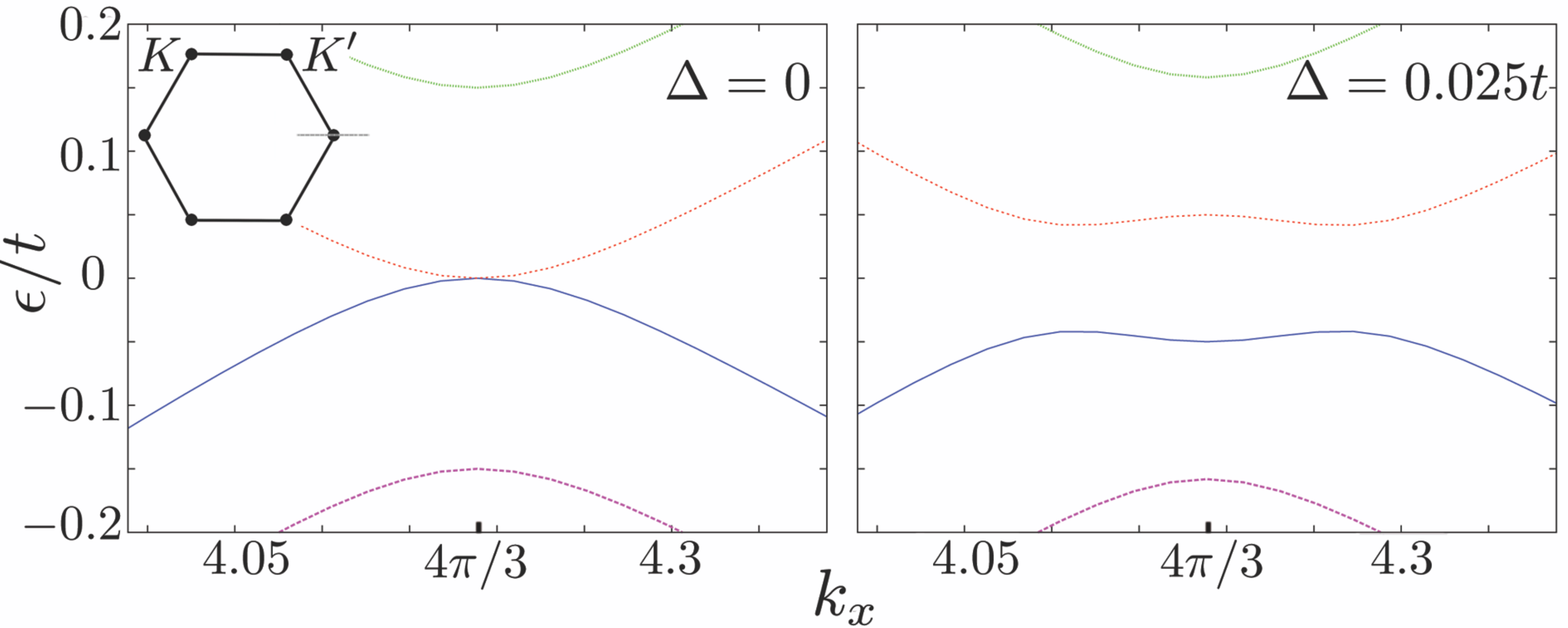} \caption{(a) Bilayer graphene in a dual-gate configuration. (b) Schematic diagram of a plaquette-centered (large, red) adatom impurity on the top layer of bilayer graphene. (c) Cross-section of the dispersion relation for unbiased (Left) and biased (Right) graphene close along $k_y=0$ through the K-point ($\Delta=0$ and $\Delta=0.025t$, respectively). Inset: The two unique K-points and the cross-sectional cut are indicated in the Brillioun zone.} \label{model} 
\end{figure}
Our work goes beyond the preliminary work of \citet{Ding:2009}, which studied local moment formation for site-centered adatoms on bilayer graphene, in several important respects. (i) We consider adatoms that are positioned at the center of a hexagonal plaquette on one of the layers. The study of this configuration is motivated by a recent ab initio study of adatoms in monolayer graphene that indicates plaquette centered impurities are generally more energetically favourable than on-site impurities \citep{Chan:2008}. We expect a similar situation to hold in bilayer graphene. (ii) An applied electric field is shown to directly tune the impurity energy. This is because an impurity position will, in general, be located closer to the top layer of bilayer graphene. Accounting for this impurity energy shift allows us to identify regions of the phase diagram where local moment formation can be turned on and/or off by the application of a perpendicular electric field. (iii) For a particular impurity level chosen so that its $renormalized$ (with self-energy corrections) energy level lies in the middle of gap in presence of the bias, we construct phase diagrams at zero, positive and negative bias by sweeping the chemical potential. The resulting phase diagram exhibits the onset of a Coulomb-blockade phase where any arbitrarily small $U$ results in the formation of local moments. (iv) As a consequence of the chiral wavefunctions of bilayer graphene and the fact that the plaquette centered impurity adatom couples to many sites, the coupling between the impurity and the quasiparticles of bilayer graphene has strong momentum and band dependence. This affects many of the details of the phase diagram. For instance, the self-energy develops a large real part that has nontrivial frequency dependence, and substantially renormalizes the position of the impurity spectral peak in a manner that depends on the chemical potential and the applied bias. We provide a detailed physical explanation for how this affects the resulting phase diagrams, which were not provided in reference \citet{Ding:2009}. Furthermore, to better illustrate the effect of the wavefunctions and chirality of bilayer graphene on the phase diagrams, the bilayer graphene system is contrasted with a fictitious system of non-chiral fermions with the same DOS and dispersion relation. (v) We go beyond the issue of local moment formation to address the tunable RKKY interactions between such local moments on bilayer graphene.

We begin in Section \ref{Sect:Formalism} by introducing the Anderson impurity model specific to bilayer graphene. Section \ref{Sect:MFT} summarizes the Anderson mean field theory formalism. Armed with this background, in Section \ref{Sect:PhaseDiagrams} we construct the impurity model phase diagrams for plaquette-centered adatoms (shown schematically in Fig.~\ref{model}b). To highlight some of the unusual features of these phase diagrams, we contrast it with an impurity model of a fictitious system of electrons that have an identical dispersion but a band-independent coupling to the adatom. Finally, in Section \ref{Chapt:RKKY}, we discuss the RKKY interaction, and its tunability, for local moments on bilayer graphene.

\section{Adatom model in bilayer graphene} \label{Sect:Formalism}

Consider an adatom on bilayer graphene, described by the Anderson impurity model \citep{Anderson:1961}, 
\begin{eqnarray}
	H_{\rm BLG} &=& \sum_{\bk,s,\sigma} (\epsilon^\pdg_{\bk s} - \mu) c^\dg_{\bk s \sigma} c^\pdg_{\bk s \sigma}, \\
	H_{\rm imp} &=& \sum_\sigma (\epsilon^\pdg_d - \mu) d^\dg_\sigma d^\pdg_\sigma + U n^\pdg_{d \upa} n^\pdg_{d \dna}, \\
	H_{\rm mix} &=& - \sum_{{\bf{r}} \sigma} \chi^\pdg_{\bf{r}} (c^\dg_{{\bf{r}} \sigma} d^\pdg_\sigma + d^\dg_\sigma c^\pdg_{{\bf{r}} \sigma}). 
\end{eqnarray}
Here $\epsilon^\pdg_{\bk s}$ is the bilayer graphene electron dispersion for electrons labelled by momentum $\bk$ and band index $s$ with creation (annihilation) operators are $c^\dg_{\bk s \sigma}$($ c^\pdg_{\bk s \sigma}$). In $H_{imp}$ and $H_{mix}$, $c^\dg_{\br \sigma}$($ c^\pdg_{\br  \sigma}$) and $d^\dg_\sigma $($d^\pdg_\sigma $) are the electron creation (annihilation) operators at lattice site $\br$ and the impurity site, respectively, with spin $\sigma$.    We assume a minimal model for the bilayer graphene dispersion that includes a nearest-neighbor hopping amplitude, $t$, to sites on the same layer, and an interlayer hopping amplitude, $t_\perp$, between the two sites that sit one on top of the other. Henceforth, we set $t=1$ and note that $t \approx 3$ eV and $t_\perp/t \approx 0.15$ in bilayer graphene. In $H_{\rm imp}$, we denote the impurity energy by $\epsilon_d$, while $U$ denotes the electron-electron repulsion on the impurity site. Conduction electrons at sites ${\bf{r}}$ can hop on or off the adatom impurity with an amplitude $\chi^\pdg_{{\bf{r}}}$. We assume a common equilibrium chemical potential $\mu$ for the impurity and the conduction electrons. The complete Hamiltonian for unbiased bilayer graphene is then given by $H=H_{\rm BLG}+H_{\rm imp}+H_{\rm mix}$.

Electronic structure studies of transition metal adatoms on monolayer graphene suggest that the low-energy configuration of many types of impurities corresponds to the adatom residing at the center of a hexagonal plaquette \citep{Chan:2008}. We therefore fix the adatom position to be at the plaquette center on the top layer (labelled $\ell=1$) of bilayer graphene, as shown in the schematic diagram on the right in Fig.\ \ref{model}. For simplicity, we assume that $\chi^\pdg_{\bf{r}}\!\!=\!\!\chi$ for the set of sites $\{{\bf{r}}_n\}$, which includes the six nearest neighbor plaquette sites in layer-$1$ and the site on layer-$2$ that lies directly below the adatom, and $\chi_{\bf{r}}\!\!=\!\!0$ for all other sites. This simplifying assumption about the impurity model allows us to focus on unconventional features of local moment formation intrinsic to bilayer graphene. Future density functional studies would be useful in incorporating details of the impurity atomic orbitals. Turning to the mixing Hamiltonian $H_{\rm mix}$ that allows the impurity electrons to hybridize with the bilayer graphene electrons, let us set 
\begin{equation}
	\label{eq_Vkn} V^\pdg_{\bk s} \equiv \chi \sum_{{\bf{r}}=\{\!{\bf{r}}_n\!\}} \phi^\pdg_{\bk s}({\bf{r}}), 
\end{equation}
where $\phi^\pdg_{\bk s}({\bf{r}})$ denotes the wave function at site ${\bf{r}}$ for electrons in band-$s$ and momentum $\bk$. We then obtain 
\begin{eqnarray}
	H=\sum_{\bk,s,\sigma} \left( (\epsilon^\pdg_{\bk s} - \mu) c^\dg_{\bk s \sigma} c^\pdg_{\bk s \sigma} +V^\pdg_{\bk s} c^\dg_{\bk s \sigma} d^\pdg_\sigma + V^*_{\bk s} d^\dg_\sigma c^\pdg_{\bk s \sigma}\right) \nonumber \\
	+\sum_{\sigma}(\epsilon^\pdg_d - \mu) d^\dg_\sigma d^\pdg_\sigma + U n^\pdg_{d \upa} n^\pdg_{d \dna} . 
\end{eqnarray}
While the impurity model Hamiltonian in bilayer graphene looks similar to that in conventional systems or monolayer graphene, there are two important new ingredients in the impurity physics of bilayer graphene with plaquette centered impurities.
\begin{figure}
	[tb] \centering 
	\includegraphics[width=.9 
	\textwidth]{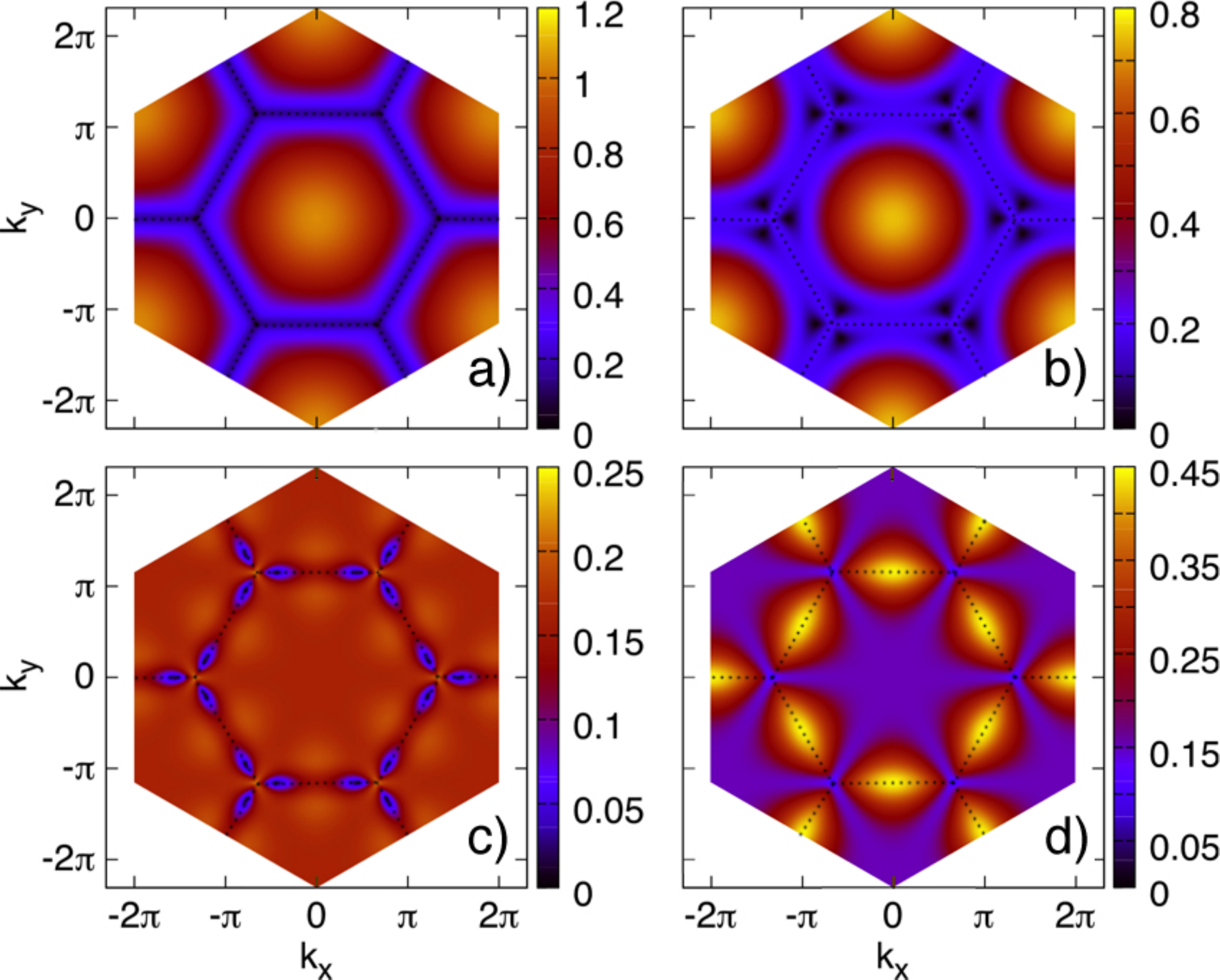} \caption{ Coupling of the impurity to the four bands (ordered from lowest to highest energy and scaled by system size), (a) $|V_{\bk 1}|$, (b) $|V_{\bk 2}|$, (c) $|V_{\bk 3}|$, and (d) $|V_{\bk 4}|$, with impurity hopping strength $\chi = 0.3t$. Dotted line indicates the Brillouin zone.} \label{Vkn} \vskip -0.2in 
\end{figure}

First, for bilayer graphene (or multilayer graphene), as opposed to monolayer graphene, one can tune the density of states by applying an electric field perpendicular to the layers. Let $\Delta$ denote the potential imbalance between the top and bottom layer induced by the electric field. Assuming that the adatom is at the same height as the top layer, this leads to an extra term in the Anderson Hamiltonian 
\begin{equation}
	\label{bias} H_{\rm bias} = - \frac{\Delta}{2} \sum_{\ell,{\bf{r}}_\ell,\sigma} (-1)^\ell c^\dg_{{\bf{r}}_\ell \sigma} c^\pdg_{{\bf{r}}_\ell \sigma} + \frac{\Delta}{2} \sum_\sigma d^\dg_\sigma d^\pdg_\sigma 
\end{equation}
where ${\bf{r}}_\ell$ denotes the sites in the top ($\ell=1$) and bottom ($\ell=0$) layers. In writing this modification to the Hamiltonian, we have assumed that $\chi$ and $t$ remain unchanged in the presence of an electric field. If intercalation of the impurity occurs, this will reduce the shift of the impurity energy, but will always be nonzero on grounds of the crystal symmetry. Incorporating the bias in this way thus has three effects: (i) a renormalization of the bilayer graphene dispersion; (ii) a modification of the hybridization $V_{\bk s}$ through a change in the bilayer graphene quasiparticle wavefunctions; and (iii) a shift the impurity energy to $\epsilon^\pdg_d +\Delta/2$. We will refer to the renormalized bilayer graphene dispersion and the hybridization as $\epsilon^\pdg_{\bk s}(\Delta)$ and $V_{\bk s}(\Delta)$ respectively. It is well-known that such a bias in bilayer graphene can open a band gap and significantly change the low-energy density of states; what is perhaps not appreciated is that this also effectively tunes the impurity energy in multilayer graphene. The last term in Eqn.\ \ref{bias} describing this effect was not present in reference \citet{Ding:2009} and it will be shown to have a remarkable effect on local moment formation in presence of a bias.

A second important difference arises from the tunneling matrix elements, $V_{\bk s}$, for the four bands of the bilayer. As shown in Fig.\ \ref{Vkn}, these matrix elements display strong band- and momentum-dependence, which does not appear for the site-centered impurities discussed in reference \citet{Ding:2009}. The rich structure of the coupling between the chiral bilayer graphene quasiparticles and the impurity site leads to a number of differences in the impurity model phase diagram when compared with conventional non-chiral fermions with a similar density of states, where we simply replace $\phi_{\bk s}({\bf{r}}) \sim \exp(i\bk\cdot{\bf{r}})$ in Eqn.\ \ref{eq_Vkn}.

\section{Mean field theory}\label{Sect:MFT}

A mean field treatment of the adatom impurity model is obtained, following Anderson \citep{Anderson:1961}, by setting 
\begin{equation}
	U n^\pdg_{d \upa} n^\pdg_{d \dna} = U \sum_{\sigma=\pm} (\frac{1}{2} \rho^\pdg_d - \sigma m^\pdg_d) n^\pdg_{d\sigma} 
\end{equation}
where $\rho_d = \sum_\sigma \la n^\pdg_{d\sigma} \ra$, and $m_d= \frac{1}{2} \sum_\sigma \sigma \la n^\pdg_{d\sigma} \ra$. Let us then define 
\begin{eqnarray}
	\xi^\pdg_{d\sigma} &\equiv& \epsilon_d - \mu + U (\frac{\rho^\pdg_d}{2} - \sigma m^\pdg_d) \\
	\xi^\pdg_{\bk s}(\Delta) &\equiv& \epsilon_{\bk s}(\Delta) - \mu. 
\end{eqnarray}
With this mean field approximation, the entire Hamiltonian splits into two single particle impurity Hamiltonians, one for each spin, with 
\begin{eqnarray}
	H^\sigma_{\rm imp} &=& ( \xi^\pdg_{d\sigma} + \frac{\Delta}{2}) d^\dg_\sigma d^\pdg_\sigma \\
	H^\sigma_{\rm BLG} &=& \sum_{\bk,s} \xi^\pdg_{\bk s}(\Delta) c^\dg_{\bk s \sigma} c^\pdg_{\bk s \sigma} \\
	H^\sigma_{\rm mix} &=& - \sum_{\bk} (V^\pdg_{\bk s}(\Delta) c^\dg_{\bk s \sigma} d^\pdg_\sigma \!+\! V^*_{\bk s}(\Delta) d^\dg_\sigma c^\pdg_{\bk s \sigma}). 
\end{eqnarray}
These are coupled together by the self-consistency conditions that fix $\xi^\pdg_{d\sigma}$ via $m^\pdg_d$ and $\rho^\pdg_d$. The single particle Green function for the impurity is given by 
\begin{equation}
	G_{dd}^\sigma(i \omega_n) = \frac{1}{i\omega_n - (\xi^\pdg_{d\sigma} + \frac{\Delta}{2}) - \Sigma^\pdg_{d} (i\omega_n)}, 
\end{equation}
where the impurity self-energy is given by 
\begin{equation}
	\label{eq_selfenergy} \Sigma_{d} (i\omega_n) = \sum_{\bk s} \frac{|V_{\bk s}(\Delta)|^2}{i\omega_n-\xi_{\bk s}(\Delta)}. 
\end{equation}
We can analytically continue this to the real frequency axis by setting $i\omega_n \!\to\! \omega + i 0^+$ to obtain the real and imaginary parts of the self-energy $\Sigma_d(\omega)$. We can then compute at $T=0$ 
\begin{eqnarray}
	\rho^\pdg_d \!\!&=&\!\! -\frac{1}{\pi} \int_{-\infty}^0 \!\!d\omega \sum_\sigma {\cal I}m~G^\sigma_{dd} (i\omega_n \!\to\! \omega + i 0^+), \\
	m^\pdg_d \!\!&=&\!\! -\frac{1}{2 \pi} \int_{-\infty}^0 \!\!d\omega \sum_\sigma \sigma {\cal I}m~G^\sigma_{dd} (i\omega_n \!\to\! \omega + i 0^+). 
\end{eqnarray}
Within this mean field approach, the presence of a local moment on the impurity is signalled by a self-consistent solution with a nonzero $m^\pdg_d$.

Alternatively, it is possible to self-consistently solve the mean field Hamiltonian using exact diagonalization for small system sizes. All of the phase diagrams in the next section were checked for consistency using this method.  Another approach that provides more insight into the Kondo resonance at the Fermi level is to use a slave-rotor representation of the impurity states \citep{Florens:2002,Florens:2004}.  Although, this method has yet to be applied to adatoms in bilayer graphene, an outline of the formalism and implementation of a mean field treatment of the slave-rotor model is provided in Appendix \ref{Append1}.  Preliminary numerical works has been initiated, but remains to be completed.

\section{Local moment formation}\label{Sect:PhaseDiagrams}

Using the above mean field theory enables us to study local moment formation on an impurity atom residing on bilayer graphene. Since the bilayer graphene band structure can be tuned by the electric field, we choose to define $\Gamma_0\! \equiv \! \pi \chi^2/t$ as a rough scale for the impurity level broadening in the absence of interactions. Thus, $\Gamma_0$ remains fixed for a given $\chi$ even as the electric field and chemical potential are varied. In this section, we begin by discussing the case when $\Delta=0$ (i.e.\ without an applied electric field perpendicular to the layers). Phase diagrams are constructed by varying $\epsilon_d$ and $U$ for fixed $\chi=0.3 t$ (which implies $\chi \sim 1$ eV in conventional units) with various choices of the chemical potential. Next, we consider how varying $\Delta$ can be used to tune the phase diagrams. After which, we discuss an alternative phase diagram for an impurity with a fixed bare energy level (although the actual energy level will be modified in the presence of a bias) with various choices of $\Delta$. To construct these phase diagrams, $\mu$ and $U$ are varied, while $\epsilon_{d}$ and $\chi$ ($=0.3t$) are kept fixed.

We have checked that varying $\chi$ modestly makes no qualitative changes to various features in the phase diagram, although it does shift the phase boundaries as expected. We ascribe the complexities of the impurity model phase diagram in bilayer graphene to the effective momentum- and band-dependent mixing $V_{\bk s}$. As we discuss below, the strong variation of this coupling between different bands results in particle-hole asymmetry of the impurity model phase diagram via the impurity self-energy. This is despite the fact that in the simplest tight-binding parameterization, which we have considered, the bilayer graphene band dispersion itself is particle-hole symmetric for $\mu=0$. 

\subsection{Phase diagram in the unbiased case: $\Delta=0$} 
\begin{figure}
	[t] \centering 
	\includegraphics[width=0.9 
	\textwidth]{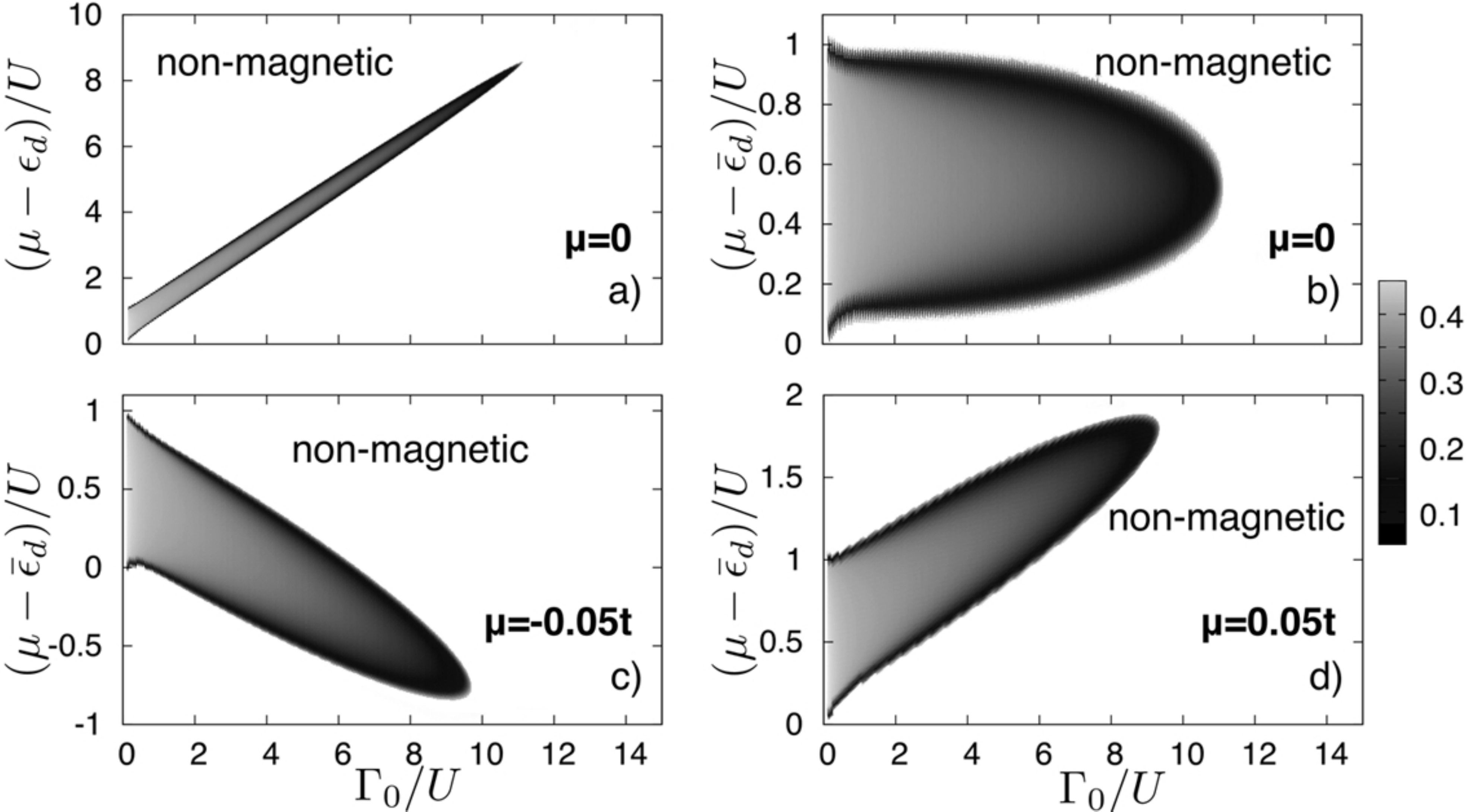} \caption{Phase diagram of local moment formation on plaquette centered impurities in terms of $\epsilon_{d}$ for (a) $\mu=0$, and in terms of $\overline{\epsilon}_{d}=\epsilon_{d} +\Sigma'_{d}(0)$ for (b) $\mu=0$, (c) $\mu=-0.05t$, and (d) $\mu=0.05t$ d). $\chi=0.3t$ in all figures. Grey-scale measures the local moment $m_d$.} \label{plaquette} 
\end{figure}

The $T\!=\!0$ mean field phase diagram for a plaquette-centered impurity embedded in `intrinsic' ($\mu=0$) bilayer graphene with $\Delta=0$ is shown in Fig.\ \ref{plaquette}(a). The phase diagram shares some qualitative features with that of local moment formation in a typical host metal. Namely, there exists a critical ratio of $\Gamma_0/U$ before the onset of mean field magnetization and a clear Coulomb staircase in the small $\Gamma_0/U$ limit. Despite these similarities, there are two unusual aspects to this phase diagram. We next start by highlighting these novel features and then clarify their physical origin.

(i) As seen from Fig.\ \ref{plaquette}(a), there is an extreme skewing of the magnetic regime from being centered at $(\mu-\epsilon_{d})/U \! \sim \! 0.5$ for small $\Gamma_0/U$ to being centered around large positive values of $(\mu-\epsilon_{d})/U$ with increasing $\Gamma_0/U$. This strong particle-hole asymmetry arises from the fact that the impurity couples asymmetrically to the two layers of bilayer graphene, leading to a significant real part of the impurity self-energy $\Sigma_{d}'(\omega)$. The effect of which is to strongly renormalize $\epsilon_d$, which causes the observed skewing. In order to eliminate this large skewing in later plots, we split the real part of the impurity self-energy as 
\begin{equation}
	\Sigma_{d}'(\omega) = \Sigma_{d}'(0) + (\Sigma_{d}'(\omega) - \Sigma_{d}'(0)) 
\end{equation}
and absorb $\Sigma_{d}'(0)$ into the impurity energy, defining a renormalized impurity energy $\bar{\epsilon}_{d}=\epsilon_{d}+\Sigma'(0)$. The resulting renormalized self-energy $\tilde{\Sigma}_{d}'(\omega) = (\Sigma_{d}'(\omega) - \Sigma_{d}'(0))$ then vanishes at $\omega=0$, and remains small but nonzero away from $\omega=0$. Plotting the impurity model phase diagram in terms of the renormalized impurity energy $\bar{\epsilon}_{d}$, to a large degree but {\it not completely}, removes the strong particle-hole asymmetry for $\mu=0$; this can be seen in Fig.\ \ref{plaquette}(b). Of course, strong particle-hole asymmetry continues to exist away from $\mu=0$ even after accounting for the impurity energy renormalization, as shown in Fig.\ \ref{plaquette}(c),(d); this can be ascribed to the particle-hole asymmetry of the bilayer graphene dispersion at nonzero $\mu$.

(ii) As seen from  Fig.\ \ref{plaquette}(a), there is a dramatic elongation of the magnetic region to large values of $\Gamma_0/U \sim 10$, which one can partially attribute the small density of states at $\mu=0$. However, the phase diagram is also influenced by the {\it wavefunctions} of the bilayer graphene quasiparticles. A close inspection of the phase boundaries reveals that they are not symmetric about $\mu=0$ even after accounting for the self-energy correction discussed above. We understand that this residual particle-hole symmetry breaking arises from the asymmetric broadening of impurity level caused by the disparate effective hybridizations with the different bands. This effect is also seen in the phase diagrams for systems by comparing the $\mu=0.05t$ and $\mu=-0.05t$ phase diagrams. While one might na\"ively expect that the symmetry between the valence and conduction dispersions would lead to symmetric phase diagrams for positive and negative chemical potential, subtle differences between the two regions again reflect the influence of the wavefunctions of the electrons that hybridize with the impurity level.

It is, in fact, extremely instructive to compare the complete impurity phase diagram of bilayer graphene with a fictitious system of electrons obtained by setting $\phi_{\bk s}({\bf{r}}) = \exp(i\bk\cdot{\bf{r}})/\sqrt{N_s}$ in Eqn.\ \ref{eq_Vkn}, where $N_s$ is the total number of sites in the bilayer. These fictitious electrons are chosen to have the same dispersion as the bilayer graphene quasiparticles, but their coupling to the impurity does not account for the chirality or the band dependence of the quasiparticle wavefunctions. We find that some of the unusual features of the bilayer graphene impurity phase diagram, discussed above, are eliminated upon making this change.
\begin{figure}
	[t] \centering 
	\includegraphics[width=.5 
	\textwidth]{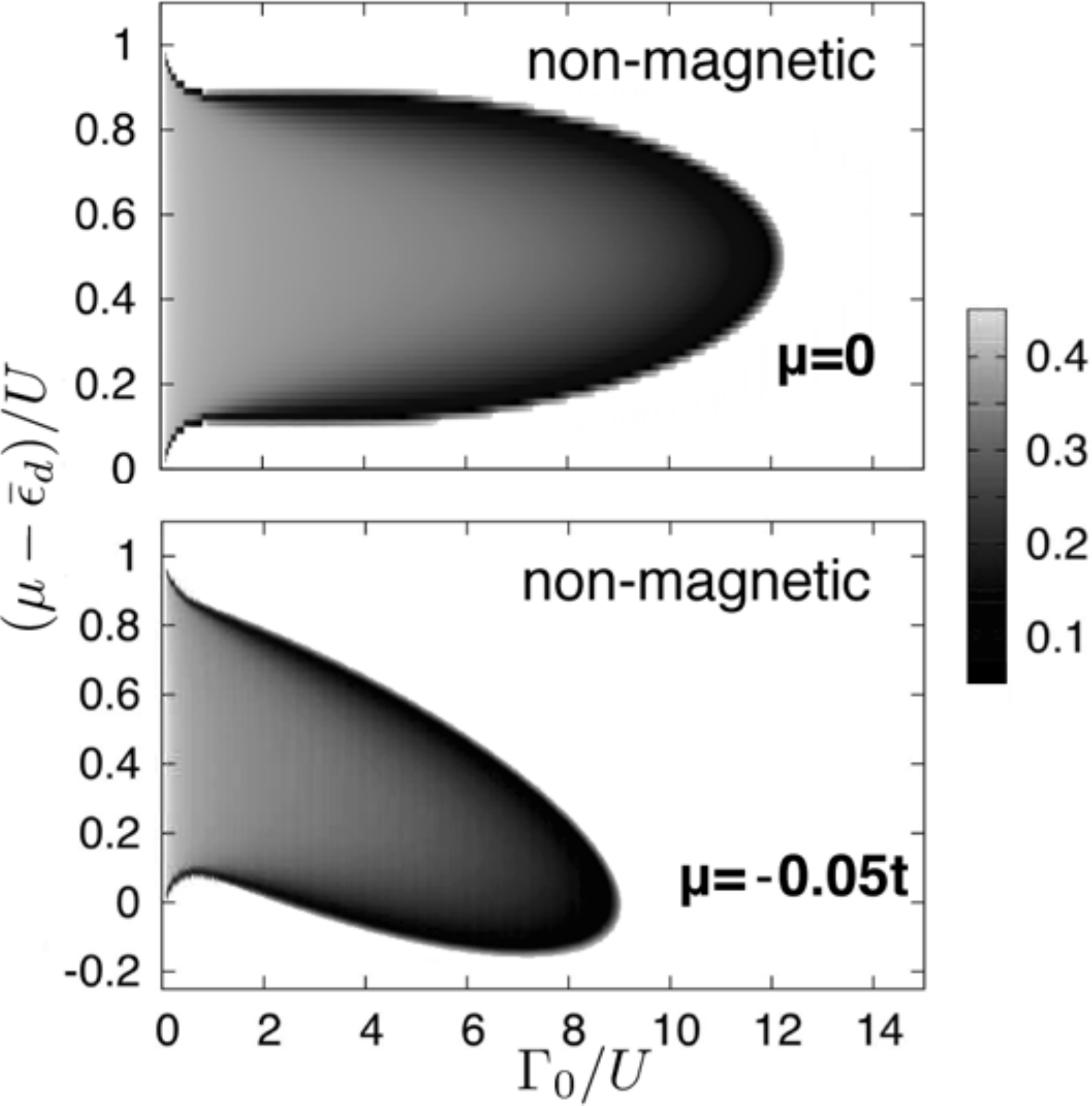} \caption{The phase diagram of local moment formation for fictitious fermions with the same dispersion as bilayer graphene for (above) $\mu=0$, and (below) $\mu=-0.05t$, plotted in terms of $\bar{\epsilon_d}$. (Note, $\bar{\epsilon}_d=\epsilon_{d}$ when $\mu=0$.) The phase diagram for $\mu=0.05t$ is related to that of $\mu=-0.05$ by a reflection about $\mu-\bar{\epsilon}_{d}/U=0.5$. } \label{fiction} 
\end{figure}

Most noticeably, the phase diagram of the fictitious fermions is not skewed when $\mu=0$ even when plotted in terms of the unrenormalized impurity energy, indicating that $\bar{\epsilon}_d=\epsilon_d$, so that $\Sigma_d(0)=0$. This stems from a symmetry $\Sigma_d(-\omega)=-\Sigma^*_d(\omega)$ in the expression for the self-energy in Eqn.\ \ref{eq_selfenergy} upon assuming a band-independent $V_{\bk s}$. Moreover, it also follows that the phase diagram for the fictitious fermions is {\it exactly} particle-hole symmetric, in contrast to the case of bilayer graphene.

Similar arguments also exactly relate the non-chiral phase diagrams of systems with corresponding chemical potential $\mu$ and $-\mu$ by noting that the self-energy at finite chemical potential can be obtained by $\Sigma_{d}(\omega+\mu)$ of the self-energy at $\mu=0$. Consequently, the phase diagram of the $-\mu$ system is obtained by reflecting the phase diagram of the $\mu$ system. This is again in contrast to the phase diagram of bilayer graphene where there is no such relation between systems with positive and negative $\mu$. In bilayer graphene, particle-hole excitations in a system with positive chemical potential favour different bands than those of a system with negative chemical potential. Since each band has a unique effective coupling to the impurity in bilayer graphene, the hybridization of the impurity states will depend on the sign of the chemical potential and so the phase diagrams will be different. Finally, the other major distinction between the finite chemical potential phase diagrams of the two systems is that, once again, the bilayer graphene phase diagram is more strongly skewed, even when plotted in terms of $\bar{\epsilon}_{d}$. This confirms that the band- and momentum-dependence of the hybridization to the bilayer graphene quasiparticles is responsible for sizeable shift in the impurity energy via a sizeable real self-energy.

\subsection{Phase diagram in the biased case: $\Delta \neq 0$}

We now turn our attention towards a bilayer graphene system in a dual-gate configuration. This setup allows one to continuously tune the layer bias and the average chemical potential independently by applying an external electric field perpendicular to the layers. In the presence of a symmetric interlayer bias, the chemical potential remains fixed while a band gap opens in the bulk electronic spectrum of bilayer graphene. In the context of local moment formation, this modification to the density of states is expected to substantially change the extent to which an impurity state hybridizes with the bilayer graphene electrons. In addition to this, the impurity energy levels also shift up or down depending on the potential of the layer in which it resides. This remarkable ability to alter the energy of an impurity level with respect to the chemical potential through the application of an external electric field is unique to multilayer systems, and has no analog in monolayer graphene.

In the first part of this section, we explore how biasing the layers affects local moment formation by reconstructing phase diagrams similar to those above, but for gated systems with different layer bias and fixed $\mu=0$. Doing so allows us to identify regions of impurity parameters where local moment formation can be turned on and/or off by the electric field. In the subsequent part of this section, we consider the ability to tune both the chemical potential and bias by constructing alternative phase diagrams where $\mu$ and $U$ are varied and it is the bare impurity energy which is fixed. This is again done for a selection of values for the bias. 

\subsubsection{Impurity energy variation}
\begin{figure}
	[t] \centering 
	\includegraphics[width=.5 
	\textwidth]{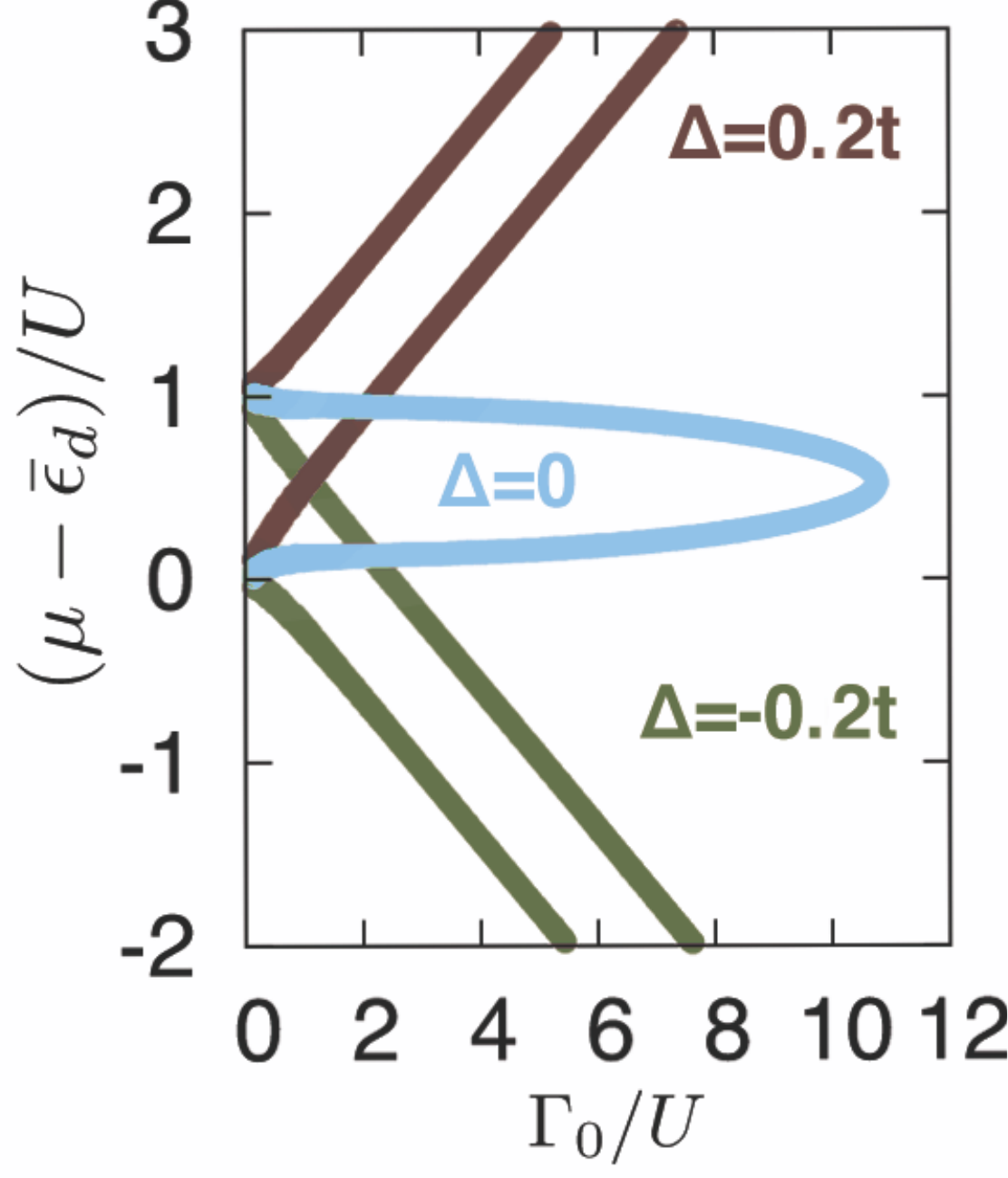} \caption{Phase diagram of local moment formation in plaquette centered impurities with $\Delta=0.0t$, $0.2t$, $-0.2t$ as a function of $\overline{\epsilon}_{d}=\epsilon_{d} +\Sigma'_{d} \left(\omega=0, \Delta=0\right)$. In both figures $\chi=0.3t$ and $\mu=0$.} \label{phaseV} 
\end{figure}

Figure \ref{phaseV} is the phase diagram of the impurity model for experimentally accessible values of $\Delta$ plotted in terms of the redefined impurity energy $\bar{\epsilon}_{d}$ introduced above ($\chi=0.3t$ and $\mu=0$). The bias has two effects on the impurity model: (i) it opens a band gap $\sim \Delta$ in the bilayer graphene dispersion, and (ii) it shifts the impurity energy by $\Delta/2$. Let us discuss, in turn, the impact of these two effects on the phase diagram.

{\bf (i)} First consider restricting the effect of turning on a bias to opening a gap in the bilayer graphene spectrum so that the impurity energy level remains unaltered. Then, the dominant effect of a large $\Delta$ is the elongation of the phase boundary to large $\Gamma_0/U$, regardless of the parity of $\Delta$. This occurs because the bias induces a large band gap and, when the impurity spectral peak lies in this gap, the coupling between the impurity and the extended states becomes negligible because the density of states vanishes. (We have to be careful that the {\it renormalized} impurity energy, taking self-energy corrections into account, should lie in the gap; this renormalization is small if the band gap is large compared to $\Gamma_0$.) Hence, the impurity spectral functions become simple delta functions and if we vary $\epsilon_d$ for fixed $\Gamma_0/U$ the local moment phase boundary resembles that of a simple Coulomb staircase in the atomic limit.

{\bf (ii)} The effect of shifting the energy of the impurity level is similar to the effect of the real part of the self-energy in the phase diagram; it dramatically skews the local moment phase about $\epsilon_{d}/U=0.5$. The direction of the skewing depends on the parity of the bias, as this determines the direction of the impurity energy shift.

If the impurity energy shift and opening of a band gap are taken together, both skewing and elongation of the local moment phase boundary occur. As the electric field is increased from zero to large field strengths, the local moment phase continuously elongates and `peels' away from the zero bias boundary. Although slight, it is important to note that the phase diagrams with opposite bias parity are not symmetric but have slight differences that arise from the breaking of layer symmetry by the impurity. One of the key new results is the identification of regions in the impurity parameter space where local moments can be turned either on and/or off by adjusting the electric field. The region where local moments survive both in the presence and absence of the electric field are simply where the phases overlap.

\subsubsection{Chemical potential variation.}

Now we explore the possibility of tuning the chemical potential of the system to control local moment formation both in the unbiased and biased cases. To do this, we construct phase diagrams for a given $\epsilon_d$ and $\Delta$, and we now vary $\mu$ and $U$. We do this for $\Delta=0,\pm 0.2 t$, for a choice of the bare impurity energy such that the noninteracting impurity spectral peak appears in the midgap when $\Delta= -0.2 t$, which we do by choosing $\epsilon_{d} + \Delta/2 = - \Sigma(\omega=0,\Delta)$.

\begin{figure}
	[t] \center 
	\includegraphics[width=.7 
	\textwidth]{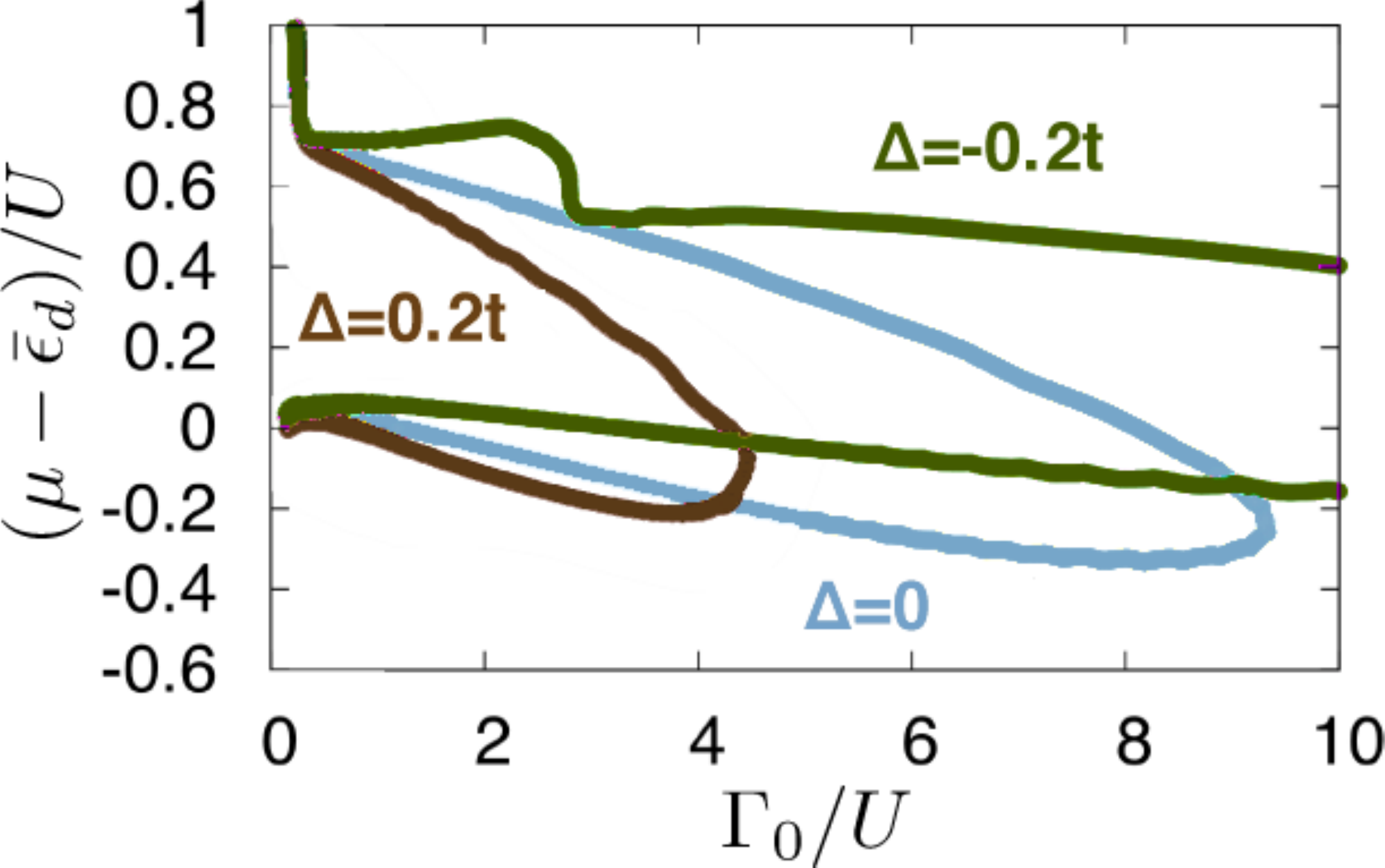} \caption{Local moment phase diagram for biased bilayer graphene. The impurity level at $\Delta=-0.2t$ bias was chosen so that its spectral peak lies in the middle of the gap.} \label{altphase} 
\end{figure}

In Fig.\ \ref{altphase}, the phase diagram is plotted in terms of a redefined impurity energy $\bar{\epsilon}_{d} = \epsilon_{d} + \Sigma(\omega=0,\Delta=0)$. It is important to emphasize that the location of the spectral peaks mostly do not correspond to $\bar{\epsilon}_{d}$. The real part of the self-energy has significant frequency dependence that shifts the location of the spectral peak, whose effect must also be accounted for in order to fully understand the phase diagrams.

When the system is unbiased (i.e.\ $\Delta=0$) the impurity energy level lies within the conduction bands (see reference \citet{Castro-Neto:2009} or reference \citet{Castro:2007} for details on the band structure). In this case, the phase diagram is qualitatively similar to that of a single site impurity (see reference \citet{Ding:2009}). When a positive bias is in place, $\Delta=0.2 t$, a band gap opens and the impurity energy shifts deeper into the conduction band. Consequently, the phase boundaries for local moment formation are significantly reduced because of the enhanced broadening due to the increase in the density of states at higher energy in the conduction bands.

The more interesting case is when $\Delta=-0.2t$ and the impurity spectral peak shifts down into the middle of the gap. Then, if $U$ is small enough so that the doubly occupied state also lies at sub-gap energies, both the singly and doubly occupied states can no longer hybridize with the bilayer graphene states and the impurity spectral function reduces to delta functions. Hence, we again recover local moment formation very similar to the atomic limit, but now in the large $\Gamma_{0}/U$ limit. However, in this limit the upper and lower phase boundaries of the coulomb-staircase are not separated by $(\mu-\bar{\epsilon}_{d})/U=1$ because of level repulsion. The doubly occupied state shifts down in energy due to $\Sigma'_{d}(\omega)$. 

Thus, this phase diagram is unusual in the sense that it has two regimes resembling the atomic limit at large and small $\Gamma_{0}/U$. Separating these regimes is the part of the phase diagram where the doubly occupied state's energy lies beyond the band edge and hybridizes with the conduction states. This occurs at about $\Gamma_{0}/U \sim 2.5$ when $U\sim 0.1t$, precisely where the unusual `hump'-like feature is seen in the upper phase boundary. The cause of the feature can again be attributed to level repulsion, as it becomes very strong for states close to the the gap edge and $\Sigma'(\omega)$ exhibits a large peak.

\section{Controlling interactions between local moments.} \label{Chapt:RKKY}

In this closing chapter, we explore the RKKY coupling between local moments \citep{Ruderman:1954, Kasuya:1956, Yosida:1957} and study how it can be tuned by varying the band gap and chemical potential using a dual-gate configuration. We have seen in the previous section that such variations will, in general, modify the local moment. Here, we focus on changes to the RKKY coupling induced purely by changes in the bulk band structure and filling.

We consider two classical local moments that couple to the set of sites $\{{\bf{r}}\}$ and $\{{\bf{r}}'\}$, respectively, 
\begin{equation}
	\label{moments} H'=\sum_{\{{\bf{r}}\}}J^{(1)}_{{\bf{r}}} \bS_{1}\cdot {\bf s}_{{\bf{r}}}+\sum_{\{{\bf{r}}'\}}J^{(2)}_{{\bf{r}}'} \bS_{2}\cdot {\bf s}_{{\bf{r}}'}, 
\end{equation}
where $J^{(a)}_{{\bf{r}}}$ is the strength of the exchange coupling of an electron's spin, ${\bf s}_{{\bf{r}}}$, at site ${\bf{r}}$ with the magnetic impurity $\bS_a$. Upon integrating out the itinerant electrons and retaining only those terms that are second order in $J^{(a)}_{\bf r}$, one obtains a reduced Hamiltonian for the local moments, 
\begin{equation}
	H_{eff}=J_{RKKY}\bS_1\cdot \bS_2. 
\end{equation}
The coupling $J_{RKKY}$ is given by 
\begin{eqnarray}
	\label{J} J_{RKKY}=\! \frac{1}{2N} \! \sum_{\substack{ \bq \bk i j n m }} M_{i j}(\bq) \, \phi^{*n}_{\bk}(i) \phi^{n}_{\bk}(j) \phi^{*m}_{\bk + \bq}(j) \phi^{m}_{\bk + \bq}(i)\, e^{i\bq\cdot({\bf{r}}_{1}-{\bf{r}}_{2})} \nonumber \\
	\times \frac{n_{F}(\xi^{m}_{\bk+\bq})-n_{F}(\xi^{n}_{\bk})}{\xi^{m}_{\bk+\bq}-\xi^{n}_{\bk}}, 
\end{eqnarray}
where $m/n$ are band indices, $i/j$ are the combined sublattice and layer label, $n_{F}$ is the Fermi distribution, and $M_{ij}(\bq)$ is a matrix describing the Fourier transform between different sites weighted by $J^{(1)}_{\bf r} J^{(2)}_{\bf r'}$. The explicit form of $M_{ij}$ for the case of interest is provided below.

For monolayer graphene, it has been shown that a perturbative treatment in the continuum low-energy theory \citep{Saremi:2007} produces approximate results that match closely with exact diagonalization \citep{Black-Schaffer:2010a} and lattice Green's functions methods \citep{Sherafati:2010}, as long as an appropriate high-energy cutoff scheme is applied. In the above perturbative treatment, the entire band structure is used in the calculation so as to avoid any cutoff dependence and the RKKY coupling is accurately reproduced for monolayer graphene. We therefore expect this perturbative calculation to also be a reasonable approach to study the RKKY coupling in bilayer graphene in the dual-gate configuration.
\begin{figure}
	\center 
	\includegraphics[width=.5 
	\textwidth]{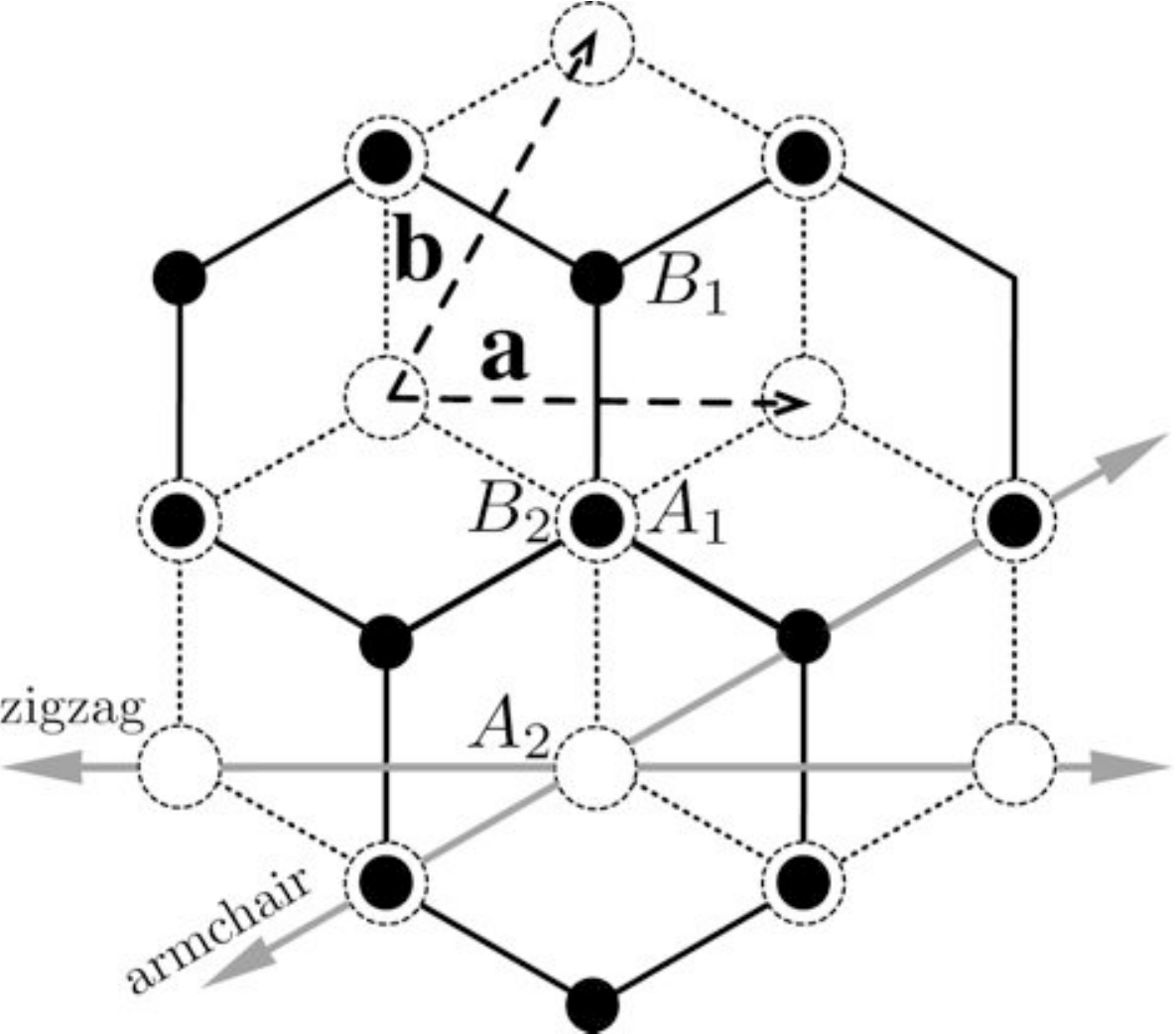} \caption{Crystal structure of bilayer graphene and site labelling convention. The primitive lattice vectors are ${\bf a}$ and ${\bf b}$ and the armchair and zigzag directions are indicated by the arrows. The local moments considered here are plaquette centered and reside on the same layer above an $A_2$ atom.} \label{crystal} 
\end{figure}

We analyzed various moment configurations such as single site (AA, BB, AB) and plaquette coupled moments both along the zigzag and armchair directions (see Fig.\ \ref{crystal} for labelling conventions). The effects of varying the chemical potential and layer bias were seen to be qualitatively similar for each case, so we have chosen to present only the results for plaquette centered moments that lie along the zigzag direction. The impurity atom is taken to lie above an $A_2$ in the center of a hexagonal plaquette in layer 1. For simplicity, we assume the coupling to each of the seven sites is equal so that $J_{\bf r}^{(1)}=J_{\bf r}^{(2)}\equiv J$, although we have checked that the results are qualitatively unaffected if there is an unequal coupling to the site below the impurity on the other layer ($A_2$). In this case, the components of $M_{ij}$ are 
\begin{eqnarray}
	M_{A_{1}A_{1}}\!=J^2 \Big( 4+2\left( \cos(q_{a}) + \cos(q_{b}) +\cos(q_{a}-q_{b})\right) \Big) \nonumber \\
	M_{A_{1}B_{1}}\!=J^2\Big( 2 + 2\left( \cos(q_{a}) + \cos(q_{b}) +e^{i(q_{a}-q_{b})} + e^{-i(2q_{a}-q_{b})} \right) \nonumber \\
	M_{A_{1}A_{2}}=J^2\Big(1+e^{iq_{b}}+e^{i(q_{a}-q_{b})}\Big) \nonumber \\
	M_{B_{1}A_{2}}\!=J^2\Big(e^{iq_{b}}+e^{i(q_{a}-q_{b})}+ e^{i(q_{a}-2q_{b})}\Big) \nonumber \\
	M_{A_{2}A_{2}}\!=J^2, \phantom{\Big( 2} M_{i j}=M^{*}_{j i}, 
\end{eqnarray}
where $q_a={\bf q} \cdot {\bf a}$, $q_b={\bf q} \cdot {\bf b}$ and ${\bf a}={\bf \hat{x}}$ and ${\bf b}={\bf \hat{x}}/2 + \sqrt{3}{\bf \hat{y}}$/2.

\begin{figure}
	[tb] \center 
	\includegraphics[width=.55 
	\textwidth]{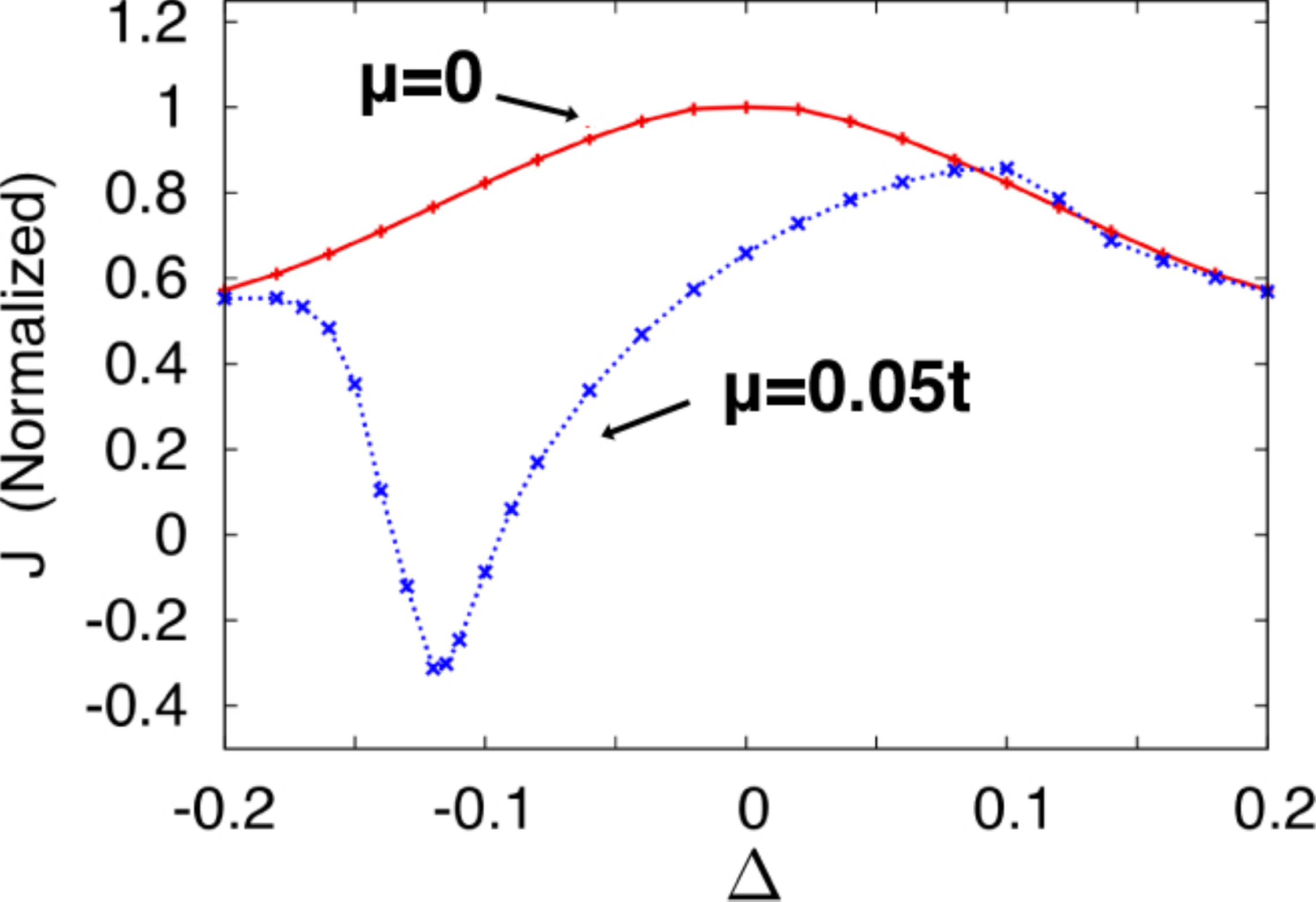} \caption{Normalized RKKY coupling strength between to classical plaquette centered moments at a distance of $10$ lattice spacings along the zigzag direction. Both moments are located on the same layer and the chemical potential is chosen to be $\mu=0$ and $\mu=0.05t$ at a temperature $T=0.002t$ ($\sim70$ K).} \label{JVvariation} 
\end{figure}

To demonstrate the ability to tune the RKKY interaction using the dual-gate configuration, the $J_{RKKY}$ coupling, normalized to its value at $\mu=0$ and $\Delta=0$, is plotted in Fig.\ \ref{JVvariation} as a function of the interlayer bias $\Delta$ for two moments separated by 10 lattice spacings. This is done for $\mu=0$ and $\mu=0.05t$. For experimental considerations, one must keep in mind that the RKKY coupling is quite small in bilayer graphene. As an example, a bare exchange term equal to $J^{(1)}=J^{(2)}=0.2t$ produces an effective coupling $J_{RKKY}=1.3\times10^{-4}t$ ($\sim 4.4$ K) at 4 lattice spacings, and just $J_{RKKY}=7.5\times10^{-6}t$ ($\sim 0.3$ K) at $10$ lattice spacings. However, similar to monolayer graphene, electron interactions are expected to make the coupling strength more long ranged \citep{Black-Schaffer:2010}. At shorter distances, the RKKY interaction is enhanced, but the tunability is reduced.

Before describing the tunable features of the RKKY coupling, it is important to first understand that the wavefunctions of a given band are sensitive to the parity of the bias between the layers, even though the dispersion is not. Their dependancy on the parity can significantly influence how $J_{RKKY}$ changes with bias, as explained below. When a positive bias is present, states in the upper two bands are more heavily weighted to layer 1 sites, while states in the lower band are more heavily weighted to the layer 2 sites. This weighting is reversed when the parity of the bias is negative. In contrast, when there is no bias the weighting of the wavefunction is the same for each layer.

With this background, it is possible to explain the symmetry/asymmetry between the two curves. When $\mu=0$, the chemical potential lies between the valence and conduction bands, and so particle-hole excitations can only occur between them. This corresponds to one of the states being localized to layer 1 and the other localized to layer 2. It follows that the coupling strength $J_{RKKY}$ is parity invariant and so it is symmetric for positive and negative biases. 

In contrast, when $\mu \neq 0$, the coupling is sensitive to the parity of the bias. If $\mu=0.05t$, $\Delta>0$ and $\mu$ is greater than the band gap, the chemical potential lies in the third band where the states tend to localize to layer 1, the layer in which the moments reside. The finite chemical potential causes some of the particle-hole excitations between the lower and upper two bands to be suppressed by Pauli-blocking, and also introduces low-energy excitations between the two upper bands where the wavefunctions are weighted to layer 1. If however, $\mu=0.05t$, $\Delta<0$ and $\mu$ is greater than the band gap, the chemical potential lies in the third band, but now these states tend to localize to layer 2. Although the energetics of the scattering processes remain the same, the matrix elements do not. The excitations between the upper two bands now have matrix elements whose weighting on layer 1 is much less. Thus, the coupling is dependent on the $relative$ parity o f $\mu$ to $\Delta$. Hence, if we consider a system with $\mu=-0.05t$, the $J_{RKKY}$ curve will be reflected about $\Delta=0$.

In addition to the effects described above, the density of states about the chemical potential tends to increase for small biases, as the band edge flattens and is pushed closer to the chemical potential. At large bias strengths, the dispersion close to the band-edge resembles that of a `mexican-hat', leading to further complexity in the density of states. Furthermore, at finite chemical potential, the Fermi points extend out to form a Fermi-surface symmetric about the $K$-points. The combination of all these effects lead to the non-trivial changes seen the RKKY coupling in Fig.\ \ref{JVvariation}.

Interestingly, at a distance of 10 lattice spaces, the coupling remains antiferromagnetic when $\mu=0.05t$ and $\Delta >0$ and tends to increase with bias strength. However, when $\Delta$ becomes increasingly negative, the antiferromagnetic coupling strength is reduced to zero then switches to a ferromagnetic coupling about $\Delta \sim -0.12$. Regardless of the sign of the bias, once the band gap exceeds the chemical potential, the $\mu=0.05t$ curve begins to merge with the $\mu=0$ curve, as expected at low temperature. The ability to dynamically tune the strength of the RKKY interaction, as well as its sign from being antiferromagnetic to being ferromagnetic, for two fixed moments using just an electric field (rather than doping) is perhaps the most interesting feature of this system.

\chapter{Part II: Closing Remarks} 
Part \ref{Part:Adatom} of this thesis focused on local moment formation of adatoms on bilayer graphene and the RKKY interaction between local moments.  Just as there are predictions of interesting physics regarding local moments in single layer graphene, we also demonstrated there to be novel new features that appear for bilayer graphene, where the underlying fermions are also chiral.  Moreover, the local moments were shown to be even more widely tunable than in the single layer because the bilayer has an additional tuning parameter --- the bandgap.  Since much of the physics of local moments is mediated by the underlying sea of chiral fermions, which have gate tunable properties, many interesting nuances were found in both the phase diagram for local moment formation and in the RKKY coupling.  The ability to turn on and off local moments, and the ability to tune the sign and magnitude of the RKKY coupling using electric fields constitutes physics beyond what has been discussed for monolayer graphene.  In the remainder of this closing chapter, we address the role of many-body interactions, discuss recent developments regarding local moments in graphene, and close with some thoughts for future research.  

\section{Effects of many-body interactions}
We have, in our discussion, ignored the effects of electron-electron interactions among the bilayer graphene electrons. These are known to be important for quadratic band touching \citep{Sun:2009, Vafek:2010, Zhang:2010, Nandkishore:2010, Nandkishore:2010a}, leading to symmetry breaking and many-body gaps for zero doping and zero electric field. However, so long as there is nonzero doping or the presence of an electric field that gaps out the low-energy bilayer graphene states, the perturbative effects of electron-electron interactions are benign and will not lead to qualitatively new many-body effects. Using a clean substrate may also be a viable route to mitigating the effects of rippling and disorder. Moreover, the substrate may partially screen the bilayer graphene electron-electron interactions and reduce the many-body gap recently observed in suspended bilayer graphene \citep{Weitz:2010}. The competition between impurity physics and many-body interactions in bilayer graphene deserves a careful separate investigation.      

\section{Recent developments} 

Since the completion of this study, there have been tremendous efforts to understand the magnetism of various adatoms on graphene.  In a recent (unpublished) review article, \citet*{Fritz:2012} identify three key questions to be addressed.  We paraphrase them here:    
\begin{list}{}{}
\item i)\,\, What is the lattice position  of the adatom? 
\item ii)\, What is its spin state of the adatom?
\item iii)\, How does the adatom hybridize with the conduction electrons?
\end{list}     

There have been many \emph{ab-initio} calculations that have attempted to answer these three questions, but the results are inconclusive \citep{Wehling:2010a,Wehling:2011,Jacob:2010a}.  For cobalt, most of the studies suggest that it resides at the center of the hexagonal plaquette in an $S=3/2$ spin state.  However, the situation remains unresolved because of the sensitivity to the precise value of the on-site Hubbard interaction, which is unknown.

In addition to these theoretical studies, there are also two important experiments involving cobalt adatoms worth mentioning.  The first experiment employed STM measurements to show that the ionization of a cobalt adatom could be controlled by external gates \citep{Brar:2011}.  Despite demonstrating that the electronic structure of the adatom can be controlled and the direct observation of a screening cloud, they found no evidence for local moment physics.  The second experiment was performed on a SiC substrate \citep{Manoharan:2011}.  They reported the observation of key signatures in their STM data that they identified to be rooted in a highly unconventional two-channel Kondo effect.  Unfortunately, this report remains unpublished at the time of writing.

It should also be mentioned that other groups measured the magnetotransport \citep{Chen:2011} and magnetization \citep{Nair:2012} of irridated graphene with local defects and interpreted their results as arising from Kondo screening and paramagnetic moments, respectively.  This is consistent with theoretical studies that show there to be localized states residing on such defects. 

At present, the outlook of local moment physics does appear promising.  However, it is very likely that only a combination of more in-depth theoretical studies and further experimentation is necessary before the observation of magnetic adatoms is unequivical.
   
\section{Future direction}
The experimental realization of such tunable local moments in bilayer graphene is a compelling prospect. It would be interesting to study such a system using scanning tunnelling spectroscopy and to probe the quantum dynamics of interacting local moments in experiments. We expect that thermal fluctuations will only slightly alter the phase boundaries as long as the temperature is below the Hubbard gap. Quantum fluctuations are expected to lead to Kondo screening or valence fluctuations in points of the phase diagram --- this is an interesting direction for future research.  Finally, it would be interesting to consider more careful how the Kondo effect behaves as a function a bias induced bandgap.          

%
\appendix 
\chapter{Slave Rotor Formulation}\label{Append1}
The Anderson impurity model is given by
\begin{eqnarray}
H_{\rm imp} &=& \sum_\sigma 
\xi^\pdg_d d^\dg_\sigma d^\pdg_\sigma + U 
(n^\pdg_{d \upa}-\frac{1}{2}) (n^\pdg_{d \dna}-\frac{1}{2}) \\
H_{\rm BLG} &=& \sum_{\bk,s,\sigma} \xi^\pdg_{\bk s} c^\dg_{\bk s \sigma} c^\pdg_{\bk s \sigma} \\
H_{\rm mix} &=& - \sum_{\{\br\}\sigma} 
\chi^\pdg_\br (c^\dg_{\br \sigma} d^\pdg_\sigma + d^\dg_\sigma 
c^\pdg_{\br \sigma}),
\end{eqnarray}
where $\{\br\}$ is the set of sites with which the impurity adatom has 
nonzero hybridization, with strength
$\chi^\pdg_{\br}$, in this tight binding
parametrization. Note that
$\xi_d = (\epsilon_d - \mu)$ and $\xi_{\bk s} \equiv (\epsilon^\pdg_{\bk s} - \mu)$ measure the
impurity site energy and the BLG electron band dispersion relative to the chemical potential. 
Let us set 
\begin{equation}
V^\pdg_{\bk s} \equiv \sum_{\{\br\}} \chi^\pdg_\br \psi_{\bk s}(\br),
\end{equation}
where $\psi^\pdg_{\bk s}(\br)$ denotes the wave function at site $\br$ for electrons in band-$s$ and 
momentum $\bk$.
We then obtain
\begin{equation}
H_{\rm mix} = - \sum_{\bk s \sigma} 
(V^*_{\bk s} c^\dg_{\bk s \sigma} d^\pdg_\sigma + V^\pdg_{\bk s} d^\dg_\sigma 
c^\pdg_{\bk s \sigma}).
\end{equation}
To solve this model, we will employ the
slave-rotor mean field theory introduced by Florens and Georges.
The
Hilbert space at the impurity site has four electronic states,
$|0\ra, |\upa\ra, |\dna\ra, |\upa\dna\ra$. That is, it could be
empty, or be occupied by a spin up electron, or a spin down
electron, or be doubly occupied. In the slave rotor representation,
the electron charge degree of freedom is described by a charged
rotor, with its angular momentum operator $L$ counting the total
charge, and the spin degree of freedom is described by a spin-1/2
fermionic spinon. Each of the four physical states then corresponds
to a direct product of the rotor state and the spinon state, 
\begin{eqnarray}
|0\ra &\equiv& |1\ra|0\ra \\
|\upa\ra&\equiv& |0\ra|\upa\ra\\
|\dna\ra &\equiv& |0\ra|\dna\ra\\
|\upa\dna\ra &\equiv& |-1\ra|\upa\dna\ra
\end{eqnarray} 
Here on the r.h.s.,
the first ket $|\ell \rangle$ is the eigenstate of $L$ (rotor charge or angular momentum)
with eigenvalue $\ell =0,\pm 1$, and the second ket is eigenstate
of spinon occupation number, $n_{f,\sigma}=0, 1$ for
$\sigma=\uparrow,\downarrow$. Notice we have chosen a background
charge $+1$ for the state with no electrons and each added electron
contributes charge $-1$. The enlarged rotor-spinon Hilbert space
contains unphysical states such as $|1\ra|\upa\ra$. These unphysical
states are avoided by imposing the operator constraint 
\begin{equation}\label{constraint}
n^f_{\upa} + n^f_{\dna} + L = 1. 
\end{equation} 
In the slave
rotor representation, the impurity electron number is equal to the spinon
number, i.e., 
\begin{equation} 
n^{d}_{\sigma} = n^f_{\sigma}. 
\end{equation} 
The impurity
electron creation (annihilation) operator 
\begin{eqnarray}
d^\dagger_{\sigma}&=&f^\dagger_{\sigma} {\rm e}^{-i\theta} \label{cre},\\
d^{\vphantom\dagger}_{\sigma}&=&f^{\vphantom\dagger}_{\sigma}
{\rm e}^{+i\theta}, 
\label{ann} 
\end{eqnarray} 
where $f_{\sigma}$ is the
spinon annihilation operator, and the rotor creation (annihilation)
operator , ${\rm e}^{+i \theta}$ (${\rm e}^{-i\theta}$), is
defined by 
\begin{eqnarray} 
{\rm e}^{\pm i\theta} |\ell \ra&=&|\ell \pm 1\ra . 
\end{eqnarray} 
We can rewrite the impurity electron dependent parts of
the Hamiltonian in terms of the spinon and rotor field
operators as 
\begin{eqnarray}
\!\!H_{\rm imp} \!\!&=\!\!& (\xi^\pdg_d + \lambda) \sum_\sigma n^f_\sigma
 +\frac{U}{2} (L^2 - \frac{1}{2}) + \lambda L \\
\!\!H_{\rm mix} \!\!&=\!\!&
- \! \sum_{\{\br\}\sigma} 
\chi^\pdg_\br (c^\dg_{\br \sigma} f^\pdg_\sigma {\rm e}^{i\theta} \!+\! 
f^\dg_\sigma c^\pdg_{\br \sigma} {\rm e}^{-i\theta})
\end{eqnarray}
where $\lambda$ is a Lagrange multiplier which imposes the
constraint in Eq.~(\ref{constraint}) on average, i.e., it fixes
$\la n^f_{\upa} \ra + \la n^f_{\dna} \ra + \la L \ra = 1$.
It is convenient to set $(\xi_d + \lambda) = \xi_f$. At this stage,
we resort to the approximation that the electronic states of this
system can be decoupled into a rotor sector and a sector which has
spinons coupled to the BLG electrons. The resulting rotor and 
BLG-spinon Hamiltonians take the form
\begin{eqnarray}
H_\theta &=& \lambda L + \frac{U}{2} L^2 - \Gamma ({\rm e}^{i\theta} 
+ {\rm e}^{-i\theta}) \\
\!\!H_{c,f} &=& \sum_{\bk,s,\sigma} \xi^\pdg_{\bk s}
c^\dg_{\bk s \sigma} c^\pdg_{\bk s \sigma} + \sum_\sigma \xi_f n^f_\sigma
\nonumber \\
&-& \Phi \sum_{\bk s \sigma} 
(V^*_{\bk s} c^\dg_{\bk s \sigma} f^\pdg_\sigma +
V^\pdg_{\bk s} f^\dg_\sigma c^\pdg_{\bk s \sigma}).
\end{eqnarray}
Here $\Phi = \la {\rm e}^{-i\theta}\ra_R$ 
(which we assume to be real without loss of generality), and
\begin{equation}
\Gamma = \frac{1}{2} \sum_{\bk s \sigma} (
V^*_{\bk s} \la c^\dg_{\bk s \sigma} f^\pdg_\sigma\ra + 
V^\pdg_{\bk s} \la f^\dg_\sigma c^\pdg_{\bk s \sigma} \ra),
\end{equation}
are parameters which are determined self-consistently, which
thus couples together $H_\theta$ and $H_{cf}$.
Physically, a ground state solution where $\Phi=0, \ell=0$ corresponds 
to an unscreened
Kondo moment, $\Phi=0,\ell=\pm 1$ corresponds to the absence of a local
moment, and $\Phi\neq 0$ (with fluctuating $\ell$) 
corresponds to the presence of a local moment with
Kondo screening.

\subsubsection{Solution}
The impurity model of spinons and electrons, $H_{c,f}$ 
can be solved exactly, and the solution is most conveniently
expressed by defining a ``self-energy''
\begin{equation}
\Sigma(i\omega_n) = \sum_{\bk,s} \frac{|V_{\bk s}|^2}{i\omega_n-
\xi_{\bk s}}.
\end{equation}
In terms of this ``self-energy'', the required Green functions for $H_{cf}$
are given by
\begin{eqnarray}
G_{ff}(i\omega_n) \!\!&=&\!\! 
\frac{1}{i\omega_n \!- \xi_f \!- \Phi^2 \Sigma(i\omega_n)} \\
G_{fc}(i\omega_n, \bk s) \!\!&=&\!\! 
\frac{- \Phi V_{\bk s}}{(i\omega_n - \xi_{\bk s}) 
(i\omega_n \!- \xi_f \!- \Phi^2 \Sigma(i\omega_n))}.
\end{eqnarray}
Knowing these, we can define retarded and advanced Green functions via
$G^{R/A}(\omega) = G(i\omega_n \to \omega \pm i 0^+)$, in terms of which
we find
\begin{eqnarray}
\!\!\!\!\la c^\dg_{\bk s \sigma} f^\pdg_\sigma \ra \!\!&\!=\!&\!\! 
\int_{-\infty}^{+\infty} \!\!\frac{d\omega}{2\pi i} n_F(\omega)\!
\left[ G^A_{fc} (\omega, \bk s) \!-\! G^R_{fc}(\omega,\bk s)\right] \\
\!\!\!\!\la f^\dg_{\sigma} f^\pdg_\sigma \ra \!\!&\!=\!&\!\! 
\int_{-\infty}^{+\infty} \!\!\frac{d\omega}{2\pi i} n_F(\omega)\!
\left[ G^A_{ff} (\omega) \!-\! G^R_{ff}(\omega)\right]
\end{eqnarray}
Using $+ (-)$ to denote retarded (advanced) functions, we can obtain the following compact
expressions for $\Gamma$ and $\la n_f \ra$:
\begin{eqnarray}
\!\!\!\!\!\!\!\!\!\Gamma\!\!&\!\!=\!\!&\!2\Phi \!\! \int_{-\infty}^{+\infty}\! \!\!\frac{d\omega}{2\pi i} n_F(\omega)\!
\!\sum_{g=\pm}\! g \frac{\Sigma_g(\omega)}{\omega\!-\!\xi_f \!-\! \Phi^2 \Sigma_g(\omega)\!+\!i g 0^+} \\
\!\!\!\!\!\!\!\!\!\la n_f \ra\!\!&\!\!=\!\!&\! - 2 \!\! \int_{-\infty}^{+\infty}\! \!\!\frac{d\omega}{2\pi i} n_F(\omega)\!
\!\sum_{g=\pm} \! g \frac{1}{\omega\!-\!\xi_f \!-\! \Phi^2 \Sigma_g(\omega)\!+\!i g 0^+}
\end{eqnarray}

To solve the rotor Hamiltonian, we have to resort to numerically
matrix diagonalization with an upper cutoff on the rotor angular
momentum $\ell$.

\addcontentsline{toc}{chapter}{Bibliography}


\bibliographystyle{apsrmp}

\bibliography{ut-thesis}

\end{document}